\documentclass[12pt, a4paper, oneside]{article}
\usepackage{setspace}
\usepackage[stable]{footmisc}
\long\def\symbolfootnote[#1]#2{\begingroup %
\def\thefootnote{\fnsymbol{footnote}}\footnote[#1]{#2}\endgroup}
\usepackage{tabularx}
\usepackage{parskip}
\usepackage{float}

\usepackage[singlelinecheck=true]{caption}
\usepackage{natbib}
\usepackage[ansinew]{inputenc}
\usepackage[T1]{fontenc}                        
\usepackage{textcomp}                           

\usepackage{longtable}
\usepackage{colortbl}
\usepackage{booktabs}
\usepackage{tabularx}
\newcolumntype{L}[1]{>{\raggedright\arraybackslash}p{#1}} 
\newcolumntype{C}[1]{>{\centering\arraybackslash}p{#1}} 
\newcolumntype{R}[1]{>{\raggedleft\arraybackslash}p{#1}} 

\usepackage{graphicx}
\usepackage{epic}
\usepackage{overpic}
\graphicspath{{graphics/}}
\usepackage{import}
\usepackage{tikz}
\usetikzlibrary{arrows,calc}
\usepackage{booktabs}       
\usepackage{dcolumn}        
\usepackage[labelfont=bf]{caption}
\usepackage{latexsym}
\usepackage{amsmath}
\usepackage{amsfonts}
\usepackage{amssymb}
\usepackage{url}
\usepackage{amsbsy}
\usepackage{verbatim}
\usepackage{chngpage}
\usepackage{rotating}
\usepackage{lscape}
\usepackage{fancyhdr}
\usepackage{lmodern}
\usepackage{tabulary}
\usepackage{bm}
\usepackage{multirow}
\usepackage{rotating}
\usepackage[left]{eurosym}
\usepackage[title]{appendix}
\usepackage{geometry}
\renewcommand{\baselinestretch}{1.1}
\usepackage{array,multirow}
\usepackage{hhline}
\usepackage{adjustbox}
\usepackage{rotating}

\def\path{FigsTables/}


\usepackage{adjustbox}
\usepackage{enumerate}
\usepackage{wrapfig}
\usepackage{setspace}
\usepackage{multirow}
\usepackage[super]{nth}
\usepackage[colorlinks=true,allcolors=blue]{hyperref}
\usepackage{booktabs}
\usepackage{siunitx}
\usepackage{threeparttable}
\usepackage{mathpazo}
\usepackage{pdflscape}
\usepackage{needspace}
\usepackage{dsfont}
\usepackage{subcaption}
\usepackage{bm}
\usepackage{setspace}
\usepackage{comment}
\usepackage{siunitx}
\usepackage{dcolumn}
\newcolumntype{d}[1]{D{.}{.}{#1}}
\usepackage{ragged2e}

\begin{document}  
\addtolength{\footskip}{5cm}

\begin{titlepage}
\title{\vspace{-2cm}\Huge Peace Dividends: \\
\LARGE The Economic Effects of Colombia's Peace Agreement\thanks{Email: \href{mailto:m.fajardo-steinhauser@lse.ac.uk}{m.fajardo-steinhauser@lse.ac.uk}. I thank Michael Callen for his guidance and support throughout the project. I thank Nava Ashraf, Gharad Bryan, Robin Burgess, Omar Hamoud-Gallego, Edward Wiles and seminar participants at the London School of Economics, the ACES Summer School, PacDev 2023, EEA-ESEM Congress 2024, LSE-NYU PSPE 2024, RES Annual Conference 2024, German Development Economics Conference, NICEP 2024 Conference, 2024 ESOC Annual Meeting, and AYEW Webinar for helpful comments. I acknowledge financial support from the Economic and Social Research Council and STICERD.}}
\author{ \Large Miguel Fajardo-Steinhäuser 
\\ \large London School of Economics}
\date{\today}
\maketitle
\begin{abstract}
\noindent The last decades have seen a resurgence of armed conflict globally, renewing the need for durable peace agreements. In this paper, I evaluate the economic effects of the peace agreement between the Colombian government and the largest guerrilla group in the country, the FARC, ending one of the lengthiest and most violent armed conflicts in recent history. Using a difference-in-difference strategy comparing municipalities that historically had FARC presence and those with presence of a similar, smaller guerrilla group, the ELN, before and after the start of a unilateral ceasefire by the FARC, I establish three sets of results. First, violence indicators significantly and sizeably decreased in historically FARC municipalities. Second, despite this substantial reduction in violence, I find precisely-estimated null effects across several economic indicators, suggesting no effect of the peace agreement on economic activity. In addition, I use a sharp discontinuity in eligibility to the government's flagship firm and job creation program for conflict-affected areas to evaluate the policy's impact, also finding precisely-estimated null effects on the same economic indicators. Third, I present evidence that suggests the reason why historically FARC municipalities could not reap the economic benefits from the reduction in violence is a lack of state capacity, caused both by their low initial levels of state capacity and the lack of state entry post-ceasefire. These results indicate that peace agreements require complementary investments in state capacity to yield an economic dividend. \\
\vspace{0in}\\

\bigskip
\end{abstract}
\end{titlepage}
\pagebreak \newpage

\pagenumbering{Roman}
\setcounter{page}{1} \addtolength{\footskip}{-5cm}
\pagestyle{fancy}
\renewcommand{\headrulewidth}{0pt}
\fancyhead{}
\fancyfoot{}
\fancyfoot[C]{\thepage}
\fancypagestyle{plain}{%
\fancyhead{}
\fancyfoot{}
\fancyhf{} 
\fancyfoot[C]{\thepage} 
}
\newcommand{\citedouble}[4]{(\citealp{#1}, #2; \citealp{#3}, #4)}
\newcommand{\citetriple}[6]{(\citealp{#1}, #2; \citealp{#3}, #4; \citealp{#5}, #6)}


\pagenumbering{arabic}
\setcounter{page}{1} 
\renewcommand{\baselinestretch}{1}
\interfootnotelinepenalty=0

\begin{flushright}
\textit{``My country celebrates and welcomes these (Sustainable Development) Goals, \\
because we are aware that they are also necessary conditions for building peace and, \\
in turn, peace in Colombia will have very high economic, social and \\
environmental dividends. It will be a virtuous circle.''} 
\\
-- Juan Manuel Santos, then President of Colombia, to the UN in 2015.
\end{flushright}

\section{Introduction}

Over the past two decades, armed conflict has substantially increased globally. \citet{von2017civil} show that since 2007, the number of major civil wars has almost tripled and that there has been a six-fold increase in battle fatalities, with 2014 and 2015 being the deadliest years since the end of the Cold War. The increase in armed conflict begs the question of how best to proceed \textit{after} a conflict ends. While there are many ways in which armed conflicts can end, peace agreements between warring parties are a common conflict-ending mechanism. Over one-third of all conflicts waged between 1989 and 2018 were resolved by peace agreements, with over 350 agreements in total \citep{pettersson2019organized}.\footnote{Examples include the 1998 Good Friday Agreement in Northern Ireland, the 2015 agreement between the Malian government and the CMA, the 2018 agreement between the Ethiopian government and the Ogaden National Liberation Front, and the 2020 agreement in South Sudan.} Despite comprehensive peace agreements being considered ``the gold standard in international peacemaking''  \citep{pospisil2022dissolving}, they almost always entail lengthy negotiations, and their effectiveness is unclear, both in terms of reducing the likelihood of conflict re-emerging and bringing economic prosperity. Around 60\% of conflicts resolved in the early 2000s relapsed within five years \citep{von2017civil}. Given the popularity of peace agreements, a deeper understanding of how they work is critical to bring economic prosperity and avoid violence reemerging.


This paper investigates the economic effects of a recently signed peace agreement, the 2016 peace agreement between the Colombian government and the Revolutionary Armed Forces of Colombia (FARC). The agreement dissolved the FARC as an insurgent group, ending over 50 years of conflict, the lengthiest ongoing conflict at the time \citep{fisas2012anuario}. The conflict in Colombia has been bloody and violent: since 1985, Colombia's Victims' Unit has registered over 9 million people as victims of conflict, with over 8 million forced displaced, 1 million murdered, around 200.000 forced disappeared, and almost 90.000 terrorist acts/fights (as of May 2022). One of the main goals of the peace agreement was to promote the economic development of the areas most affected by the conflict, as frequently mentioned by government officials during the negotiations \citep[e.g. the statement by Colombia's President at the time,][]{cancilleria}. Expectations for the agreement were high, with then-president Juan Manuel Santos receiving the Nobel Peace Prize for his ``resolute efforts to bring the country's more than 50-year-long civil war to an end''.

I use a difference-in-difference strategy, leveraging the fact that the 2016 peace agreement involved only \textit{one} insurgent group in the country, the FARC. Other criminal groups, including the National Liberation Army (ELN), a smaller guerrilla group, remained operational. More specifically, I compare the evolution of municipalities under FARC control to those under ELN control, before and after the start of a (credible and sustained) unilateral ceasefire by the FARC as part of the peace negotiations. The FARC and ELN share a broadly similar history and evolution. While the FARC was always larger, the ELN was not an insignificant power. Despite this, it did not participate in the peace discussions and continued operating after the demobilisation of the FARC. Identifying which group has (historically) controlled a municipality is challenging, with much of their authority exerted through soft power and intimidation. However, there is one common element in the way the FARC and ELN manifested their control -- the use of violence. \citet{arjona2011presencia} highlight that, for guerrilla groups in Colombia, violent activity by a given group is highly indicative of the group's presence. Thus, I use detailed administrative data on the number of criminal actions by insurgent groups to create two measures of municipalities most affected by the FARC and ELN. I provide evidence that both measures capture and distinguish areas historically associated with either FARC or ELN presence, and the results are robust to using either measure.

My analysis proceeds in three parts. \citet{hsr2012sexual} show that, between 1950 and 2004, violence relapsed within five years of ceasefires and peace agreements in 38.2\% and 32.4\% of cases, respectively. Thus, I first evaluate whether the start of the unilateral ceasefire by the FARC in December 2014 as part of the peace negotiations translated into a reduction of violence in FARC-controlled municipalities. Using official data from the Victims' Unit, the Ministry of Defense, and other sources, I consistently find large, significant reductions in measures of crime and violence in FARC-controlled municipalities after the start of the unilateral ceasefire relative to ELN-controlled municipalities. In my preferred specification, I find sizeable reductions in forced displacement (50\% of the pre-treatment mean across FARC and ELN municipalities), forced disappearances ($\sim$40\%), theft ($\sim$40\%), homicides ($\sim$15\%) and clashes between armed actors (50\%), among other indicators. Moreover, these reductions were not short-lived, most persisting until 2019, and there is no evidence of subsequent entry by other criminal groups in former FARC municipalities. These results suggest that the unilateral ceasefire and subsequent agreement drastically reduced crime and violence.

Second, given that the ceasefire successfully reduced violence in municipalities historically affected by the FARC, I evaluate whether this new-found peace brought economic improvements to these areas. As conflict-affected municipalities tend to be hard-to-reach, small and poor, measuring economic activity there is difficult. I use a battery of economic indicators to assess the economic effects of the peace agreement. These include official value added estimates from the National Statistical Office, nighttime light intensity data, a measure of agricultural productivity, firm entry, formal employment from administrative data, and the share of the urban population in a municipality, as well as many alternative measures of economic activity. Overall, I find precisely-estimated null effects across the different measures. For example, using a summary measure of these variables, I find an insignificant \textit{reduction} of economic activity of just 0.004 SD, with tight SEs that would allow me to reject the null of an increase as small as 0.06 SD. Moreover, event-study regressions show no upward trajectory for these measures up to five years post-ceasefire, ruling out that this overall effect of the ceasefire is masking an improvement over time. Studying the short- and middle-term economic responses to reductions in violence is particularly important since many conflicts relapse and peace agreements unravel quickly \citep{von2017civil,hsr2012sexual}. Governments have a small ``window of opportunity'' to make these work, and post-conflict economic recovery likely plays a key role. In an additional exercise, I evaluate a government program that specifically targeted economic activity in conflict-affected areas by providing substantial fiscal incentives for firms to operate in those areas, again finding no improvement across the same economic indicators. Thus, conflict-affected municipalities did not seem to have benefitted economically from the peace agreement, even those aided with large economic government programs.

In the third part of my analysis, I investigate mechanisms behind this puzzling set of results. Reports suggest that the implementation of the peace agreement has lagged, with the state absent from areas previously under FARC control \citep[e.g.,][]{isacson2021long, piccone2019peace, un2021report}. Theory suggests that state capacity plays an influential role in spurring economic growth and, importantly, that very low levels of state capacity can lead to self-perpetuating poverty traps \citep{besley2010state}. I present three pieces of evidence along the lines of this theory, with a lack of state capacity and presence preventing FARC-controlled municipalities from economically benefiting from the reduction in violence post-ceasefire. First, FARC-controlled municipalities had much lower levels of state capacity than the rest of the country at the outset. Second, and in line with the reports, I find no improvement in a wide array of state capacity and presence indicators such as tax revenue per capita, dependence on the national government, and indicators of municipal government performance in FARC-controlled municipalities after the start of the ceasefire. Third, there is suggestive evidence of minor economic improvements in the few areas the state entered (even if weakly). These results \textit{suggest} that the ceasefire did not bring economic improvements due to low initial levels of state capacity and a lack of state entry. Furthermore, I present evidence to suggest that alternative mechanisms, including credit constraints for agricultural producers, coca production, migration of Venezuelan migrants, a shift from production towards education, land restitution issues, and a lack of support/trust in the agreement, are unlikely to explain the results. 

Finally, I show that my results are robust to multiple checks. These include using different definitions of FARC- and ELN-controlled municipalities, employing an alternative summary index, controlling for municipalities' pre-intervention characteristics, using placebo tests in the pre-intervention period, extending the timeframe until 2021, a permutation test randomly assigning municipalities to treatment and control groups, and estimating the different parts using the synthetic difference-in-difference estimator developed by \citet{arkhangelsky2021synthetic}. I also check the robustness of the results to violations of the parallel trend assumption, following \citet{rothpre}, and to spillovers on the control municipalities, following \citet{butts2021difference}, among other exercises. 

This study contributes to several literatures. First, while the literature on the effects of armed conflict is abundant, few studies have studied the effects of \textit{peace}.\footnote{In their review of (civil) war's causes, conduct and consequences, \citet{blattman2010civil} state that in terms of post-conflict recovery policy, ``most of that literature comes in the form of best practices summaries, case studies, and other literature produced by international aid organisations, governments, and NGOs. Academic research remains limited, and where it exists, it tends to focus on high-level analysis'', highlighting the need for rigorous, high-quality research on post-conflict themes. An exception is \citet{honig2021impact} in Nigeria.} Conflict has been shown to have large negative effects on economic growth \citep{abadie2003economic}, educational attainment \citep{akbulut2014children}, health outcomes \citep{bendavid2021effects}, human capital \citep{waldinger2016bombs}, house prices \citep{besley2012estimating}, among others \citep[see][for a review]{rohner2021elusive}. However, little suggests that the end of conflict is enough to bring about the converse effects. While large-scale destruction can happen quickly, recovery can take long or not even materialise again. This paper is one of the first to provide rigorous evidence on the causal effects of peace on economic activity. I find that, while the peace agreement did lead to a considerable reduction in violence in FARC-controlled municipalities, this new-found peace did not translate into improvements in economic indicators in the short and medium run. This suggests that, contrary to physical capital shocks that usually dissipate quickly \citep{brakman2004uthe, davis2002bones,miguel2011long}, long-drawn, violent conflicts can have persistent effects even after their end. 


This study also contributes to the literature evaluating the relationship between state capacity and economic growth \citep[see][for a review]{besley2014causes}. Using a theoretical model, \citet{besley2010state} show that state capacity is an important determinant of economic growth. This theoretical prediction has received some empirical support. For example, \citet{dincecco2016state} find that European governments that undertook fiscal centralisation reforms experienced faster economic growth over the long run. \citet{aneja2022strengthening} show that the strengthening of the US Post Office in the late \nth{19} century facilitated long-distance innovation, a key determinant of economic growth. The results of this study \textit{suggest} that FARC municipalities did not benefit economically from the reduction in violence caused by the peace agreement because the state did not enter these long-neglected areas, with no improvement in indicators of state capacity and presence. Given these municipalities' low initial state capacity levels and the suggestive evidence of minor economic improvements in areas where the state did enter, the results fit the theoretical prediction that a basic level of state capacity is needed for peace to translate into economic growth. 

A large body of political science literature has studied how to make peace agreements work. One strand of the literature focuses on understanding how support for peace agreements emerges \citep[see, e.g.][]{matanock2018considering, garbiras2021using, haas2020if}. Closer to this work, another strand of the literature tries to determine what factors lead to the success of peace agreements, such as \citet{matanock2022does}, \citet{gartner2011signs}, \citet{white2020perils} and \citet{joshi2017implementing}. For armed conflicts driven by financial motives, like in Colombia, post-conflict economic growth is likely a necessary condition for lasting peace. This study contributes to this literature by analysing the economic effects of a recent comprehensive peace agreement.  It suggests that reducing violence is not a sufficient condition for designing effective peace agreements. Complementary investments in state capacity are needed for the economic benefits of such agreements to materialise and, therefore, for their ultimate success. Indeed, qualitative evidence indicates that conflict is re-emerging in Colombia \citep[see, e.g.][]{icg2021,icg2022}, likely due to a lack of economic opportunities for former combatants and local communities that never received the promised peace dividends.

Finally, this paper also contributes to a literature studying the Colombian armed conflict in general, and the peace agreement with the FARC specifically. From a historical perspective, \citet{lopez2018agrarian} show that the rise of the FARC is associated with historical dispossession of peasants' lands by landlords. More recently, \citet{acemoglu2020perils} and \citet{prem2021rise} study how policies designed by the Colombian government to address different aspects of the armed conflict backfired, while \citet{acemoglu2013monopoly} suggest that criminal groups (in their case, paramilitaries) have important sway in high-stake elections. Lastly, papers have looked at the effect of the peace agreement with the FARC on the killing of social leaders \citep{prem2021selective}, credit access \citep{de2021forgone}, demography \citep{guerra2021peace}, entrepreneurship dynamics \citep{bernal2022peaceful}, among others. Using a different identification strategy, this paper adds to this literature by studying the economic impacts of the peace agreement, evaluating a government policy aimed at spurring economic activity in conflict-affected areas, and highlighting the role that state capacity and presence have in shaping any (economic) peace dividends. 

This study has important policy implications for Colombia and other countries. \citet{milian2021peace} mention that peace negotiation processes were ongoing in 37 countries in 2021. In Colombia, the idea of negotiating a peace agreement with the ELN has been around for years, with the current president Gustavo Petro starting preliminary peace dialogues with the ELN in October 2022. He has even opened the doors to similar dialogues with other criminal organisations. The results in this paper provide a cautionary tale for those peace efforts and highlight the importance of a strong government presence post-agreement in previously-disputed areas.

The paper proceeds as follows. Section \ref{cont_data} provides more details about the history of the armed conflict in Colombia and the peace negotiation process, and describes the different data sources. Section \ref{id_strategy} introduces the identification strategy, followed by the results on violence and economic activity in Section \ref{results}. In Section \ref{sec_sc}, I present evidence in support of one potential mechanism that might be driving the results in Section \ref{results}: a lack of state capacity and presence in previously-affected municipalities. Section \ref{rob_checks} presents different robustness checks, and Section \ref{conclusion} concludes.

\section{Context \& Data}
\label{cont_data}

\subsection{Conflict in Colombia}

Colombia has had a long history of armed conflict and violence. During the \nth{19} and early \nth{20} centuries, traditional political parties used violence to settle disputes and fight for political power. The fight between the Liberal and Conservative parties, the two largest traditional parties in the country, reached its highest point between 1946 and 1958 (a period called ``La Violencia''). This led to the bipartisan ``National Front'' creation in 1958. During the years of the ``National Front'' (1958-1974), the presidency was rotated between these parties every four years. Even though one of the reasons behind the ``National Front'' was to reduce the competition between the two parties and, by extension, the violence in the country, violence against agrarian, worker and left-wing urban movements continued \citep{sanchez2014grupo}. 

In the mid-1960s, the decades of political violence and growing urban-rural inequality led to the emergence of the FARC and the ELN. The FARC started as a Marxist-Leninist peasant self-defence organisation in distant, rural regions, with close connections to Colombia's Communist Party. Similarly, the ELN was Marxist-Leninist at its outset, although with a more visible military orientation. It was primarily composed of students, rural organisers, and religious leaders inspired by the Cuban Revolution. In addition to their similar ideologies, the two groups share a similar trajectory: until around 1982, they remained small and isolated, then between 1982 and 1996, they experienced considerable growth, both in numbers and territory, with this growth coinciding with the development of Colombia's drug trafficking business. \citet{bejarano1997inseguridad} and \citet{arias2014costos} estimate that the FARC went from 7 fronts and 850 fighters to 66 fronts and over 16.000 fighters between 1978 and 2000, while the ELN grew from 3 fronts and 350 fighters to 35 fronts and 4.000 fighters from 1982-2000. The most violent period of the conflict took place between 1996 and 2005, when both groups continued expanding and the government focused on a military solution to the armed conflict. This military campaign significantly weakened both groups, particularly from 2005, although they remained sizable and with a continued presence throughout the country. Both the FARC and ELN have used violence and terror extensively as part of their strategy to control areas and exert pressure on the government \citep{sanchez2014grupo,feldmann2018revolutionary}. While they sometimes fought each other, most notably in the department of Arauca, their main enemies have been far-right paramilitary groups and the state \citep{razonpublica2019,isc2021}.

There had been several attempts to broker a peace agreement between the government and the FARC. In the 1990s, both discussed a possible agreement several times, with the most serious effort taking place between 1998 and 2002, which ultimately failed. Contacts were reestablished in secret in 2011, with negotiations starting in Cuba in 2012 and the story leaking later that year. While the ELN indicated a desire to participate in the talks, they ultimately did not. The negotiations led to the FARC announcing a unilateral ceasefire in December 2014, with the final agreement reached in 2016. This agreement was put to a popular vote in a plebiscite in October 2016, which was narrowly rejected. Following the rejection of the referendum, a revised agreement was signed in November 2016 and approved by Congress. As part of the agreement, the FARC committed to hand over their weapons, temporarily concentrate its troops in remote areas where they would be safe, and participate in a commission to understand the history of the conflict in Colombia. They also transitioned into a political party and were given seats in Congress for two terms. 

One of the main goals of the peace deal was to promote the economic development of the areas that suffered the most from the armed conflict. As stated by the official government institution in charge of implementing the peace agreement, part of the Peace Treaty's first objective was the ``Integral Rural Reform (RRI)''. It was ``developed to reverse the effects of the conflict and guarantee the sustainability of the peace agreement,'' which ``aims to increase the well-being of rural habitants and spurring the integration of regions and social and economic development, promoting opportunities for rural Colombia and especially for the populations most affected by the conflict and poverty'' (see \href{https://www.portalparalapaz.gov.co/publicaciones/811/explicacion-puntos-del-acuerdo/}{\textit{here}}). Moreover, the first point of the final peace agreement remarks that ``an integral rural development is key to spur the integration of the regions and the equitable social and economic development of the country''. It goes on to admit that ``while access to land is a necessary condition for the transformation of the countryside, it is not sufficient. Thus, national plans, financed and promoted by the government, ought to be designed that target rural development and provide public goods and services such as education, health, recreation, infrastructure, [...], that provide wellbeing to the rural population'' (see \href{https://www.portalparalapaz.gov.co/publicaciones/809/texto-del-acuerdo/}{\textit{here}}, pages 10 and 11).

At the time of the agreement's signing, several organisations forecasted its economic impact. The most optimistic estimates came from Colombia's National Planning Department, which predicted a 1.1pp-1.9pp increase in GDP growth \citep{gaviria2015dividendo}. Less optimistically, the Ministry of Finance suggested that GDP growth would only increase by 0.3pp in the 15 years following the signature of the agreement, very close to estimates from Bank of America - Merril Lynch \citep{villar2017informe}. In between these two, \citet{clavijo2017dividendos} estimated that the reduction in conflict and drug trafficking would translate into 0.5\%-1\% higher GDP growth over the decade after 2016 and that the implementation of the peace agreement would cost around 5\% of GDP per year during the same time. Thus, while all studies agreed that the peace agreement would bring economic growth, the magnitude of such effects was believed to be small \textit{for the whole country}. Other than one contemporaneous and related study to this \citep[][who find that the peace agreement had no overall effects on entrepreneurship dynamics]{bernal2022peaceful}, there have been no attempts to quantify whether the peace agreement realised the expansion of economic activity it aspired to, especially in the areas most affected by the conflict, a void that this study helps to fill.  

\subsection{Data}

The data used in this paper come from multiple sources. Most data are provided by CEDE at Universidad de los Andes, which provides a municipality-level panel on many variables since the 1990s. CEDE collects data mostly from government agencies. Appendix \ref{data_detail} contains a variable-by-variable description of data sources.  

One of the main difficulties in evaluating the impact of Colombia's peace agreement in municipalities previously affected or controlled by the FARC is the lack of administrative data on the presence of insurgent groups across municipalities over time. This is a general difficulty in the literature studying armed conflicts. However, \citet{arjona2011presencia} suggest that the infliction of violence by an insurgent group tends to be a good predictor of their presence/control in Colombia. The literature analysing the effects of Colombia's peace agreement has therefore used an insurgent group's criminal activity as a proxy for its presence \citep[see, e.g.][among others]{de2021forgone, bernal2022peaceful, guerra2021peace}. Thus, I develop two measures of FARC/ELN presence based on the location of their violent actions.

Both measures are based on administrative data from the National Police and Administrative Department of Security, which contain disaggregated data on different types of criminal acts committed at the municipality-year-insurgent group level. I focus on the period between 1996 and 2008, the highest point of the conflict and before the start of the peace negotiations. My preferred measure, used for all the baseline results, is meant to capture the ``extensive margin'' of conflict and identify municipalities constantly exposed to FARC/ELN actions in that timeframe. More specifically, in the baseline definition, I classify a municipality as having been affected by, or under the control of, an insurgent group (I use these terms interchangeably) if the municipality has at least one criminal act committed by the guerrilla group in at least 60\% of the years between 1996-2008 (see Appendix \ref{data_detail} for the specific criminal acts used). I exclude the top 2\% largest municipalities, which never were under FARC or ELN control. 

The second measure is meant to capture the ``intensive margin'' of conflict and aims to identify the municipalities that were hit hardest by the activities of a particular guerrilla group. First, I calculate the average number of criminal acts per 100.000 inhabitants for each municipality over time. In the baseline definition of this variable, I classify a municipality as having been affected/under the control of the FARC/ELN if its average number of events per 100.000 inhabitants between 1996 and 2008 is at the top 20\% of the average across all municipalities. Thus, while the first measure identifies municipalities exposed to a specific guerrilla group for an extended period, the second identifies municipalities that were hit particularly hard during this timeframe. While the main results will use the ``extensive margin'' measure, in Section \ref{rob_checks}, I show that all the results are virtually identical using the ``intensive margin'' measure and when changing the cut-off levels used to create the variables. The fact that both rely on a long time horizon also reduces misclassification concerns due to mismeasurement or one-off events by either group that are likely in this context.

\begin{figure}[h!]
\caption{Baseline Treatment and Control Group Municipalities}
\label{maps}
\begin{subfigure}[b]{0.5\textwidth}
\includegraphics[width=\textwidth]{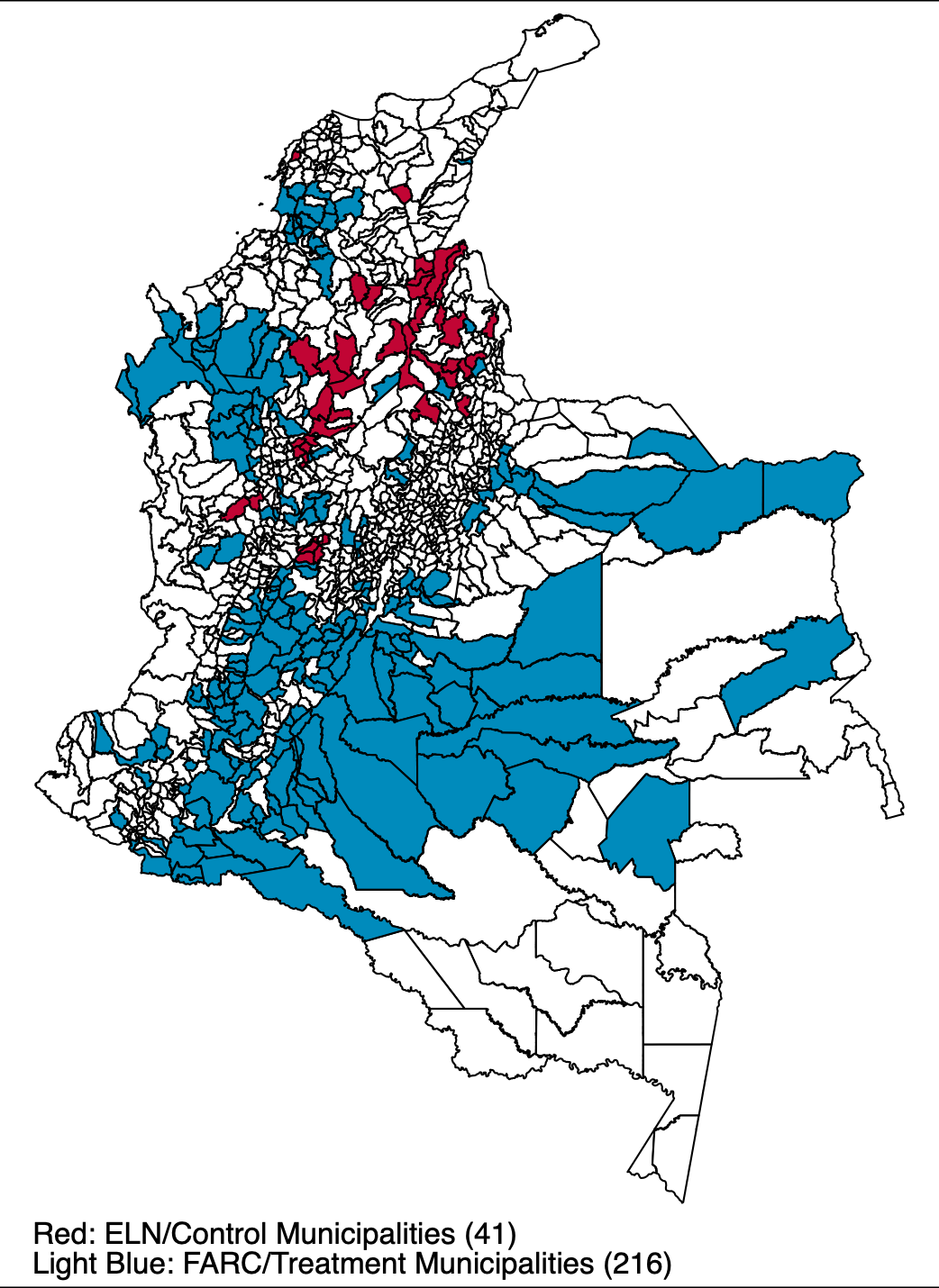}
\subcaption{Extensive Margin, over 60\% of Years}
\end{subfigure}
\begin{subfigure}[b]{0.5\textwidth}
\includegraphics[width=\textwidth]{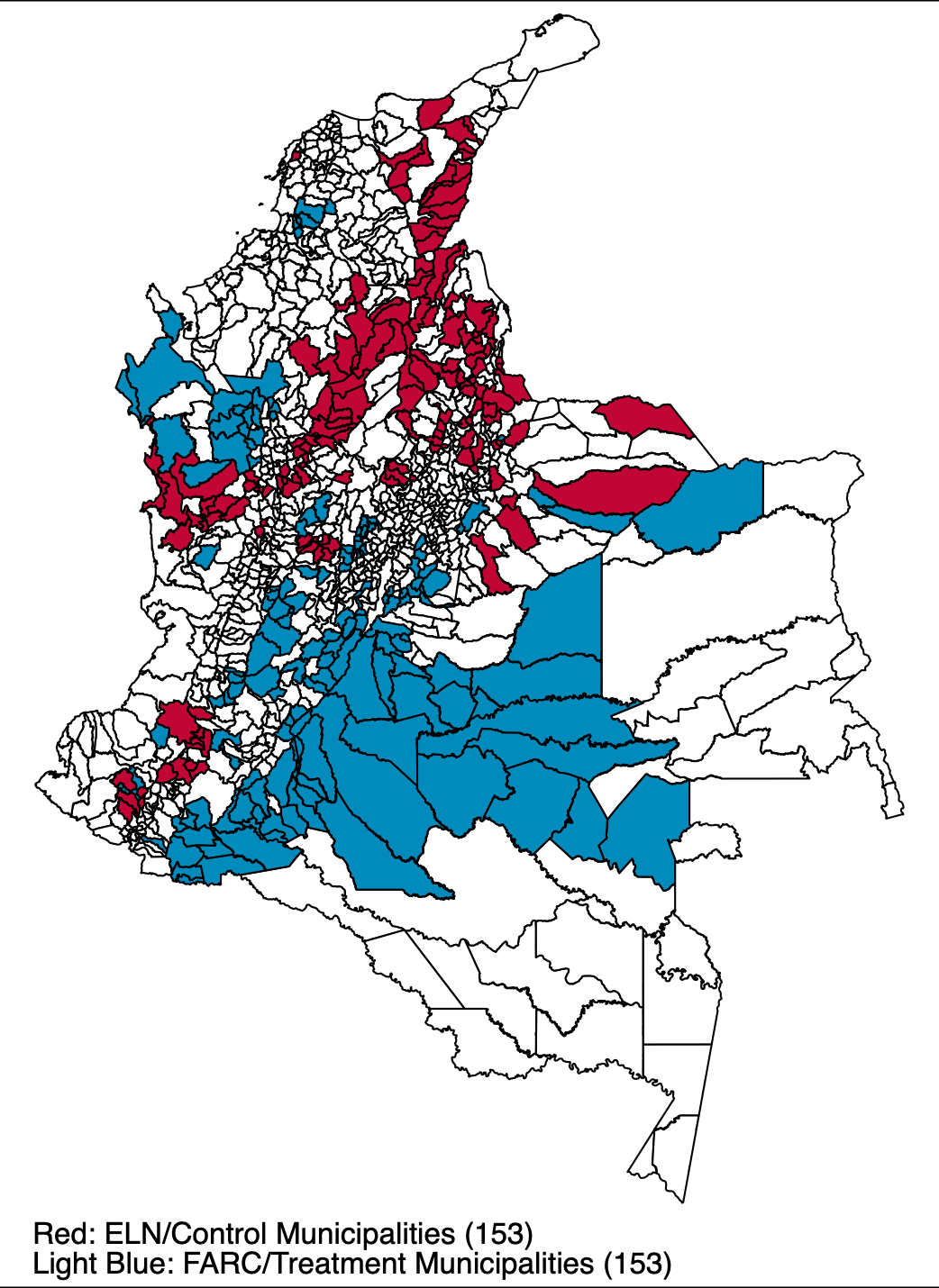}
\subcaption{Intensive Margin, Top 20\%}
\end{subfigure}
\justifying 
\footnotesize{\textbf{Notes:}  Maps show the distribution of treated (i.e. historically FARC, in blue) and control (i.e. historically ELN, in red) municipalities for the two measures of presence of these groups.}
\end{figure}

Figure \ref{maps} shows the distribution of municipalities categorised as having been affected/under the control of the FARC (treatment) and the ELN (control) according to the ``extensive margin'' measure in Panel A and to the ``intensive margin'' measure in Panel B. Note that municipalities that are classified as both under the FARC and ELN control are excluded from the analysis. Reassuringly, these broadly correspond to the areas in which the FARC and ELN operated according to other sources \citep[for example,][]{echandia1998indagacion, pares2015}, and there is significant overlap between the two maps. They also correspond to areas that have been historically associated with each of the two groups, with the FARC operating in the central departments (e.g. Tolima, Huila, Meta and Caquetá), and the ELN further north, in Santander, Norte de Santander, and Cesar. Moreover, over 80\% and 95\% of the municipalities studied in \citet{blair2022preventing}, which were selected due to having had ``historical FARC presence'' where ``other armed groups (including ELN and paramilitaries) were historically present as well'', are correctly identified by the intensive and extensive margin measures.\footnote{Several additional pieces of evidence suggest that these measures capture FARC/ELN municipalities. First, all municipalities that were part of El Caguán's Demilitarized Zone (an area where the FARC has historically had control) as part of the negotiations with the FARC between 1999 and 2002 are classified as having been under FARC's presence. Second, after the signature of the peace agreement, FARC members were located in 26 camps (\textit{Zonas Veredales Transitorias de Normalización} and \textit{Puntos Transitorios de Normalización}, ZVTN and PTN, respectively) across 25 municipalities in areas in which the FARC used to operate. For the extensive (intensive) margin measure of presence, 23 (22) out of 25 municipalities that were either PTN or ZVTN are classified under FARC/ELN control. Third, qualitative evidence suggests that criminal groups recruit mainly where they operate (see, e.g. \href{https://razonpublica.com/vinculacion-de-menores-de-edad-a-las-guerrillas-colombianas/}{\textit{Razón Pública, 2013}} and \href{https://www.semana.com/on-line/articulo/asi-reclutan-farc/79954-3/}{\textit{Semana, 2006}}). Table \ref{demob_CEDEExt_p60} shows the number of demobilised members of FARC and ELN that state they live in FARC/ELN/rest of the country municipalities after demobilising. Demobilised FARC (ELN) members consistently name municipalities classified as FARC (ELN) municipalities as their municipality of residence (first four columns for FARC, last four columns for ELN). Importantly, FARC (ELN) members are located in ELN (FARC) and the rest of the country municipalities in similar proportions, again suggesting that these are indeed ELN (FARC) municipalities. The same pattern holds when looking at captures of FARC/ELN members using data between 2010 and 2017.} Thus, both measures seem to identify the presence of these insurgent groups accurately. 

%

\section{Identification Strategy}
\label{id_strategy}

Having identified which municipalities were most affected/under control of the FARC in the pre-agreement period, the next question is how to estimate the agreement's causal effects. Simply comparing the evolution of FARC municipalities against that of the rest of the municipalities in the country in a difference-in-difference setting (which is what most of the literature has done) is unlikely to recover the causal effects of the agreement. The municipalities affected by the FARC are radically different from the rest of the country, and it is unlikely that the parallel trends assumption needed for the difference-in-difference estimation is satisfied. Table \ref{SumStats_CEDEExt_p60} shows how different FARC (column 2) and non-FARC municipalities (column 1) are along many characteristics in the pre-negotiation period. FARC municipalities are much less populated, farther away from their department's capitals, and more likely to have experienced conflict for long. They also tend to have much lower levels of state capacity (broadly defined), institutional performance, and provision of public goods than non-FARC municipalities (Panel B). Looking at economic indicators in Panel C, FARC municipalities have much lower levels of development and nighttime light intensity while having higher levels of poverty. They tend to be more rural (with more cultivated land and fewer manufacturing firms), yet their agricultural productivity is lower. Lastly, survey data from the Gran Encuesta Integrada de Hogares shows that salaries, ownership of assets, and socioeconomic status are lower in these areas. Overall, FARC municipalities have consistently worse economic, social, and performance indicators than non-FARC municipalities. It is especially hard to believe that violence in the rest of the country, which includes large cities like Medellín, Bogotá and Cali, would have evolved in the same way as in FARC-controlled municipalities in the absence of the peace agreement. On the other hand, FARC and ELN (column 3) municipalities are very similar across these dimensions. Even if this does not prove that ELN municipalities are a good counterfactual, it does suggest that, to begin with, these two groups are similar and much more so than other, non-FARC municipalities.

\begin{table}[h!] \centering
\newcolumntype{R}{>{\raggedleft\arraybackslash}X}
\newcolumntype{L}{>{\raggedright\arraybackslash}X}
\newcolumntype{C}{>{\centering\arraybackslash}X}

\caption{Summary Statistics: Extensive Margin, Events in Over 60\% of Years}
\label{SumStats_CEDEExt_p60}
\begin{adjustbox}{max width=\linewidth, max height=\textwidth}\begin{tabular}{l|ccc|ccc}

\toprule
& \multicolumn{3}{c|}{\textbf{Mean}} & \multicolumn{3}{c}{\textbf{p-value of Difference}} \tabularnewline & Rest Country & FARC & ELN & Rest-FARC & Rest-ELN & FARC-ELN \tabularnewline 
{}&{(1)}&{(2)}&{(3)}&{(4)}&{(5)}&{(6)} \tabularnewline
\midrule \addlinespace[\belowrulesep]
\midrule \textbf{Panel A. General Characteristics}&&&&&& \tabularnewline
Population&42.65&25.15&27.36&0.36&0.72&0.66 \tabularnewline
Area (km\(^2\))&783.28&1842.32&711.46&0&0.88&0.04 \tabularnewline
Distance Dept. Capital (km)&81.12&79.73&101.34&0.76&0.04&0.02 \tabularnewline
Distance Bogot\'a (km)&317.89&303.70&345.42&0.34&0.39&0.11 \tabularnewline
Conflict in 1901/30&0.03&0.07&0.11&0.02&0&0.30 \tabularnewline
Spanish Occupation&0.46&0.18&0.07&0&0&0.08 \tabularnewline
&&&&&& \tabularnewline
\midrule \textbf{Panel B. State Capacity}&&&&&& \tabularnewline
Gov. Transfers (pc)&0.07&0.05&0.03&0&0&0.13 \tabularnewline
Tax Revenue (pc)&0.09&0.05&0.05&0&0.06&0.67 \tabularnewline
Savings Capacity&33.81&29.26&32.15&0&0.54&0.30 \tabularnewline
Fiscal Performance&62.45&60.61&61.29&0&0.39&0.62 \tabularnewline
Overall Performance&60.09&55.29&59.41&0&0.78&0.11 \tabularnewline
Aqueduct Coverage&59.5&56.58&57.18&0.27&0.68&0.92 \tabularnewline
Garbage Collection&44.59&47.33&43.04&0.30&0.78&0.45 \tabularnewline
&&&&&& \tabularnewline
\midrule \textbf{Panel C. Economic Conditions}&&&&&& \tabularnewline
Multidimensional Poverty&68.01&73.41&72.80&0&0.07&0.79 \tabularnewline
Municipal Development&67.75&63.84&65.31&0&0.24&0.47 \tabularnewline
Nighttime Light Intensity&0.12&--0.33&--0.20&0&0.06&0.06 \tabularnewline
Cultivated Land (per HA)&0.25&0.28&0.31&0.47&0.27&0.26 \tabularnewline
Agricultural Productivity&7.61&6.88&5.21&0.40&0.15&0.41 \tabularnewline
\# Manufacturing Firms (EAM, pc)&2.42&1.11&0.85&0.01&0.16&0.64 \tabularnewline
Gross Salary Last Month (GEIH)&788.89&716.86&624.29&0&0&0.28 \tabularnewline
Asset Ownership Index (GEIH)&--0.15&--0.68&--0.60&0&0&0 \tabularnewline
Low Socioecon. Stratum (GEIH)&0.64&0.86&0.77&0&0&0 \tabularnewline
\bottomrule \addlinespace[\belowrulesep]

\end{tabular} \end{adjustbox}
\\ \parbox{\linewidth}{\scriptsize \justifying \noindent \(Notes\): Variables are measured in the last pre-treatment year for which data are available (2008) for all variables except multidimensional poverty (from 2005 Census) and those from the Gran Encuesta Integrada de Hogares (GEIH, 2009) and the Encuesta Anual Manufacturera (EAM, 2009). Conflict in 1901/1930 denotes whether the municipality experienced social conflict between 1901-1930. Distance variables are measured in kilometers. Population in 1000s. Government transfers and tax revenue are per capita (in thousand COP). Savings capacity, fiscal performance, overall performance and municipal development are indices created by different government agencies. Agricultural productivity is defined as total tonnes produced divided by the total area cultivated in hectares. The two agricultural measures are based on data from 271 different crops. The number of manufacturing firms is per 10000 inhabitants, from the EAM, and is conditional on having at least one firm surveyed (most municipalities don't have any surveyed manufacturing firm). Gross salary last month is in 1000 COP, and asked only to employed individuals. Asset ownership index is based on ownership of the following assets: phone, TV, fridge, washing machine, microwave, motorbike, car, bicycle, computer, and access to internet, using as weights those from Colombia's 2015 DHS survey. Low socioeconomic stratum is based on whether the respondent said they are classified as belonging to one of the bottom two strata (out of 6) based on the energy bill. These three variables come from the individual-level GEIH, which only covers a subset of municipalities in the country and is not representative at the municipality level.}
\end{table}

Instead of comparing FARC and all non-FARC municipalities, I use a difference-in-difference strategy comparing municipalities with a long-lasting FARC presence (treatment) and those with ELN presence (control), before and after the start of the unilateral ceasefire announced by the FARC in December 2014. There are several reasons to believe this comparison would satisfy the parallel trends assumption needed to recover the agreement's causal effect. The FARC and ELN were both left-wing guerrilla organisations created around the same time for similar reasons, and they have shared the same trajectories, as described in Section \ref{cont_data}. They both have used violence and terror as a way of imposing their control, kidnapped and extorted as a way of funding their operations, and fought the state and paramilitary forces for decades. Their terror acts have fluctuated over time in a similar way \citep{feldmann2018revolutionary}. While the ELN was always smaller than the FARC and mainly focused on extracting resources from oil-producing regions, it has followed the FARC in trafficking drugs in recent times. The FARC and ELN even joined other smaller guerrilla groups to consolidate under a single organisation between 1987 and 1994, the Coordinadora Guerrillera Simón Bolívar. Moreover, while not identical to FARC's municipalities, Table \ref{SumStats_CEDEExt_p60} shows that municipalities with FARC and ELN presence were much more similar before the start of the peace negotiations than FARC and non-FARC municipalities.

I estimate three different regression equations. First, I estimate the following two-way fixed-effects regression to recover the overall effects of the start of peace on municipalities with historical FARC presence relative to those with ELN presence:
\begin{equation}
\label{eq_did}
y_{mt} = \beta \textit{Ceasefire}_{t} \times \textit{FARC Presence}_m + \eta_m + \mu_t + \varepsilon_{mt}
\end{equation}
\noindent where $y_{mt}$ is the outcome of municipality $m$ in year $t$, \textit{Ceasefire}$_t$ is a dummy that equals one for the post-ceasefire years (from 2015 onwards), \textit{FARC Presence}$_m$ is a dummy that equals one for municipalities that are classified as having had FARC's presence according to either of my two measures of presence, and zero for municipalities classified as having had ELN's presence, $\eta_m$ are municipality fixed-effects, $\mu_t$ are year fixed-effects, and $\varepsilon_{mt}$ is an error term. Municipalities classified as having had both FARC and ELN presence are excluded from the analysis, as it is unclear which group they should belong to. Standard errors are clustered at the municipality level. For most variables, the regression is estimated using data from 2009 (the year after the period used to create the presence measures) to 2019. Note that while there is now an extensive literature highlighting problems in the estimation of TWFE regressions \citep[for reviews, see][]{de2022two,roth2022trending}, these are not relevant in this setting as the treatment is not staggered. 

Second, I estimate event-study-like TWFE regressions to study the dynamics of the treatment effects, which also allows the evaluation of parallel trends before the start of the ceasefire. Formally, I estimate:
\begin{equation}
\label{eq_es}
y_{mt} = \sum_{j \in T \notin 2014} \alpha_j ( \mathds{1}[t = j] \times \textit{FARC Presence}_m ) + \kappa_m + \lambda_t + \upsilon_{mt}
\end{equation}
\noindent with indicator variables for each year between 2009 and 2019, $T$, with the last pre-treatment period (2014) omitted. The parameters $\alpha_j$ measure the difference in the outcome variable in municipalities with FARC's presence and municipalities with ELN's presence, in year $j$ relative to 2014, the last year before the ceasefire started. Standard errors are clustered at the municipality level. 

Third, I estimate triple difference-in-difference equations to study heterogeneous treatment effects by interacting the post-ceasefire and FARC dummies with a dummy for a given measure of heterogeneity:
\begin{equation}
\label{eq_het}
\begin{split}
y_{mt} =& \rho \textit{Ceasefire}_{t} \times \textit{FARC Presence}_m + \theta \textit{Ceasefire}_{t} \times \textit{Heterogeneity}_m \\
 &+ \gamma \textit{Ceasefire}_{t} \times \textit{FARC Presence}_m \times \textit{Heterogeneity}_m + \xi_m + \zeta_t + \iota_{mt}
\end{split}
\end{equation}
\noindent where the coefficient of interest is $\gamma$. It measures whether the growth rate of $y_{mt}$ in FARC relative to ELN municipalities differs for those municipalities that belong to the heterogeneity group (i.e. those with $Heterogeneity_m = 1$) and those who do not (i.e., $Heterogeneity_m = 0$).

\section{Results}
\label{results}

In this Section, I present the main results of this study. First, in Subsection \ref{sec_viol}, I show that the start of the ceasefire led to a large reduction in violence indicators in municipalities with FARC presence relative to ELN's. Then, in Subsection \ref{sec_econ}, I analyse whether this reduction in violence translated into improvements in economic indicators. I show this is not the case, with precisely-estimated null effects on economic activity. Using a different identification strategy, I evaluate a government program specifically designed to spur economic activity in conflict-affected municipalities by granting tax incentives to firms opening in those areas, also finding precisely-estimated null effects. 

\subsection{Violence}
\label{sec_viol}


Table \ref{DID_final_Vio_CEDE_Ext_p60} shows the results of estimating Equation \eqref{eq_did} for a battery of violence-related indicators, defining the treatment and control groups using the extensive margin measure of presence (note that the violence data used to create these groups are not used in this analysis). Panel A shows the results using measures from the Victims' Unit, while Panel B uses measures from multiple sources, mainly from the Ministry of Defense. Appendix \ref{data_detail} contains a complete list of sources and definitions for each variable. All measures are standardised per 1000 inhabitants. 

\begin{table}[h!] \centering
\newcolumntype{R}{>{\raggedleft\arraybackslash}X}
\newcolumntype{L}{>{\raggedright\arraybackslash}X}
\newcolumntype{C}{>{\centering\arraybackslash}X}

\caption{DiD Violence Analysis: Extensive Margin, Events in Over 60\% of Years}
\label{DID_final_Vio_CEDE_Ext_p60}
\begin{adjustbox}{max width=\linewidth, max height=\textwidth}\begin{tabular}{lccccccc}

\toprule
{}&{(1)}&{(2)}&{(3)}&{(4)}&{(5)}&{(6)}&{(7)} \tabularnewline
\midrule \addlinespace[\belowrulesep]
\midrule \textbf{Panel A. UARIV}&Terror. Act&Threats&Disapp.&Sex Crimes&Child Recruit.&Torture&Prop. Loss \tabularnewline
\midrule Ceasefire \( \times \) FARC&--0.330*&--0.865**&--0.046**&--0.029*&--0.029***&--0.005&--0.608*** \tabularnewline
&(0.195)&(0.342)&(0.019)&(0.016)&(0.006)&(0.003)&(0.135) \tabularnewline
&&&&&&& \tabularnewline
Treated Munic.&216&216&216&216&216&216&216 \tabularnewline
Control Munic.&41&41&41&41&41&41&41 \tabularnewline
Mean Dep. Var.&0.649&2.508&0.107&0.068&0.037&0.014&0.689 \tabularnewline
Observations&2,827&2,827&2,827&2,827&2,827&2,827&2,827 \tabularnewline
\(R^2\)&0.252&0.523&0.308&0.608&0.453&0.360&0.363 \tabularnewline
&&&&&&& \tabularnewline
\midrule \textbf{Panel B. Other}&Kidnap.&Homicides&Theft&Mines&Forced Mig.&Clash/Att.&\textbf{\textbf{And. Index}} \tabularnewline
\midrule Ceasefire \( \times \) FARC&0.001&--0.074**&--0.427**&--0.292***&--10.210***&--0.019***&--0.289*** \tabularnewline
&(0.004)&(0.032)&(0.210)&(0.050)&(3.198)&(0.006)&(0.104) \tabularnewline
&&&&&&& \tabularnewline
Treated Munic.&216&216&216&216&216&216&216 \tabularnewline
Control Munic.&41&41&41&41&41&41&41 \tabularnewline
Mean Dep. Var.&0.013&0.443&1.100&0.460&19.870&0.036&0.079 \tabularnewline
Observations&2,827&2,827&2,827&2,827&2,827&2,570&2,827 \tabularnewline
\(R^2\)&0.148&0.574&0.769&0.618&0.546&0.332&0.431 \tabularnewline
\bottomrule \addlinespace[\belowrulesep]

\end{tabular} \end{adjustbox}
\\ \parbox{\linewidth}{\scriptsize \justifying \noindent \(Notes \): Results from estimating Equation \eqref{eq_did}. Standard errors clustered at the municipality level. Includes municipality and year fixed effects. Considers only years from 2009 and before 2020. Mean Dep. Var.: Mean of the dependent variable in the pre-intervention period for municipalities in the sample (FARC and ELN). Defines the post period as 2015. All variables are measures in 1000's of inhabitants. Prop. Loss: Property Loss. Clash/Att.: Number of clashes and attacks between government and paramilitary and guerrilla groups. Panel B shows variables from other data sources, with the first three columns coming from the Ministry of Defense, the fourth from the agency against anti-personnel mines (DAIMA), the fifth from the Victims' Unit, and the sixth from Juan Vargas. And. Index is a summary measure created following Anderson (2008) that summarizes all the different outcomes variables. It is the weighted average of the standardized outcomes, weighted by their inverted covariance matrix.}
\end{table}

A consistent picture emerges from this Table: most indicators show a significant and large decline in violence after the start of the ceasefire for municipalities with FARC presence relative to those with ELN presence. Interpreting the coefficients, municipalities with previous FARC presence experienced decreases of 0.046 forced disappearances, 0.608 property losses, 0.019 clashes between different groups, 0.292 mine events, and ten forced displacements per year (among others) after the start of the ceasefire relative to municipalities with ELN presence, with the decreases being between 40\%-50\% of the pre-treatment mean in FARC and ELN municipalities. 

I create an index following \citet{anderson2008multiple}'s approach to summarise the different measures in a single variable and increase power. In brief, it is an inverted-covariance-weighted mean of the various standardised measures, with several attractive properties relative to other summary indices. Appendix \ref{data_detail} details how this index is created. Column 7 of Panel B shows the result for this summary measure. It also shows a significant, negative and sizable decrease in overall violence. 

To understand these effects' dynamics and check whether trends look parallel for pre-intervention periods, I follow \citet{freyaldenhoven2021visualization}' suggestions and estimate Equation \eqref{eq_es}. I only present results for the Anderson Index in Figure \ref{viol_es_ext} for brevity. Each measure's results can be found in Appendix \ref{app_figures}. There are several takeaways from this Figure. First, violence decreased over time after the ceasefire, suggesting that at least until 2019, the decrease in violence was sustained. Following the advice in \citet{rothpre} and \citet{freyaldenhoven2021visualization}, I test the joint significance of all the pre-intervention coefficients. The $p$-value of such test is shown at the bottom left of each figure, and I can't reject the null of no joint pre-intervention effects. When using the pre-trends test suggested by \citet{borusyak2021revisiting} (second line at the bottom of the figure), I can only marginally reject the joint test. However, there is no individually significant coefficient and no discernible trend in the pre-intervention period, with all coefficients around the same magnitude. Finally, as suggested by \citet{freyaldenhoven2021visualization}, the sup-t confidence band is also plotted around the coefficients, which is a more adequate way of testing for the event-time path of the outcomes. Reassuringly, this sup-t confidence band covers 0 for all the pre-intervention periods. Overall, while the parallel trends assumption is untestable, these results for the pre-intervention period support the assumption's validity. 

\begin{figure}[h!]
\begin{center}
\caption{Violence in FARC Municipalities vs. ELN Municipalities -- Extensive Margin, Events in Over 60\% of Years}
\label{viol_es_ext}
\includegraphics[width=0.8\textwidth]{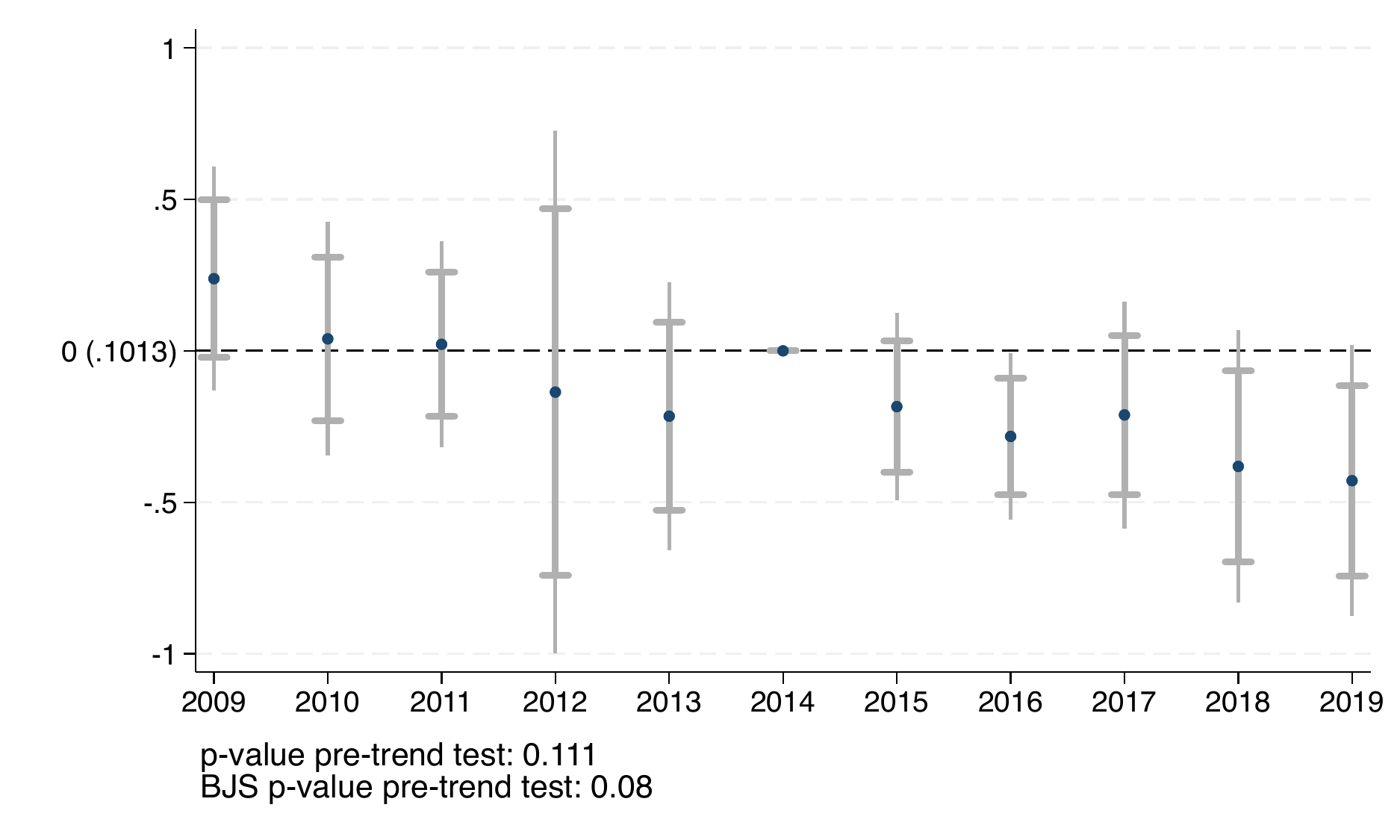}
\end{center}
\justifying 
\footnotesize{\textbf{Notes:}  Event study plots from estimating Equation \eqref{eq_es}, including 95\% confidence intervals (based on standard errors clustered at the municipality level). The index is created following \citet{anderson2008multiple} and is based on the violence measures in Table \ref{DID_final_Vio_CEDE_Ext_p60}.}
\end{figure}

Two open questions remain: Did these reductions in violence materialise in all FARC municipalities? And, if the FARC stopped operating in these municipalities, did other criminal groups start operating there? To address the first question, I check if the violence effect varies by whether the municipality is especially attractive for armed groups. Three important sources of income for armed groups in Colombia are illegal gold mining, coca production and extorting oil companies. In Figure \ref{viol_cocaGold}, I show that there is no heterogeneity in the reduction in violence in municipalities that produced oil or gold or that are more suitable for growing coca \citep[from][]{mejia2013bushes}, or both suitable to grow coca and with the presence of chemical gold anomalies. Thus, the decrease in violence seems to have happened across FARC municipalities. I use data from \citet{osorio2019mapping} to address the second question. They use NLP tools to identify the armed group responsible for human rights violations (and geolocate these) based on reports from a Colombian NGO, CINEP. CINEP gathers information about HR incidents related to the armed conflict in Colombia since 1987. Table \ref{DID_violSpill_CEDE_Ext_p60} presents results of estimating Equation \eqref{eq_did} for each armed group. Consistent with the results above, the Table shows that FARC-associated HR violations decreased significantly. There are also large reductions in activity by the government (although insignificant) and other criminal groups. The last column shows, as before, that HR violations decreased after the start of the ceasefire. Importantly, there is no increase in criminal activity by other groups (ELN, criminal bands, paramilitaries), suggesting that other criminal groups did not replace the FARC. While there is an increase in HR violations by FARC dissident groups, this comes mostly in 2018, the magnitude is very small, and they were very minor players until 2020 (see Section \ref{rob_checks} for a discussion of these groups). 

\subsection{Economic Activity}
\label{sec_econ}

The results from the previous Subsection suggest that the start of the ceasefire reduced violence in FARC municipalities relative to ELN ones, and that this effect has persisted over time. While the reduction in violence following the ceasefire is a worthy achievement in itself, one of the agreement's stated goals was to bring economic prosperity to areas affected by the conflict. Thus, in this Subsection, I evaluate whether this decrease in violence translated into improved economic indicators in precisely these areas, the intended beneficiaries of the agreement. Whether Colombia \textit{as a whole} experienced an economic dividend is an open question beyond the scope of this paper. The analysis here follows in two parts. First, I evaluate the economic impacts of the ceasefire using the same difference-in-difference strategy as in Subsection \ref{sec_viol}. Then, I focus on municipalities that received a fiscal incentive program for firms and analyse whether this government program targeting conflict-affected municipalities succeeded in improving economic indicators using a different identification strategy.

\subsubsection{Difference-in-Difference}

Measuring economic activity in municipalities that experienced the long-running presence of armed groups is difficult given that i) these tend to be small, distant, rural communities, ii) there is a lack of survey- or individual-level data in Colombia representative at the municipality level with good geographical coverage, and iii) any data collection effort is made more difficult by the insecurity created by these armed groups. Thus, I use a variety of indicators of economic activity to provide a general perspective. While none of these measures is perfect, they paint a consistent picture together.

Table \ref{DID_Econ_CEDE_Ext_p60} shows the results from estimating Equation \eqref{eq_did} on different economic indicators using the extensive margin measure of presence. First, following a large recent literature \citep[see for example, ][]{donaldson2016view,henderson2012measuring,henderson2018global}, I use nighttime light intensity as a proxy for economic development in column 1.\footnote{The data come from \citet{li2020harmonized}. Although there are concerns about the usage of nighttime data in rural areas, these have been used frequently in research about Colombia \citep[see, e.g.][]{de2021forgone, ch2018endogenous, ch2019paras, prem2022landmines}. Evidence suggests that, in Colombia, the data I use are good proxies for economic activity at the municipality level, even in rural areas \citep{perez2021night}.} For comparability, I standardise this measure for each year. Column 2 uses official estimates of municipality value added from the National Statistical Office (DANE), available between 2011 and 2020. It is based on department-level figures of value added created following international standards and best practices. This is then distributed across municipalities within the department according to a municipality's measure of economic importance based on 30 socioeconomic variables. Given that the analysed municipalities tend to be primarily rural agricultural municipalities, in column 3, I use data from the Ministry of Agriculture and analyse agricultural productivity (tonnes/area cultivated) based on the cultivation of 270+ crops. These data are collected through a bottom-up, multi-stage, rigorous process, are used to create official statistics, and provide the best spatial and temporal coverage. Column 4 uses the share of the population living in urban areas based on census data and official projections from DANE, which can capture any structural transformation taking place due to movement restrictions being lifted after the ceasefire or increased economic opportunities in urban areas. Column 5 uses firm entry data from the \textit{Registro Único Empresarial y Social} (RUES) collected by the Confederation of Chambers of Commerce. Colombian firms must obtain a license from their local Chamber of Commerce within a month of starting their commercial activity for many regular business activities. However, this license does not imply formalisation, as registration with tax authorities is a separate process, meaning that this measure will capture at least some part of the informal sector.\footnote{For more information on RUES, see \citet{bernal2022peaceful} or \citet{londono2022reparations}.} Column 6 uses data from the Ministry of Health's \textit{Planilla Integrada de Liquidación de Aportes} (PILA), which captures formal employment. PILA records for all formal wage-earners and self-employed individuals in the country their monthly contributions to healthcare, pension funds and workers' compensations. I estimate the average number of active contributors across months in each municipality as a measure of formal employment. The last column presents the results of the Anderson Index composed of the measures in columns 1 to 6.

\begin{table}[h!] \centering
\newcolumntype{R}{>{\raggedleft\arraybackslash}X}
\newcolumntype{L}{>{\raggedright\arraybackslash}X}
\newcolumntype{C}{>{\centering\arraybackslash}X}

\caption{DiD Economic Outcomes: Extensive Margin, Events in Over 60\% of Years}
\label{DID_Econ_CEDE_Ext_p60}
\begin{adjustbox}{max width=\linewidth, max height=\textwidth}\begin{tabular}{lccccccc}

\toprule
& Nighttime & Value Added (pc) & Agricultural & Share Urban & Firm  & Formal & \textbf{Anderson} \tabularnewline & Light & DANE & Productivity & Population & Entry & Empl. & \textbf{Index} \tabularnewline 
{}&{(1)}&{(2)}&{(3)}&{(4)}&{(5)}&{(6)}&{(7)} \tabularnewline
\midrule \addlinespace[\belowrulesep]
\midrule Ceasefire \( \times \) FARC&--0.007&--2.348*&0.264&--0.001&0.177&0.002&--0.004 \tabularnewline
&(0.021)&(1.208)&(0.491)&(0.004)&(0.329)&(0.004)&(0.038) \tabularnewline
&&&&&&& \tabularnewline
Treated Munic.&216&216&216&216&216&216&216 \tabularnewline
Control Munic.&41&41&41&41&41&41&41 \tabularnewline
Mean Dep. Var.&--0.314&12.007&6.654&0.430&6.078&0.107&--0.097 \tabularnewline
Observations&2,827&2,313&2,827&2,827&2,827&2,827&2,827 \tabularnewline
\(R^2\)&0.950&0.863&0.975&0.981&0.783&0.971&0.952 \tabularnewline
\bottomrule \addlinespace[\belowrulesep]

\end{tabular} \end{adjustbox}
\\ \parbox{\linewidth}{\scriptsize \justifying \noindent \(Notes \): Results from estimating Equation \eqref{eq_did}. Standard errors clustered at the municipality level. Includes municipality and year fixed effects. Considers only years from 2009 and before 2020. Mean Dep. Var.: Mean of the dependent variable in the pre-intervention period for municipalities in the sample (FARC and ELN). Defines the post period as 2015. Nighttime light intensity defined so that grid cells in the border of multiple municipalities are assigned in proportion to the share of the grid cell in each municipality (weighted). Value added per capita comes DANE (National Department of Statistics), in millions of COP. Agricultural productivity is defined as total tonnes produced of 271 agricultural crops divided by total area cultivated in hectares, based on data from the Ministry of Agriculture. Firm entry comes from RUES and is measured per 1000 inhabitants. Formal employment is measured as the average number of individuals paying contributions to healthcare, pension funds and workers' compensations across the year in the municipality per 18-60 years old, from PILA. Anderson Index is a summary measure created following Anderson (2008) that summarizes the measures in columns 1-6. It is the weighted average of the standardized outcomes, weighted by their inverted covariance matrix.}
\end{table}


The results across different proxies of economic development are consistent and show that economic indicators did not improve in areas previously affected by the FARC (if anything, the only significant coefficient is actually \textit{negative}, on value added per capita).\footnote{The insignificant effects on firm entry are similar to those of \citet{bernal2022peaceful} when looking at the whole post-ceasefire period (their Table 2). Unlike them, this variable has no differential result when analysing the post-ceasefire, pre-agreement, and post-agreement periods separately.} Moreover, this does not seem to be due to a lack of power or imprecisely estimated treatment effects, as the null effects are precisely estimated. For example, for nighttime light intensity data, the estimates suggest a reduction in light intensity of 0.007 SD, with SEs that would identify significant effects of around 0.03 SD, a small effect in magnitude. Similarly, for firm entry and formal employment, the SEs on the estimates are around $1/20$ and $1/25$ of the overall pre-intervention mean across FARC and ELN municipalities. Thus, even small effects in magnitude would have been picked up. The results using the Anderson Index in the last column are also insignificant and precisely-estimated. Overall, while each of these measures only imperfectly captures economic activity in this context, each captures important facets of these municipalities' economy (the labour market, the agricultural sector, and value-added) and together paint a consistent and comprehensive picture: the ceasefire, while proceeded by a decrease in violence, did not bring economic improvements to areas previously affected by the FARC.\footnote{It is possible that those municipalities that experienced the largest reduction in violence benefitted economically from the ceasefire. To study this, I employ an IV-DiD strategy, in which the first stage corresponds to the violence results shown in column 7, Panel B of Table \ref{DID_final_Vio_CEDE_Ext_p60}, and the second stage regresses the economic outcomes on the (instrumented) Anderson Index of the violence outcomes. The underlying assumption is that the ceasefire affects economic outcomes only through its effect on conflict. Under the assumption that this holds (unlikely), the results of this exercise are shown in Table \ref{DIDIV_Econ_CEDE_Ext_p60}. While noisier, the coefficients on all economic outcomes remain insignificant and small in magnitude, suggesting that economic effects are lacking regardless of the reduction in violence.}

I look at a battery of other development-related outcomes in Table \ref{DID_Other_CEDE_Ext_p60}, finding similar results. Each of these has important data quality or coverage limitations, so the results are suggestive at best. Still, the fact that they mimic the main set of outcomes is reassuring. The Table's notes describe the variables employed. Panel A looks at the robustness of the value added, nighttime and agricultural productivity outcomes (columns 1 to 4), as well as two measures based on satellite images: One measuring urban built-up (column 5), which has been shown to perform well in settings like Colombia \citep{valdiviezo2018built}, and one measuring the UN Human Development Index, following \citet{sherman2023global} (column 6). Panel B uses measures of agricultural trade (agricultural deliveries to each of the main agricultural markets across the country) based on DANE's System of Price Information (SIPSA in Spanish), which is only available since 2013. While there is a small, marginal increase in the number of markets supplied, there is no increase in the number of products, deliveries, average price, or total value and quantity delivered. Given the importance of the agricultural sector in these areas, this is further evidence that they did not experience much growth after the ceasefire. Panel C looks at measures of tourism and migration from various sources. While there is no increase in the provision of tourism accommodation or employment (columns 1 and 2), there is a significant increase in the number of national tourists visiting former FARC municipalities (data are only available for four years) and no increase in travel expenditures (columns 3-5).\footnote{Data on international tourists is only available since 2015, but Figure \ref{internationalTourists} shows no apparent increase in FARC areas relative to ELN ones after 2015.} Columns 6 and 7 perform a long DiD using Census data from 2005 and 2018 to assess whether people have migrated into former FARC municipalities in the past year (for the whole population and those aged over 20), finding no increase in immigration. Panel D looks at different health and education variables, finding no improvement in measures of children's health or more official schools, teachers or students. Panel E uses survey data from the Gran Encuesta Integrada de Hogares, Colombia's largest survey used to estimate the official unemployment rate and provide a high-frequency snapshot of the labor market. It is a monthly, repeated cross-section that is not representative at the municipality level.\footnote{Most FARC/ELN municipalities are never surveyed given that the emphasis is on the largest municipalities in the country, but Table \ref{SumStatsGEIH_CEDEExt_p60} shows that surveyed FARC/ELN municipalities are not too different from non-surveyed FARC/ELN municipalities in terms of baseline characteristics.} As a measure of house prices, column 1 uses the hypothetical price for which homeowners would be willing to sell their house. Column 2 creates an asset index based on ownership of different assets (using weights from Colombia's 2015 DHS survey). Columns 3 to 7 use labour market outcomes, including the intensive and extensive margin of employment, salary, and length of unemployment spells. Overall, there is no improvement in labour market conditions or asset ownership. Across this wide range of outcomes, results support those of the main outcomes, indicating a lack of improvement in economic conditions in former FARC areas.\footnote{Results using the baseline intensive measure of presence are very similar and presented in Table \ref{DID_Other_CEDE_Int_p80}. This represents the universe of data sources on economic activity in Colombia. Other datasets are either not publicly available or only survey a non-representative sample of municipalities. The Annual Manufacturing Survey (EAM) surveys all manufacturing firms above a certain amount of employees and revenue. There are very few of these firms in FARC/ELN municipalities, and the municipalities where the survey takes place are unsurprisingly much better off than those without reported surveys. However, there has been no increase in the number of manufacturing firms in FARC municipalities or their performance regarding the number of employees, total salaries, investment in fixed assets, energy consumption, or the value of production or sales (results available upon request). The ELCA is a longitudinal survey of farmers conducted in 2010, 2013, 2016, and 2019 (DANE did the last one as ELCO). Unfortunately, it only surveys a tiny sample of FARC/ELN municipalities (especially in 2010 and 2013), so it is of limited use. Other agricultural datasets have even less temporal/spatial coverage. There are no municipality-level data on foreign investment.}

\begin{figure}[htp]
\caption{Economic Activity in FARC vs. ELN Municipalities -- Extensive Margin, Events in Over 60\% of Years}
\label{dyn_econ_ext}
\begin{center}
\begin{subfigure}[b]{0.44\textwidth}
\includegraphics[width=\textwidth]{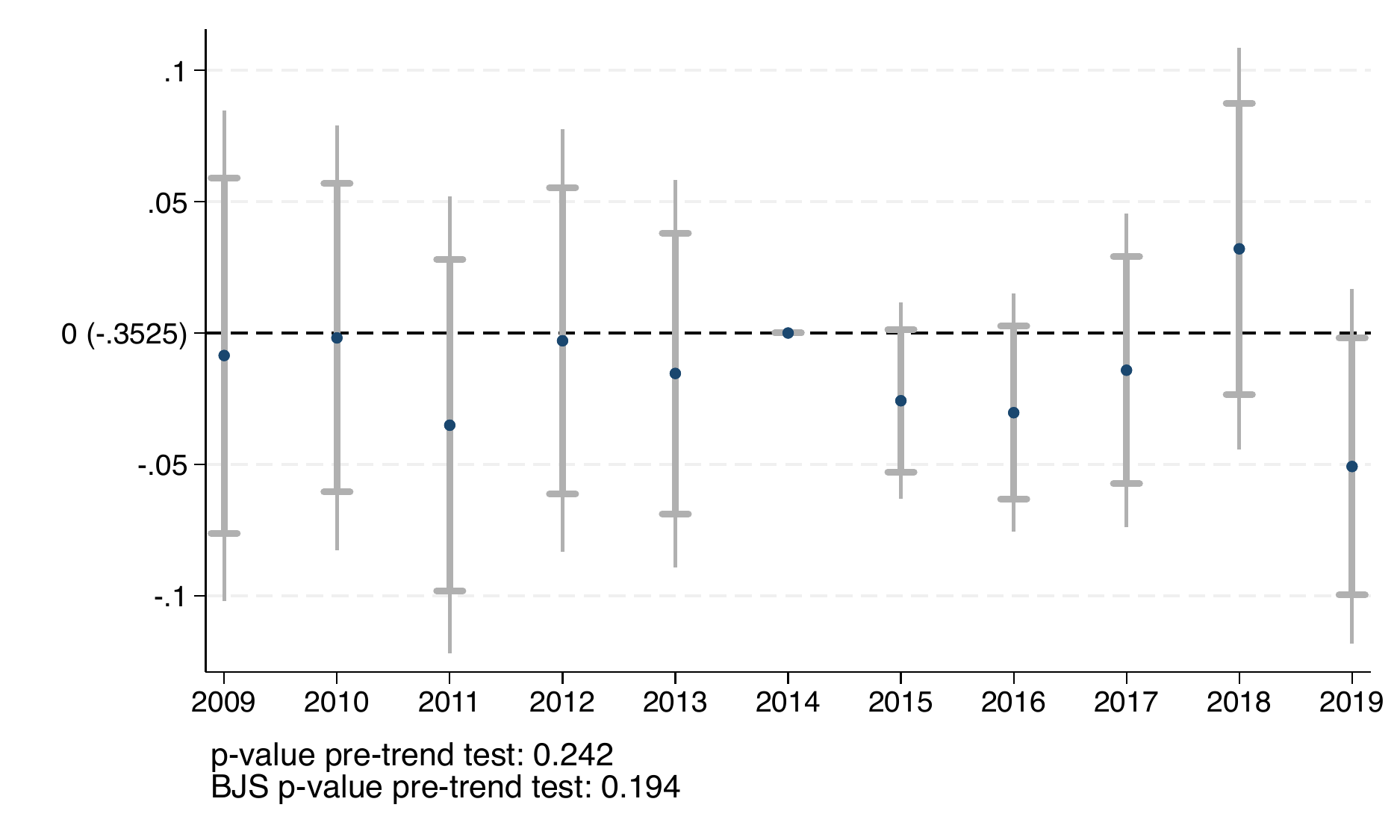}
\subcaption{Nighttime Light Weighted}
\end{subfigure}
\begin{subfigure}[b]{0.44\textwidth}
\includegraphics[width=\textwidth]{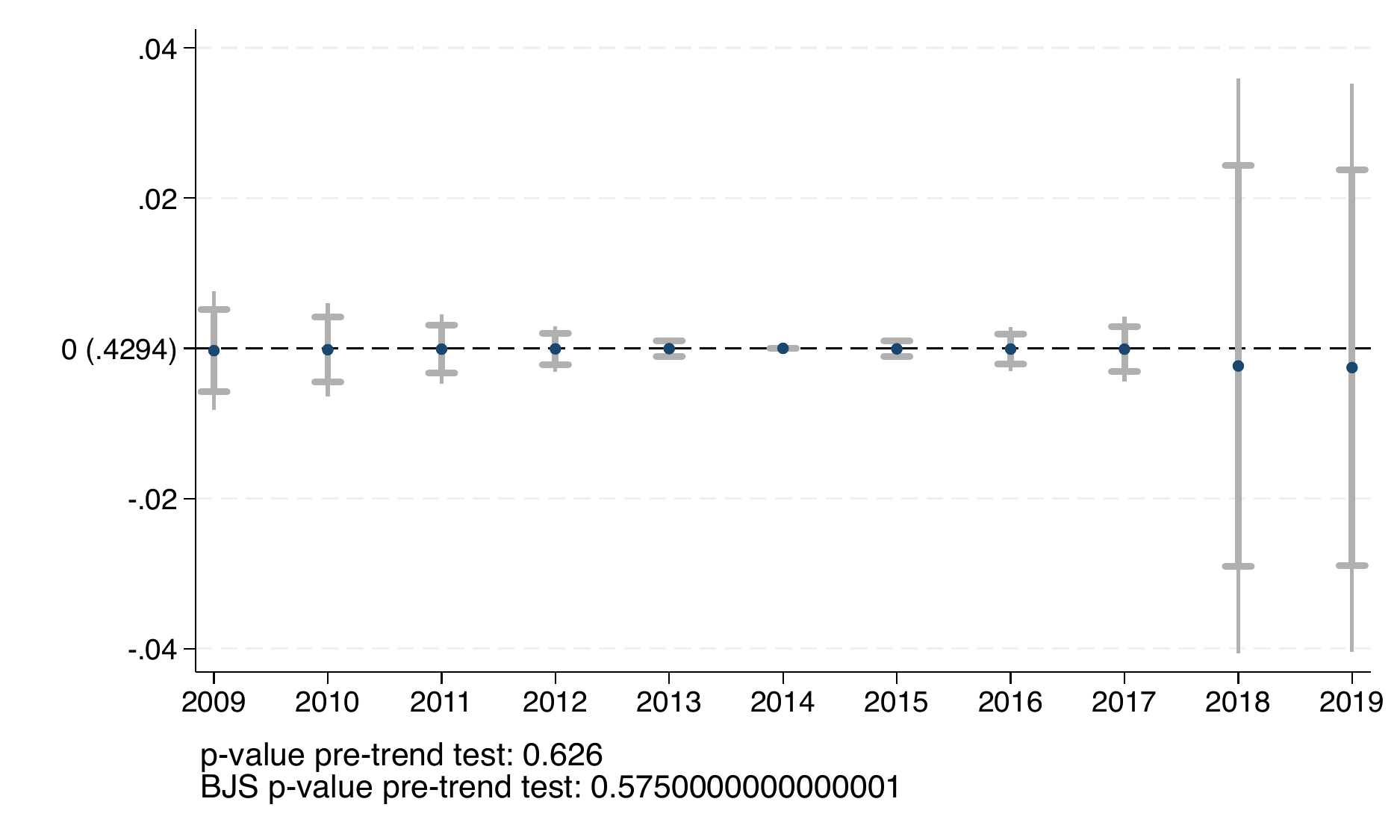}
\subcaption{Share Urban Population}
\end{subfigure}
\begin{subfigure}[b]{0.44\textwidth}
\includegraphics[width=\textwidth]{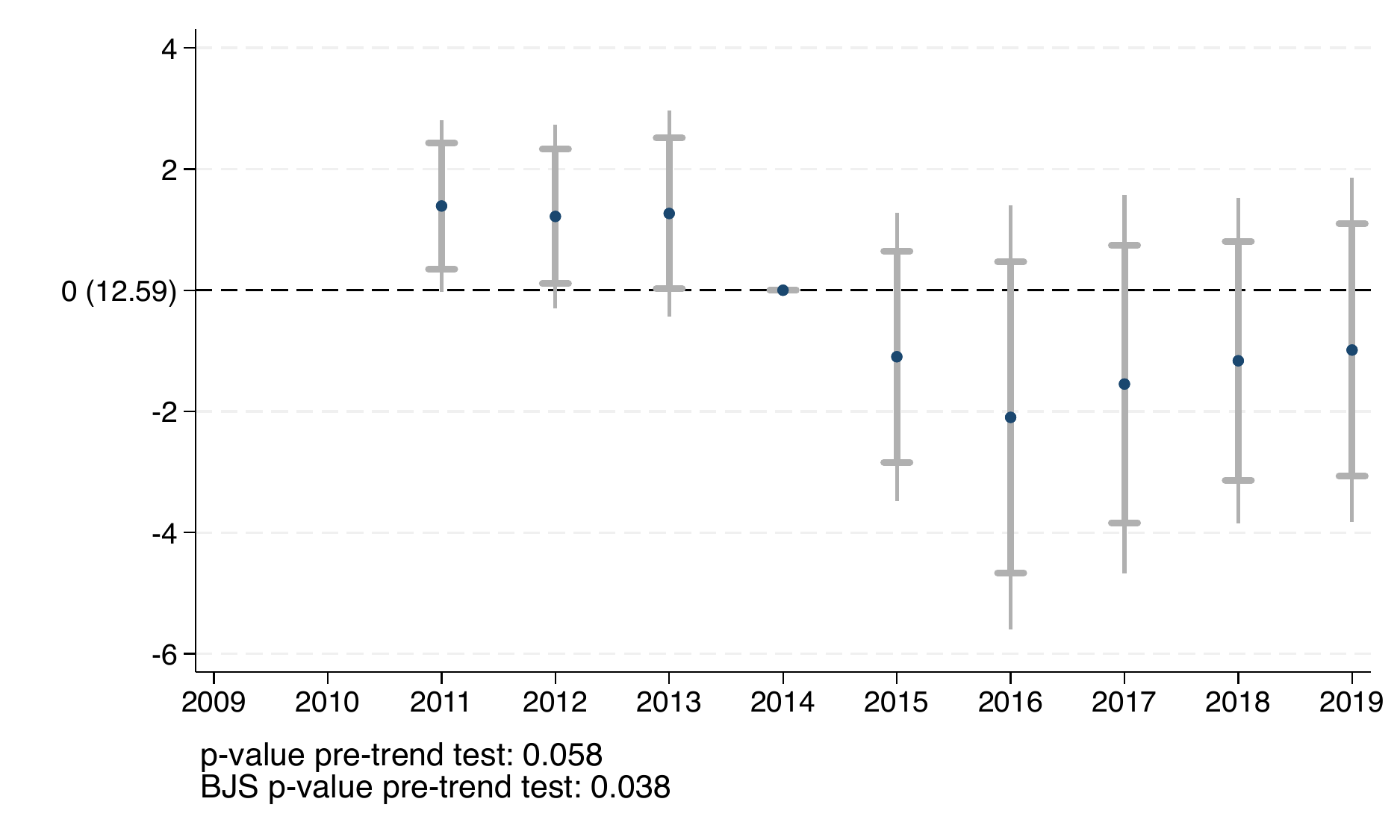}
\subcaption{Value added (pc, DANE)}
\end{subfigure}
\begin{subfigure}[b]{0.44\textwidth}
\includegraphics[width=\textwidth]{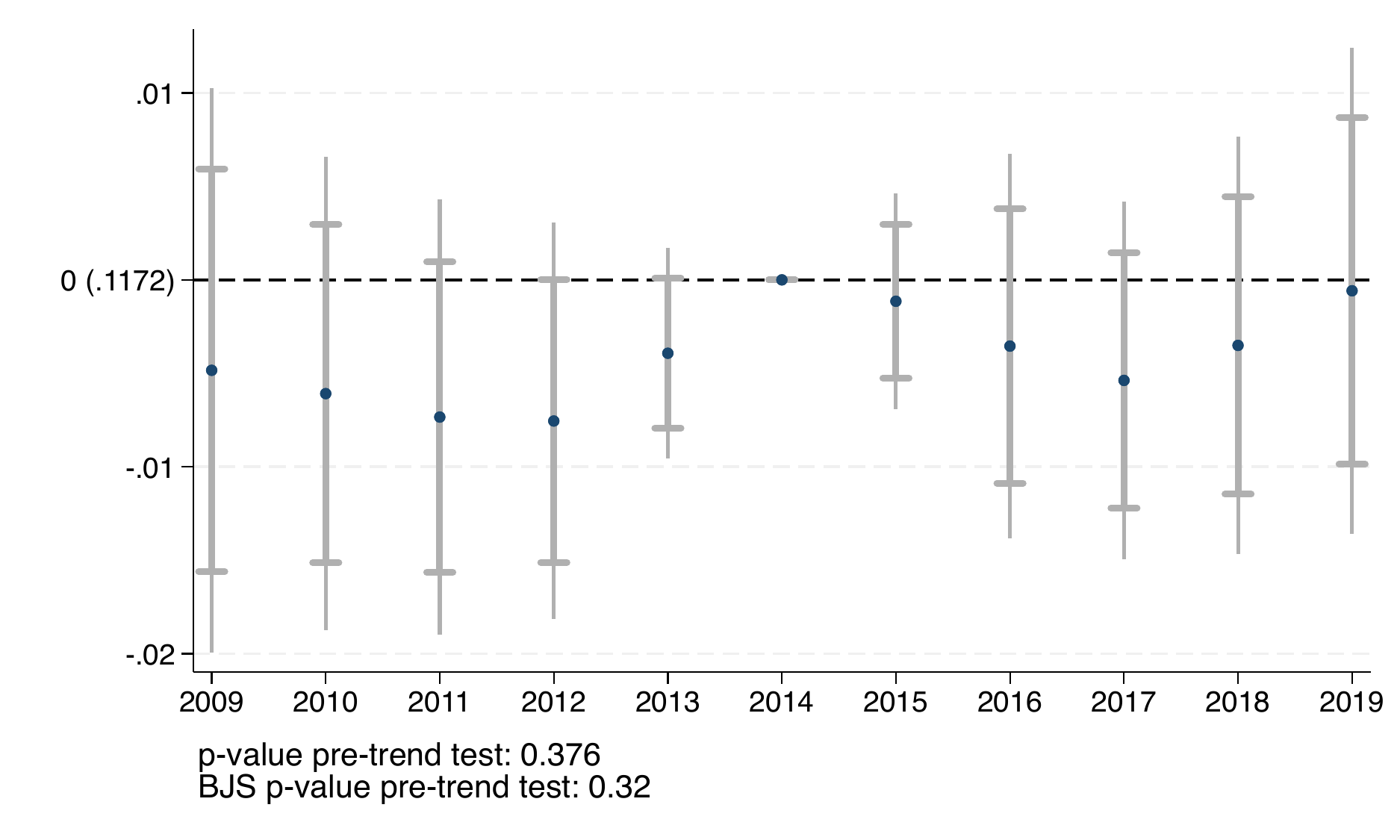}
\subcaption{Formal Employment}
\end{subfigure}
\begin{subfigure}[b]{0.44\textwidth}
\includegraphics[width=\textwidth]{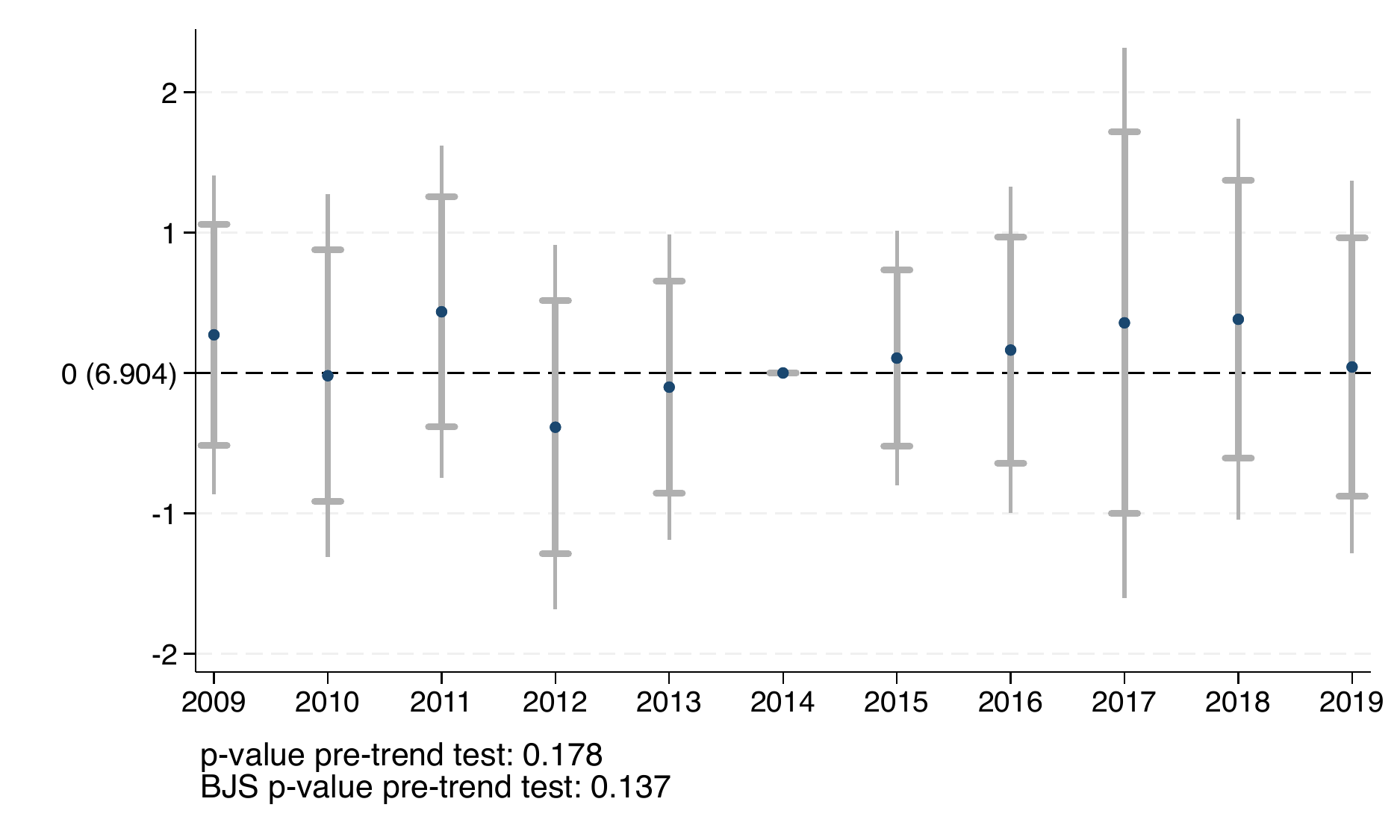}
\subcaption{Firm Entry}
\end{subfigure}
\begin{subfigure}[b]{0.44\textwidth}
\includegraphics[width=\textwidth]{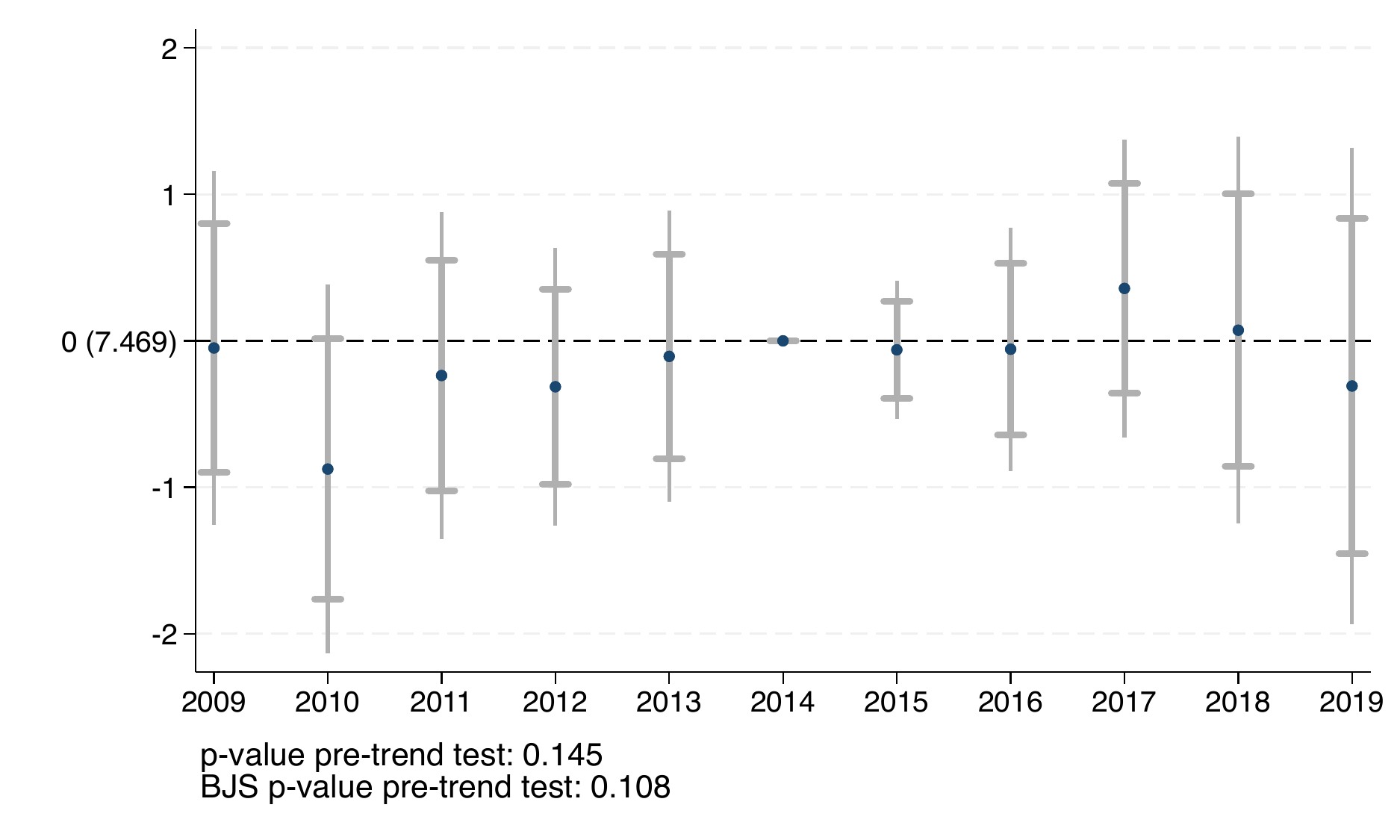}
\subcaption{Agricultural Productivity}
\end{subfigure}
\begin{subfigure}[b]{0.44\textwidth}
\includegraphics[width=\textwidth]{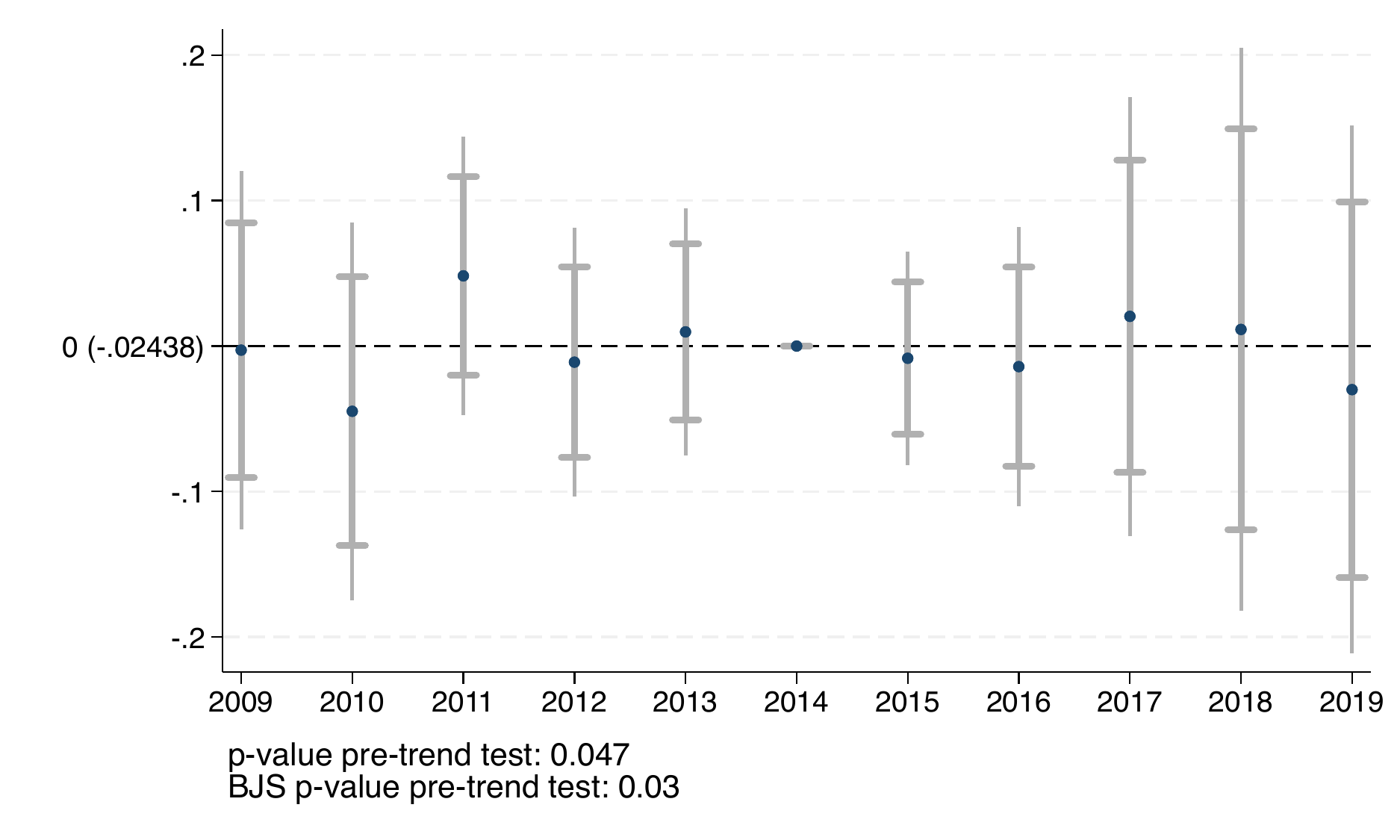}
\subcaption{Anderson Index}
\end{subfigure}
\end{center}
\justifying 
\footnotesize{\textbf{Notes:}  Event study plots from estimating Equation \eqref{eq_es}, including including 95\% confidence intervals (based on standard errors clustered at the municipality level). The Index is created following \citet{anderson2008multiple} and is based on all the other variables.}
\end{figure}

A natural concern with the results in Table \ref{DID_Econ_CEDE_Ext_p60} is that the analysis might simply ignore dynamics. On the one hand, it could be the case that the economic returns from peace do not happen immediately but take some time to materialise. On the other hand, the initial period post-ceasefire is likely the most critical one since there is a small ``window of opportunity'' to make the agreement work. To analyse the dynamic evolution of the different indicators, I estimate Equation \eqref{eq_es} following \citet{freyaldenhoven2021visualization}'s suggestions. Figure \ref{dyn_econ_ext} shows the event study figures for the different measures of economic activity.

There are several takeaways from Figure \ref{dyn_econ_ext}. First, for almost all joint significance tests of pre-intervention periods, the null hypothesis of no effect is not rejected, using either a test of joint significance of all pre-treatment coefficients or the pre-trends test suggested by \citet{borusyak2021revisiting}. There is no trend pre-treatment for the two graphs for which the null is rejected (the pre-intervention coefficients are simply flat), and the sup-t confidence bands cover 0. Moreover, the significant joint test for the Anderson Index comes entirely from the firm entry measure, as shown in Figure \ref{rob_indices}. However, the main coefficient is robust to the composition of the index. Thus, there seems to be no violation of the parallel trend assumption in the pre-intervention period, providing supportive evidence of the validity of the identification strategy. Second, and most importantly, it does not seem to be the case that the economic effects of the ceasefire are materialising in the medium run. The coefficients remain close to zero for the entire post-intervention period, with no apparent improvement over time or a differential effect before or after the actual ratification of the peace agreement in 2016. 

Lastly, it could be that the null result is masking considerable heterogeneity, with some municipalities benefiting from the ceasefire and others not. I focus on two sets of characteristics and how they affect the economic activity index, shown in Figure \ref{econHeterogeneity}. First, in Panel A, I show estimates of $\gamma$ from estimating Equation \eqref{eq_het} using different characteristics of the municipalities that are either time-invariant (e.g. distance) or from the 2005 census. I create heterogeneity measures by whether the municipality is above or below the median of said variable. Second, in Panel B, I show heterogeneity by the presence of resources usually exploited by guerrilla and other criminal groups, which could be more likely to be ``recaptured''. The figures show that, across heterogeneity variables, the triple interaction terms $\hat{\gamma}$ are insignificant at the 5\% level. This suggests that the lack of economic improvements was widespread. 

\subsubsection{Difference-in-Discontinuities -- ZOMAC Program}

The results from the previous Subsection show no improvement in economic indicators when comparing FARC and ELN municipalities, before and after the start of the ceasefire. In this Subsection, I go one step further and analyse a government policy designed to spur economic activity in conflict-affected areas. Importantly, the inclusion of municipalities in this program was based on set thresholds of some economic and violence indicators. I briefly introduce the program, identification strategy and results in this Subsection, with a more detailed exposition in Appendix \ref{zomac_sec}. 

To incentivise business and employment creation in areas affected by the conflict (ZOMAC municipalities), the government started a tax incentive program for firms in 2017. The main incentive is a progressive business tax tariff for 10 years beginning in 2017, which varies depending on the firm's size, as shown below in Table \ref{taxbenefits}. The reduction in business tax rates is sizeable. For firms to benefit from the tax reduction, they must i) have been created after December 29, 2016, ii) have their primary address in a ZOMAC municipality, iii) perform their whole productive processes in ZOMAC municipalities, and iv) satisfy some investment and job-creation requirements. These investment and job-creation requirements vary depending on the sector and the firm's size.\footnote{For example, a micro firm in the agricultural sector must invest 40 monthly minimum wages (SMLMV, around 30M COP) and generate two jobs to receive the incentives. A large firm in the same sector must invest 7800 SMLMV (1.5M USD) and generate 49 direct jobs to benefit. The table in the appendix of \citet{zomac2015} stipulates the requirements for each firm type and industry.} Informal firms that formalise and meet these criteria can also benefit from these incentives. 

\begin{table}[h!]
\begin{center}
\caption{Percentage of Business Tax Rate Paid -- ZOMAC Program}
\label{taxbenefits}
\begin{tabular}{l|cccc}
\multicolumn{1}{c|}{\textbf{Firm Size}} & \textbf{2017-2021} & \textbf{2022-2024} & \textbf{2025-2027} & \textbf{2027-} \\ \hline
Micro \& Small                          & 0\%                & 25\%               & 50\%               & 100\%          \\
Medium \& Large                         & 50\%               & 75\%               & 75\%               & 100\%         
\end{tabular}
\end{center}
\end{table}

Participation in the program was based on several different socioeconomic indicators with precise cutoffs. I exploit the discontinuity in participation caused by an index of incidence of the armed conflict (IICA)\footnote{The IICA score is the average across six violence-related variables between 2002 and 2013, and the participation threshold is set at 0.019, with municipalities with a score above that included in the program, conditional on some additional variables.} by embedding a regression-discontinuity design based on this variable in a diff-in-diff set-up. This approach was first formalised by \citet{grembi2016fiscal} and called the ``difference-in-discontinuity'' estimator. It allows me to compare the evolution of those municipalities just below and above the inclusion threshold over time. Three assumptions need to be satisfied to recover this estimator: first, all the potential outcomes must be continuous at the discontinuity; second, an assumption similar in spirit to the parallel trends assumption in traditional difference-in-difference settings; and third, the effect of the treatment at the discontinuity does not depend on any confounding policy. I describe these assumptions in detail and present evidence in support of their validity in Appendix \ref{zomac_sec}. A recent paper \citep{picchetti2024difference} shows that even in the absence of anything else changing at the cutoff, the difference-in-discontinuity estimator has lower bias and better coverage than the traditional RDD because it leverages the time dimension.

Following \citet{grembi2016fiscal}, I estimate the regression:
\begin{equation}
\label{eq_diff_in_disc}
\begin{aligned}
y_{it} =& \delta_0 + \delta_1 IICA_{m}^{*} + \delta_2 \textit{IICA Treatment}_{m} + \delta_3 IICA_{m}^{*} \times \textit{IICA Treatment}_{m}
\\
&+ \delta_4 Post_t + \delta_5 Post_t \times IICA_{m}^{*}  + \delta_6 Post_t \times \textit{IICA Treatment}_{m} 
\\
&+ \delta_7 Post_t \times \textit{IICA Treatment}_{m} \times IICA_{m}^{*} + u_{mt}
\end{aligned}
\end{equation}

\noindent where $IICA_{m}^{*}$ is the normalized IICA score ($IICA_{m}^{*} = IICA_{m} - 0.0191$) of municipality $m$ in year $t$, \textit{IICA Treatment}$_{m}$ is a dummy for municipalities with an $IICA$ score above 0.0191 (i.e. ZOMAC municipalities), and $Post_t$ is an indicator for the post-treatment period. As the ZOMAC program started in 2017, I denote the post-treatment years as those from 2017 in these regressions. Standard errors are clustered at the municipality level. The difference-in-discontinuity estimator of interest is the coefficient $\delta_6$ and identifies the treatment effect of receiving the fiscal incentives for firms. 

Table \ref{RDD_Econ_IICA} shows the estimates $\hat{\delta}_6$ from estimating Equation \eqref{eq_diff_in_disc} on the measures of economic activity. I follow \citet{calonico2014robust} and select the bandwidth to minimise the Mean Square Error, but the results are similar if the bandwidth minimises the Coverage Error Rate instead. Following \citet{grembi2016fiscal}, the bandwidth is the average of those in the pre- and post-treatment periods. The $p$-value of the diagnostic test suggested by \citet{picchetti2024difference}, described in Appendix \ref{zomac_sec}, is shown at the bottom of Panel A and is insignificant for all but one outcome. Given the lack of theoretical estimators for the case with multiple time periods, I also follow \citet{picchetti2024difference} and estimate in Panel B the same model collapsing the data in the pre- and post-treatment periods.

\begin{table}[h!] \centering
\newcolumntype{R}{>{\raggedleft\arraybackslash}X}
\newcolumntype{L}{>{\raggedright\arraybackslash}X}
\newcolumntype{C}{>{\centering\arraybackslash}X}

\caption{Difference-in-Discontinuities Analysis: IICA Cutoff, Economic Outcomes}
\label{RDD_Econ_IICA}
\begin{adjustbox}{max width=\linewidth, max height=\textwidth}\begin{tabular}{lccccccc}

\toprule
& Nighttime & Value Added (pc) & Agricultural & Share Urban & Firm  & Formal & \textbf{Anderson} \tabularnewline & Light & (DANE) & Productivity & Population & Entry & Empl. & \textbf{Index} \tabularnewline 
{}&{(1)}&{(2)}&{(3)}&{(4)}&{(5)}&{(6)}&{(7)} \tabularnewline
\midrule \addlinespace[\belowrulesep]
\midrule \textbf{Panel A. Diff-in-Disc}&&&&&&& \tabularnewline
IICA Treatment \( \times \) Post&0.031&0.692&--0.341&--0.027**&--1.064&--0.001&--0.022 \tabularnewline
&(0.047)&(1.269)&(1.606)&(0.014)&(0.966)&(0.011)&(0.109) \tabularnewline
&&&&&&& \tabularnewline
Observations&1,859&936&1,100&1,067&1,584&1,166&1,353 \tabularnewline
\(R^2\)&0.010&0.044&0.024&0.054&0.059&0.029&0.073 \tabularnewline
Treated Municipalities&65&46&43&42&60&47&49 \tabularnewline
Control Municipalities&104&58&57&55&84&59&74 \tabularnewline
\(p\)-value Joint Test Pre-Treat Years&0.949&0.998&0.964&0.999&0.001&0.226&0.792 \tabularnewline
&&&&&&& \tabularnewline
\midrule \textbf{Panel B. Collapsing Pre-Post Years}&&&&&&& \tabularnewline
IICA Treatment \( \times \) Post&0.006&--1.057&--0.966&--0.010&--1.114&--0.006&--0.124 \tabularnewline
&(0.037)&(1.572)&(1.154)&(0.012)&(0.852)&(0.009)&(0.104) \tabularnewline
&&&&&&& \tabularnewline
Observations&506&348&304&322&384&366&348 \tabularnewline
\(R^2\)&0.012&0.011&0.050&0.013&0.082&0.027&0.040 \tabularnewline
Treated Municipalities&85&68&63&65&75&72&68 \tabularnewline
Control Municipalities&168&106&89&96&117&111&106 \tabularnewline
\bottomrule \addlinespace[\belowrulesep]

\end{tabular} \end{adjustbox}
\\ \parbox{\linewidth}{\scriptsize \justifying \noindent \(Notes \): Results from estimating Equation \eqref{eq_diff_in_disc}. Standard errors clustered at the municipality level. Includes municipality and year fixed effects. Considers only years from 2009 and before 2020. Defines the post period as 2017. Panel A shows results estimated following the approach in Grembi et al. (2016), while Panel B collapses the data for the pre- and post-intervention period and follows the approach in Picchetti et al. (2024). Bandwidth optimally estimated using Calonico et al. (2014) approach, minimizing the Mean Squared Error. Nighttime light intensity defined so that grid cells in the border of multiple municipalities are assigned in proportion to the share of the grid cell in each municipality (weighted). Value added per capita comes DANE (National Department of Statistics), in millions of COP. Agricultural productivity is defined as total tonnes produced of 271 agricultural crops divided by total area cultivated in hectares, based on data from the Ministry of Agriculture. Firm entry comes from RUES and is measured per 1000 inhabitants. Formal employment is measured as the average number of individuals paying contributions to healthcare, pension funds and workers' compensations across the year in the municipality per 18-60 years old, from PILA. Anderson Index is a summary measure created following Anderson (2008) that summarizes the measures in columns 1-6. It is the weighted average of the standardized outcomes, weighted by their inverted covariance matrix. The last row of Panel A shows the \(p\)-value of the diagnostic test suggested in Picchetti et al. (2024), which uses a stacked regression of each of the pre-treatment years and tests the equality of all the triple interaction terms between year dummies, the (placebo) dummy, and the normalised IICA score.}
\end{table}

The results align with those found in the previous Subsection: the ZOMAC program to incentivise firm and employment creation in areas affected by the conflict did not improve a wide array of economic indicators. All the coefficients are statistically insignificant, but for the share of the urban population, which shows a modest \textit{decrease}. The estimates are precisely-estimated, and the null result is not due to large standard errors. Appendix \ref{zomac_sec} shows robustness to a battery of tests. The main drawback of this evaluation is the short post-intervention period considered, with only three years of data available after the start of the ZOMAC program. Still, results are very similar when extending the timeframe until 2021 (Table \ref{timeframes_CEDEExt_p60}). However, given that the incentives were larger for early movers, it would have been optimal for businesses considering this opportunity to start as soon as possible to enjoy the tax reductions for longer. Even for firm entry (column 5) and formal employment (column 6), outcomes targeted explicitly by this program, there is no significant improvement. These are especially relevant since, to benefit from the tax incentives, firms had to formalise and pay social security contributions to their employees, which these two variables capture. This shows that even government programs specifically designed to spur economic activity in these areas failed to improve economic conditions. 

\section{Mechanisms: State Capacity \& Entry}
\label{sec_sc}

The results from the previous Section are puzzling. On the one hand, the results in Subsection \ref{sec_viol} show that the start of the ceasefire led to a significant decrease in violence in municipalities that have historically had FARC presence relative to those with ELN presence. On the other hand, the results in Subsection \ref{sec_econ} show that the start of the ceasefire and its subsequent reduction in violence did not lead to improvements in a wide array of economic indicators in the short- and medium-run, even in municipalities targeted explicitly by the government with economic policies. In this Section, I investigate one potential channel that could reconcile these two seemingly contradictory results: a general lack of state capacity in these areas due to their long-term neglect during the most violent periods of the conflict and a lack of state entry post-ceasefire, left them unable to benefit economically from their new-found peace. 

In Colombia, areas affected by the armed conflict have been long ignored by the government due to corruption, difficulties reaching those distant, rural areas, and the inherent difficulties of imposing government authority in areas where armed groups are present. \citet{colchester2020implementing} have noted the failure of the Colombian government to engage effectively in conflict-affected areas, saying ``state institutions in Colombia have suffered from a lack of capacity and effectiveness, especially in the marginalised regions which were particularly affected by the armed conflict leading to ineffective implementation responses.'' Nor was this an unknown factor, with several different organisations and reports during the peace negotiations emphasising the importance of the state consolidating its presence in areas previously under the control of the FARC for the agreement to succeed \citep{sanchez2014grupo,meacham2014colombia,isacson2014ending,pares2015}. However, in recent years, concerns have grown that implementation of the agreement has lagged, especially under President Duque's government, which opposed the agreement, with the state not entering previous FARC areas. Evaluating the implementation of the agreement in 2021, five years after its ratification, the \citet{isacson2021long} stated that, as the rural reform at the heart of the agreement falls behind, ``so does Colombia's effort to govern its own territory -- especially its effort to govern democratically, with the whole state, not just the security forces -- in parts of the countryside where that has never happened before.'' \citet{piccone2019peace} similarly observed that ``the heavy demands of addressing multiple challenges simultaneously -- (...) building a state presence for rural development -- are taxing, if not overwhelming, the government's capacity to keep the process on track.'' Several programs designed to increase the state's presence in affected municipalities have disappointed due to their flawed implementation, or lack thereof \citep{garcia2020implementacion}.

I present three pieces of evidence that \textit{suggest} that the lack of economic benefits from the reduction in violence is due to i) these areas having very low initial levels of state capacity, likely due to their sustained exposure to illegal armed groups, and ii) a lack of state entry and improvement of local state capacity after the start of the ceasefire. Intuitively, economic investments will not follow in areas without state or where it can't provide basic public goods, especially if other armed groups are still operating, since the inherent uncertainty of recouping such investments persists. In areas where there is at least a tentative state entry, there is evidence to suggest that former FARC municipalities did experience economic improvements. This would align with the theoretical model by \citet{besley2010state}, which shows that low state capacity can lead to self-reinforcing low-income traps. Consequently, breaking from this poverty trap requires investments in state capacity. Given the many meanings and definitions of ``state capacity'', I will interpret it broadly, presenting multiple measures commonly associated with state capacity.\footnote{There is a long literature discussing what state capacity is and how it can be measured. Measuring state capacity at the subnational level is even more difficult due to a general lack of indicators at a disaggregated level \citep{hanson2021leviathan}. Earlier work originally referred to it as the power of the state to raise revenue \citep{north1981structure,tilly1985war}. Based on Mann's concept of infrastructural power \citep[the capacity of the state to penetrate society and ``to implement logistically political decisions throughout the realm'',][]{mann1984autonomous}, another strand of the literature has emphasised it as the state's ability to provide the basic infrastructure necessary to sustain economic activity, including the provision of public goods \citep{hanson2021leviathan}. Yet another strand of the literature follows Weber's definition of the state not just by its monopoly on the legitimate use of force but also the effectiveness in which an organised bureaucracy uses this monopoly and thus focuses on measures of bureaucratic quality \citep{hendrix2010measuring}. As each definition captures important components of a state's capacity, I use a broad range of different state capacity indicators based on and relating to these literatures.}

\subsection{Evidence 1: Low-Levels of Baseline State Capacity}

Table \ref{SC_SumStats_CEDEExt_p60} shows the mean value of a multitude of state capacity measures in FARC (column 2) and ELN municipalities (column 3), as well as across the rest of the country (column 1) in 2008, before peace negotiations started. Municipalities with FARC and ELN presence, while similar to each other in terms of initial levels of state capacity, have much lower levels than the rest of the country to begin with. This is likely due to the decades of armed conflict they experienced and a historical lack of state presence in these areas. The second row shows that total tax revenue per capita (perhaps the most common measure of state capacity) is, on average, around 102.000 COP across the rest of the country, while in FARC and ELN municipalities, it is considerably lower at around 70.000 COP.\footnote{100.000 COP are around 26.5 USD according to the exchange rate on June 2, 2022.} This difference is statistically significant between the two groups and the rest of the country (but not when comparing FARC and ELN municipalities). A similar pattern holds for other financial measures, although power is low when comparing with ELN municipalities due to the few municipalities categorised as ELN. These municipalities also receive lower transfers from the central government than the rest of the country (\nth{3} row). The results suggest that conflict-affected municipalities have much lower institutional quality and performance (rows 6-9) and provide fewer basic public goods such as aqueduct provision and garbage collection (rows 10-11).

\begin{table}[h!] \centering
\newcolumntype{R}{>{\raggedleft\arraybackslash}X}
\newcolumntype{L}{>{\raggedright\arraybackslash}X}
\newcolumntype{C}{>{\centering\arraybackslash}X}

\caption{Baseline State Capacity Summary Statistics: Extensive Margin, Events in Over 60\% of Years}
\label{SC_SumStats_CEDEExt_p60}
\begin{adjustbox}{max width=\linewidth, max height=\textwidth}\begin{tabular}{l|ccc|ccc}

\toprule
& \multicolumn{3}{c|}{\textbf{Mean}} & \multicolumn{3}{c}{\textbf{p-value of Difference}} \tabularnewline & Rest Country & FARC & ELN & Rest-FARC & Rest-ELN & FARC-ELN \tabularnewline 
{Variable}&{(1)}&{(2)}&{(3)}&{(4)}&{(5)}&{(6)} \tabularnewline
\midrule \addlinespace[\belowrulesep]
\midrule 1. Total Revenue&811.02&703.91&611.84&0&0&0.13 \tabularnewline
2. Tax Revenue&102.95&71.40&68.75&0&0.07&0.78 \tabularnewline
3. Gov. Transfers&92.11&65.94&52.24&0&0.01&0.13 \tabularnewline
4. Operational Expenditures&122.58&95.91&81.05&0&0&0.08 \tabularnewline
5. Tax Rev. / Nighttime Light&5232.20&77.27&--181.69&0.60&0.81&0.36 \tabularnewline
6. Savings Capacity&0.05&--0.20&--0.02&0&0.63&0.22 \tabularnewline
7. Fiscal Perf.&0.03&--0.17&0.03&0&0.99&0.16 \tabularnewline
8. Administrative Perf.&0.07&--0.23&0&0&0.64&0.14 \tabularnewline
9. Rule Compliance&0.01&--0.12&--0.09&0.05&0.49&0.82 \tabularnewline
10. Aqueduct Coverage&59.74&54.15&54.52&0.03&0.34&0.94 \tabularnewline
11. Garbage Collection&44.91&43.84&43.91&0.67&0.85&0.98 \tabularnewline
\bottomrule \addlinespace[\belowrulesep]

\end{tabular} \end{adjustbox}
\\ \parbox{\linewidth}{\scriptsize \justifying \noindent \(Notes\): All the variables are measured in 2008 (the last period before the panel used in the main analysis) but for the rule compliance variable that is only available from 2010. The financial measures are in per capita terms (in thousand COP), and the performance measures are standardised (by year).}
\end{table}

\subsection{Evidence 2: Lack of (General) State Entry Post-Ceasefire}

A great deal of emphasis was given during the peace negotiations to the importance of the state establishing a firm presence in former FARC municipalities to avoid other criminal organisations taking over this territory. I analyse whether the start of the ceasefire led to an improvement in state capacity measures in former FARC municipalities in Table \ref{DID_SC_CEDE_Ext_p60}, which shows the results of estimating Equation \eqref{eq_did}. I focus on a range of broad state capacity and presence measures: Tax revenue per capita (column 1), which captures both the capacity to collect taxes and the legitimacy of local governments. While it could also capture more economic activity, the results in Subsection \ref{sec_econ} suggest that this is unlikely. Operational costs (per capita), which measures how much is spent in the functioning of the local government (personnel, general functioning, etc). This is a rough indicator of the presence of bureaucrats, civil servants, and the state in general. The ratio of government transfers to total municipality revenue (excluding transfers from the central government, column 3) indicates how much the local government depends on the central government to fund their operations (higher values indicate lower state capacity). The last three columns look at indicators of institutional quality and performance developed by different government entities using administrative data to monitor municipal governments. Fiscal performance (column 4) is composed of six indicators that measure the financial management of the local government, an important indicator of how efficiently resources are being spent. Administrative performance (column 5) comprises four broad indicators that measure how well municipalities execute their programs and administrative capacity. Rule compliance (column 6) measures how well local governments comply with rules set by the national government and have adopted the monitoring tools it requires. It consists of the sum of two indices, one measuring whether the municipality's contracting is transparent and how well it provides information on important social programs, and the other one is how it implements internal monitoring tools and follows adequate accounting standards. Each measure captures different and important aspects of state capacity and presence, from tax capacity to local governments' size to institutional performance.

\begin{table}[h!] \centering
\newcolumntype{R}{>{\raggedleft\arraybackslash}X}
\newcolumntype{L}{>{\raggedright\arraybackslash}X}
\newcolumntype{C}{>{\centering\arraybackslash}X}

\caption{DiD State Capacity Analysis: Extensive Margin, Events in Over 60\% of Years}
\label{DID_SC_CEDE_Ext_p60}
\begin{adjustbox}{max width=\linewidth, max height=\textwidth}\begin{tabular}{lccccccc}

\toprule
& Tax & Operational  & Ratio Gov. Trans. & Fiscal & Administrative & Rule  & \textbf{Anderson} \tabularnewline & Revenue (pc) & Costs (pc) & to Revenue & Performance  & Performance & Compliance & \textbf{Index} \tabularnewline 
{}&{(1)}&{(2)}&{(3)}&{(4)}&{(5)}&{(6)}&{(7)} \tabularnewline
\midrule \addlinespace[\belowrulesep]
\midrule Ceasefire \( \times \) FARC&--0.009&0.005&--0.050&0.012&--0.058&0.122&0.067 \tabularnewline
&(0.011)&(0.005)&(0.539)&(0.108)&(0.090)&(0.100)&(0.054) \tabularnewline
&&&&&&& \tabularnewline
Treated Municip.&216&216&216&216&216&216&216 \tabularnewline
Control Municip.&41&41&41&41&41&41&41 \tabularnewline
Mean Dependent Var.&0.100&0.113&7.149&--0.065&--0.132&--0.087&0.014 \tabularnewline
Observations&2,822&2,822&2,815&2,823&2,827&2,567&2,827 \tabularnewline
\(R^2\)&0.778&0.813&0.676&0.548&0.483&0.433&0.595 \tabularnewline
\bottomrule \addlinespace[\belowrulesep]

\end{tabular} \end{adjustbox}
\\ \parbox{\linewidth}{\scriptsize \justifying \noindent \(Notes \): Results from estimating Equation \eqref{eq_did}. Standard errors clustered at the municipality level. Includes municipality and year fixed effects. Considers only years from 2009 and before 2020. Mean Dep. Var.: Mean of the dependent variable in the pre-intervention period for municipalities in the sample (FARC and ELN). Defines the post period as 2015. Column 1 uses tax revenue. Column 2 uses the municipality's operational costs. Both these measures are in millions of COP per capita. Column 3 uses the ratio of transfers from the central government to total municipality revenue (excluding transfers from the central government). These three measures have been winsorized for observations ten SDs away from the mean for the whole country. Column 4 uses an indicator of fiscal performance created by the National Department of Planning and measures how well the municipality spends its resources. Column 5 uses an indicator of the municipality's overall administrative performance created by the National Department of Planning. It takes into account the local administration's efficiency and efficacy, compliance with rules, and the municipality's administrative capacity. Column 6 combines two indicators created by the Office of the Inspector General, which measure how well municipalities report information, implement internal monitoring tools, and the transparency of contracting practices. These last three are standardised. Column 7 summarizes these variables into a single index following Anderson (2008). It is the weighted average of the standardized outcomes, weighted by their inverted covariance matrix.}
\end{table}

The results in Table \ref{DID_SC_CEDE_Ext_p60} using these state capacity indicators are consistent: after the start of the ceasefire, indicators of state capacity and presence did not improve in former FARC municipalities relative to ELN municipalities, suggesting that the state did not enter these areas to fill in the vacuum left by the FARC. The estimates are precisely estimated, suggesting that the insignificance is not due to noisy estimates. For example, column 1 shows that tax revenue \textit{decreased} by 9.000 COP per capita per year on average in FARC municipalities relative to ELN municipalities, which is only 9\% of the pre-intervention mean across FARC and ELN municipalities. The SEs would pick up effects of around 20.000 COP per year, which are small in magnitude. In terms of operational costs, column 2 shows an increase of around 5.000 COP per capita per year, suggesting that there was no meaningful increase in the physical size of the state in FARC areas post-ceasefire. While there is an (insignificant) decrease in the reliance on government transfers (column 3), this is very small in magnitude. Similar results hold when looking at institutional performance indicators: Fiscal performance increases by only 0.012 SDs (column 4), while administrative performance decreases by 0.058 SDs (column 5), both of which are insignificant. Lastly, there is also no significant improvement in compliance with the rules set out by the national government (column 6). Column 7 shows the results of combining these measures into a single index, following \citet{anderson2008multiple}, finding an insignificant effect of around 0.067 SDs. These results are robust to using the ratio of tax collection to a municipality's economy or not windsorising the measures of state capacity, see Table \ref{rob_DID_SC_CEDE_Ext_p60}.\footnote{Other measures of state capacity and presence in Colombia are difficult to come by. As far as I can tell, there is no publicly available data at the municipality-year level on the presence of police, courts, prosecutors or other government staff (these are at least captured in part by the measure in column 2). Contrary to improving state capacity, other results available upon request suggest that the ceasefire led to more extortion, lower coverage of garbage collection, and no improvement in the provision of other public goods or participation in elections at any level. However, all of these have important data caveats.}

\begin{figure}
\caption{State Capacity Outcomes in FARC vs. ELN Municipalities -- Extensive Margin, Events in Over 60\% of Years}
\label{dyn_stateCap_ext}
\centering
\begin{subfigure}[b]{0.48\textwidth}
\includegraphics[width=\textwidth]{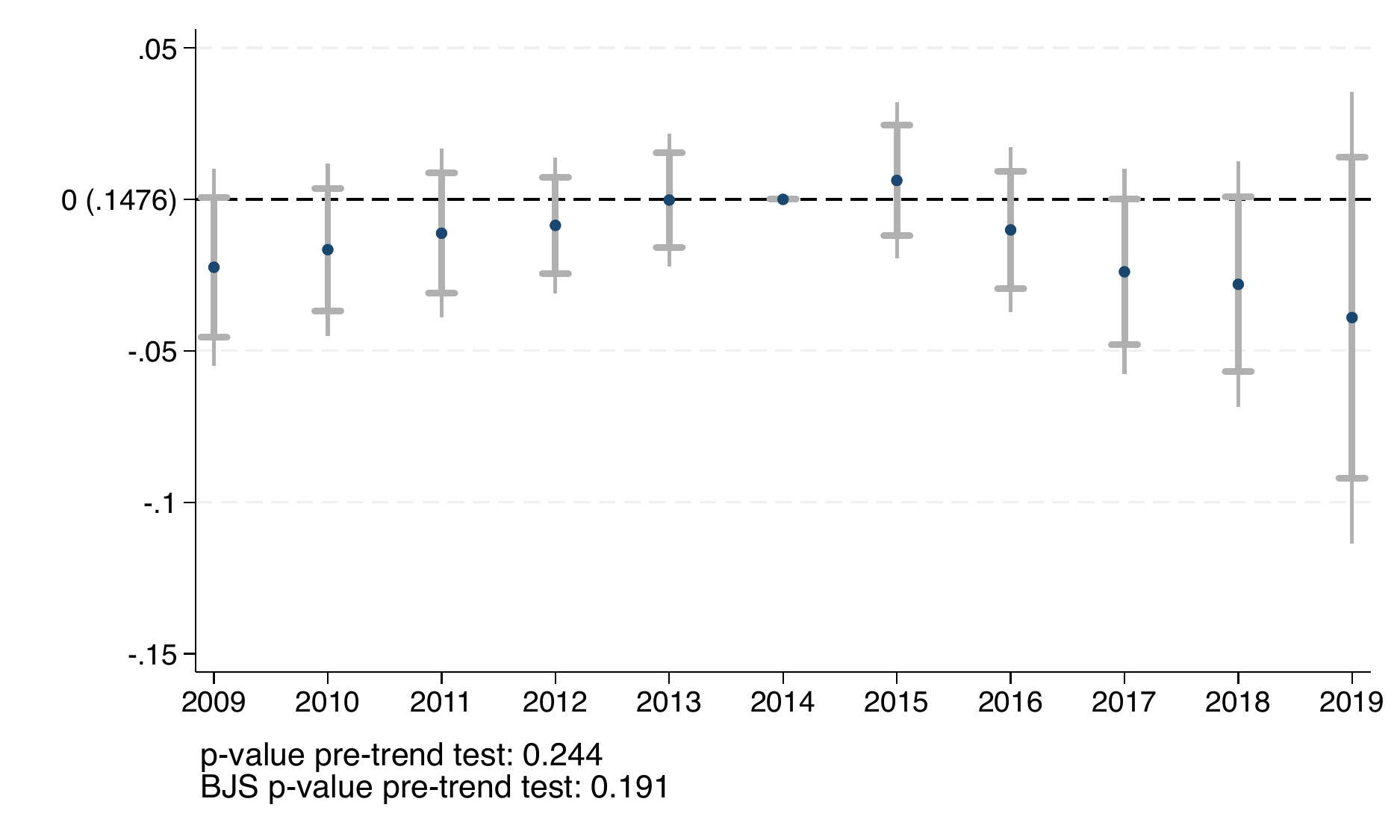}
\subcaption{Tax Revenue per Capita}
\end{subfigure}
\begin{subfigure}[b]{0.48\textwidth}
\includegraphics[width=\textwidth]{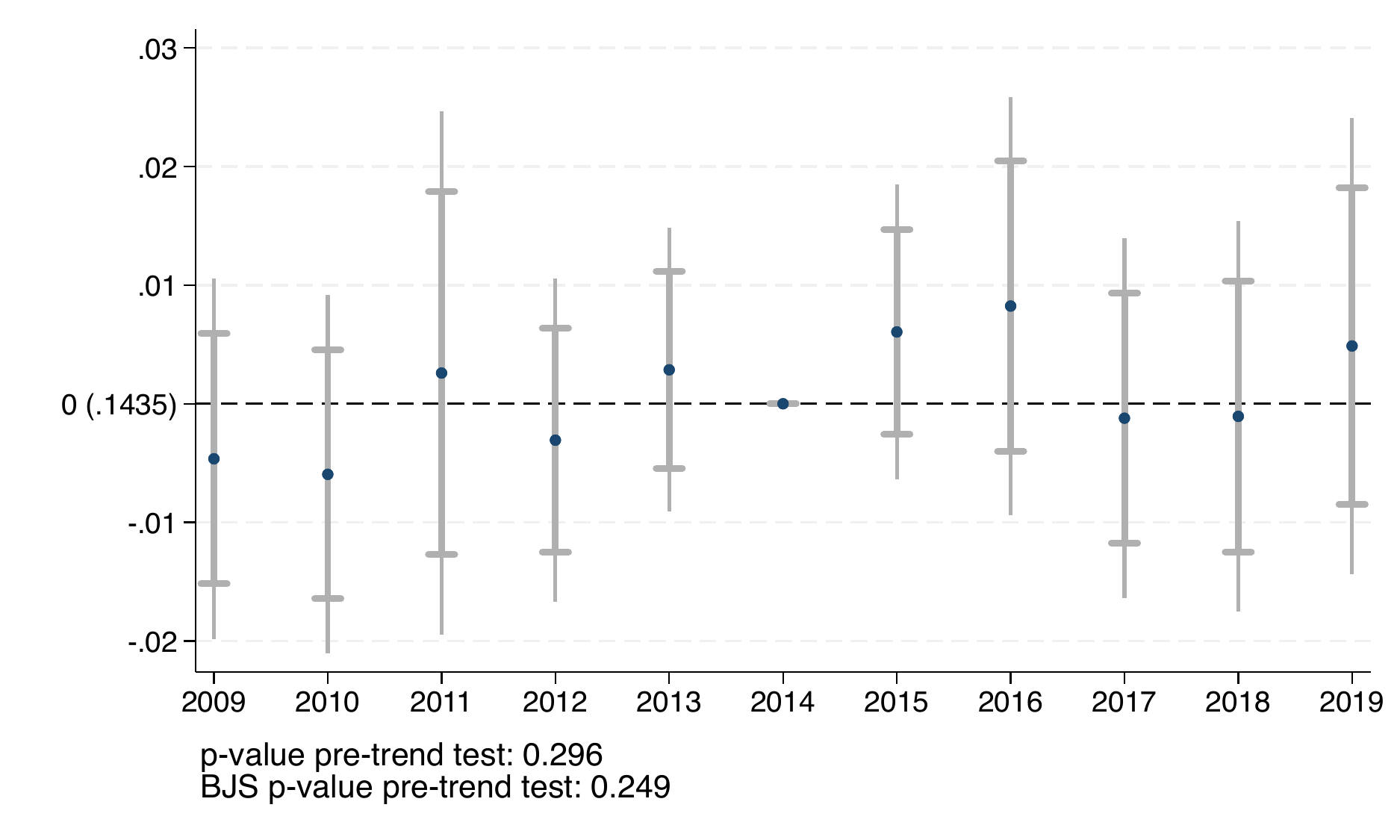}
\subcaption{Operational Costs (pc)}
\end{subfigure}
\begin{subfigure}[b]{0.48\textwidth}
\includegraphics[width=\textwidth]{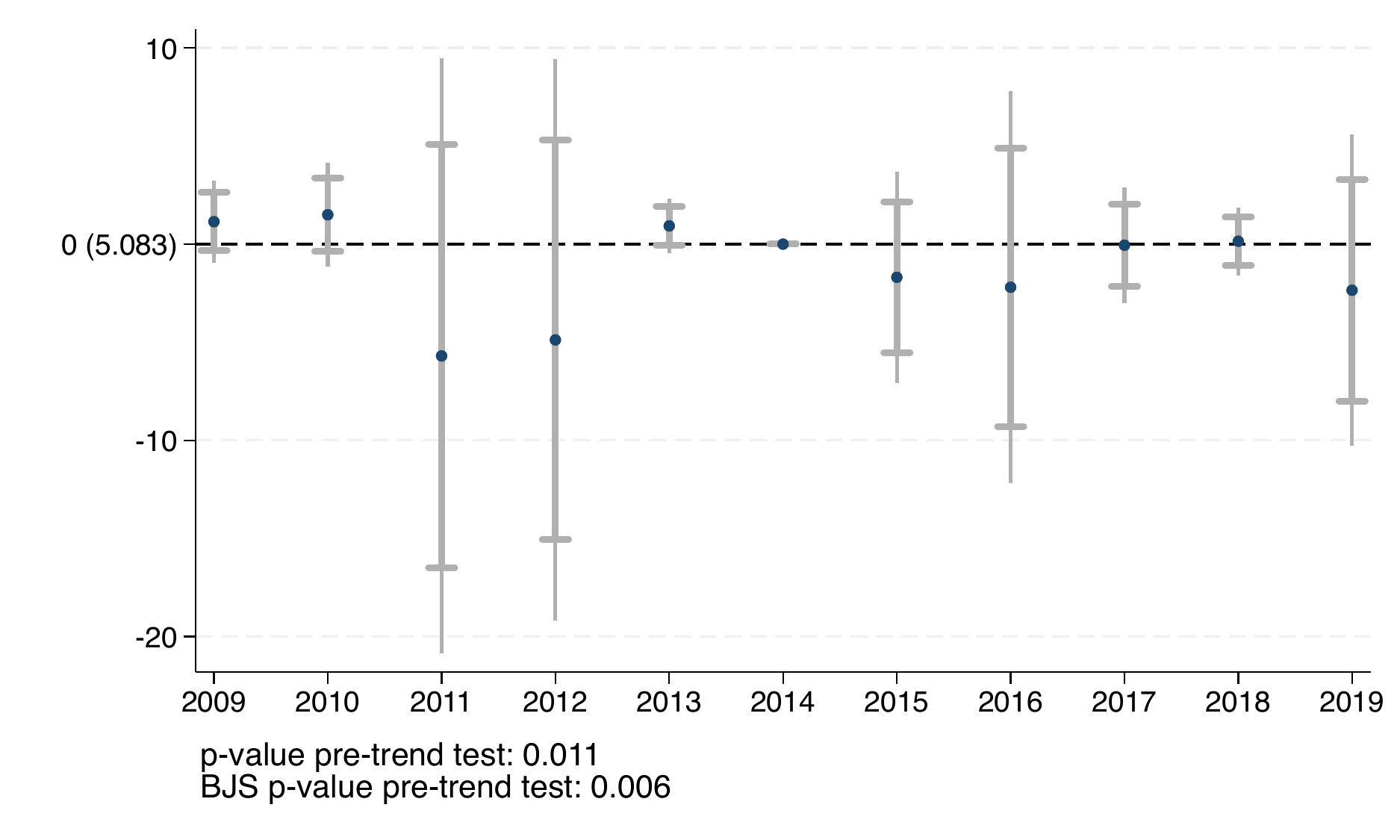}
\subcaption{Ratio Gov. Trans. to Revenue}
\end{subfigure}
\begin{subfigure}[b]{0.48\textwidth}
\includegraphics[width=\textwidth]{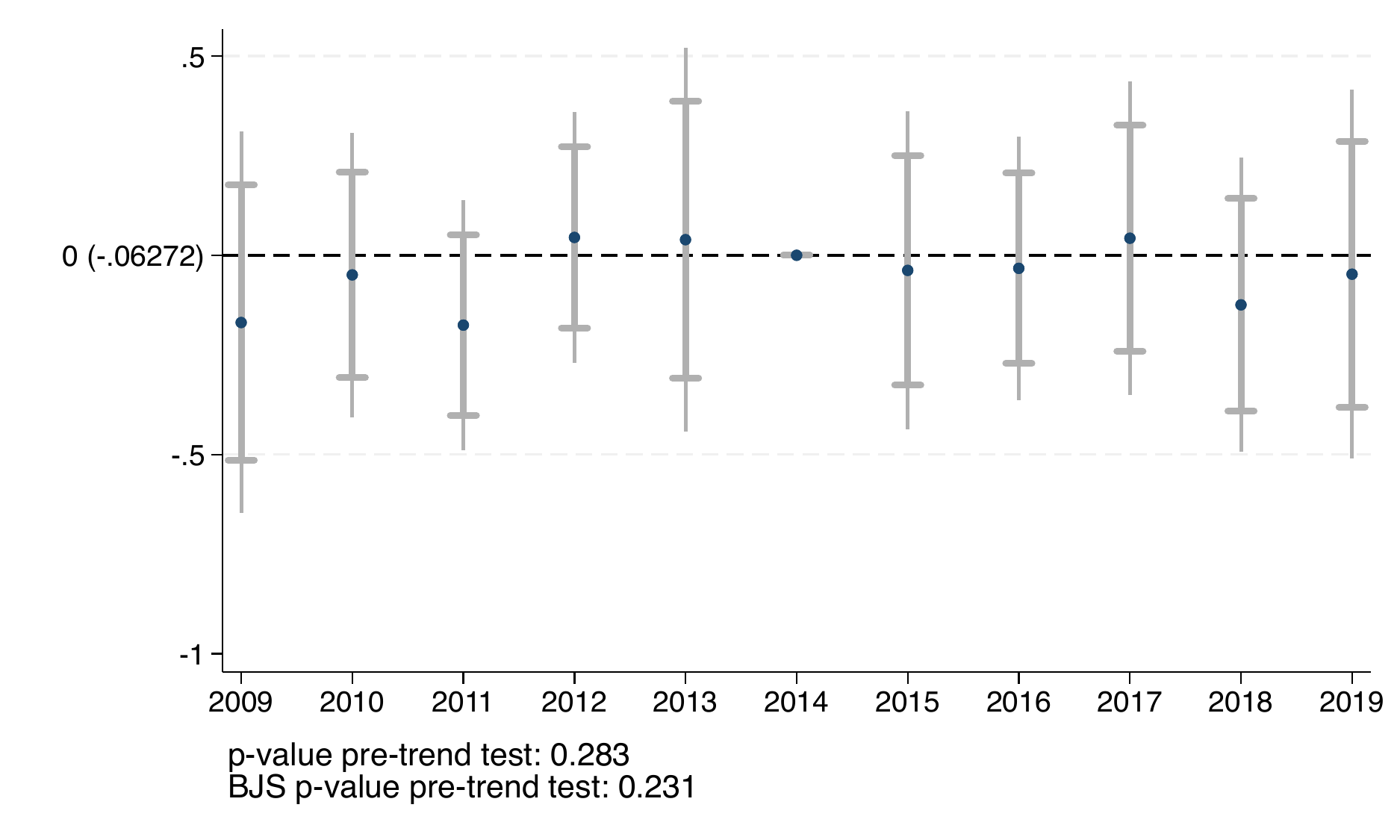}
\subcaption{Fiscal Performance}
\end{subfigure}
\begin{subfigure}[b]{0.48\textwidth}
\includegraphics[width=\textwidth]{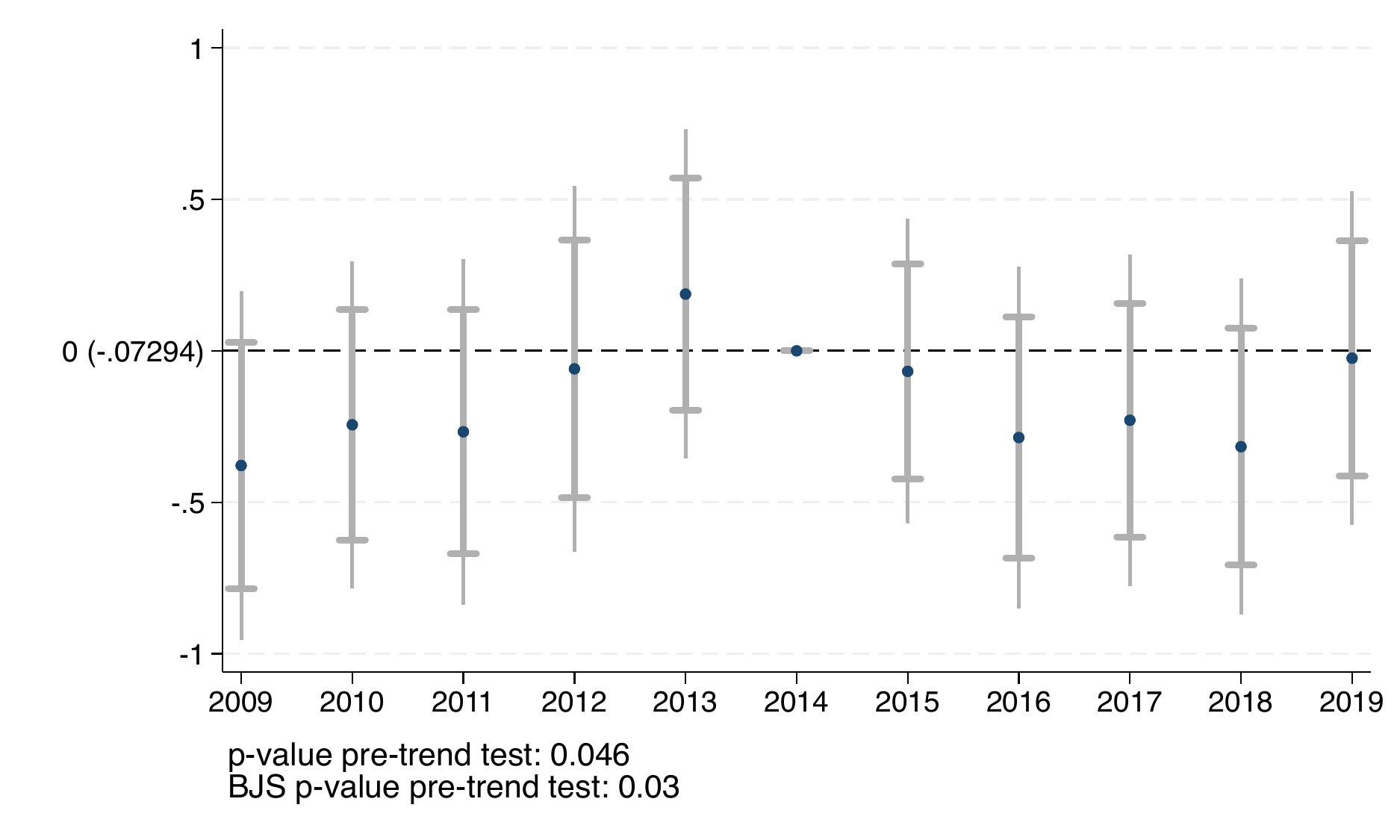}
\subcaption{Administrative Performance}
\end{subfigure}
\begin{subfigure}[b]{0.48\textwidth}
\includegraphics[width=\textwidth]{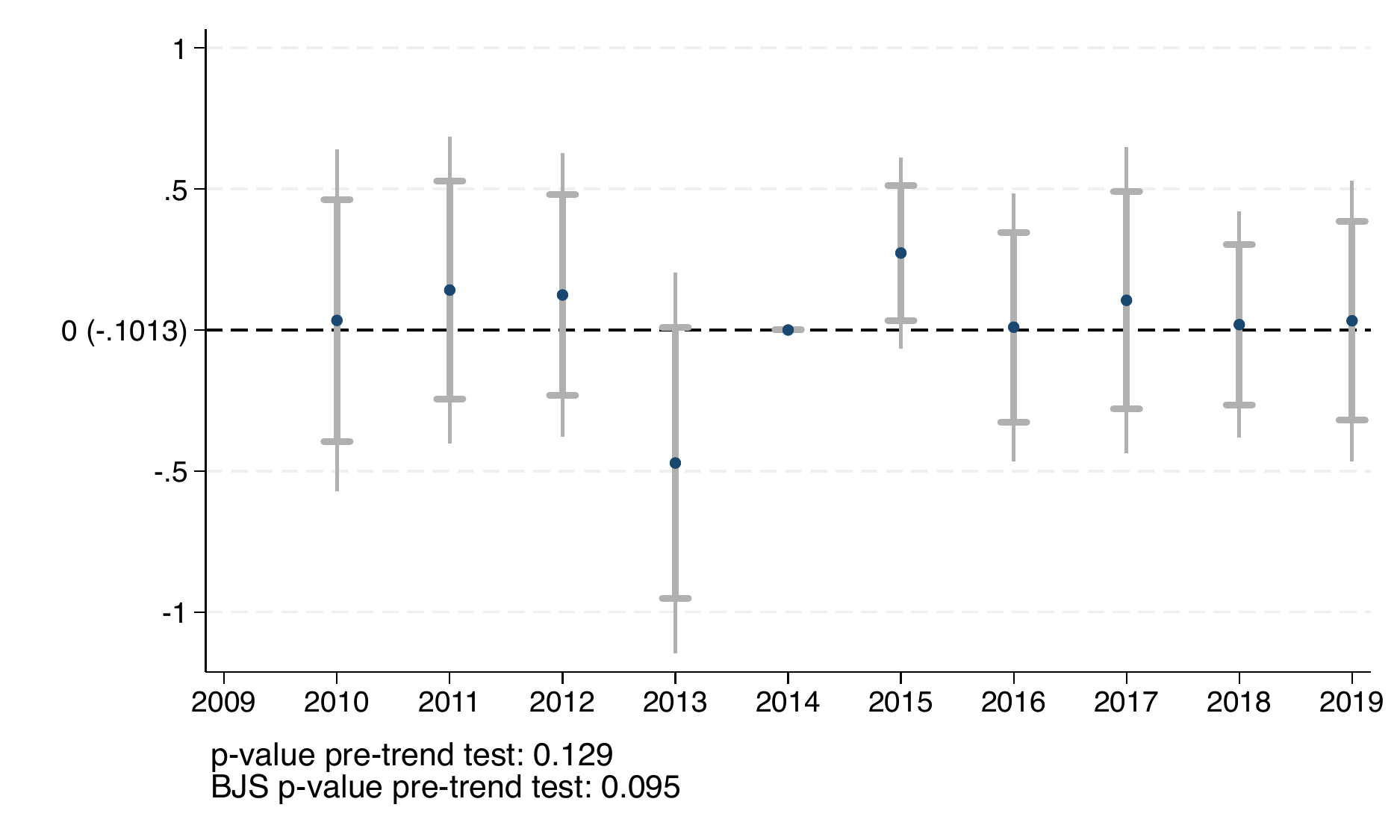}
\subcaption{Information Openness}
\end{subfigure}
\begin{subfigure}[b]{0.48\textwidth}
\includegraphics[width=\textwidth]{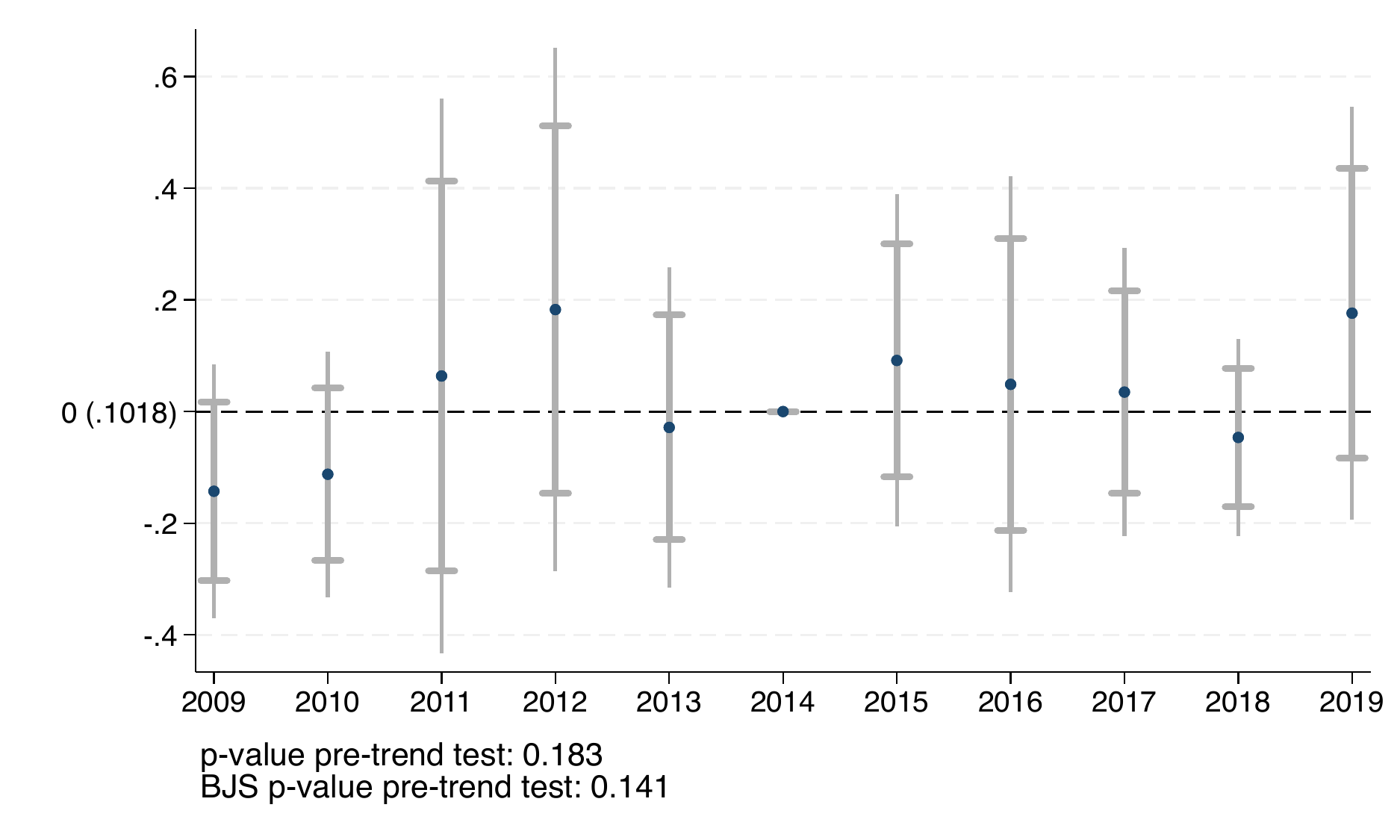}
\subcaption{Anderson Index}
\end{subfigure}
\justifying 

\footnotesize{\noindent \textbf{Notes:}  Event study plots from estimating Equation \eqref{eq_es} of the state capacity measures, including 95\% confidence intervals (based on standard errors clustered at the municipality level). The panels show results using tax revenue, operational costs, the ratio of transfers from the central government to total municipality revenue (excluding transfers from the central government), an indicator of fiscal performance, an indicator of the municipality's overall administrative performance (both created by the National Department of Planning, DNP), and an indicator of rule compliance created by the Office of the Inspector General, respectively. The last panel shows the results using a summary index of all these variables following \citet{anderson2008multiple}.}

\end{figure}

Event-study plots are displayed in Figure \ref{dyn_stateCap_ext}. The main takeaway is that there is no apparent upward trend in these measures post-ceasefire, which would be expected if improvements took some years to materialise. Pre-intervention trends also look parallel, with the sup-t confidence bands covering the zero in all figures and no individual coefficient being significant at the 5\% level across the figures.

\subsection{Evidence 3: State Entrance is Associated with Economic Improvements}

One of the main difficulties in assessing whether areas in which the state entered experienced economic improvements is that, at least until 2020, there was a general lack of state entry/presence post-ceasefire \citep[see, e.g.][]{isacson2021long}. One possible proxy for state entry could be participation in \textit{Territorially Focused Development Programs} (PDET in Spanish). This was a key program designed by the government as part of the agreement's Integral Rural Reform. PDETs consist of ``plans, developed in consultation with local leaders, to address long-neglected governance and development priorities'' \citep{isacson2021long}. Sixteen areas encompassing 170 municipalities were identified across the country. Thus, these are areas identified by the government in which discussions with local leaders were conducted to address the needs of their communities, arguably the closest it comes to the state establishing a presence. However, the program's implementation has been extremely slow, with little to show in terms of progress. By August 2018, there were plans for only 2 of the 16 regions. Only around 22\% of the funding that should have been spent by 2021 had been spent. 

I evaluate whether FARC municipalities that benefited from the PDET before 2020 experienced economic improvements compared to those that did not. Given that most municipalities classified as PDET had not seen any state entrance by 2020 due to logistical issues, I focus instead on those classified as PDET that had an economic project finished before 2020.\footnote{Data on PDET projects and investments come from the Agency for Territorial Renovation (ART), accessed from \href{https://centralpdet.renovacionterritorio.gov.co/inversion-nacional/}{\textit{here}} on March 22, 2022. The data include the state of the project (planning, execution, finished), the year of completion, and the project's sector. I categorise economic projects as those in the agricultural, commercial, industrial, mining, jobs, transport, and housing sectors, while non-economic projects are those in sectors like education, sports and culture.} Table \ref{pdetDescriptives} provides an overview of all PDET projects and those finished by 2019, divided by the sector and whether they are economic-related or not. Economic projects have a mean (median) value of 2B COP (213M COP) and tend to be much larger/expensive than non-economic ones, which have a mean (median) value of 488M COP (67M COP). Almost 40\% of economic projects relate to transport and target road improvements to connect conflict-affected municipalities to the regional economy, while the largest projects relate to energy and mines and target electricity provision and connectivity, key for economic activity. Thus, they capture meaningful, visible investments by the state in these municipalities.

\begin{figure}[h!]
\begin{center}
\caption{Heterogeneity by State Entry}
\label{tripleInter}
\includegraphics[width=0.7\textwidth]{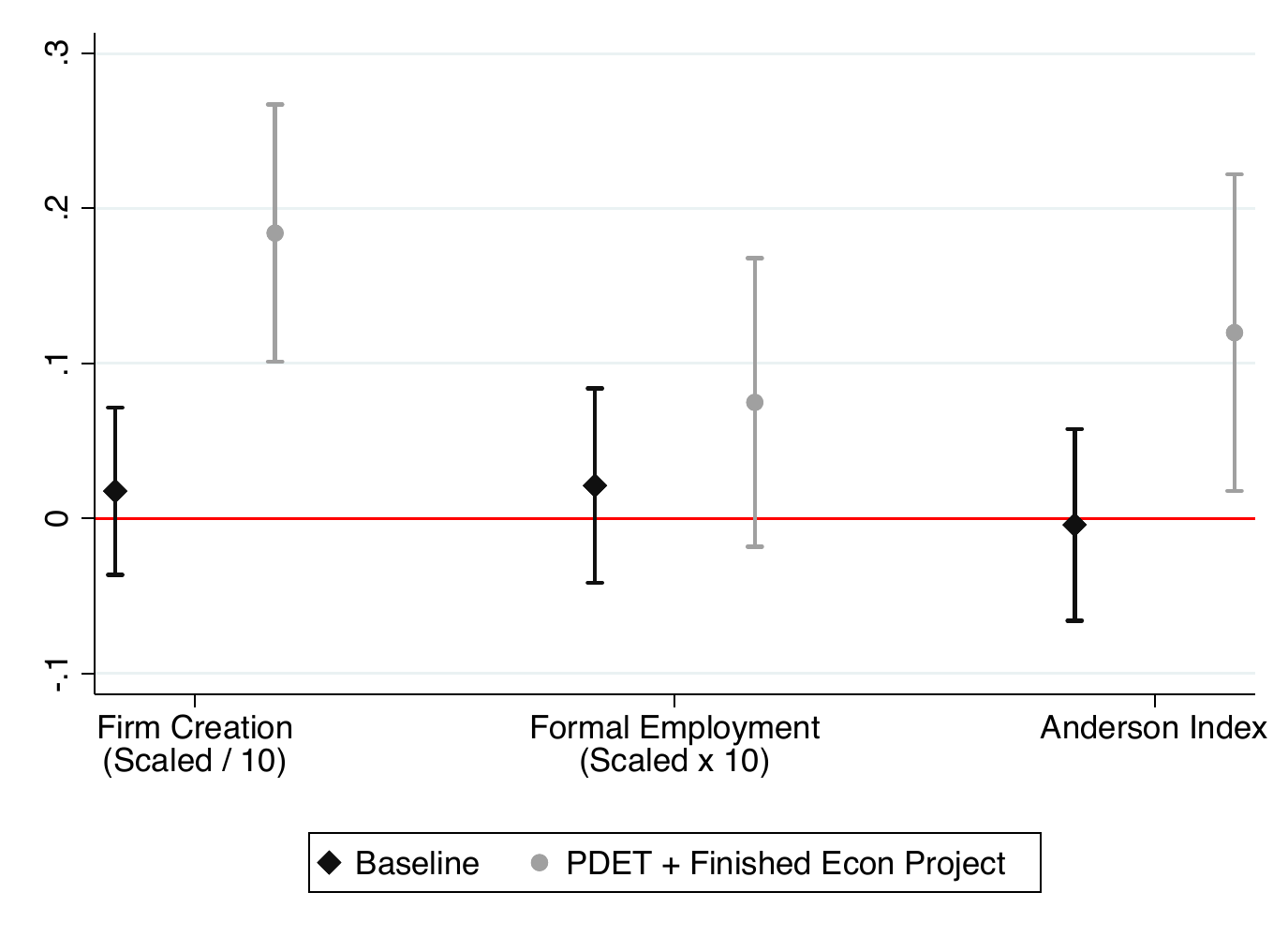}
\end{center}
\justifying 
\footnotesize{\textbf{Notes:}  Estimates of $\gamma$ and their 90\% confidence intervals from estimating Equation \eqref{eq_het}. The black coefficients are the baseline estimates for FARC municipalities, while the grey coefficients are the estimates of the triple interaction term, $\hat{\gamma}$.}
\end{figure}

I estimate Equation \eqref{eq_het}, where the heterogeneity variable is a dummy that equals one if the municipality was denoted as PDET and had an economic project finished before 2020 and zero otherwise.\footnote{Another test of this hypothesis would be to check whether municipalities with higher initial levels of state capacity experienced improvements in economic indicators post-ceasefire. However, baseline state capacity levels are much lower in FARC and ELN municipalities compared to the rest of the country (as seen in Table \ref{SC_SumStats_CEDEExt_p60}), and there is little variation in these measures in the sample.} Figure \ref{tripleInter} shows in black the baseline estimates for three economic outcomes (firm creation, formal employment, and the Anderson Index of the economic variables) and in grey the estimates of the triple interaction term between post-ceasefire, FARC presence, and having finished an economic project as part of the PDET program. There is a clear trend for the coefficient on the triple interaction to be larger than the baseline one and marginally significant. While the effect is small, this is probably because the PDET program started relatively late (2017), and its implementation has been very messy. Note that given the non-random nature of the areas in which the state executed PDET programs, these results are only suggestive (but not necessarily causal) of the presence of the state being a necessary condition for an economic peace dividend and need to be interpreted with caution.

\subsection{Alternative Mechanisms}

While the results so far have focused on state capacity as being the mechanism driving the results, there could be alternative mechanisms that explain why the reduction in violence did not translate into economic improvements in FARC areas. In Appendix \ref{app_alt_mechanisms}, I present evidence suggesting that several alternative mechanisms are unlikely to be causing the results. More specifically, I show that i) FARC municipalities did not produce more coca than ELN municipalities, nor were they disproportionately targeted by the government's eradication program; ii) FARC municipalities did not receive more Venezuelan migrants (which might shock these small labour markets) than ELN municipalities; iii) small agricultural producers in FARC municipalities started receiving more credit after the signature of the peace agreement, suggesting that credit constraints do not explain the results; iv) there was no shift from productive activities to education in FARC municipalities; v) claims for land restitution have evolved similarly in FARC and ELN municipalities; and vi) residents of FARC municipalities were supportive of the peace agreement, believed that it would benefit them (both economically and in terms of security) and were knowledgeable about the content of the agreement. 

Overall, the results in this Section are supportive of state capacity being a, if not the, potential mechanism to explain the results in the previous Section. Municipalities that have historically had FARC/ELN presence have much lower levels of state capacity than the rest of the country to begin with, probably as a consequence of the long-running conflict. Moreover, contrary to what was emphasised during the peace negotiations, the central government does not seem to have entered former FARC municipalities to build a state presence and the capacity of local governments, leading to no improvement in state capacity indicators in former FARC municipalities after the start of the ceasefire. This lack of state capacity could have hindered these municipalities from reaping economic benefits from the large reduction in violence observed after the start of the ceasefire. Indeed, suggestive evidence shows that the few municipalities with at least a weak state entrance seem to experience economic improvements. This is consistent with the theory in \citet{besley2010state}, in which low levels of state capacity can lead to self-perpetuating poverty traps.

\section{Robustness Checks}
\label{rob_checks}

\begin{figure}[htp]
\caption{Robustness to Alt. Thresholds of Presence Measures -- Extensive Margin}
\label{rob_thresholds_ext}
\centering
\begin{subfigure}[b]{0.65\textwidth}
\includegraphics[width=\textwidth]{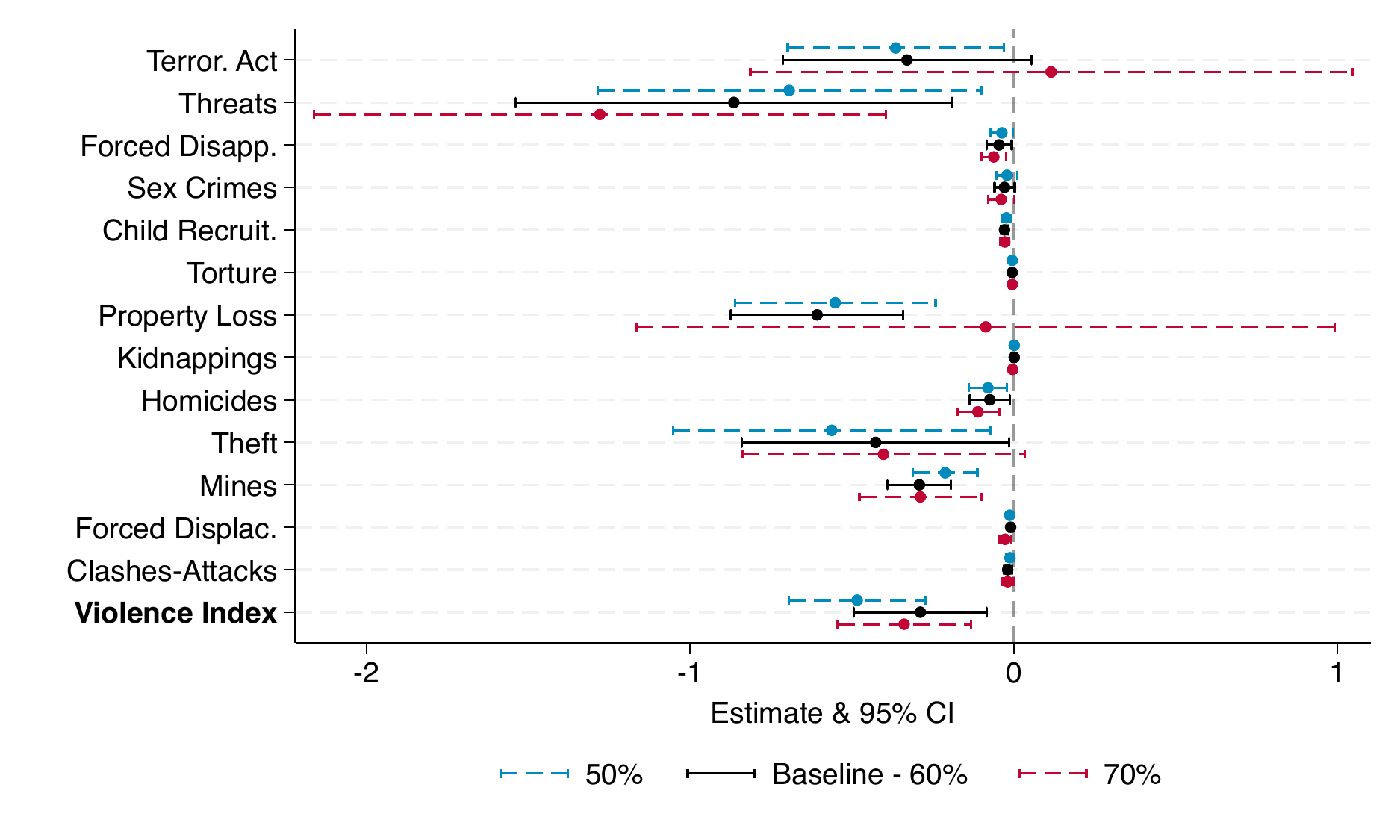}
\subcaption{Violence Measures}
\end{subfigure}
\begin{subfigure}[b]{0.65\textwidth}
\includegraphics[width=\textwidth]{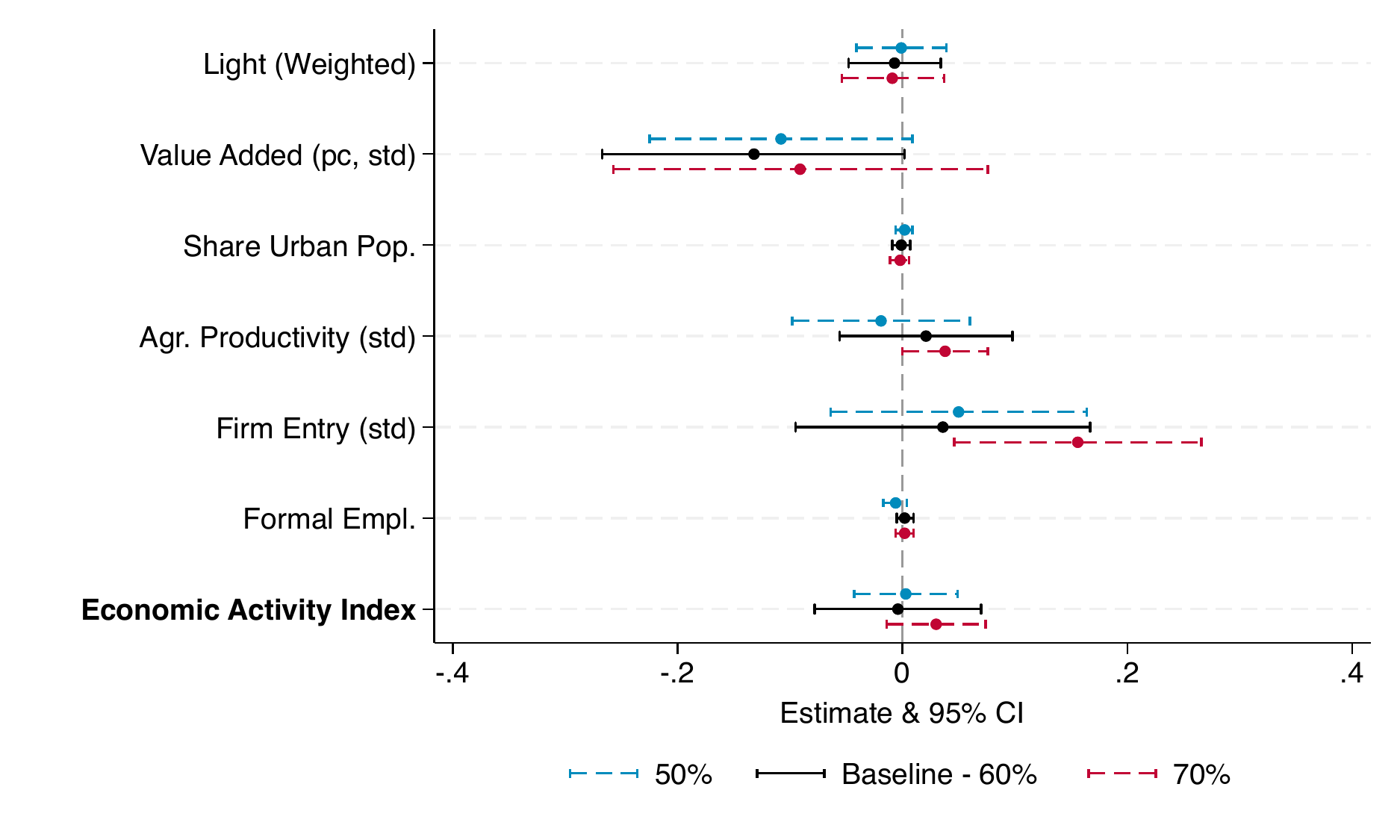}
\subcaption{Economic Indicators}
\end{subfigure}
\begin{subfigure}[b]{0.65\textwidth}
\includegraphics[width=\textwidth]{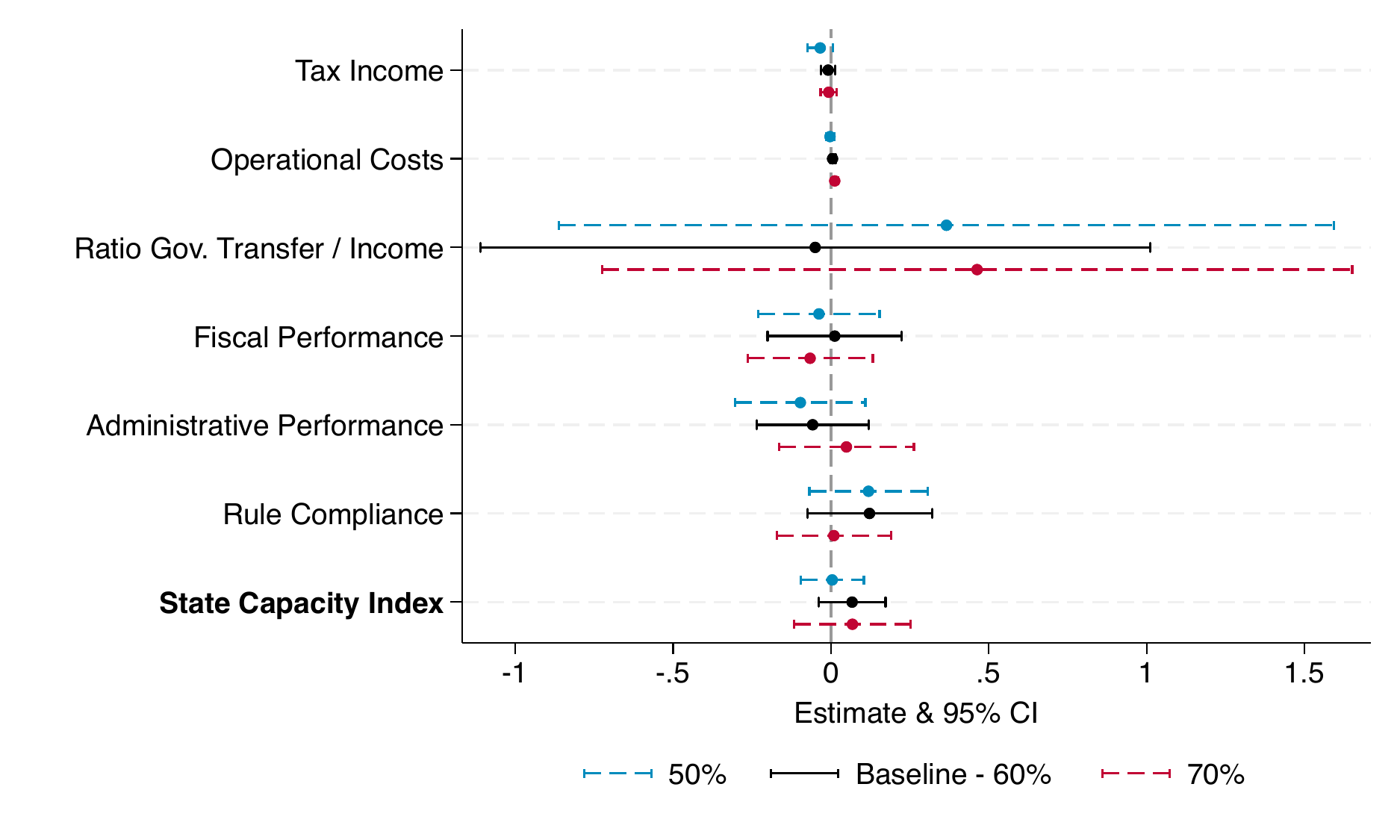}
\subcaption{State Capacity Measures}
\end{subfigure}

\justifying
\noindent\footnotesize{\textbf{Notes:} In Panel A, all variables are in 1000s of inhabitants, except for the migration ones (forced displaced and forced migration), which are measured in per capita terms. In Panel B, the measures of value added per capita, agricultural productivity and firm entry have been standardised for comparability.}
\end{figure}

In this Section, I present a multitude of robustness checks to the results presented in Sections \ref{results} and \ref{sec_sc}. First, the baseline extensive margin measure defines a group as being present in a given municipality if the municipality experiences events from the insurgent group in at least 60\% of the years. I test the robustness by using two alternative cutoffs, 50\% and 70\% of the years, to define affected municipalities. Figure \ref{rob_thresholds_ext} presents the results of varying the definition cutoff for the three sets of variables, with the top panel using the violence variables. Regardless of the thresholds used to define the presence measure, the results indicate a consistent reduction in violence in former FARC municipalities relative to ELN municipalities. Results for the economic indicators are presented in the middle figure, which show that the estimates remain consistent when using different thresholds. Finally, the bottom figure confirms that the state capacity results also remain qualitatively unchanged when using these different thresholds.

Second, while the baseline results use the extensive margin measure of presence of insurgent groups, the results are robust to using instead the intensive margin measure described in Section \ref{cont_data}, which captures the municipalities in which a particular guerrilla group was the most violent (in per capita terms). The baseline intensive margin measure classifies a municipality as under the control of a given armed group if it belongs to the top 20\% of municipalities that had the most events per capita by that armed group on average between 1996 and 2008. Tables \ref{DID_final_Vio_CEDE_Int_p80}, \ref{DID_Econ_CEDE_Int_p80}, \ref{SC_SumStats_CEDEInt_p80}, and \ref{DID_SC_CEDE_Int_p80} use the intensive margin measure and mimic Tables \ref{DID_final_Vio_CEDE_Ext_p60}, \ref{DID_Econ_CEDE_Ext_p60}, \ref{SC_SumStats_CEDEExt_p60}, and \ref{DID_SC_CEDE_Ext_p60}. Table \ref{DID_final_Vio_CEDE_Int_p80} shows the results using the different violence measures. As with the extensive margin measure, it shows significant, large reductions in violence across multiple violence indicators. Figure \ref{viol_es_int} shows that the improvement in violence seems to have happened gradually and persisted at least until 2019, with the joint test of pre-treatment coefficients being insignificant. Moreover, the results in Table \ref{DID_Econ_CEDE_Int_p80} also show that this large reduction in violence did not translate into improvements in economic indicators, as almost all coefficients show precisely-estimated null effects. Figure \ref{dyn_econ_int} shows no improvement over time in these economic indicators. Turning to the mechanisms, Tables \ref{SC_SumStats_CEDEInt_p80} and \ref{DID_SC_CEDE_Int_p80} show that FARC and ELN municipalities have much lower initial levels of state capacity (although the two groups are less similar than when using the extensive margin measure) and that FARC municipalities did not see improvements in these after the start of the ceasefire relative to ELN municipalities, even when looking at the event studies in Figure \ref{dyn_stateCap_int}. To show that the results do not depend on the specific cutoff used to create the intensive margin measure, Figure \ref{rob_thresholds_int} shows that the intensive margin results are robust to setting the inclusion threshold to the top 30\% or 10\% municipalities with the most violence by FARC or ELN.\footnote{Other papers studying the peace agreement in Colombia have used DiD strategies comparing some definition of FARC municipalities against the rest of the country \citep[e.g.,][]{prem2021human, guerra2021peace, bernal2022peaceful, de2021forgone}. While it is difficult to match the strategies across those papers exactly due to differences in the data used to create the different groups, controls and weights used, and how exactly FARC municipalities are defined, Figure \ref{robNonFARCControl} shows the results of the main indices using different variations of what others have done. Based on Ministry of Defence (MoD) data, it shows the baseline results in black, which compare FARC and ELN municipalities. In red results using MoD data comparing FARC municipalities against the rest of the country. The other two use CINEP data between 2011 and 2014 (when peace negotiations between the FARC and the government were ongoing) to compare the FARC and the rest of the country municipalities. In blue, FARC municipalities are those with at least one FARC event in that timeframe. In green, FARC municipalities are those in the top 75\% of FARC events. The remaining municipalities are the control group. The results are similar: Violence decreases significantly and sizeably, while there is either no change or a slight deterioration in economic activity and state capacity. Two things are worth highlighting here. First, CINEP has few FARC events in the 2011-2014 period: Out of the 142 municipalities with at least a FARC event, 62 (44\%) have precisely one event, while 100 (70\%) have at most three. Whether these result from mismeasurement, fortuitous events by the FARC, or something else is difficult to say, but it is unlikely that this captures prolonged exposure to the FARC, as the baseline measures do. Second, concerns about SUTVA violations when looking only at ELN municipalities would naturally translate when using the whole country as the control group, since the latter will also include ELN municipalities.}


While the historical evidence suggests that municipalities under ELN's control could be a valid counterfactual for those under FARC's control, the majority of tests for joint significance of pre-intervention coefficients presented here fail to reject the null, the sup-t confidence bands recommended by \citet{freyaldenhoven2021visualization} cover zero, and no apparent pre-trend are observed, I present results using the synthetic diff-in-diff estimator developed by \citet{arkhangelsky2021synthetic}. This estimator tries to find a group of ``donor'' municipalities that mimics the treated ones in the pre-intervention period, with the intuition that if their pre-trends were parallel, then their post-treatment counterfactual evolution would also have been in parallel.\footnote{The authors recommend restricting the donor pool somewhat to avoid having very different units as potential donors. While there is little guidance on how to do this exactly, I take a conservative approach and reject only a few municipalities that are very different to the ones with historical FARC presence. Briefly, for some variables of interest, I pick the most extreme observations (the max or the min) for the treatment group in the pre-intervention period and drop those municipalities with even more extreme values. I select six variables to restrict the donor pool: total population, distance to the department's capital, rurality index, fiscal performance index, overall performance index and child mortality rate (below 1 year). These variables should help identify municipalities that are very different to historical FARC municipalities along economic and social measures. For example, I check what is the largest population in my treated groups. Then, ignore those municipalities with a minimum population in the pre-intervention period larger than the maximum population in the treated groups. } Appendix Tables \ref{SDnD_CEDEExt_p60} and \ref{SDnD_CEDEInt_p80} show the results using this method and the extensive and intensive margin measures of insurgent group presence, respectively. For brevity, I only present the results for violence using the Anderson Index and a different index created following the procedure suggested by \citet{kling2007experimental}. The results echo the results in the previous Section using historical ELN municipalities as the control group: there is a substantial reduction in violence in treated municipalities, not followed by increases in economic indicators or improvements in state capacity levels. Appendix Figures \ref{synt_Viol}, \ref{synt_Econ} and \ref{synt_SC} show the corresponding diff-in-diff plots.

The results so far have used \citet{anderson2008multiple}'s approach to summarising different variables into a single index. Results using the summary measure proposed by \citet{kling2007experimental} (KLK Index) are very similar to the ones using the Anderson Index (see Table \ref{robCheck_CEDEExt_p60}), suggesting that the construction of a specific index variable does not drive the results. As described in Appendix \ref{data_detail}, the two indices are created similarly. Nonetheless, the Anderson Index has several advantages over the KLK Index, and that's why the baseline results use this index. First, the Anderson Index assigns more weight to variables that are less correlated to other component variables, as these might carry more relevant information not captured by the other component variables. Intuitively, uncorrelated indicators represent ``new''  information and thus receive more weight. Second, it also assigns less value to indicators with missing observations rather than imputing missing observations using the mean of the treatment assignment group, which might artificially decrease the index's variance. 

Another cause of concern is potential spillovers from the treated to the control municipalities, which would lead to a violation of the SUTVA and thus bias the coefficients. For example, the start of the ceasefire could have affected the level of violence in ELN municipalities if the government started focusing its efforts on ELN municipalities. Assuming that spillovers are local (meaning that spillovers only happen within a certain distance $d$ of the treated units) and under a slightly modified parallel trends assumption requiring that the counterfactual trend among control units outside the spillover distance $d$ is the same as that of all treated units, \citet{butts2021difference} shows that one can estimate two different effects: the ``total'' effect, which captures the effect of going from a world with no treatment to a world with the realised level of treatment (including spillovers on treated municipalities), and the spillovers on control municipalities.

Table \ref{spatSpill_CEDEExt_p60} shows the results of estimating the total effect (odd columns) and the spillovers on the control units (even columns) for the three Anderson Indices of each category of variables. I do this for different values of $d$, based on the distribution of distances between municipalities and their department's capital. Specifically, I restrict spillovers from the closest treated municipality up to the \nth{35}, \nth{50} or \nth{65} percentile of the distribution. The results are similar to the baseline ones, with a general lack of spatial spillovers. The total effects when taking into account spillovers to the control units are negative and significant for the violence index, and insignificant for the state capacity index, regardless of the distance considered. They also show no significant spillovers on nearby control municipalities. For the economic index, both the total effect and the spillovers are insignificant at the \nth{50} and \nth{75} percentiles, although they are negative and marginally significant at the \nth{35} percentile. Overall, these results suggest that violations to SUTVA are unlikely to be a cause of concern in this setting and seem to rule out the most likely avenue in which SUTVA violations could manifest themselves, which is through spillovers to nearby units.

\begin{table}[h!] \centering
\newcolumntype{R}{>{\raggedleft\arraybackslash}X}
\newcolumntype{L}{>{\raggedright\arraybackslash}X}
\newcolumntype{C}{>{\centering\arraybackslash}X}

\caption{Spatial Spillovers: Extensive Margin, Events in Over 60\% of Years}
\label{spatSpill_CEDEExt_p60}
\begin{adjustbox}{max width=\linewidth, max height=0.6\textwidth}\begin{tabular}{lcccccc}

\toprule
& \multicolumn{2}{c}{\textbf{\nth{35} Percentile}} & \multicolumn{2}{c}{\textbf{\nth{50} Percentile}} & \multicolumn{2}{c}{\textbf{\nth{65} Percentile}} \tabularnewline \cmidrule(lr){2-3} \cmidrule(lr){4-5} \cmidrule(lr){6-7} \vspace{-4mm} \tabularnewline  & Total & Spillover & Total & Spillover & Total & Spillover \tabularnewline 
{}&{(1)}&{(2)}&{(3)}&{(4)}&{(5)}&{(6)} \tabularnewline
\midrule \addlinespace[\belowrulesep]
\midrule 1. Violence Index&--0.466***&--0.203&--0.341*&--0.056&--0.476***&--0.198 \tabularnewline
&(0.146)&(0.130)&(0.191)&(0.181)&(0.169)&(0.160) \tabularnewline
&&&&&& \tabularnewline
2. Economic Index&--0.096*&--0.105**&--0.108&--0.112&--0.127&--0.130 \tabularnewline
&(0.055)&(0.053)&(0.084)&(0.083)&(0.107)&(0.106) \tabularnewline
&&&&&& \tabularnewline
3. State Capacity Index&--0.025&--0.105&0.064&--0.003&0.066&--0.001 \tabularnewline
&(0.077)&(0.066)&(0.087)&(0.077)&(0.106)&(0.097) \tabularnewline
\bottomrule \addlinespace[\belowrulesep]

\end{tabular} \end{adjustbox}
\\ \parbox{\linewidth}{\scriptsize \justifying \noindent \(Notes \): Considers only years 2009 to 2019. Defines the post period as 2015. All panels use indices created following Anderson (2008). The first panel uses the index based on the different violence variables. The second panel uses the index based on weighted nighttime light intensity, value added per capita (from DANE), share of urban population, agricultural productivity, formal employment and firm entry measures. The third panel uses the index based on measures of state capacity. The  columns show the total effect of treatment, the  columns show the spillovers on control municipalities within a distance \(d\) of the closest treated municipality, where \(d\) corresponds to the \nth{35} (51kms, columns 1 and 2), \nth{50} (67.8kms, columns 3 and 4), or the \nth{65} (87.3kms, columns 5 and 6) percentile of the distribution of the distances between municipalities and their department's capital. Standard errors are estimated following Conley (1999), allowing for spatial correlation up to \(d\)kms away from a municipality's centroid.}
\end{table}

A recent literature has developed tools to measure the extent of violations of pre-trends and how these would affect the results \citep{rambachan2019honest,rothpre}. Appendix Table \ref{Roth_baseline_slope} shows the results of implementing the test suggested by \citet{rothpre}. Panel A shows the results for the extensive margin measure of presence, while Panel B shows the results for the intensive margin measure of presence. The first column of each panel shows the slope of a linear violation of the parallel trend assumption that would be detected with a power of 80\%. This slope is small in magnitude for all measures, suggesting that the tests are sufficiently powered to detect a small violation of the parallel trend assumption. For example, the slope in Panel A for the Anderson Index of violence measures says that a pre-trend with a size of 0.079 SDs is likely to generate at least one statistically significant pre-period event study coefficient. In columns 2 and 3, I calculate the unconditional and conditional (on surviving the pre-test) biases that would be caused by a trend of that size for the average of the post-treatment coefficients.\footnote{To calculate the unconditional bias, one simply calculates the sum of the linear interpolation of the slope over the post-treatment periods. I follow \citet{rothpre}'s Proposition 3.1 to estimate the conditional bias. For details on implementing each of these, see \citet{miller2021medicaid}.} For example, focusing on the second column of Panel A, an unconditional bias of 0.199 SDs means that if a trend of a size of up to 0.066 SDs is indeed present (although not detectable), it would generate on average across all post-treatment periods a bias of at most 0.199 SDs, around 65\% the size of the mean post-treatment coefficients. Appendix Figure \ref{roth_pretrend_fig} shows the event study graphs for the Anderson Index of the violence measures, with the linear violation of the parallel trend assumption that would be detected at 80\% power. 

The baseline results consider the years between 2009 and 2019. These are probably the most critical years since most conflicts relapse quickly, and a small ``window of opportunity'' exists to make a peace agreement work. However, the results look similar if the timeframe is extended until 2021 (beyond that, too many variables are missing), as shown in Table \ref{timeframes_CEDEExt_p60}. Panel A presents estimates of the DiD Equation \ref{eq_did} and Panel B of the difference-in-discountinuity Equation \eqref{eq_diff_in_disc}. For each of these, the baseline set of results for 2009 to 2019 is shown for reference on top, and the corresponding coefficient for the extended timeframe from 2009 to 2021 is at the bottom. 

The estimates for the extended timeframe need to be interpreted with caution. While FARC dissident groups started emerging after 2017, they were relatively small until the end of 2019, when they began coordinating and consolidating \citep{pares2024}, mostly in former FARC areas (see Table \ref{DID_violSpill_CEDE_Ext_p60}). Thus, they are less of a concern when evaluating data until 2019. Since then, they have seen rapid growth, and the country has experienced a worsening of the security situation due to the dissidents and other criminal groups \citep[see, e.g.,][]{defensoria2024, icg2022, indepaz2022, insightcrime2024}. Thus, the extent to which ``peace'' persists in former FARC municipalities post-2019 is debatable. In line with this evidence, the coefficient on the violence index in column 1 of Table \ref{timeframes_CEDEExt_p60} shrinks towards 0 when including post-2019 years. The coefficient on the economic index remains insignificant and precisely-estimated if one ignores municipalities with any actions by FARC dissidents (column 8). Figure \ref{timeframes_CEDEExt_p60} shows the event study plots for the three main indices when extending the timeframe. The reversal in the violence evolution is apparent (especially in 2021), while the lack of upward trajectory in the economic and state capacity indices persists. While the extended results need to be taken with a grain of salt, even then, they are qualitatively the same as in the baseline analysis. Whether the lack of economic dividends played a role in the rise of FARC dissident groups remains an open question.

I present several additional robustness checks in Appendix \ref{app_addRobChecks}. First, I perform a ``permutation'' test for the violence result, randomly assigning municipalities to the treatment and control groups, finding a coefficient larger in magnitude in only 0.5\% of iterations than the baseline one. Second, I use an alternative inference method, an alternative summary index (the KLK Index), add pre-treatment controls, add municipality-specific time trends, and use alternative measures of FARC/ELN presence (each of these checks separately), with the baseline results robust to these alternative specifications/checks. Then, I show that the Anderson Index's results for the different groups of variables are robust to excluding each of the indices' component variables individually, showing that the results are not driven by the specific set of variables used to construct these indices. Moreover, I show that the lack of economic improvements in FARC municipalities is not due to the difference-in-difference estimates masking improvements in FARC and ELN municipalities with respect to the rest of the country but instead due to a lack of catching up in those areas. Another concern is that the FARC and the ELN, while sharing a similar history and modus operandi, did operate in different areas of the country and focused on different sources of income (for the FARC, mainly drug trade, while for the ELN, mostly smuggling and extorting oil companies). While some of this will be captured by the municipality and the year fixed effects, Table \ref{robCheckControls_CEDEExt_p60} shows that the baseline results (Panel A) remain qualitatively unchanged when including controls for geographic and demographic characteristics (Panel B), measures of connectivity (Panel C), various measures related to illegal activities performed by the FARC and ELN (Panel D), and all these together (Panel E), interacted with the post-ceasefire dummy. Finally, results also hold when using a continuous rather than discrete measure of FARC presence, as shown in Table \ref{contMeasAlt_CEDEExt_p60}. 

\section{Conclusion}
\label{conclusion}

The recent resurgence of armed conflicts globally has led to the frequent use of peace agreements to end conflicts. Yet, while the economic literature on the effects of war is abundant, there is comparatively little evaluating the impacts of \textit{peace} and the determinants of successful peace agreements. This study addresses these questions by evaluating the peace agreement in Colombia, signed between the Colombian government and the FARC, the largest insurgent group in the country at the time, ending what was then the lengthiest and one of the most violent conflicts in the world. 

Comparing municipalities that had historically had FARC presence and those that had had ELN presence (another similar but smaller guerrilla group) before and after a unilateral ceasefire by the FARC as part of the peace negotiations, the study finds three sets of results. First, several violence indicators, including forced displacement, combats, disappearances, and terrorist acts, among others, significantly and sizeably decreased after the ceasefire in FARC municipalities. This suggests that the ceasefire did lead to the desired reduction in violence. Second, despite this significant reduction in violence, affected municipalities did not experience improvements across different economic indicators, from agricultural production to nighttime light intensity. Importantly, these are precisely-estimated null effects. Even considering only conflict-affected municipalities that received a government program that provided substantial fiscal incentives for firms to operate in these areas, these do not experience any economic improvement. Third, I evaluate what might explain these two seemingly contradictory results. In line with qualitative evidence, my results suggest that the reason historically FARC municipalities could not reap the economic benefits from their new-found peace following the ceasefire is a lack of state capacity and presence, caused both by their low initial levels of state capacity probably from decades of conflict and the lack of state entry post-ceasefire. In areas where the state entered, even if only tentatively, there is \textit{suggestive} evidence of an economic improvement. These results align with the theoretical model of \citet{besley2010state}, which posits that low levels of state capacity can lead to self-perpetuating poverty traps. 

The disappointing lack of economic improvements in historically FARC areas does not mean the peace agreement was unsuccessful. The significant reduction in violence following the ceasefire is in itself a welcome and worthy development, and whether \textit{the country as a whole} benefitted economically from it remains a possibility. However, this contradicts the government's promises during the negotiations and signifies a missed opportunity to help long-neglected areas. More speculatively, the lack of state entry, capacity building, and economic prosperity following the ceasefire could explain why peace started to unravel throughout the country after six years of the ratification of the agreement. While the FARC transitioned to a political party as part of the agreement, amid accusations of drug trafficking by FARC leaders and problems in the implementation of different elements of the accord, it splintered in 2019, and a faction returned to arms, leading to the consolidation of FARC dissident groups. Recent reports have associated these factions with increasing violence levels in the country \citep{bohorquez2022,torrado2022,fip2022ni},\footnote{The increase in violence is not only associated with FARC dissidents but also with increased activity from other criminal groups \citep{defensoria2024, icg2022, pares2024, insightcrime2024}.} and concerns are growing that gains achieved by the agreement will be quickly lost. 

There are several avenues for future research in this area. First, it is important to understand whether a necessary condition for peace agreements to succeed is that they bring economic improvements in areas previously affected by the conflict. It is common for conflicts to start for economic reasons (lack of opportunities, rising inequality, an ignored countryside, etc), and agreements that fail to resolve these challenges are more likely than not to fail in time, as suggested by the results here in the case of Colombia. Governments have only a small ``window of opportunity'' to make these agreements work, and the short- and middle-term economic recovery post-conflict likely plays a key role in this. Second, as my results suggest that the capacity of local governments is a crucial factor to capitalise economically from peace, understanding what types of programs governments can implement to fill the power vacuum left after peace agreements quickly and to build state capacity in conflict-affected areas \citep[along the lines of][]{blair2022preventing} is crucial for the successful implementation of future peace agreements. Finally, in countries like Colombia, where multiple criminal organisations co-exist, evaluating how crime and conflict reorganise after the disbandment of a criminal organisation would help understand the dynamics of large criminal organisations. Do new criminal organisations emerge, or do already-existing organisations fight to take over the territory and businesses of the removed criminal organisation, or given that the state has one criminal organisation less to worry about, can it concentrate on existing ones, leading to their weakening? Understanding these general equilibrium effects is important to guide governments' post-peace military strategy.


\clearpage
\phantomsection

\bibliography{References}



\renewcommand{\thefigure}{A\arabic{figure}}
\setcounter{figure}{0}
\renewcommand{\thetable}{A\arabic{table}}
\setcounter{table}{0}

\clearpage
\appendix
\phantomsection
\chead{\textsc{Online Appendix}}
\addcontentsline{toc}{section}{Online Appendix}
\section*{Online Appendix}
\label{sec:appendix}
\renewcommand{\thesubsection}{\Alph{subsection}}
\setcounter{subsection}{0}

\subsection{Data Sources}
\label{data_detail}

First, to create the measures of historical presence of insurgent groups, I use detailed data from the National Police and the Administrative Department of Security. These data record how many of these crimes were committed by a given insurgent group in each municipality-year pair for different types of crimes. The criminal acts used for the creation of the two measures are: attacks and assaults against the population, incursions to populated centers, incendiary terrorist acts, explosive terrorist acts, offensive actions, kidnappings of politicians, members of armed forces and civilians, attacks against transport infrastructure, illegal road checkpoints, armed harassment, attacks against official entities, illegal roadblocks, assaults against private property, political attacks, ambushes, clashes, armed contacts, homicides, armed forces wounded and killed, and killings of members of the insurgent group. The creation of the two measures is described in detail in Section \ref{cont_data}. 

Population estimates come from the latest projections (January 2022) from the Colombian National Administrative Department of Statistics (DANE), available \href{https://www.dane.gov.co/index.php/estadisticas-por-tema/demografia-y-poblacion/proyecciones-de-poblacion}{\textit{here}}, not from CEDE. 

In Table \ref{data_sources_tab}, I summarise the data sources of the different variables used throughout the paper. A * denotes those that come from CEDE. For the Victims' Unit measures, ``events'' means the number of occurrences of such an act. Therefore, a person can suffer several ``events'' of the same type over time.

\begin{longtable}[]{| p{0.2\textwidth} | p{0.65\textwidth} | p{0.2\textwidth} |}
\caption{Data Sources}
\label{data_sources_tab}
\\
\hline
\multicolumn{1}{|c|}{\textbf{Variable Name}} & \multicolumn{1}{|c|}{\textbf{Description}} & \multicolumn{1}{|c|}{\textbf{Source}} \\ \hline \hline
\multicolumn{3}{|c|}{\textbf{Panel A. Violence Measures}} \\ \hline  \hline
Fights & Number of terrorist acts, fights, combats, clashes and attacks (events). & Victims' Unit \\ \hline
Threats & Number of threats. & Victims' Unit \\ \hline
(Forced) Disappearances & Number of forced disappearances (events). & Victims' Unit \\ \hline
(Forced) Displaced & Number of forced displaced individuals (events). & Victims' Unit \\ \hline
Homicides & Number of homicides (events). & Ministry of Defence*\\ \hline
Kidnaps & Number of kidnappings (events). & Ministry of Defence* \\ \hline
Torture & Number of tortures (events). This variable shows some extreme outliers. Thus, values 10 SDs above the mean are windsorised. & Victims' Unit \\ \hline
Property loss & Number of losses of real estate property (events). & Victims' Unit \\ \hline
Terrorist acts & Number of terrorist acts. & Ministry of Defence*  \\ \hline
Mines & Number of mine-related events. & Ministry of Defence* \\ \hline
Theft & Total number of thefts. & Ministry of Defence* \\ \hline
Clashes/Attacks & These data were kindly provided by Juan Vargas and have been used in a multitude of papers \citep[see, e.g.][]{de2021forgone,dube2013commodity,guerra2021peace}. The data were originally collected by \citet{restrepo2003dynamics} and have been updated by Universidad del Rosario. It covers the period from 1996 to 2018. \citet{de2021forgone} say the following about these data ``records conflict events (i.e. clashes, attacks) involving the different agents in the conflict (left-wing guerrillas, right-wing paramilitaries, government forces). For each event, the dataset records of the location and the date of occurrence. The data is based on news reports from over 20 major newspapers, complemented with additional reports from NGOs and the Catholic church''. Also see \citet{dube2013commodity} for more information. I simply add all the clashes and attacks between the three different groups to come to a single, aggregate measure. & Juan Vargas \\ \hline
Forced migration & I received data from the Victims' Unit on the amount of forced displaced coming into a municipality and the mount of individuals forced to leave a municipality. Based on these in- and out-migration flows, I create this measure of total forced migration as ((total forced migrated into municipality - total forced to outmigrate from municipality) / the municipality's population) * 1000, as a measure of total forced migration. & Victims' Unit \\

\hline \hline

\multicolumn{3}{|c|}{\textbf{Panel B. Economic Indicators}} \\

 \hline  \hline

Nighttime light intensity & Nighttime light intensity from two sources (the Defense Meteorological Satellite Program (DMSP), and the Visible Infrared Imaging Radiometer Suite (VIIRS)) are combined, cleaned, and harmonized to create a single panel dataset by \citet{li2020harmonized}. & \citet{li2020harmonized} \\ \hline
Value added per capita & One measure of value added per capita at the municipal level comes from DANE, \href{ https://www.dane.gov.co/index.php/estadisticas-por-tema/cuentas-nacionales/cuentas-nacionales-departamentales}{\textit{here}}, available only from 2011 to 2020. The other measure is created following the procedure proposed by \citet{sanchez2012urbanizacion}. This is done in the following way: first, sum the total tax intake (property plus industry and commerce) for a municipality $m$ in department $d$ in year $t$. Then, calculate the total tax intake in $d$ in year $t$. Calculate the share of $m$'s tax intake relative to the $d$'s intake. This gives $m$'s relative importance. Then assign this share of $d$'s value added to municipality $m$. Department GDP comes from DANE, tax revenue from the National Planning Department (DNP). & DANE and DNP* \\ \hline
Urban built-up & This is generated using the Band Ratio for Built-Up Area (BRBA) index developed by \citet{waqar2012development}. It is measured as $BRBA = TM3/TM5$, where TM are different (spectral) bands from Landsat TM satellite images. This measure has been shown to perform very well in identifying urban built-up areas in settings like Colombia \citep{valdiviezo2018built}. I estimate the index using Google Earth Engine. I first mask clouds and water bodies (identified using the MNDWI index), as water bodies affect the performance of the BRBA index. To classify a pixel as urban built-up, I use a threshold of 0.48 based on visual inspection of different threshold values. Values above 0.48 are classified as urban built-up. Then, I calculate the share of pixels within each municipality classified as urban.  & Landsat satellite images via Google Earth Engine \\ \hline
Agricultural productivity & Defined as total tonnes produced of 271 agricultural crops in a given municipality divided by total area cultivated in hectares in the municipality. An alternative measure focuses only on those crops produced in at least 2\% of the municipality-year pairs for which data are available. & Based on data from the Ministry of Agriculture* \\ \hline
Agricultural production & Defined as total tonnes produced of 271 agricultural crops per capita. & Based on data from the Ministry of Agriculture* \\ \hline
Share of urban population & Estimates of urban population based on population projections. & DANE* \\ \hline
Firm entry & Number of (new) businesses that requested a mercantile license from a Chamber of Commerce in a given year, and listed their commercial activities as taking place in a given municipality, & Confecámaras \\ \hline
Formal employment & Average number of individuals (wage earners and self-employed) making contributions to healthcare (mandatory), pension funds and/or workers' compensations in a given municipality-year. Note that for some of the earlier years, data are available for only some months.  & Ministry of Health and Social Protection, \textit{PILA} \\ \hline
Human Development Index & Estimate of the United Nations Human Development Index (HDI) based on machine learning tools and satellite imagery. & \citet{sherman2023global} \\ \hline
Agricultural trade outcomes & The data come from two sources. First, from Agronet, the data portal of the Ministry of Agriculture,  I obtained historical prices for each agricultural product sold in each of the main wholesale markets in Colombia. I drop prices  above or below 3 SDs of the mean price in a given year to deal with outliers. Second, from the System of Price and Supply of the Agricultural Sector (SIPSA), I obtained data on agricultural deliveries to each of these wholesale markets. These data are only available since 2013, and include information on the municipality of origin of the delivery, the quantity delivered and the market of destination, with information at the delivery level. I then matched these two datasets. However, some of the categories in SIPSA are either too broad, do not show up or match the categories in Agronet's price data, in which case they are excluded. In total, I could merge 45\% of the tonnes delivered in the SIPSA data to the corresponding price data from Agronet. Using this dataset, I create measures at the municipality of origin-year level of the number of  agricultural products were shipped, the number of wholesale markets reached, the number of deliveries, the average price of products (weighted by the total amount of each product delivered), the total value of all deliveries to all markets, and the total quantity delivered to all markets. & Agronet (Ministry of Agriculture) \& SIPSA \\ \hline
GEIH Survey Data & Seven outcomes are analysed from the Gran Encuesta Integrada de Hogares (GEIH). This is a monthly, repeated cross-section survey, not representative at the municipality level. The outcomes are: the hypothetical price a homeowner would be willing to sell their house for; an index based on ownership of phone, TV, fridge, washing machine, microwave, motorbike, car, bicycle, computer, and access to internet, with weights for each of these assets based on Colombia's 2015 DHS survey, that creates a similar index; the length of the unemployment spell in weeks (for those unemployed); months in the current work, months worked in the past 12 months, average weekly hours worked in that job and the gross salary in the previous month (for those employed). & Gran Encuesta Integrada de Hogares \\ \hline
\# of Beds & Number of beds for touristic purposes. & Registro Nacional de Turismo \\ \hline
Acc. employees & Number of employees working in touristic accommodations. & Registro Nacional de Turismo \\ \hline
Tourists & Number of national tourists that report visiting the municipality. & Encuesta de Gasto Interno en Turismo \\ \hline
Travel expenses & Expenditures on travel-related items. & Encuesta de Gasto Interno en Turismo \\ \hline
Immigrated & Number of individuals who report having moved to a given municipality in the past 12 months. There is a version that focuses only on those people aged at least 20. & 2005 and 2018 Census \\ \hline
International tourists & Number of international tourists that report that they are visiting a given municipality. & Ministry of Commerce, Industry and Tourism \\

\hline \hline

\multicolumn{3}{|c|}{\textbf{Panel C. State Capacity Measures}} \\

 \hline  \hline
 
Total revenue & Sum of current revenue and capital income. & DNP* \\ \hline
Tax revenue & Total tax revenue, including property tax, industry and commerce tax, and gasoline tax, among others. & DNP* \\ \hline
Government transfers & Transfers to the municipality from other government levels. & DNP* \\ \hline
Total expenditures & Sum of current expenditures and capital expenditures. & DNP* \\ \hline
Operational expenditures & Expenditures on the operation of the municipality administration. & DNP* \\ \hline
Total deficit & Difference between current income and current expenditures. & DNP* \\ \hline
Credit & Net income from internal and external credits (received - paid). & DNP* \\ \hline
Savings capacity & Current savings over current income. & DNP* \\ \hline
\% of Expenditures invested (also called ``Investment'' in some tables) & Share of expenditures used for investments. & DNP* \\ \hline
Fiscal performance & Index created by the DNP to measure municipalities' fiscal performance. Goes from 0 (lowest) to 100 (highest). It is composed of six different indicators: share of current income spent on the operation of the local government, total debt, share of income coming from transfers from other levels of the government, capacity to generate own income, savings capacity and size of investments. & DNP* \\ \hline
Overall performance & From 2016, based on score on municipal performance from DNP. Before 2016, based on the Index of Integral Performance, also by the DNP. Both try to capture the effectiveness, efficiency and administrative capacity of the municipal administration. & DNP* \\ \hline
Information openness & Index of Open Government created by the Office of the Inspector General of Colombia, available only since 2010. Created using measures of internal control, organization of information and document management and others. For a precise definition of this variable, see the CEDE data catalogue.  & Office of the Inspector General of Colombia* \\ \hline
Aqueduct coverage & Total aqueduct coverage. Data from CEDE covers only from 2010 to 2016. Data from the 2018 Census is used to complement this, from DANE. & Sistema Único de Información de Servicios Públicos* and DANE \\ \hline
Garbage collection & Total coverage of garbage collection system. Data from CEDE covers only from 2010 to 2016. Data from the 2018 Census is used to complement this, from DANE. & Sistema Único de Información de Servicios Públicos* and DANE \\ \hline
Sewage coverage & Total coverage of sewage system. Data from CEDE covers only from 2010 to 2016. Data from the 2018 Census is used to complement this, from DANE. & Sistema Único de Información de Servicios Públicos* and DANE  \\ \hline
Share voting & Voting data for elections since 2006 (both national, department, and municipal) come from CEDE. & CEDE \\ 

\hline \hline

\multicolumn{3}{|c|}{\textbf{Panel D. Other Variables}} \\

 \hline  \hline
 
Population & & DANE \\ \hline
Area &  & DANE* \\ \hline
Dist. Capital & Distance to department's capital. & DANE* \\ \hline
Dist. Bogotá & Distance to Bogotá. & DANE* \\ \hline
Small credit & Value of credit to small producers. & Agronet* \\ \hline
Total credit & Total value of credit to producers. & Agronet* \\ \hline
Conf. 1901/30 & Dummy for whether the municipality experienced a land-related conflict between 1901 and 1931. & CEDE \\ \hline
Spanish occup. & Dummy for whether the municipality was occupied by the Spanish between 1510 and 1561. & CEDE \\ \hline
GINI &  & CEDE based on census data from DANE \\ \hline
MDP & Municipal Poverty Incidence. & CEDE based on census data from DANE \\ \hline
NBI & Unsatisfied Basic Needs. & DANE*  \\ \hline
Social leaders killed & Number of social leaders killed. & Somos Defensores \\ \hline
Extortion & Number of total extortion acts. & Ministry of Defense* \\ \hline
Harbers15 & Tax revenue divided by nighttime light intensity (standardized). & Based on \citet{harbers2015taxation} \\ \hline
Distance to border to Venezuela & Distance to border to Venezuela in kms. & From \citet{martinez2017transnational} \\ \hline
Venezuelan Migrants with PEP & Numbers of Venezuelan migrants that have received a Permiso Especial de Permanencia and registered at a given municipality. & From Colombia's Migration Office, can be accessed \href{https://public.tableau.com/app/profile/migraci.n.colombia/viz/PermisoEspecialdePermanencia-PEP/Inicio}{\textit{here}} \\ \hline
Coca production & Area cultivated with coca crops (in HAs). & Ministry of Justice \\ \hline
Coca eradication areas & Total area of coca crops eradicated (manually and by plane). & Ministry of Justice \\ \hline
PNIS program & Number of municipalities and beneficiaries from the PNIS coca cultivation program. & UNODC, from \href{https://www.unodc.org/documents/colombia/2021/Febrero/INFORME_EJECUTIVO_PNIS_No._23.pdf}{\textit{here}} \\ \hline
Credit to agricultural producers (Banco Agrario) & Total value of credits to different types of agricultural producers given by the Banco Agrario. & Agronet* \\ \hline
Credit to agricultural producers (FINAGRO) & Total value of credits to different types of agricultural producers given by FINAGRO. & FINAGRO \\ \hline
First-year students & Number of students enrolled in the first year of higher education programs (technical, university or post-graduate degrees). & Ministry of Education \\ \hline
Number of higher education institutions & Number of institutions providing higher educations programs (technical, university or post-graduate degrees). & Ministry of Education\\ \hline
Demobilized individuals & Number of demobilized members of insurgent groups (FARC/ELN) that have registered as living at a given municipality. & Agency for Reincorporation and Normalization \\ \hline
Variables related to land restitution processes & Number of land restitution claims presented to UAEGRTD (including the number of claimants and plots involved), number of requests handled and denied by land restitution courts, and the number of beneficiaries and plots returned by courts to claimants. & Land Restitution Unit (URT) \\ \hline
Survey data & Data come from the AmericasBarometer surveys (since 2011) and other occasional surveys (``Muestra Especial Zonas Conflicto'', for 2013, 2015 and 2017) conducted by the Observatorio de la Democracia. Can be found \href{https://obsdemocracia.org/encuestas/}{\textit{here}}. & Observatorio de la Democracia \\ \hline
Child health outcomes & Weight and size at birth, Apgar score 1 and 5 minutes after birth, and child mortality (calculated as the number of under-1-year olds dead relative to the number of newborns in a municipality in a given year). & Estadísticas Vitales, DANE \\ \hline
Education outcomes & Number of students, teachers, and official schools. & CEDE \\ \hline
Distance to other municipalities and wholesale markets & \citet{gafaro2022trade} calculate travel distances based on Google Maps data across all municipalities in 2014. These travel distances are used to compute the number of municipalities and wholesale markets within a certain amount of kms of each municipality. & \citet{gafaro2022trade}  \\ \hline
International Brent spot price & In USD per barrel.& US Energy Information Administration \\ \hline
Coca Suitability Index & & \citet{mejia2013bushes}  \\ \hline
Coca prices & Average price of coca leaves and coca paste at production sites, and average price of cocaine in main cities (in COP). UNODC do not report these prices in 2019, so I use instead values from 2020 from the National Police, at 2019's exchange rate. & UNODC. \& National Police   \\ \hline
Presence of gold geochemical anomalies & This is a dummy that equals one if the municipality has gold geochemical anomalies. & Ministry of Energy   \\ \hline
Oil royalties & Both an indicator for whether the municipality paid any oil-related royalties to the central government, and the total amount paid, to control for oil production. & Agencia Nacional de Hidrocarburos \\ \hline
Years of schooling & Average years of schooling of population aged 15+. & DANE, based on Census* \\ \hline
Attending education & Share of population aged 5 to 24 attending an educational institution. & DANE, based on Census* \\ \hline
Conflict incidence (2002-2013) & Index of incidence of armed conflict between 2002 and 2013 (used to determine ZOMAC eligibility). & DNP \\ \hline

\end{longtable}

\subsubsection{Creation of Summary Indices}

In several of the sections of this study, I analyse the effects of the start of the ceasefire on a wide array of outcome variables for one particular indicator (e.g. violence, economic activity or state capacity). For brevity and compactness, I use two approaches to summarise all the different outcome variables into single measures that capture the relevant information. One is based on \citet{kling2007experimental}, and the other is based on \citet{anderson2008multiple}. Here, I explain how these are created:

\begin{enumerate}
	\item \citet{kling2007experimental}: results using the index suggested by \citet{kling2007experimental} are called ``KLK index''.  KLK argue that ``the aggregation improves statistical power to detect effects that go in the same direction within a domain''. To create this index, the following steps are followed:
	\begin{enumerate}
		\item First, each of the measures is standardised by the pre-intervention values of the variable in the control group (note: in the original paper, there is no time dimension, so they just standardise by the control group. However, for the purposes here, only pre-intervention periods are used to avoid treatment contamination).
		\item Some variables have missing values for some years. For variables with data missing for a year before (after) the treatment, I follow their imputation method and assign to the control/treatment group the mean (standardised) value of the control/treatment group in the pre-(post-)treatment periods. 
		\item All variables are aligned in the same direction, so higher values indicate ``better'' outcomes.
		\item The final index is the equally-weighted average of $z$-scores of the index's component variables, created in the first two steps.
		\item The final measure is then standardised for easiness of interpretation.
	\end{enumerate}
	\item[] This index has been used, for example, by \citet{blumenstock2022modernizing} and \citet{casey2012reshaping}.
	\item \citet{anderson2008multiple}: the second index has been used by \citet{egger2019general,gilligan2014civil,  haushofer2016short} and \citet{karlan2010expanding}, for example. It is implemented using the ``swindex'' command in Stata created by \citet{schwab2020constructing}. This one is very similar to the KLK index, but assigns more weight to component variables that are less correlated to other component variables, as these might carry more relevant information not captured by the other component variables. Intuitively, uncorrelated indicators represent ``new''  information and, therefore receive more weight. It also assigns less value to indicators with missing observations. The estimation procedure, detailed in \citet{egger2019general}'s PAP, is the following:
	\begin{enumerate}
		\item For each outcome variable $y_{jk}$, where $j$ indexes the outcome group and $k$ indexes variables within outcome groups, variables are recoded so that higher values denote ``better'' outcomes.
		\item Then the covariance matrix $\hat{\Sigma}_j$ for outcomes in group $j$ is estimated, which consists of elements $\hat{\Sigma}_{jmn} = \sum_{i=1}^{N_{jmn}} \dfrac{y_{ijm} - \bar{y}_{jm}}{\sigma_{jm}^y} \dfrac{y_{ijn} - \bar{y}_{jn}}{\sigma_{jn}^y}$, where $N_{jmn}$ is the number of non-missing observations for outcomes $m$ and $n$ in outcome group $j$, $\bar{y}_{jm}$ and $\bar{y}_{jn}$ are the means for outcomes $m$ and $n$ in outcome group $j$, and the sigmas are the standard deviations in the pure control group for the same outcomes for the entire analysed period.
		\item Then the covariance matrix is inverted, and they define the weight $w_{jk}$ for each outcome $k$ in outcome group $j$ by summing the entries in the row of the inverted covariance matrix corresponding to that outcome.
		\item Each outcome variable is transformed by subtracting its mean and dividing by the control group SD, then weighting it with the abovementioned weights. Formally, $\hat{y}_{ij} = (\sum_{k \in K_j} w_{jk})^{-1} \sum_{k \in K_j} w_{jk} \dfrac{y_{ijk} -\bar{y}_{jk} }{\sigma_{jk}^y}$.  
	\end{enumerate}
\end{enumerate}


\clearpage

\renewcommand{\thefigure}{B\arabic{figure}}
\setcounter{figure}{0}
\renewcommand{\thetable}{B\arabic{table}}
\setcounter{table}{0}

\subsection{Difference-in-Discontinuities -- ZOMAC}
\label{zomac_sec}

In this Section, I explain in detail the ZOMAC program, the identification strategy and the underlying assumptions that need to be satisfied. I present evidence in support of the different assumptions and robustness checks that confirm the baseline results. 

As a way to incentivise business and employment creation in areas that have been affected by the conflict (ZOMAC municipalities), the government started a tax incentive program for firms in 2017. The main incentive is a progressive business tax tariff for 10 years beginning in 2017, which varies depending on the firm's size, as shown below in Table \ref{taxbenefits}. For firms to benefit from the tax reduction, they must i) have been created after December 29, 2016, ii) have their main address in a ZOMAC municipality, iii) perform their whole productive processes in ZOMAC municipalities, and iv) satisfy some investment and job-creation requirements. These investment and job-creation requirements vary depending on the sector and the firm's size. Informal firms that formalise and meet these criteria can also benefit from these incentives. Firms in the mining and oil sectors are excluded from the benefits. Thus, this program specifically targeted firm and job creation, and incentives were largest for early movers, who could enjoy the tax breaks for longer.

These are the conditions the government took into account to designate a municipality as a ZOMAC municipality:

\begin{enumerate}
	\item Municipalities that have a Multidimensional Poverty Index (IPM in Spanish) above 0,49, \textit{or} an Index of Fiscal Performance (DF in Spanish) below 70.
	\item Not be part of an agglomeration.
	\item Municipalities most affected by the conflict, denoted by having a score above 0.0191 in an index of incidence of the armed conflict (IICA in Spanish). The index is calculated as the average over 2002 and 2013 of six violence-related variables: armed actions, homicides, kidnappings, antipersonnel-mine victims, forced displacement, and coca crops.\footnote{The program's decree does not mention where these variables come from and how exactly they are aggregated into a single indicator. }
	\item Municipalities that belong have a Territorially Focused Development Programme (PDET in Spanish).\footnote{However, it has been reported that the implementation of these PDET has been the slowest of any of the components of the peace agreement, has been plagued by problems, and has been considerably underfunded \citep{garcia2020implementacion,isacson2021long, valencia2022}.}
	\item Municipalities below 450.000 inhabitants and more than 60 minutes away from the department's capital.
\end{enumerate}

More specifically, a municipality is denoted as ZOMAC if one of the following two criteria hold:

\begin{enumerate}
	\item It either satisfies 4) and 5), i.e. it is part of PDET and satisfies the demographic requirements, or
	\item It satisfies 1), 2), 3) and 5), i.e. it is poor \textit{or} poorly managed, affected by conflict, it is not part of an agglomeration, not a PDET, and it satisfies the demographic requirements.
\end{enumerate}
 
The selection criteria thus depend on clear thresholds for some variables. I exploit these to create a sample of treated municipalities just above the thresholds and a sample of control municipalities just below the thresholds, in a regression discontinuity (RD) approach. I focus on the second criterium, as being a PDET municipality or not is not defined by clear thresholds (it is simply a categorical variable). Even for the second criterium, there are several ways a municipality can be categorised as ZOMAC or not based on strict thresholds. In practice, among the possible selection combinations based on the second criterium, the only one with enough municipalities near the threshold is based on the IICA score. More precisely, I focus on municipalities that are just above and just below the IICA score necessary to be classified as a ZOMAC municipality and meet all the other requirements, meaning municipalities just below or above IICA $=$ 0.0191, with IPM $>=$ 0,49 \textit{or} DF $<=$ 70, not part of an agglomeration, below 450.000 population and over 60 minutes drive from department's capital, not PDET. For these municipalities, their IICA score was the sole determinant of ZOMAC classification. 

The RD based on the strict IICA score among these municipalities creates a sample of municipalities that are supposed to be equal in all respects other than being a ZOMAC municipality or not, if they are sufficiently close to the IICA cutoff. Given that I am interested in understanding the dynamic effects of the peace agreement, I evaluate the evolution of these municipalities just above and just below the IICA threshold over time, basically embedding the RD design in a difference-in-difference setting. This approach was first formalised by \citet{grembi2016fiscal}, and called the ``difference-in-discontinuity'' estimator. As this approach is new and has not been frequently used \citep[exceptions are][]{bazzi2020institutional, bergolo2021anatomy,grembi2016fiscal, hansen2020federalism}, there is no clear set of empirical guidelines or practices for using it, with a recent paper being the exception \citep{picchetti2024difference}. That paper shows the theoretical advantages of the difference-in-discontinuity estimator compared to traditional DiD and RDD specifications. However, I follow the different robustness checks and assumption tests used in other papers to argue that the assumptions needed for the validity of the difference-in-discontinuity estimator seem to be satisfied in this context. 

To recover the difference-in-discontinuities estimator, three assumptions need to be satisfied. First, all the potential outcomes must be continuous at the discontinuity. Second, suppose there is a different, confounding policy that affects municipalities above and below the discontinuity differently. In that case, the effect of the confounding policy at the discontinuity in the case of no treatment must be constant over time (so that it can be netted out using the ``difference'' part of the estimator). This is the equivalent of the parallel trends assumption in traditional difference-in-difference settings. Third, the effect of the treatment at the discontinuity does not depend on the confounding policy. In this Appendix, I present evidence in support of these three assumptions.

Following \citet{grembi2016fiscal}, I estimate the regression:
\begin{equation}
\label{eq_diff_in_disc_app}
\begin{aligned}
y_{it} =& \delta_0 + \delta_1 IICA_{m}^{*} + \delta_2 \textit{IICA Treatment}_{m} + \delta_3 IICA_{m}^{*} \times \textit{IICA Treatment}_{m}
\\
&+ \delta_4 Post_t + \delta_5 Post_t \times IICA_{m}^{*}  + \delta_6 Post_t \times \textit{IICA Treatment}_{m} 
\\
&+ \delta_7 Post_t \times \textit{IICA Treatment}_{m} \times IICA_{m}^{*} + u_{mt}
\end{aligned}
\end{equation}

\noindent where $IICA_{m}^{*}$ is the normalized IICA score ($IICA_{m}^{*} = IICA_{m} - 0.0191$) of municipality $m$ in year $t$, \textit{IICA Treatment}$_{m}$ is a dummy for municipalities with an $IICA$ score above 0.0191 (i.e. ZOMAC municipalities), and $Post_t$ is an indicator for the post-treatment period. As the ZOMAC program started in 2017, I denote the post-treatment years as those from 2017 in these regressions. Standard errors are still clustered at the municipality level. The difference-in-discontinuity estimator of interest is the coefficient $\delta_6$ and identifies the treatment effect of receiving the fiscal incentives for firms. 

\begin{figure}[h!]
\caption{Robustness Difference-in-Discontinuities Results to Bandwidth Selection}
\label{RDD_rob_bw}
\begin{center}
\begin{subfigure}[b]{0.46\textwidth}
\includegraphics[width=\textwidth]{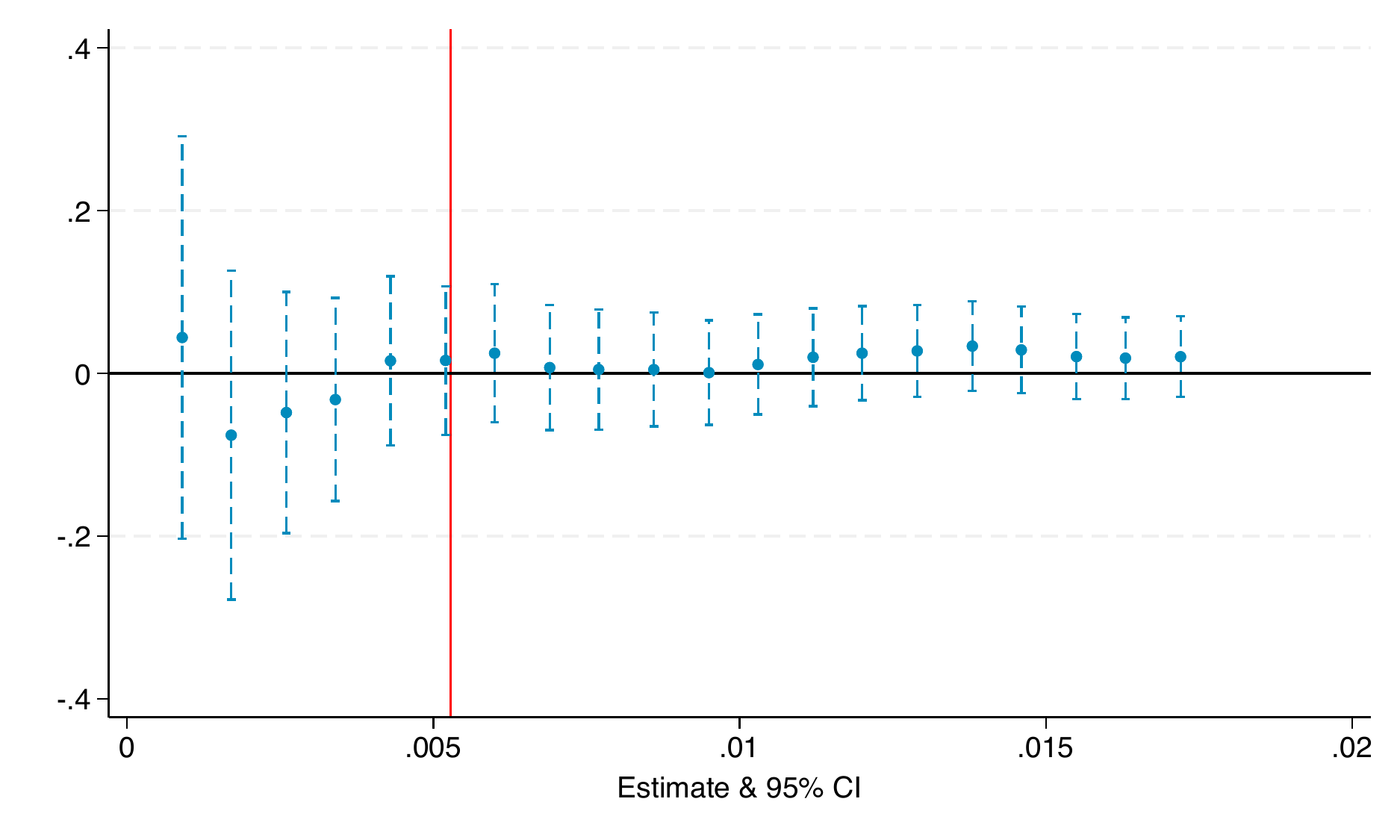}
\subcaption{Nighttime Light Intensity (Weighted)}
\end{subfigure}
\begin{subfigure}[b]{0.46\textwidth}
\includegraphics[width=\textwidth]{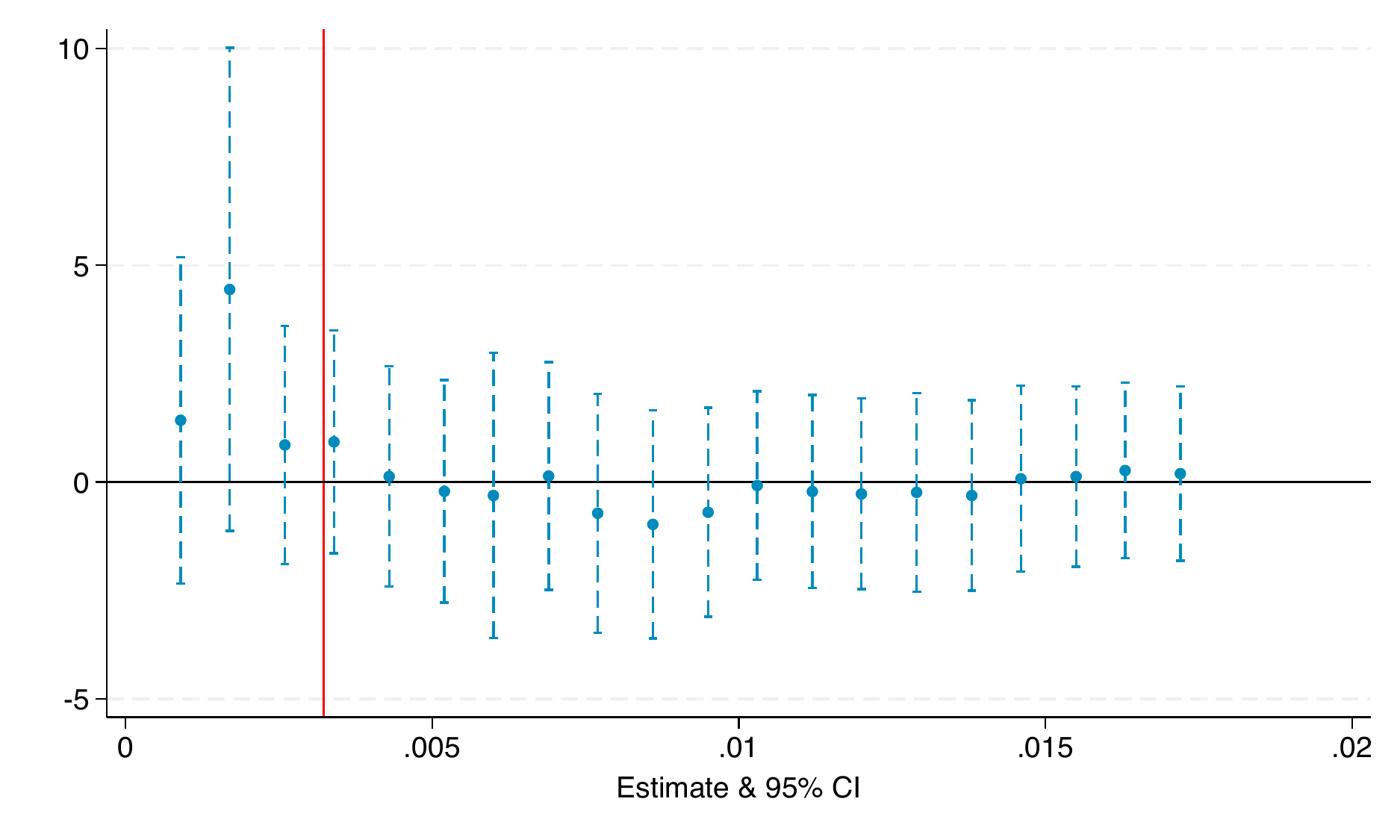}
\subcaption{Value Added (pc, DANE)}
\end{subfigure}
\begin{subfigure}[b]{0.46\textwidth}
\includegraphics[width=\textwidth]{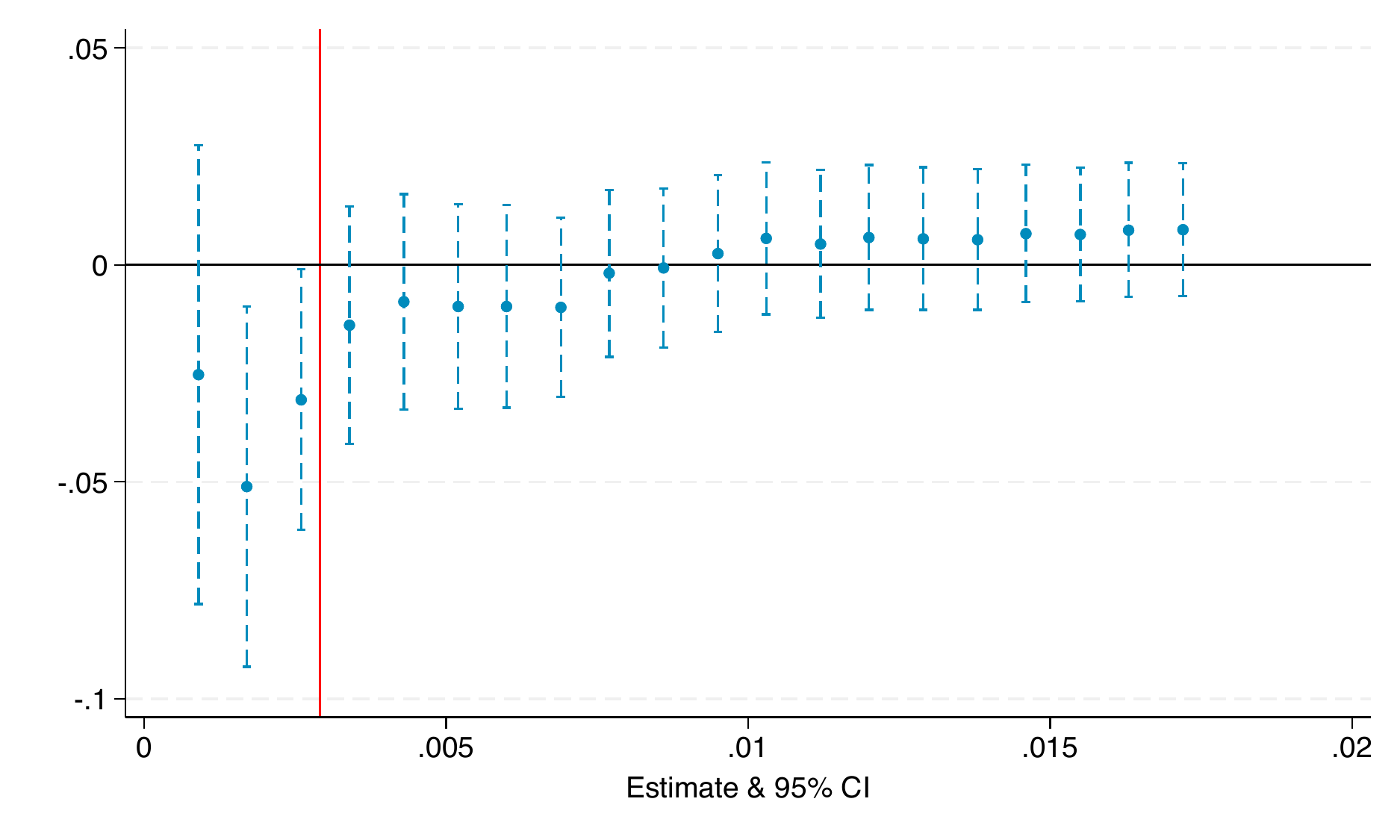}
\subcaption{Share Urban Population}
\end{subfigure}
\begin{subfigure}[b]{0.46\textwidth}
\includegraphics[width=\textwidth]{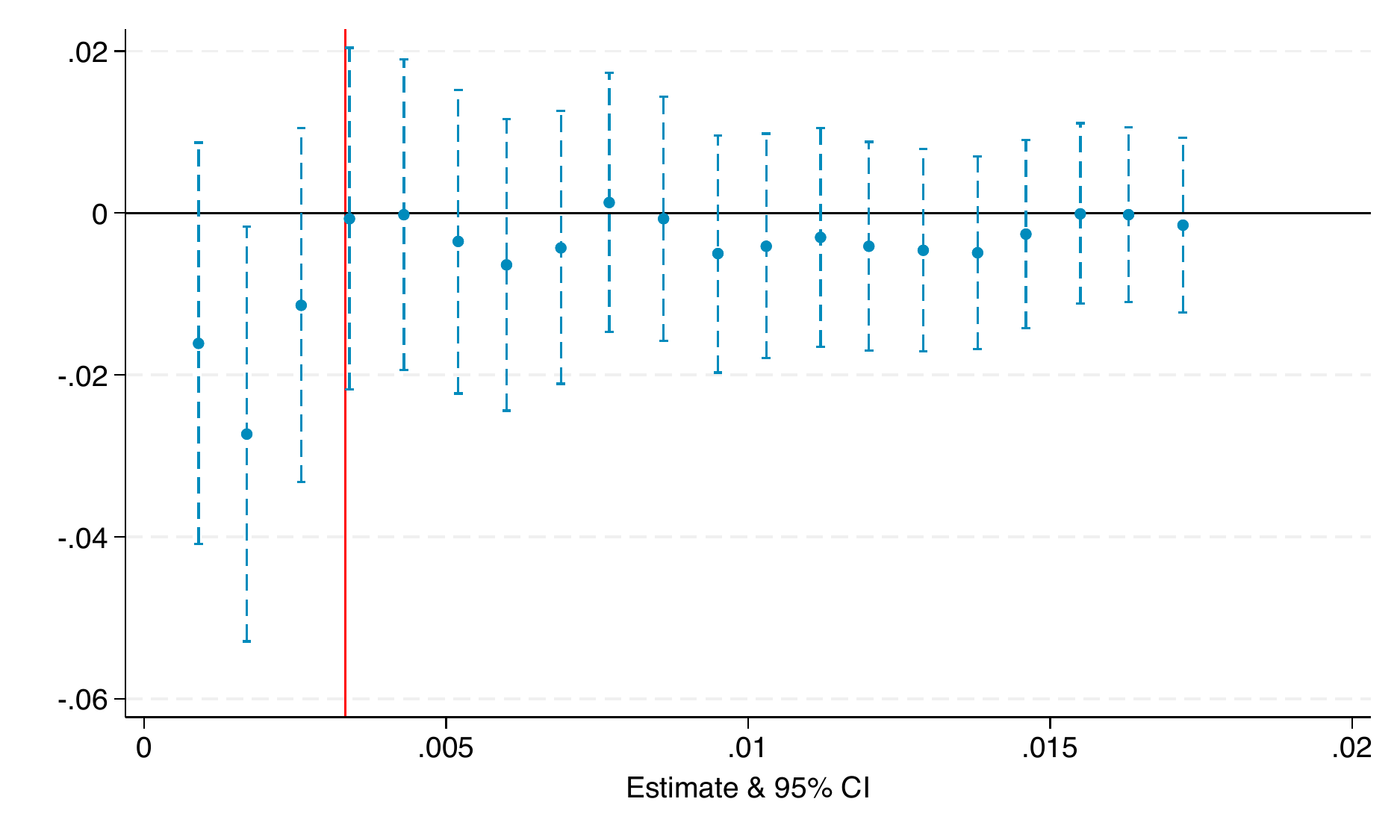}
\subcaption{Formal Employment}
\end{subfigure}
\begin{subfigure}[b]{0.46\textwidth}
\includegraphics[width=\textwidth]{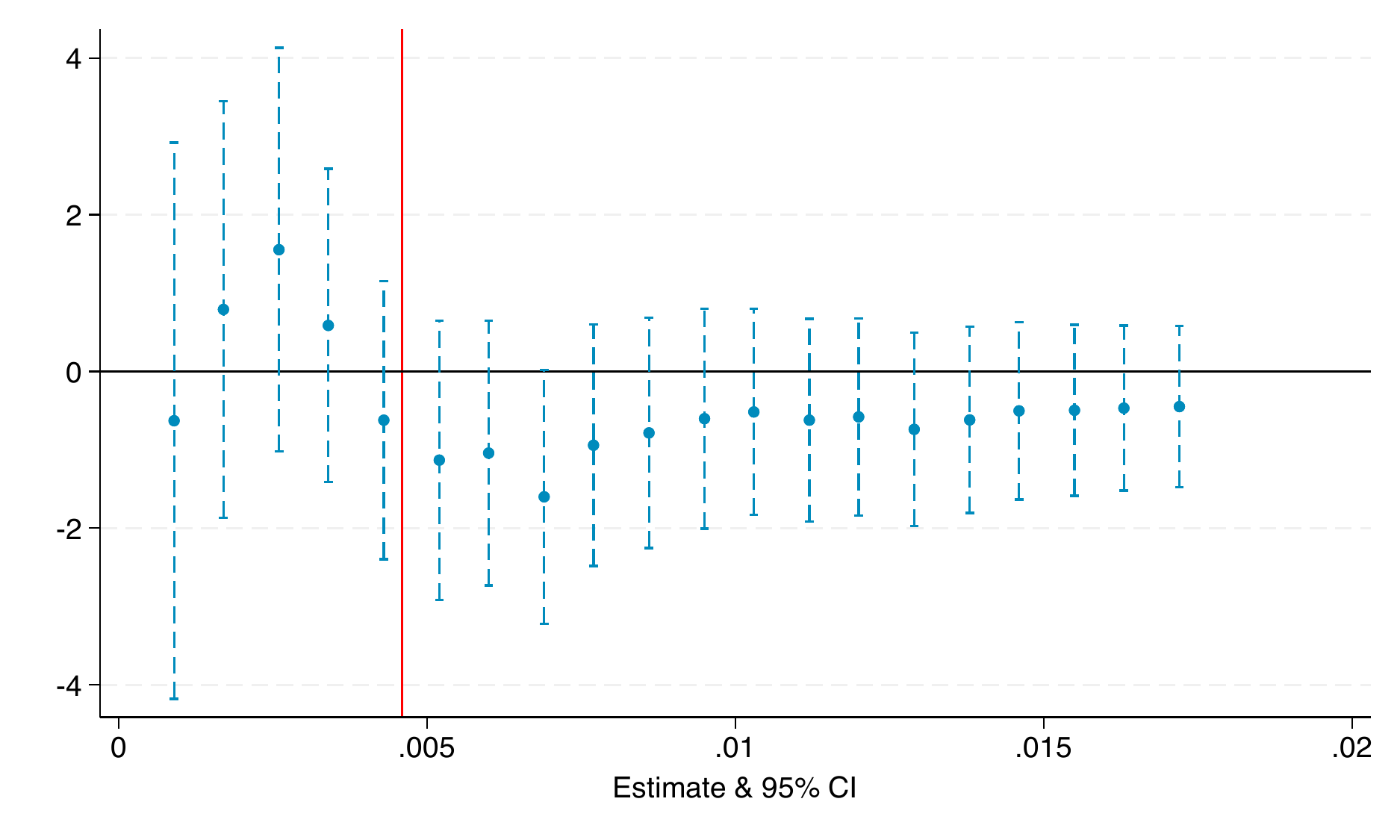}
\subcaption{Firm Entry}
\end{subfigure}
\begin{subfigure}[b]{0.46\textwidth}
\includegraphics[width=\textwidth]{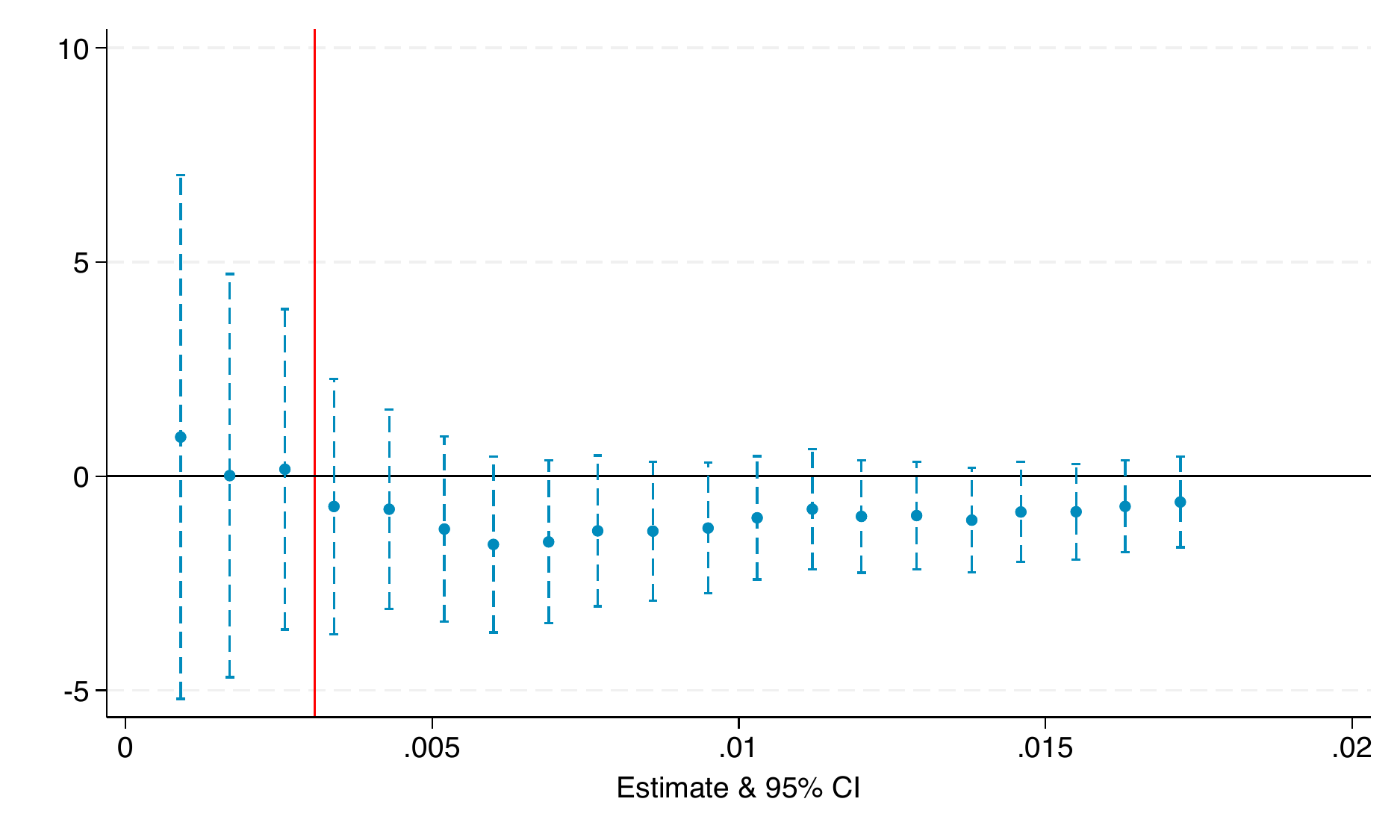}
\subcaption{Agricultural Productivity}
\end{subfigure}
\begin{subfigure}[b]{0.46\textwidth}
\includegraphics[width=\textwidth]{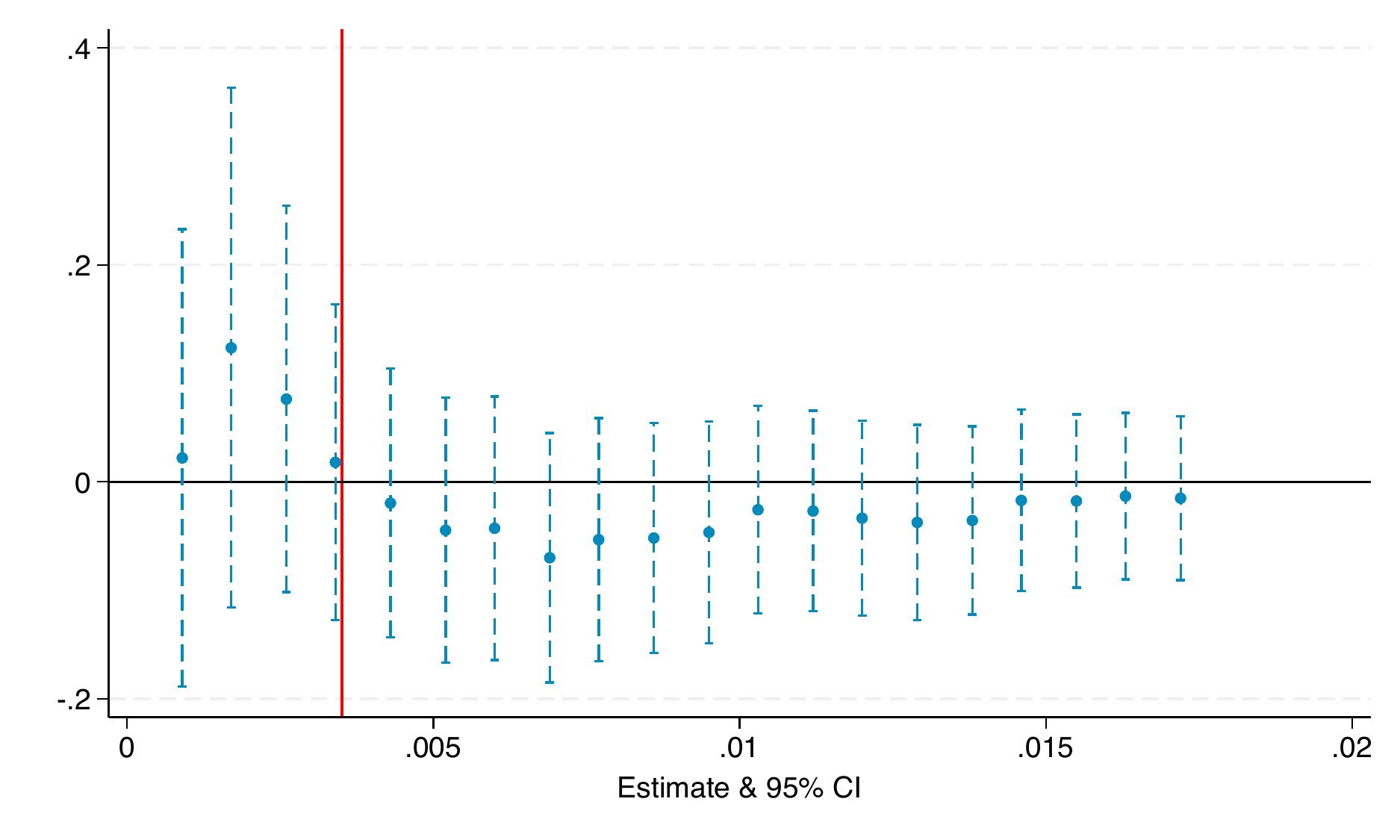}
\subcaption{Economic Activity Index}
\end{subfigure}
\end{center}
\justifying 
\footnotesize{\textbf{Notes:} Estimates of Equation \eqref{eq_diff_in_disc}, including 95\% confidence intervals (based on standard errors clustered at the municipality level). The red line is the optimal bandwidth calculated following \citet{calonico2014robust}, selected to minimise the Mean Squared Error. The blue estimates and confidence intervals are the results from this same exercise but using 20 different (evenly-spaced) bandwidths.}
\end{figure}

While Table \ref{RDD_Econ_IICA} shows the baseline results, I now assess the robustness of the results to using different functional forms, controls, definitions of the post-treatment period and bandwidths. First, in Figure \ref{RDD_rob_bw}, I plot the coefficients for each outcome variable using 20 (evenly-spaced) different bandwidths, up to one standard deviation of the running variable. The red line indicates the optimal bandwidth selected to minimise the MSE. The results are very similar regardless of the bandwidth selected. Only for the share of urban population are small bandwidths associated with significant coefficients. Second, Panel A  in Table \ref{RDD_FuncFormRob} shows the results' robustness to different inference methods. I first estimate Conley SEs \citep{conley1999gmm} that allow for correlated unobservables across municipalities within certain distances of the given municipality's centroid. For distances, I use the \nth{25}, \nth{50} and \nth{75} percentiles of municipalities' distance to their department's capital. I also show $p$-values using the wild cluster bootstrap suggested by \citet{cameron2008bootstrap} for difference-in-difference settings. The SEs and results remain unchanged. In Panel B, I estimate Equation \eqref{eq_diff_in_disc} controlling for municipality and year fixed effects, with the coefficients basically unchanged from the baseline scenario. Then, in Panels C and D, instead of using a linear function for the running variable, I use a quadratic and cubic polynomial on either side of the cutoff. The results remain qualitatively the same and similar in magnitude, although the SEs increase. Finally, although the ZOMAC program started in 2017, to rule out any anticipatory effects, in Panel E, I run a regression defining the post period as beginning in 2016 rather than 2017, which does not affect the results. 

\begin{table}[h!] \centering
\newcolumntype{R}{>{\raggedleft\arraybackslash}X}
\newcolumntype{L}{>{\raggedright\arraybackslash}X}
\newcolumntype{C}{>{\centering\arraybackslash}X}

\caption{Robustness RDD Analysis Inference, Functional Form \& Controls}
\label{RDD_FuncFormRob}
\begin{adjustbox}{max width=\linewidth, max height=\textwidth}\begin{tabular}{lccccccc}

\toprule
& Nighttime & Value Added (pc) & Agricultural & Share Urban & Firm  & Formal & \textbf{Anderson} \tabularnewline & Light & DANE & Productivity & Population & Entry & Empl. & \textbf{Index} \tabularnewline 
{}&{(1)}&{(2)}&{(3)}&{(4)}&{(5)}&{(6)}&{(7)} \tabularnewline
\midrule \addlinespace[\belowrulesep]
\midrule \textbf{Panel A. Baseline}&&&&&&& \tabularnewline
IICA Treat. \( \times \) Post&0.031&0.692&--0.341&--0.027**&--1.064&--0.001&--0.022 \tabularnewline
&(0.047)&(1.269)&(1.606)&(0.014)&(0.966)&(0.011)&(0.109) \tabularnewline
&&&&&&& \tabularnewline
Conley Perc. 25&(0.037)&(1.313)&(1.770)&(0.014)&(0.918)&(0.010)&(0.095) \tabularnewline
Conley Perc. 50&(0.039)&(1.267)&(1.612)&(0.014)&(0.616)&(0.011)&(0.103) \tabularnewline
Conley Perc. 75&(0.047)&(1.187)&(1.612)&(0.015)&(0.846)&(0.010)&(0.134) \tabularnewline
Wild Bootstrap p-value&[0.513]&[0.596]&[0.845]&[0.045]&[0.290]&[0.919]&[0.844] \tabularnewline
Observations&1,859&936&1,100&1,067&1,584&1,166&1,353 \tabularnewline
\(R^2\)&0.010&0.044&0.024&0.054&0.059&0.029&0.073 \tabularnewline
&&&&&&& \tabularnewline
\midrule \textbf{Panel B. TWFE}&&&&&&& \tabularnewline
IICA Treat. \( \times \) Post&0.031&0.692&--0.341&--0.027*&--1.064&--0.001&--0.022 \tabularnewline
&(0.049)&(1.349)&(1.690)&(0.014)&(1.015)&(0.011)&(0.114) \tabularnewline
&&&&&&& \tabularnewline
Observations&1,859&936&1,100&1,067&1,584&1,166&1,353 \tabularnewline
\(R^2\)&0.941&0.867&0.970&0.989&0.771&0.962&0.935 \tabularnewline
&&&&&&& \tabularnewline
\midrule \textbf{Panel C. Quadratic Polynomial}&&&&&&& \tabularnewline
IICA Treat. \( \times \) Post&--0.063&2.667&0.904&--0.047**&1.890&--0.027*&0.157 \tabularnewline
&(0.081)&(2.009)&(2.573)&(0.023)&(1.416)&(0.015)&(0.176) \tabularnewline
&&&&&&& \tabularnewline
Observations&1,859&936&1,100&1,067&1,584&1,166&1,353 \tabularnewline
\(R^2\)&0.036&0.062&0.048&0.095&0.094&0.040&0.075 \tabularnewline
&&&&&&& \tabularnewline
\midrule \textbf{Panel D. Cubic Polynomial}&&&&&&& \tabularnewline
IICA Treat. \( \times \) Post&--0.066&3.920&0.008&--0.039&0.366&--0.020&0.140 \tabularnewline
&(0.117)&(2.407)&(3.322)&(0.028)&(1.753)&(0.015)&(0.228) \tabularnewline
&&&&&&& \tabularnewline
Observations&1,859&936&1,100&1,067&1,584&1,166&1,353 \tabularnewline
\(R^2\)&0.054&0.079&0.060&0.129&0.104&0.040&0.113 \tabularnewline
&&&&&&& \tabularnewline
\midrule \textbf{Panel E. Alt. Post Definition}&&&&&&& \tabularnewline
IICA Treat. \( \times \) Post&--0.046&2.469&--1.869&--0.023&--0.123&--0.018&--0.027 \tabularnewline
&(0.082)&(2.094)&(3.104)&(0.016)&(1.196)&(0.014)&(0.147) \tabularnewline
&&&&&&& \tabularnewline
Observations&913&639&627&451&825&385&968 \tabularnewline
\(R^2\)&0.043&0.044&0.089&0.084&0.073&0.093&0.045 \tabularnewline
\bottomrule \addlinespace[\belowrulesep]

\end{tabular} \end{adjustbox}
\\ \parbox{\linewidth}{\scriptsize \justifying \noindent \(Notes \): Results from estimating Equation \eqref{eq_diff_in_disc}. Standard errors clustered at the municipality level. Considers only years from 2009 and before 2020. Defines the post period as 2017. Bandwidth optimally estimated using Calonico et al. (2014) approach. Conley SEs at different distance bandwidths (25th, 50th and 75th percentile of municipalities' distance to department's capital). Wild bootstrap \(p\)-value in squared brackets. Panel E defines the post-treatment period as the year 2016, after the peace agreement was approved. Nighttime light intensity defined so that grid cells in the border of multiple municipalities are assigned in proportion to the share of the grid cell in each municipality (weighted). Value added per capita comes DANE (National Department of Statistics), in millions of COP. Agricultural productivity is defined as total tonnes produced of 271 agricultural crops divided by total area cultivated in hectares, based on data from the Ministry of Agriculture. Firm entry comes from RUES and is measured per 1000 inhabitants. Formal employment is measured as the average number of individuals paying contributions to healthcare, pension funds and workers' compensations across the year in the municipality per 18-60 years old, from PILA. Anderson Index is a summary measure created following Anderson (2008) that summarizes the measures in columns 1-6. It is the weighted average of the standardized outcomes, weighted by their inverted covariance matrix.}
\end{table}

\begin{figure}[h!]
\begin{center}
\caption{Robustness of Anderson Index to Individual Component Variables}
\label{RDD_robAndIndex}
\includegraphics[width=0.7\textwidth]{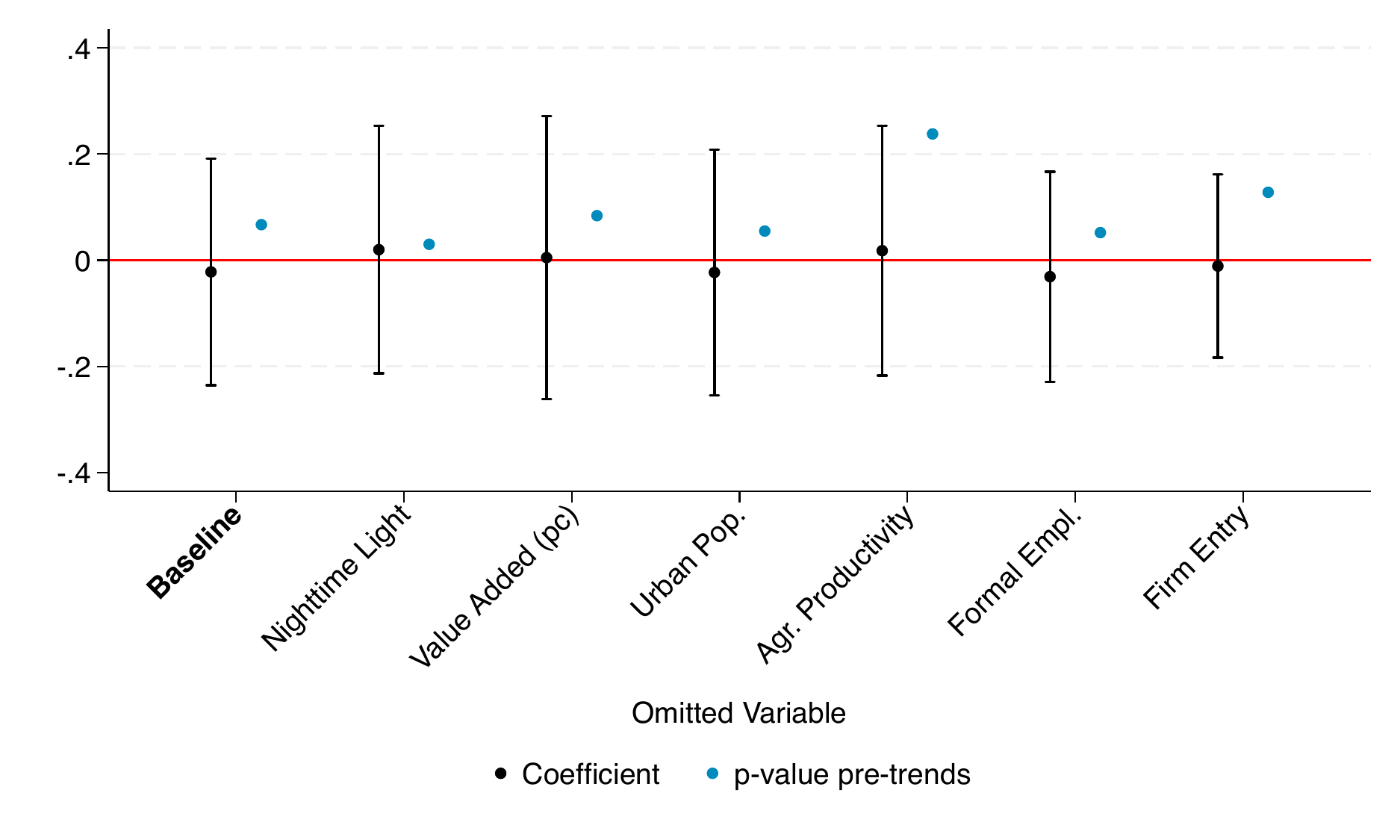}
\end{center}
\justifying 
\footnotesize{\textbf{Notes:}  Figure shows in black estimates of Equation \eqref{eq_diff_in_disc} including 95\% confidence intervals (based on standard errors clustered at the municipality level) and in blue $p$-values from a test of joint significance of all pre-treatment coefficients from estimates of Equation \eqref{eq_es} following \citet{freyaldenhoven2021visualization}. The Figure uses the index summarising the different economic indicator variables (nighttime light intensity, value added per capita (from DANE), share of urban population, agricultural productivity, firm entry and formal employment measures) created following \citet{anderson2008multiple}. The names in the $x$-axis correspond to the variable dropped from the corresponding index when estimating the results.}
\end{figure}

Any single component variable does not drive the results using the Anderson Index. Figure \ref{RDD_robAndIndex} shows in black the results of estimating Equation \eqref{eq_diff_in_disc} including 95\% confidence intervals and in blue $p$-values from a test of joint significance of all pre-treatment coefficients from estimates of Equation \eqref{eq_es} following \citet{freyaldenhoven2021visualization}, dropping each component variable individually. Regardless of the variable dropped, the results are basically identical to the baseline results, suggesting that no individual component variable drives the results. 

\begin{figure}[h!]
\begin{center}
\caption{Density Test Running Variable -- IICA Score}
\label{RDD_densityTest}
\includegraphics[width=0.7\textwidth]{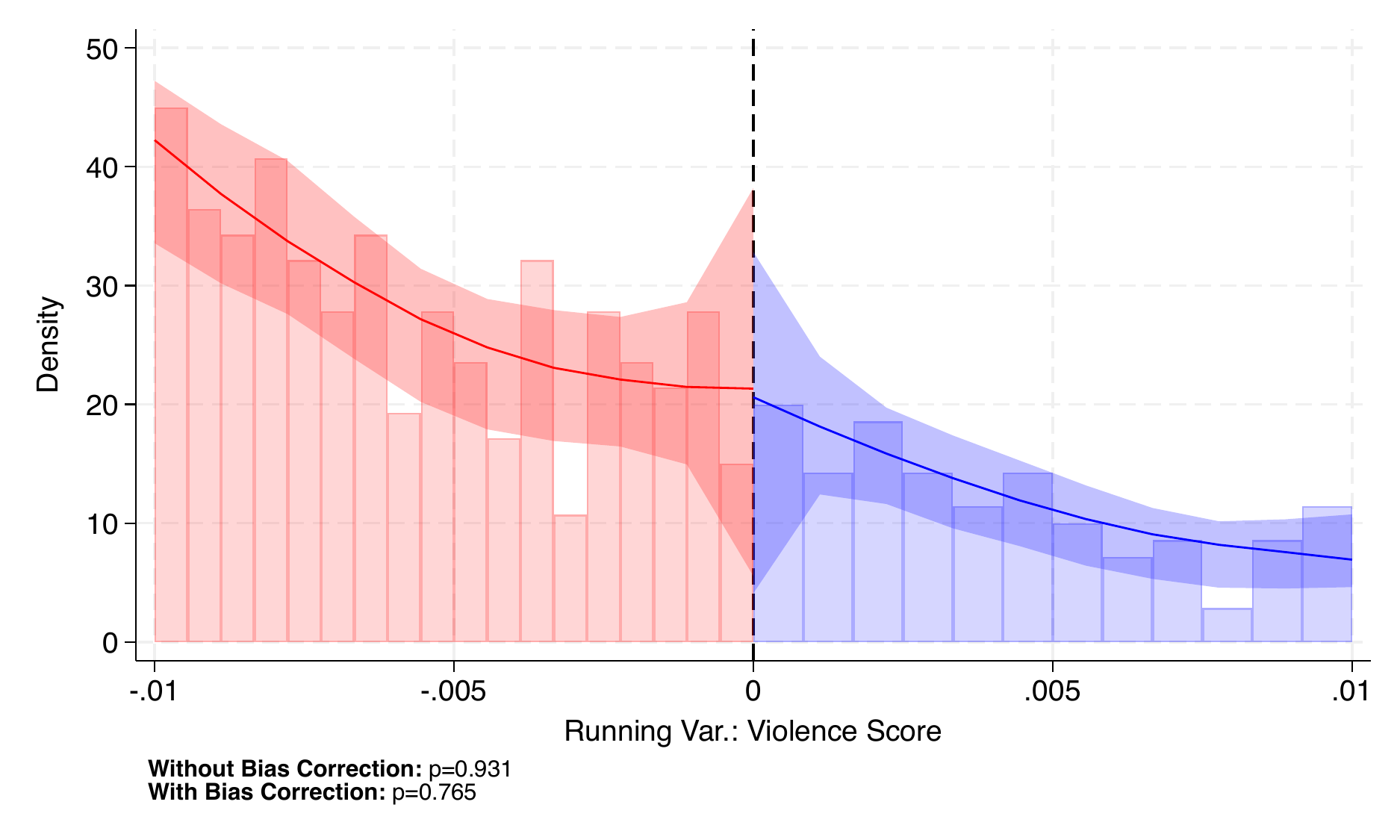}
\end{center}
\justifying 
\footnotesize{\textbf{Notes:}  Density test of the running variable, the IICA score, using local polynomial density estimators as suggested by \citet{cattaneo2020simple}. At the bottom of the Figure are the $p$-values of this test using bias correction or not.}
\end{figure}

Finally, I present evidence supporting the validity of the assumptions needed to recover the difference-in-discontinuities estimator. The first assumption requires no manipulation of the running variable, as is usual in RD designs. Figure \ref{RDD_densityTest} shows the results of a manipulation test using local polynomial density estimators as suggested by \citet{cattaneo2020simple}. Regardless of whether bias correction is used or not, the test fails to reject the null, supporting the hypothesis that there is no manipulation of the running variable. This makes intuitive sense, given that the IICA score is the average between 2002 and 2013 of different violence indicators, long before the ZOMAC program was even conceived, making manipulation very unlikely. 

\begin{figure}[h!]
\caption{Dynamic Difference-in-Discontinuities Specifications}
\label{RDD_preTrends}
\begin{center}
\begin{subfigure}[t]{0.46\textwidth}
\includegraphics[width=\textwidth]{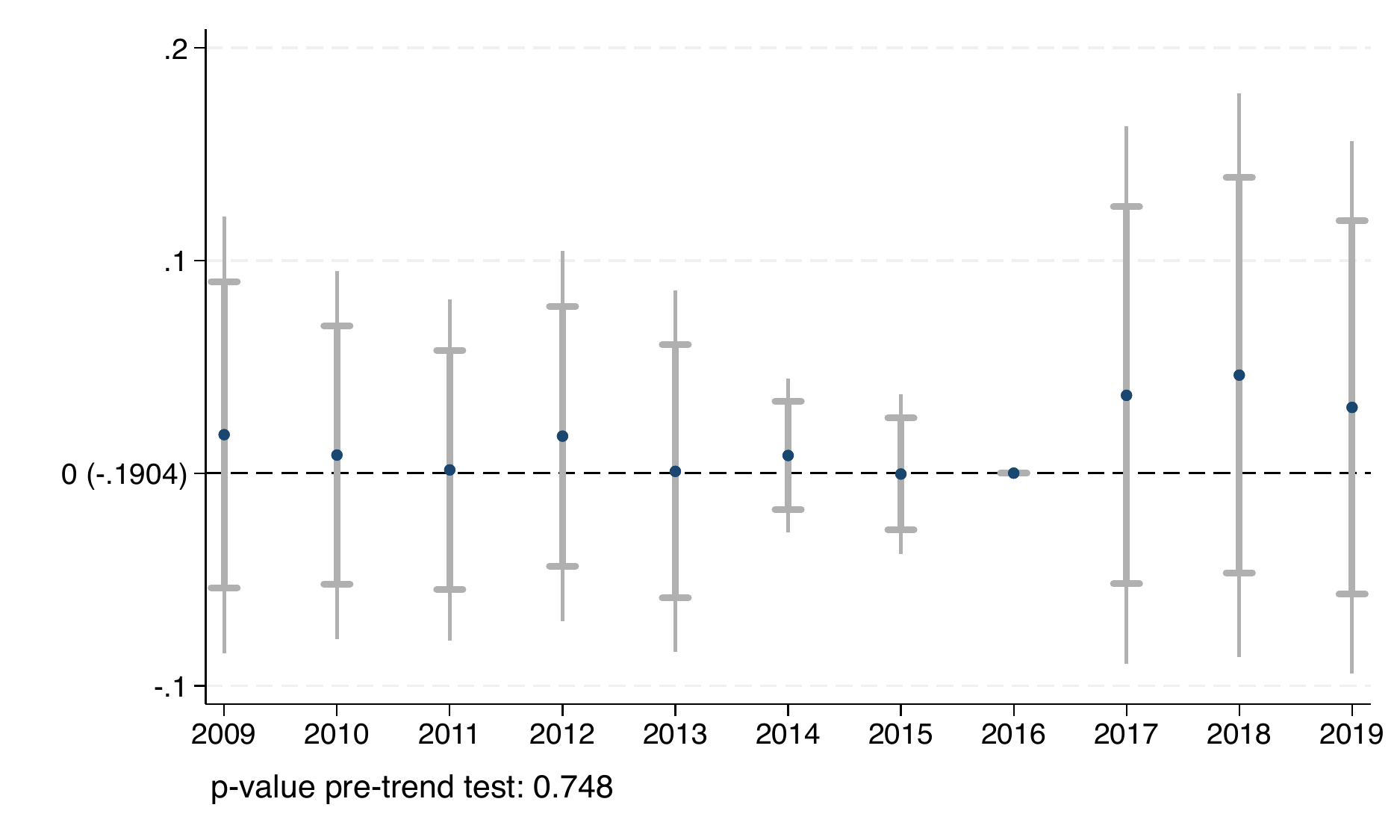}
\subcaption{Nighttime Light Intensity (Weighted)}
\end{subfigure}
\begin{subfigure}[t]{0.46\textwidth}
\includegraphics[width=\textwidth]{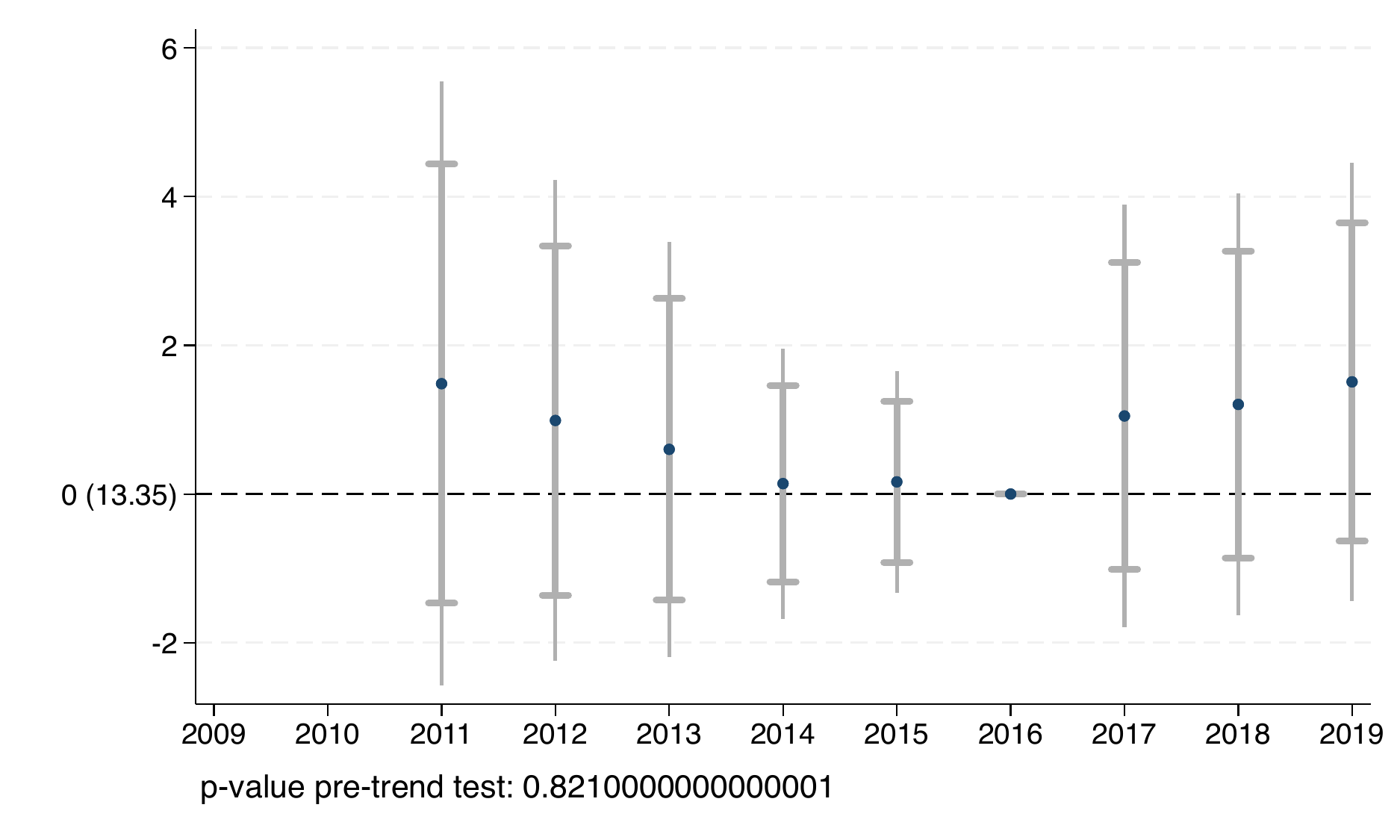}
\subcaption{Value Added (pc, DANE)}
\end{subfigure}
\begin{subfigure}[t]{0.46\textwidth}
\includegraphics[width=\textwidth]{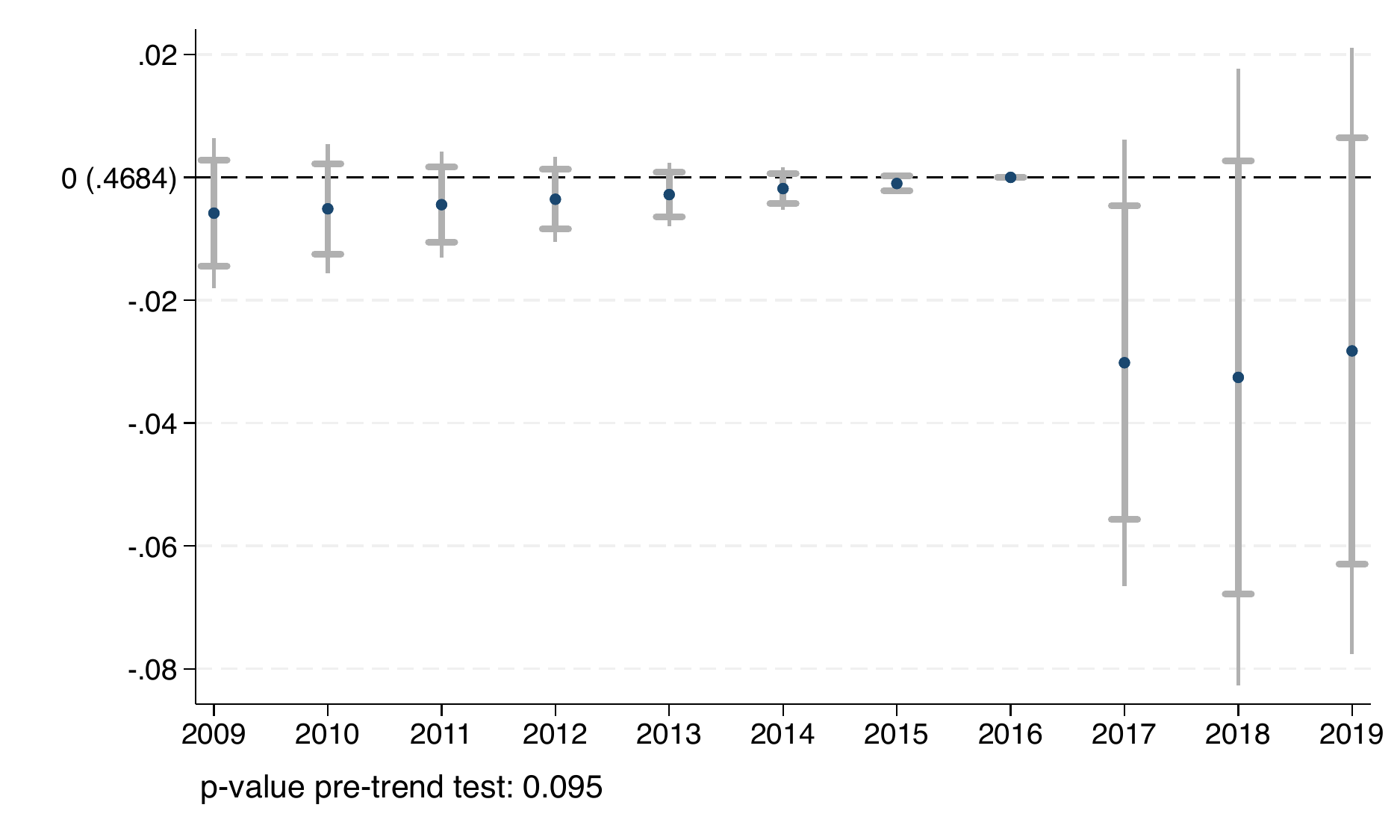}
\subcaption{Share Urban Population}
\end{subfigure}
\begin{subfigure}[t]{0.46\textwidth}
\includegraphics[width=\textwidth]{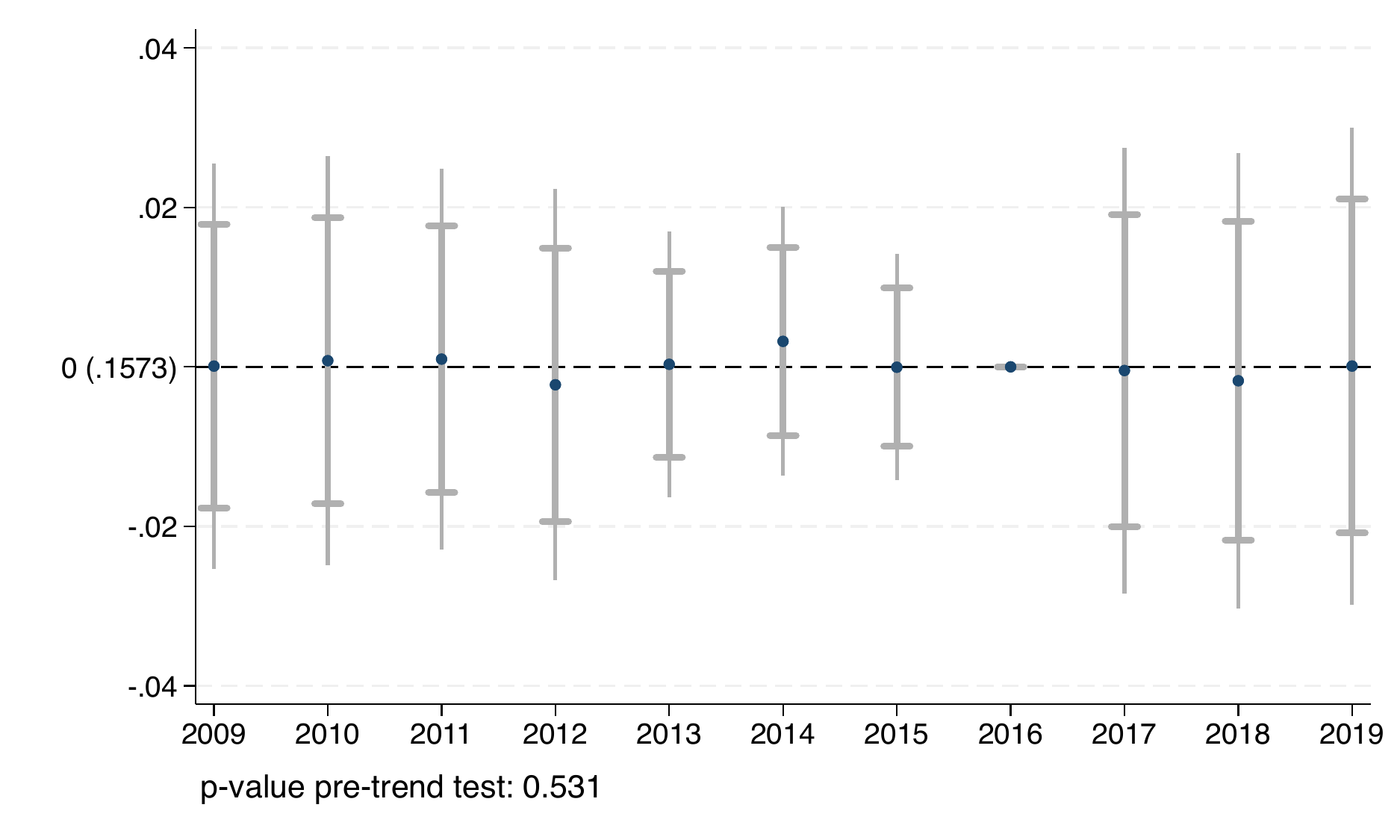}
\subcaption{Formal Employment}
\end{subfigure}
\begin{subfigure}[t]{0.46\textwidth}
\includegraphics[width=\textwidth]{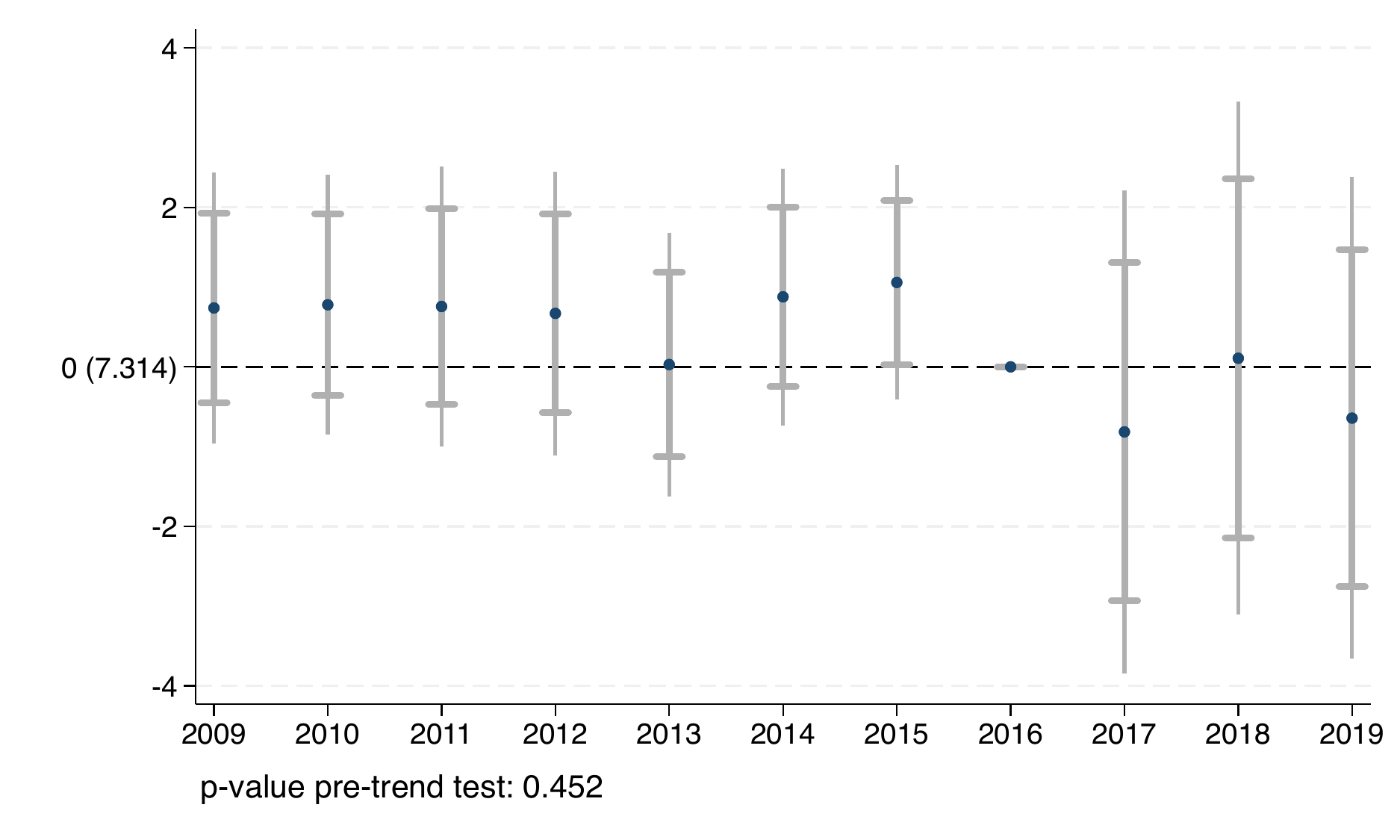}
\subcaption{Firm Entry}
\end{subfigure}
\begin{subfigure}[t]{0.46\textwidth}
\includegraphics[width=\textwidth]{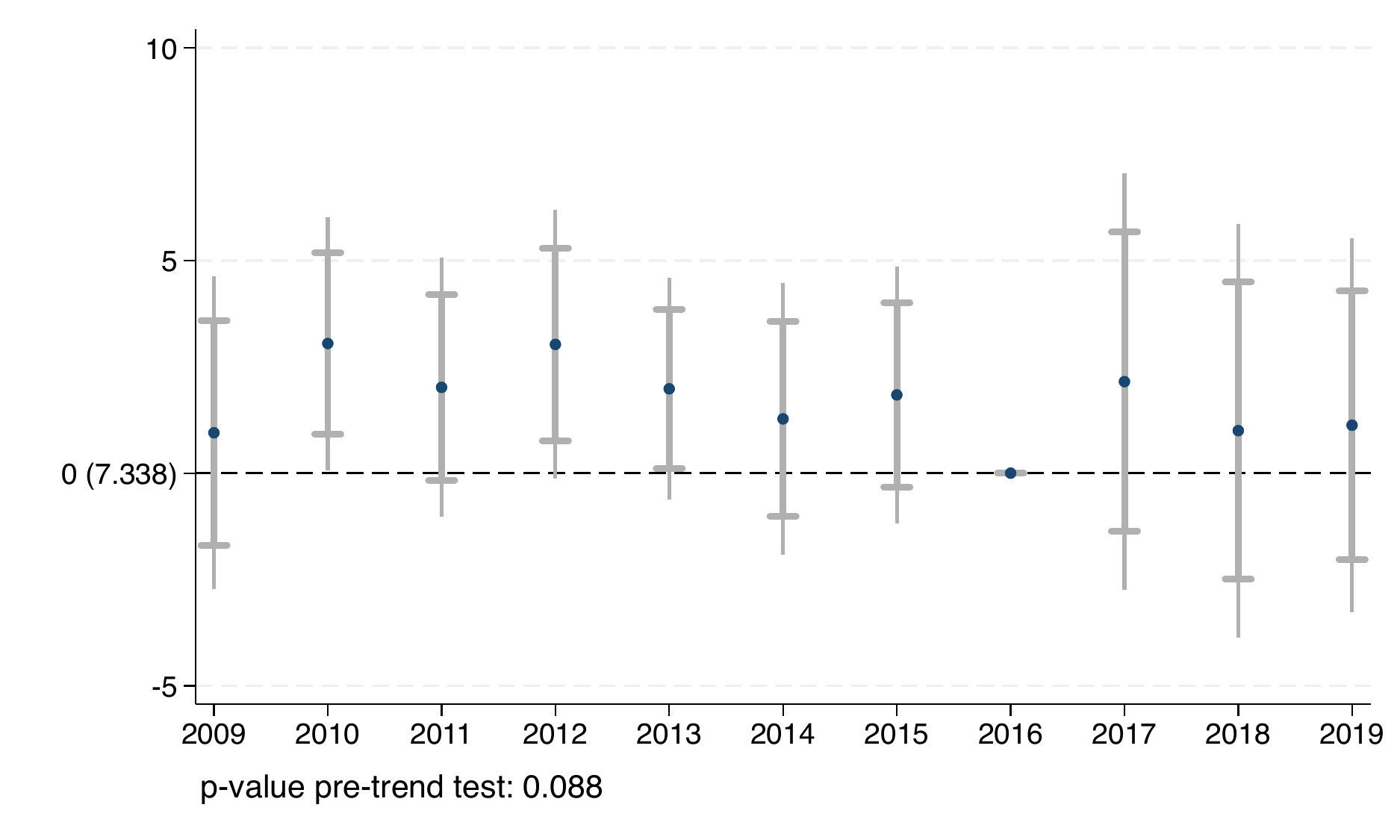}
\subcaption{Agricultural Productivity}
\end{subfigure}
\begin{subfigure}[t]{0.46\textwidth}
\includegraphics[width=\textwidth]{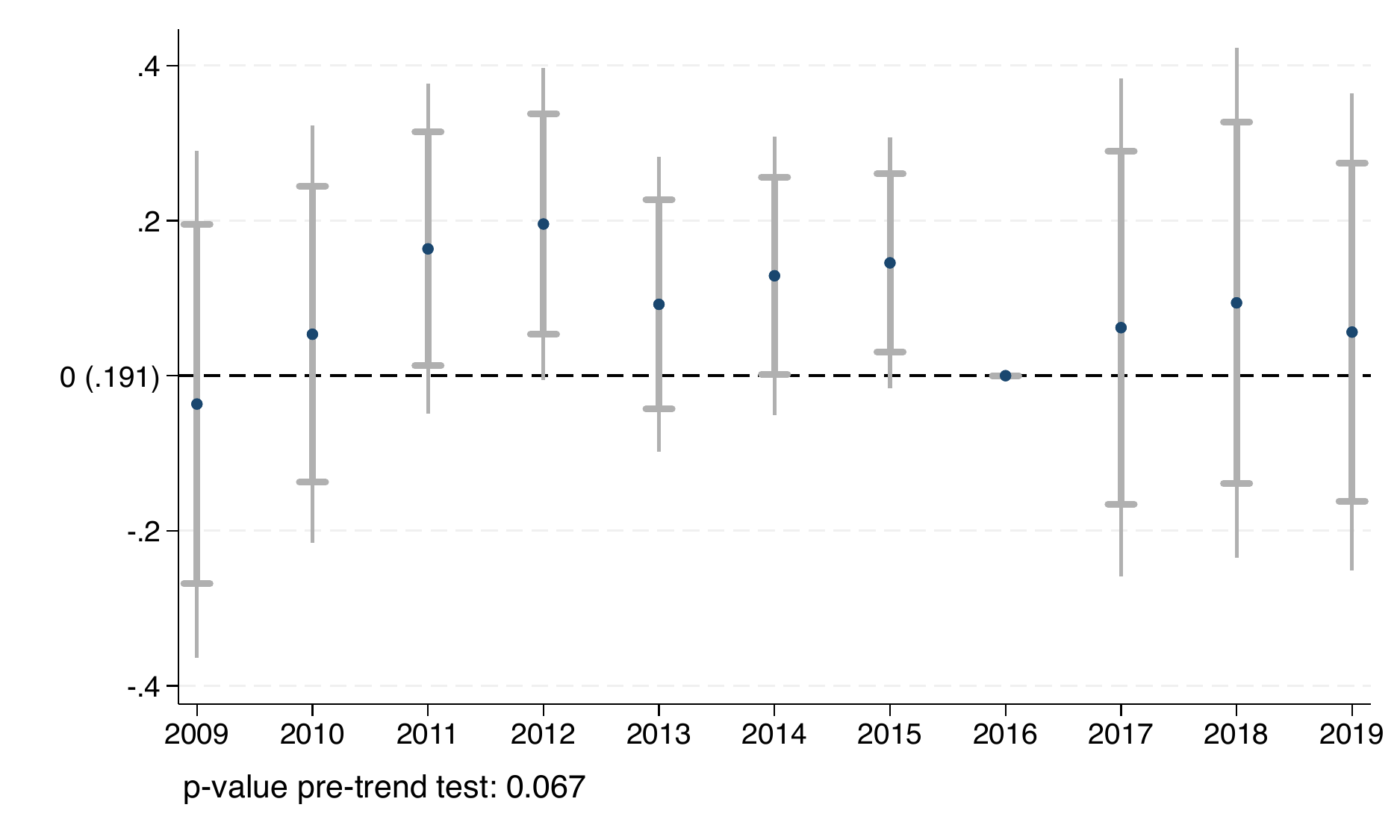}
\subcaption{Economic Activity Index}
\end{subfigure}
\end{center}
\justifying 
\footnotesize{\textbf{Notes:}  Event study plots from estimating Equation \eqref{eq_es}, including including 95\% confidence intervals (based on standard errors clustered at the municipality level). The index is created following \citet{anderson2008multiple} and is based on the other variables displayed here.}
\end{figure}

\begin{figure}[h!]
\begin{center}
\caption{Difference-in-Discontinuities: Placebo Estimates}
\label{RDD_placeboEst}
\includegraphics[width=\textwidth]{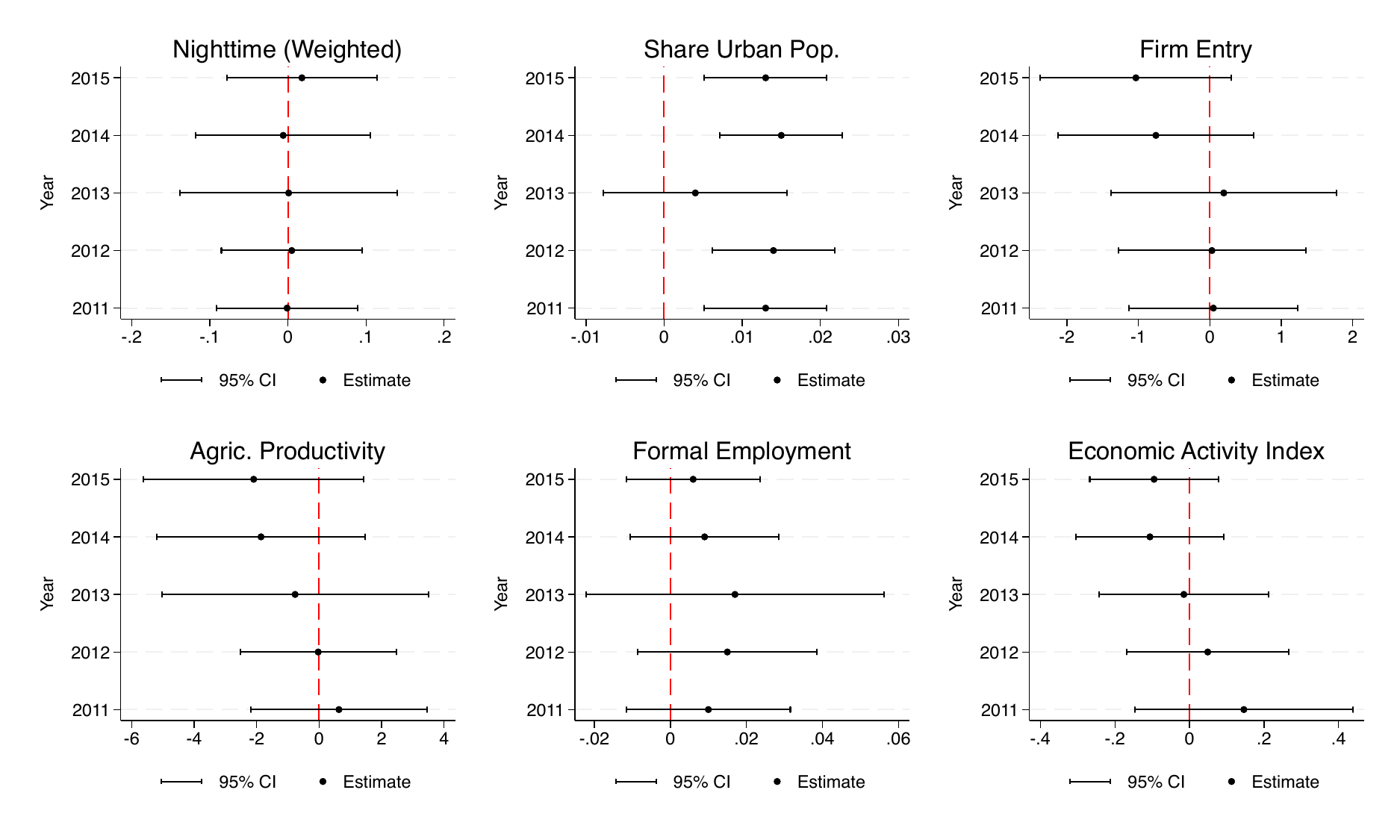}
\end{center}
\justifying 
\footnotesize{\textbf{Notes:} Estimates of Equation \eqref{eq_diff_in_disc}, including 95\% confidence intervals (based on standard errors clustered at the municipality level), using data only from the pre-treatment period (2009-2017, as the ZOMAC program started in 2017). The number in the $y$-axis corresponds to the ``placebo'' treatment year. Each figure uses a different economic outcome.}
\end{figure}

Assumptions 2 and 3 ask for municipalities above and below the cutoff to be on a (local) parallel trend in the absence of the new policy and that the effect of the treatment does not depend on other confounding policies. I provide several pieces of evidence in support of these assumptions. First, Figure \ref{RDD_preTrends} shows the results of estimating regressions akin to Equation \ref{eq_es} but for the difference-in-discontinuity setting. There seem to be no clear pre-trends for any of the variables. The joint test of pre-intervention coefficients are marginally rejected for three outcomes, but the sup-t confidence bands cover the 0 for the entire path. For the Anderson Index, the significance of the joint test comes entirely from the agricultural productivity measure, although the main coefficient remains unchanged when excluding this variable from the index (see Figure \ref{RDD_robAndIndex}). These graphs also show that there is no improvement in these economic indicators over time, as could have been expected. Second, results of running regressions of several different municipalities' characteristics pre-intervention (in 2008) on the IICA treatment dummy are presented in Table \ref{RDD_balCheck}. Out of the 20 different characteristics, only one is marginally significantly different between municipalities just below and just above the IICA threshold, suggesting that these municipalities were very similar to being with. Finally, Figure \ref{RDD_placeboEst} shows the results of running placebo versions of Equation \eqref{eq_diff_in_disc}. These placebos only use data between 2009 and 2016 (before the start of the ZOMAC policy), and assign the post-intervention period sequentially as years between 2011 and 2015. Reassuringly, the results are insignificant for all variables and placebo years, except for the share of urban population measure.  

\citet{picchetti2024difference} emphasise the importance of the confounding effect at the threshold being constant over time and propose a test to check whether this assumption is satisfied. It requires estimating a stacked RDD regression model that includes all pre-intervention time periods and test for the joint equality of all the triple interaction terms between i) the (pre-treatment) year dummies, ii) the (placebo) treatment assignment based on the cutoff, and iii) the running variable. The $p$-value of this joint test is shown at the bottom of Panel A in Table \ref{RDD_Econ_IICA}. The null is rejected only for one variable (firm entry), while for every other outcome, the $p$-value is comfortably away from traditional significance levels. 

\begin{table}[h!] \centering
\newcolumntype{R}{>{\raggedleft\arraybackslash}X}
\newcolumntype{L}{>{\raggedright\arraybackslash}X}
\newcolumntype{C}{>{\centering\arraybackslash}X}

\caption{Difference-in-Discontinuities Analysis: Pre-Treatment Balance Check}
\label{RDD_balCheck}
\begin{adjustbox}{max width=\linewidth, max height=\textwidth}\begin{tabular}{lcccccc}

\toprule
& All & Within BW & Bandwidth & Treatment & Control & Difference \tabularnewline 
{Variable}&{(1)}&{(2)}&{(3)}&{(4)}&{(5)}&{(6)} \tabularnewline
\midrule \addlinespace[\belowrulesep]
\midrule Population&38.44&25.38&0.008&23.08&26.64&0.528 \tabularnewline
Area&1017.6&568.49&0.003&521.4&605.31&0.61 \tabularnewline
Dist. Capital&81.46&75.24&0.008&73.13&76.35&0.65 \tabularnewline
Dist. Bogota&321.55&272.2&0.005&265.06&276.69&0.648 \tabularnewline
Small Credit&0.05&0.05&0.005&0.06&0.05&0.682 \tabularnewline
Total Credit&0.31&0.32&0.004&0.25&0.4&0.641 \tabularnewline
Gov. Transfers&0.07&0.05&0.006&0.06&0.05&0.873 \tabularnewline
Savings Capac.&32.91&30.46&0.009&30.02&30.66&0.777 \tabularnewline
Fiscal Perf.&62.1&61.56&0.008&62.19&61.28&0.422 \tabularnewline
Overall Perf.&58.85&58.41&0.009&57.04&59.05&0.272 \tabularnewline
Mun. Develop.&66.67&68.56&0.008&69.63&68.04&0.332 \tabularnewline
Conf. 1901/30&0.05&0.07&0.007&0.08&0.07&0.809 \tabularnewline
Spanish Occup.&0.37&0.25&0.007&0.31&0.21&0.122 \tabularnewline
Aqueduct&59.55&62.08&0.007&58.6&64.17&0.238 \tabularnewline
Garbage Coll.&45.77&51.8&0.006&52.58&51.27&0.8 \tabularnewline
Sewage&42.37&51.57&0.005&50.96&52.01&0.853 \tabularnewline
PC Expenditure&0.26&0.25&0.005&0.26&0.25&0.095 \tabularnewline
GINI&0.45&0.46&0.008&0.46&0.46&0.633 \tabularnewline
MDP&69.46&68.53&0.009&67.78&68.87&0.536 \tabularnewline
NBI&45.4&41.47&0.009&40.41&41.96&0.527 \tabularnewline
\bottomrule \addlinespace[\belowrulesep]

\end{tabular} \end{adjustbox}
\\ \parbox{\linewidth}{\scriptsize \noindent \justifying \(Notes \): All the variables are measured in 2008 (the last period before the panel used in the main analysis) but for the last four variables that are only available for 2005. Conf. 1901/1930 denotes whether the municipality experienced social conflict between 1901-1930. Distance variables are measured in kilometers. MDP is a multidimensional poverty index. Population in 1000s, value of credit in 1.000.000s COP per capita, expenditure per capita in 1.000.000s COP. The last column shows the \(p\)-value of the difference between the treatment and control groups.}
\end{table}

\renewcommand{\thefigure}{C\arabic{figure}}
\setcounter{figure}{0}
\renewcommand{\thetable}{C\arabic{table}}
\setcounter{table}{0}

\clearpage
\subsection{Alternative Mechanisms}
\label{app_alt_mechanisms}

The results in Section \ref{sec_sc} suggest that a lack of state capacity from both before the peace agreement and a lack of state entry post-ceasefire are the reasons why the large reduction in violence did not translate into economic improvements in FARC municipalities. However, there might be alternative mechanisms that could potentially explain this set of results as well. In this Section, I present evidence on some of these alternative mechanisms that suggest they are unlikely to drive the results.

\subsubsection{Migration of Venezuelans}

The worsening humanitarian situation in Venezuela due to the authoritarian regimes of Hugo Chávez and Nicolás Maduro has led to large numbers of Venezuelans migrating to Colombia since 2014. The UN estimates that about 1.8 million Venezuelans have migrated to Colombia until 2020. These large inflows of forced migrants could negatively impact local labour markets, with \citet{bahar2021give} showing that larger inflows of Venezuelan migrants reduced the employment of Colombian workers. Thus, if Venezuelan migrants were disproportionately located in FARC municipalities, as these were more attractive due to the reduction in violence, then this might explain the lack of economic improvements after the ceasefire.

Looking at Table \ref{venMig_CEDEExt_p60}, this does not seem to be the case. The Table shows the number of Venezuelan migrants that obtained a \textit{Permiso Especial de Permanencia} (PEP, a formal temporary migratory status that gives migrants the legal right to work and to access basic public services) and registered in a FARC/ELN municipality in a given year between 2017 and 2019 (first three columns) and the cumulative number of migrants that have registered in a FARC/ELN municipality since 2017 (last three columns). Unfortunately, data on the universe of Venezuelan migrants is unavailable. Thus, the numbers underestimate the true number of Venezuelan migrants in these municipalities. 

\begin{table}[h!] \centering
\newcolumntype{R}{>{\raggedleft\arraybackslash}X}
\newcolumntype{L}{>{\raggedright\arraybackslash}X}
\newcolumntype{C}{>{\centering\arraybackslash}X}

\caption{Venezuelan Migrants with Permits -- Extensive Margin, Events in Over 60\% of Years}
\label{venMig_CEDEExt_p60}
\begin{adjustbox}{max width=\linewidth, max height=\textwidth}\begin{tabular}{c|ccc|ccc}

\toprule
& \multicolumn{3}{c|}{\textbf{Flow}} & \multicolumn{3}{c}{\textbf{Stock}} \tabularnewline & FARC & ELN & p-value Diff. & FARC & ELN & p-value Diff. \tabularnewline 
{Year}&{(1)}&{(2)}&{(3)}&{(4)}&{(5)}&{(6)} \tabularnewline
\midrule \addlinespace[\belowrulesep]
\midrule 2017&1.41&2.02&&1.41&2.02& \tabularnewline
2018&17.51&29.29&&18.9&31.27& \tabularnewline
2019&1.78&1.92&&20.41&32.17& \tabularnewline
&&&&&& \tabularnewline
\textbf{\textbf{All Years}}&6.9&11.08&0.088&13.57&21.82&0.014 \tabularnewline
\bottomrule \addlinespace[\belowrulesep]

\end{tabular} \end{adjustbox}
\\ \parbox{\linewidth}{\scriptsize \justifying \noindent \(Notes\): Venezuelan migrants per 10.000 inhabitants. The first three columns show the number of Venezuelan migrants who received a special permit to live and work in Colombia (PEP) and registered in a given municipality and year (flow), while the last three columns show the total number of Venezuelan migrants with the permit that have registered in a given municipality over time. }
\end{table}

The Table shows that, if anything, Venezuelan migrants were significantly more likely to locate in ELN rather than FARC municipalities. This result is unsurprising, as ELN municipalities are much closer to the border with Venezuela than FARC municipalities: ELN municipalities are, on average, 213kms away from the Venezuelan border, while FARC municipalities are 472kms away, with the difference being significant.

\subsubsection{Coca Production and Eradication}

In 2014, during the peace negotiations with the FARC and before the ceasefire, the government announced that post-agreement it would provide farmers with material incentives to switch from growing coca to other crops. \citet{prem2021rise} show that this policy announcement led to a sharp increase in coca farming in areas suitable for coca production, as farmers expected to benefit from the government's announced substitution program. Thus, an alternative explanation to the baseline results could be that the start of the ceasefire led to a shift of economic activity from the legal sector towards coca production in FARC municipalities, which would be largely missed by the set of economic indicators used in Section \ref{sec_econ}. Moreover, it could also be that the government shifted its coca eradication programs towards FARC municipalities, as these were not controlled by an insurgent group anymore, shifting labour from productive activities towards eradication programs.

However, Table \ref{altMech_coca_CEDE_Ext_p60} suggests that this does not seem to be the case. The first three columns show results using indicator variables for whether the municipalities were producing coca or whether the government had implemented a coca eradication program in a given municipality, while the following three columns show coca production and eradication per 10.000 HAs of municipality's size. FARC municipalities were not more likely to be growing coca or to have ongoing eradication programs after the start of the ceasefire relative to ELN municipalities. There is also no increase in these measures when measured in 10.000 HAs. Thus, it seems like a shift towards coca production and eradication is also not the case in the results.

\begin{table}[h!] \centering
\newcolumntype{R}{>{\raggedleft\arraybackslash}X}
\newcolumntype{L}{>{\raggedright\arraybackslash}X}
\newcolumntype{C}{>{\centering\arraybackslash}X}

\caption{DiD Analysis -- Alternative Mechanisms -- Coca Production \& Eradication: Extensive Margin, Events in Over 60\% of Years}
\label{altMech_coca_CEDE_Ext_p60}
\begin{adjustbox}{max width=\linewidth, max height=\textwidth}\begin{tabular}{l|ccc|ccc|cc}

\toprule
& \multicolumn{3}{c|}{\textbf{Indicator Variables}} & \multicolumn{3}{c|}{\textbf{Per 10.000 HAs}} & \multicolumn{2}{c}{\textbf{PNIS Program}} \tabularnewline & Coca Production & Manual Erad. & Total Erad. & Coca Production & Manual Erad. & Total Erad. & Participating & Beneficiaries \tabularnewline 
{}&{(1)}&{(2)}&{(3)}&{(4)}&{(5)}&{(6)}&{(7)}&{(8)} \tabularnewline
\midrule \addlinespace[\belowrulesep]
\midrule Ceasefire \( \times \) FARC&--0.042&--0.009&0.009&0.721&0.197&--0.000&& \tabularnewline
&(0.037)&(0.042)&(0.036)&(0.525)&(0.321)&(0.320)&& \tabularnewline
FARC&&&&&&&0.152***&118.918*** \tabularnewline
&&&&&&&(0.036)&(26.961) \tabularnewline
&&&&&&&& \tabularnewline
Treated Munic.&216&216&216&216&216&216&216&216 \tabularnewline
Control Munic.&41&41&41&41&41&41&41&41 \tabularnewline
Mean Dep. Var.&0.323&0.320&0.364&0.559&0.474&1.135&& \tabularnewline
Observations&2,827&2,827&2,827&2,827&2,827&2,827&257&257 \tabularnewline
\(R^2\)&0.857&0.628&0.696&0.639&0.264&0.382&0.024&0.019 \tabularnewline
\bottomrule \addlinespace[\belowrulesep]

\end{tabular} \end{adjustbox}
\\ \parbox{\linewidth}{\scriptsize \justifying \noindent \(Notes\): Standard errors clustered at the municipality level. For the first six columns, includes municipality and year fixed effects. Considers only years from 2009 and before 2020. The last two columns consider only 2019, after PNIS started. Mean Dep. Var.: Mean of the dependent variable in the pre-intervention period for municipalities in the sample (FARC and ELN). Defines the post period as 2015. Coca production, manual and total eradication are per 10.000 hectares. Beneficiaries is the number of PNIS beneficiaries previously working harvesting coca per 10.000 inhabitants.}
\end{table}

The actual coca-substitution program (PNIS) that the government announced in 2014 was officially signed into decree in 2017. However, the program has been plagued by logistical and administrative problems, formally starting in 2018/2019 on a smaller magnitude than initially intended. Thus, this program started too late to explain the results. The last two columns of the Table show that i) FARC municipalities (despite not having increased their coca production relative to ELN municipalities) were more likely to be selected for the program in 2019, and ii) also had significantly more beneficiaries of the economic incentives per 10.000 inhabitants than ELN municipalities.

\subsubsection{Credit Constraints}

FARC municipalities tend to be primarily rural, distant municipalities where agriculture plays a significant role. Thus, an alternative explanation could be that FARC municipalities could not reap the economic benefits because, despite the large reduction in violence, agricultural farmers were credit-constrained and thus could not finance their projects. 

However, this seems unlikely to be the case. First, \citet{de2021forgone} show using detailed data from the largest public bank serving rural producers in Colombia (the Banco Agrario) and a similar identification strategy as here, that the number of business loans increases in FARC municipalities after the peace agreement (but not after the start of the ceasefire). Table \ref{altMech_credit_CEDE_Ext_p60} shows a similar pattern. It shows the value of credit given by two large credit-providing institutions focused on rural agricultural projects, the Banco Agrario and FINAGRO, to agricultural producers of different sizes. As is the case in \citet{de2021forgone}, credit to small agricultural producers (those most likely to be credit-constrained) increased significantly after the signing of the peace agreement (but not after the start of the ceasefire). Thus, credit constraints seemed to be eased for those more likely suffering from them in FARC municipalities after the peace agreement was ratified. 

\begin{table}[h!] \centering
\newcolumntype{R}{>{\raggedleft\arraybackslash}X}
\newcolumntype{L}{>{\raggedright\arraybackslash}X}
\newcolumntype{C}{>{\centering\arraybackslash}X}

\caption{DiD Analysis -- Alternative Mechanisms -- Credit to Agricultural Producers: Extensive Margin, Events in Over 60\% of Years}
\label{altMech_credit_CEDE_Ext_p60}
\begin{adjustbox}{max width=\linewidth, max height=\textwidth}\begin{tabular}{l|cccc|cccc}

\toprule
& \multicolumn{4}{c|}{\textbf{FINAGRO Value of Credit}} & \multicolumn{4}{c}{\textbf{Banco Agrario Value of Credit}}  \tabularnewline & Small & Medium & Large & Total Credit & Small & Medium & Large & Total Credit \tabularnewline 
{}&{(1)}&{(2)}&{(3)}&{(4)}&{(5)}&{(6)}&{(7)}&{(8)} \tabularnewline
\midrule \addlinespace[\belowrulesep]
\midrule Ceasefire \( \times \) FARC&0.000&0.008&0.027&0.035&0.007&0.008&0.028&0.043 \tabularnewline
&(0.007)&(0.007)&(0.034)&(0.038)&(0.009)&(0.007)&(0.035)&(0.038) \tabularnewline
Agreement \( \times \) FARC&0.032**&0.007&0.002&0.041&0.029*&0.001&--0.014&0.016 \tabularnewline
&(0.014)&(0.011)&(0.031)&(0.042)&(0.016)&(0.011)&(0.030)&(0.043) \tabularnewline
&&&&&&&& \tabularnewline
Treated Munic.&216&216&216&216&216&216&216&216 \tabularnewline
Control Munic.&41&41&41&41&41&41&41&41 \tabularnewline
Mean Dep. Var.&0.127&0.095&0.056&0.278&0.110&0.088&0.045&0.244 \tabularnewline
Observations&2,313&2,313&2,313&2,313&2,827&2,827&2,827&2,827 \tabularnewline
\(R^2\)&0.876&0.823&0.671&0.721&0.775&0.749&0.545&0.603 \tabularnewline
\bottomrule \addlinespace[\belowrulesep]

\end{tabular} \end{adjustbox}
\\ \parbox{\linewidth}{\scriptsize \justifying \noindent \(Notes\): Standard errors clustered at the municipality level. Includes municipality and year fixed effects. Considers only years from 2009 and before 2020. Mean Dep. Var.: Mean of the dependent variable in the pre-intervention period for municipalities in the sample (FARC and ELN). Defines the post-ceasefire period as 2015 and 2016, and the post-agreement period as the years from 2017. Value of credit to small, medium and large agricultural producers, in COP million per capita, from two different credit-giving organizations, FINAGRO and the Banco Agrario. }
\end{table}

\subsubsection{Corruption}

Another concern could be that the government did set aside funds to help FARC municipalities but corruption at the local level led to the funds being diverted to corrupt politicians or spent in non-productive investments. There are two pieces of evidence to suggest that this is not the case. First, as shown in Table \ref{DID_SC_CEDE_Ext_p60}, government transfers did not increase significantly after the start of the ceasefire in FARC municipalities (even if taken at face value, the coefficient suggests an increase of only 2.000 COP per person in FARC municipalities, less than half a dollar). Thus, it is unlikely that there were more funds to steal to begin with. Second, in Table \ref{altMech_corr_CEDE_Ext_p60} I analyse whether cases of corruption (from the General Attorney's Office, first two columns) or disciplinary actions against local government officials (from the national watchdog agency, the Procuraduría General de la Nación, last four columns) increased in FARC municipalities after the ceasefire. Panel A shows the results per 10.000 inhabitants, while Panel B uses a dummy for whether a case was opened in a given municipality. While these are only imperfect measures of corruption, they do show that neither corruption cases nor disciplinary actions increased in FARC municipalities after the ceasefire, regardless of whether one looks at the total cases (columns 2 and 6) or at more disaggregated offences (including monetary offences in column 3). Thus, it is unlikely that the lack of economic benefits post-ceasefire is simply due to stolen funds or corruption in general.

\begin{table}[h!] \centering
\newcolumntype{R}{>{\raggedleft\arraybackslash}X}
\newcolumntype{L}{>{\raggedright\arraybackslash}X}
\newcolumntype{C}{>{\centering\arraybackslash}X}

\caption{DiD Analysis -- Alternative Mechanisms -- Corruption: Extensive Margin, Events in Over 60\% of Years}
\label{altMech_corr_CEDE_Ext_p60}
\begin{adjustbox}{max width=\linewidth, max height=\textwidth}\begin{tabular}{l|cc|cccc}

\toprule
& \multicolumn{2}{c|}{\textbf{Corruption Cases}} & \multicolumn{4}{c}{\textbf{Disciplinary Actions}} \tabularnewline & Public & Total & Economic & Local Gov. & Serious & Total \tabularnewline & Administrators & Offenses & Offenses & Officials & Offenses & Offenses \tabularnewline 
{}&{(1)}&{(2)}&{(3)}&{(4)}&{(5)}&{(6)} \tabularnewline
\midrule \addlinespace[\belowrulesep]
\midrule \textbf{Panel A. Per 10.000 People}&&&&&& \tabularnewline
Ceasefire \( \times \) FARC&--0.014&--0.004&0.012&0.027&--0.016&--0.011 \tabularnewline
&(0.028)&(0.029)&(0.016)&(0.044)&(0.065)&(0.067) \tabularnewline
&&&&&& \tabularnewline
Treated Munic.&216&216&216&216&216&216 \tabularnewline
Control Munic.&41&41&41&41&41&41 \tabularnewline
Mean Dep. Var.&0.037&0.039&0.039&0.214&0.471&0.500 \tabularnewline
Observations&2,827&2,827&2,827&2,827&2,827&2,827 \tabularnewline
\(R^2\)&0.146&0.148&0.093&0.152&0.216&0.239 \tabularnewline
&&&&&& \tabularnewline
\midrule \textbf{Panel B. Dummy}&&&&&& \tabularnewline
Ceasefire \( \times \) FARC&--0.008&0.004&--0.000&0.021&0.040&0.043 \tabularnewline
&(0.027)&(0.027)&(0.015)&(0.034)&(0.042)&(0.043) \tabularnewline
&&&&&& \tabularnewline
Treated Munic.&216&216&216&216&216&216 \tabularnewline
Control Munic.&41&41&41&41&41&41 \tabularnewline
Mean Dep. Var.&0.079&0.086&0.061&0.242&0.442&0.457 \tabularnewline
Observations&2,827&2,827&2,827&2,827&2,827&2,827 \tabularnewline
\(R^2\)&0.293&0.293&0.135&0.230&0.330&0.332 \tabularnewline
\bottomrule \addlinespace[\belowrulesep]

\end{tabular} \end{adjustbox}
\\ \parbox{\linewidth}{\scriptsize \justifying \noindent \(Notes\): Standard errors clustered at the municipality level. Includes municipality and year fixed effects. Considers only years from 2009 and before 2020 Mean Dep. Var.: Mean of the dependent variable in the pre-intervention period for municipalities in the sample (FARC and ELN). Outcome variables in Panel A are per 10.000 inhabitants, while those in Panel B are dummies. Data on corruption cases come from the General Attorney's Office while that on disciplinary actions come from the national watchdog agency (Procuraduria General de la Nacion).}
\end{table}

\subsubsection{Shift Towards Education}

An alternative mechanism could be that the end of the FARC allowed young individuals to return to school and pursue an education, as they no longer needed to worry about the conflict or work for the FARC. This could have led young individuals to shift from working in productive activities to getting an education in the short run, potentially explaining the observed lack of economic improvements. 

The results in Table \ref{altMech_educ_CEDE_Ext_p60} suggest this is not the case. It shows that the number of first-year students in higher education programs (a measure of entry to higher education) in FARC municipalities did not increase significantly more than in ELN municipalities after the start of the ceasefire, regardless of the type of educational program considered (technical education programs, bachelors programs or post-graduate programs). There has also been no significant change in the number of higher education institutions in FARC municipalities, so it does not seem like a shift from production toward education is driving the results. 

\begin{table}[h!] \centering
\newcolumntype{R}{>{\raggedleft\arraybackslash}X}
\newcolumntype{L}{>{\raggedright\arraybackslash}X}
\newcolumntype{C}{>{\centering\arraybackslash}X}

\caption{DiD Analysis -- Alternative Mechanisms -- Higher Education: Extensive Margin, Events in Over 60\% of Years}
\label{altMech_educ_CEDE_Ext_p60}
\begin{adjustbox}{max width=\linewidth, max height=\textwidth}\begin{tabular}{lcccc}

\toprule
& \textbf{Number of}  & \multicolumn{3}{c}{\textbf{First-Year Students}} \tabularnewline & Learning Inst. & Technical & University & Post Grad \tabularnewline 
{}&{(1)}&{(2)}&{(3)}&{(4)} \tabularnewline
\midrule \addlinespace[\belowrulesep]
\midrule Ceasefire \( \times \) FARC&0.151&--0.061&1.766&--0.027 \tabularnewline
&(0.123)&(1.416)&(1.573)&(0.100) \tabularnewline
&&&& \tabularnewline
Treated Munic.&216&216&216&216 \tabularnewline
Control Munic.&41&41&41&41 \tabularnewline
Mean Dep. Var.&0.867&6.638&6.939&0.084 \tabularnewline
Observations&2,827&2,827&2,827&2,827 \tabularnewline
\(R^2\)&0.435&0.472&0.803&0.385 \tabularnewline
\bottomrule \addlinespace[\belowrulesep]

\end{tabular} \end{adjustbox}
\\ \parbox{\linewidth}{\scriptsize \justifying \noindent \(Notes\): Standard errors clustered at the municipality level. Includes municipality and year fixed effects. Considers only years from 2009 and before 2020. Mean Dep. Var.: Mean of the dependent variable in the pre-intervention period for municipalities in the sample (FARC and ELN). Defines the post period as 2015. Educational variables are per 10.000 inhabitants, and come from the Ministry of Education. Technical is the sum of students in technical and technological educational programs. University is the sum of students in undergraduate programs. Post grad is the sum of students enrolled in masters', specializations or PhD programs. First column shows the number of higher learning institutions in the municipality.}
\end{table}

\subsubsection{Productive Land Tied-Up in Land Restitution Processes}

Disputes over land ownership have been at the centre of the armed conflict in Colombia. For example, \citet{lopez2017agrarian} show that the historical dispossession of peasants' lands by landlords between 1914 and 1946 is associated with FARC presence in the early stages of the conflict (1974-1985). Recognising this, and as a first signal of its willingness to achieve peace, the government signed in 2011 the ``Victims and Land Restitution Law'', which provided the legal framework for conflict victims to obtain assistance and reparations from the government and a mechanism for victims to recover the lands they had lost during the conflict. 

Yet another reason for the results could be that FARC municipalities could not benefit economically because the productive land in those municipalities was either stolen by the FARC and inaccessible to farmers or tied up in the courts set up to handle land restitution processes. If this were the case, it would be expected to see large increases in land restitution claims and cases in court in FARC municipalities after the start of the ceasefire. The land restitution process proceeds in three stages: first, a person or group presents a land restitution claim to the government unit in charge of the land restitution process (UAEGRTD). In the second stage, the UAEGRTD decides whether to bring the case to the land restitution courts. Lastly, the court decides whether to restitute the land or not. The process has been criticised for its slowness: until 2019, out of the 120.000+ submitted land restitution requests, less than 10\% had been resolved (see \href{https://www.dw.com/es/restituci\%C3\%B3n-de-tierras-en-colombia-una-deuda-hist\%C3\%B3rica/a-50288146}{\textit{Deutsche Welle, 2019}}). 

However, the data do not support these hypotheses. Table \ref{altMech_landRest_CEDE_Ext_p60} shows the results of estimating Equation \eqref{eq_did} on measures related to the land restitution process. The first three columns correspond to the first step in the restitution process and show the number of requests, people and plots involved in claims brought forward to the UAEGRTD. Columns 4 and 5 show the number of requests solved and denied by the land restitution courts. The last three columns show the number of beneficiaries, plots and the total plot size of the court-approved claims. If anything, it seems like, relative to ELN municipalities, there have been fewer land restitution claims presented and resolved in FARC municipalities after the start of the ceasefire (columns 1-4), regardless of whether one measures these in per capita terms (Panel A) or not (Panel B), although the coefficients are insignificant. When looking at the requests the courts have approved, the coefficients on the number of beneficiaries, plots returned, or total area restituted are mostly negative and insignificant (columns 6-8). Thus, this alternative explanation seems unlikely to be behind the results. 

\begin{table}[h!] \centering
\newcolumntype{R}{>{\raggedleft\arraybackslash}X}
\newcolumntype{L}{>{\raggedright\arraybackslash}X}
\newcolumntype{C}{>{\centering\arraybackslash}X}

\caption{DiD Analysis -- Alternative Mechanisms -- Land Restitution: Extensive Margin, Events in Over 60\% of Years}
\label{altMech_landRest_CEDE_Ext_p60}
\begin{adjustbox}{max width=\linewidth, max height=\textwidth}\begin{tabular}{l|ccc|cc|ccc}

\toprule
& \multicolumn{3}{c|}{\textbf{Land Restitution Requests}} & \multicolumn{2}{c|}{\textbf{Court Decision}} & \multicolumn{3}{c}{\textbf{Restitutions}}  \tabularnewline & Requests & Benefic. & Plots & Solved & Denied & Benefic. & Plots & Size Plots \tabularnewline 
{}&{(1)}&{(2)}&{(3)}&{(4)}&{(5)}&{(6)}&{(7)}&{(8)} \tabularnewline
\midrule \addlinespace[\belowrulesep]
\midrule \textbf{Panel A. Per Capita}&&&&&&&& \tabularnewline
Ceasefire \( \times \) FARC&--3.378&--1.665&--2.397&--0.113&--0.056&0.247&--0.082&--0.235 \tabularnewline
&(2.311)&(1.680)&(2.084)&(0.401)&(0.037)&(0.950)&(0.251)&(0.167) \tabularnewline
&&&&&&&& \tabularnewline
Treated Munic.&216&216&216&216&216&216&216&216 \tabularnewline
Control Munic.&41&41&41&41&41&41&41&41 \tabularnewline
Mean Dep. Var.&21.309&16.004&19.136&0.500&0.020&1.033&0.344&0.206 \tabularnewline
Observations&2,313&2,313&2,313&2,056&2,056&2,056&2,056&2,056 \tabularnewline
\(R^2\)&0.566&0.618&0.585&0.472&0.273&0.475&0.478&0.472 \tabularnewline
&&&&&&&& \tabularnewline
\midrule \textbf{Panel B. Total}&&&&&&&& \tabularnewline
Ceasefire \( \times \) FARC&--5.707&--3.623&--4.190&--0.925&--0.170*&--1.123&--0.471&--14.032 \tabularnewline
&(4.357)&(3.322)&(3.796)&(0.902)&(0.101)&(2.374)&(0.590)&(11.187) \tabularnewline
&&&&&&&& \tabularnewline
Treated Munic.&216&216&216&216&216&216&216&216 \tabularnewline
Control Munic.&41&41&41&41&41&41&41&41 \tabularnewline
Mean Dep. Var.&38.137&29.154&34.364&1.157&0.032&2.920&0.860&30.783 \tabularnewline
Observations&2,313&2,313&2,313&2,056&2,056&2,056&2,056&2,056 \tabularnewline
\(R^2\)&0.528&0.520&0.541&0.521&0.412&0.516&0.549&0.590 \tabularnewline
\bottomrule \addlinespace[\belowrulesep]

\end{tabular} \end{adjustbox}
\\ \parbox{\linewidth}{\scriptsize \justifying \noindent \(Notes\): Standard errors clustered at the municipality level. Includes municipality and year fixed effects. Considers only years from 2011 (for first three columns) or 2012 (remaining columns) and before 2020. Mean Dep. Var.: Mean of the dependent variable in the pre-intervention period for municipalities in the sample (FARC and ELN). Outcome variables in Panel A are in either 10.000 inhabitants (first seven columns) or in 10.000 of HAs (last column), while those in Panel B are in totals. Data come from the Unit of Land Restitution.}
\end{table}

\subsubsection{Beliefs About the Peace Agreement}

The Democracy Observatory (Observatorio de la Democracia) at the Universidad de los Andes has been conducting surveys in Colombia as part of the broader AmericasBarometer initiative since 2004. Broadly speaking, these surveys aim to study the attitudes, experiences, values and beliefs of people across the Americas about their political institutions. While the survey contains a set of questions that are asked in every country and across time, it also includes country-specific questions. Most of the questions specific to Colombia are related to the armed conflict and the peace agreement with the FARC. On top of the AmericasBarometer surveys, the Democracy Observatory conducted three surveys focused on areas affected by the armed conflict in 2013, 2015, and 2017. The data, as well as the survey instruments, can be found \href{https://obsdemocracia.org/encuestas/}{\textit{here}}.

I combine the data from all the AmericasBarometer and the special conflict-areas surveys since 2011 to shed light on the opinions and beliefs of citizens of FARC municipalities towards the peace agreement. One concern could be that these areas did not experience economic improvements because their residents did not believe in the peace agreement, or they thought the situation would not change and, therefore, avoided making any economic investments. While these data can be useful for these purposes, it is worth emphasising that they have clear limitations. For each survey wave, only 35 to 103 municipalities are surveyed, with on average (median) 58 (36) individuals surveyed per municipality-year pair. Among FARC (ELN) municipalities, 70 out of 216 (10 out of 41) are surveyed at least once between 2011 and 2019, with 35 (9) being surveyed at most twice. While the data are meant to be representative at the municipality level, it is clear that there are limitations in terms of geographical coverage. On the other hand, these are the only surveys representative at the municipality level that are 1) conducted regularly and 2) outside the main cities of the country, and thus can provide a useful snapshot of the opinions and beliefs of people in FARC municipalities. Given the restricted sample of surveyed ELN municipalities, I focus mostly on FARC municipalities and the rest of the country.

\paragraph{Did residents of FARC municipalities believe the peace agreement would be beneficial?} First, in 2017, 2018 and 2019, people were asked whether the implementation of the peace agreement was going to improve the economic situation of their municipality on a scale from 1 (``strongly disagree'') to 7 (``strongly agree''). The share agreeing with this statement (score of 5 or above) in FARC municipalities has been constantly rising over time, from 44\% in 2017 to 47\% in 2018 to 51\% in 2019. The same pattern holds when asked whether it would improve the security situation in their municipality (going from 46\% in 2017 to 56\% in 2019) and access to land for farmers (57\% in 2017 to 64\% in 2019). Second, even in 2013, residents of FARC municipalities were 12\% more likely than people in the rest of the country to say that they believed that the end of the FARC would bring economic benefits to their municipalities. Third, when asked in 2019 if they agreed with statements saying that two key parts of the peace agreement (the PDET program and the additional seats in congress for conflict-affected areas) would benefit people like them, 59\% and 66\% of respondents agreed, higher shares than in the rest of the country. Thus, this suggests that residents of FARC municipalities believed that the peace agreement would benefit them, becoming more optimistic over time about the agreement's value. 

\paragraph{Were residents of FARC municipalities in favour of the agreement?} The answer is yes. First, residents of FARC municipalities have always been more likely to believe that the best way to end the armed conflict with guerrilla groups was through negotiations rather than military action. Before (after) the start of the ceasefire, 65\% (78\%) of citizens in FARC municipalities believed this to be the best option, compared with 57\% (72\%) in the rest of the country. This also holds for 2011, before it was leaked that the government was negotiating with the FARC. Second, the 2016 plebiscite on the final peace agreement received 50,3\% of votes in favour in FARC municipalities (i.e. would have passed), while in the rest of the country it was rejected, receiving only 49,5\% of votes in favour. Third, since 2013, the surveys have included a question asking citizens how supportive they are of the peace agreement (both before and after the actual agreement was completed). The majority of citizens in FARC municipalities have agreed with the idea of an agreement or the actual agreement in all years but one, with their support above that in the rest of the country in 5 out of 7 years for which data exist.\footnote{Across all years, 56\% of citizens in FARC municipalities and 51\% in the rest of the country have been supportive of the peace agreement. Before (after) the start of the ceasefire, these numbers were 59\% and 55\% (56\% and 50\%), respectively.} Moreover, the share in support of the agreement in FARC municipalities has been steadily increasing since 2016. The majority of people in FARC municipalities have believed (both before and after the ceasefire) that people can forgive and reconcile with former FARC fighters, higher than in the rest of the country. 

There was also broad support for different components of the peace agreement. In 2019, people were asked what percentage of people in their municipality they believed supported the implementation of the peace agreement and the Rural Reform, a key part of the agreement. In FARC municipalities, 66\% of respondents stated that they believed that at least half of their municipality supported the implementation of the Rural Reform. In contrast, 73\% believed the same to be true for the implementation of the full peace agreement, both higher than in the rest of the country. Panel A of Table \ref{suppPeaceComp} shows that, in general, citizens of FARC municipalities were more likely to agree with key elements of the peace agreement relative to people in the rest of the country. Most citizens in FARC municipalities agree with most of these components, and unpopular components fare significantly better in FARC municipalities than in the rest of the country. Interestingly, row 12 shows that, compared to citizens in other municipalities, the proposal to change the original agreement after the plebiscite was rejected was less popular in FARC municipalities. Moreover, between 2013 and 2015, when asked for their level of support for hypothetical policies similar to the ones in the actual agreement (such as penalty reductions, political participation, and so on), people in FARC municipalities were significantly more likely to support these than people in the rest of the country. However, the overall level of support was low. 

\begin{table}[h!] \centering
\newcolumntype{R}{>{\raggedleft\arraybackslash}X}
\newcolumntype{L}{>{\raggedright\arraybackslash}X}
\newcolumntype{C}{>{\centering\arraybackslash}X}

\caption{Support and Knowledge of Peace Agreement}
\label{suppPeaceComp}
\begin{adjustbox}{max width=\linewidth, max height=\textwidth}\begin{tabular}{lcccc}

\toprule
  & FARC & Rest Country & p-value & N Obs. \tabularnewline 
{}&{(1)}&{(2)}&{(3)}&{(4)} \tabularnewline
\midrule \addlinespace[\belowrulesep]
\midrule \textbf{Panel A. Support of Agreement's Components}&&&& \tabularnewline
1. Former FARC members form a political party&3.654&3.431&0&7240 \tabularnewline
2. Former FARC members run in elections&3.203&3.124&0.56&2927 \tabularnewline
3. Would accept results if FARC member wins&61.6&45.2&0&3880 \tabularnewline
4. 16 add. seats for most-affected municipalities&69.1&66&0.03&5116 \tabularnewline
5. Redistribution of land to poor farmers&81.4&81.6&0.94&2778 \tabularnewline
6. Concentration of FARC members in certain areas&34.7&30.9&0.27&1394 \tabularnewline
7. Crop substitution programs in municipality&83.4&85.2&0.39&1341 \tabularnewline
8. No jail for demobilized foot soldiers&24.8&19.8&0&2739 \tabularnewline
9. 5--8 years in jail for confessed atrocious crimes&57.6&55.4&0.53&1416 \tabularnewline
10. Over 8 years in jail for not confessed crimes&71.7&72.2&0.86&1411 \tabularnewline
11. PDET in most affected regions&76&72.8&0.05&3187 \tabularnewline
12. Changes to original agreement&45.9&46.7&0.64&4631 \tabularnewline
&&&& \tabularnewline
\midrule \textbf{Panel B. Knowledge of Agreement}&&&& \tabularnewline
1. Is this municipality a PDET?&68.4&56.8&0&1751 \tabularnewline
2. Is creation of 16 seats part of agreement?&71.7&67&0.02&2092 \tabularnewline
3. What is the max. jail penalty for FARC member?&27.4&18.3&0&1646 \tabularnewline
\bottomrule \addlinespace[\belowrulesep]

\end{tabular} \end{adjustbox}
\\ \parbox{\linewidth}{\scriptsize \justifying \noindent \(Notes\): Each row corresponds to a different component of the peace agreement. Columns 1 and 2 show the number of respondents in FARC and rest of the country municipalities that in general agree with a given component (indicated by a score higher than 5 in a 1-7 Likert-scaled question) or on a 1-10 scale in Panel A, and the number of respondents that correctly answered general questions regarding the peace agreement in Panel B. The third column shows the p-value of the difference between columns 1 and 2 (using robust standard errors). The fourth column shows the number respondents. }
\end{table}

\paragraph{Did residents of FARC municipalities understand the agreement?} One concern with the evidence so far is that it could be that people in FARC municipalities did not know the details of the peace agreement, and thus, these answers reflect a lack of understanding of what was agreed. However, this seems unlikely to be the case. Panel B of Table \ref{suppPeaceComp} shows the proportion of citizens in FARC/rest of the country municipalities that correctly answered three different questions regarding the peace agreement: whether their municipality has been designated as a PDET municipality, whether creating 16 seats for conflict-affected areas was part of the peace agreement, and what the maximum amount of years a FARC member could be sentenced to jail under the peace agreement is (8 years). Citizens in FARC municipalities are significantly more likely to get these questions right than citizens in the rest of the country, and the overall rate of correct answers is high. Moreover, Table \ref{compSurvMun_CEDEExt_p60} shows that, while surveyed and non-surveyed FARC municipalities are not identical (non-surveyed ones tend to be smaller, more rural and slightly poorer), they are fairly similar. 

Overall, the results in this Section show that citizens in FARC municipalities 1) supported the peace agreement and its key components, 2) believed that the peace agreement was going to benefit them and their municipalities (also economically), and 3) were knowledgeable regarding the content of the peace agreement, in general, more so than citizens in the rest of the country. Thus, uncertainty around, opposition against, or lack of trust in the peace agreement seem unlikely to explain why FARC municipalities did not benefit economically from the peace agreement. 

\begin{table}[h!] \centering
\newcolumntype{C}{>{\centering\arraybackslash}X}

\caption{Comparison Surveyed vs. Non-Surveyed Municipalities (Population Weighted): Extensive Margin, Events in Over 60\% of Years}
\label{compSurvMun_CEDEExt_p60}
\begin{adjustbox}{max width=\linewidth, max height=\textwidth}\begin{tabular}{lcccc}

\toprule
& \multicolumn{3}{c}{\textbf{FARC Municipalities}} & \textbf{p-value} \tabularnewline & All & Not Surveyed & Surveyed & Difference \tabularnewline
{Variable}&{(1)}&{(2)}&{(3)}&{(4)} \tabularnewline
\midrule \addlinespace[\belowrulesep]
\midrule Population&25.39&19.13&38.47&0 \tabularnewline
Urban Population&0.41&0.38&0.47&0 \tabularnewline
Area&1842.32&1498&2560.5&0.04 \tabularnewline
Distance Capital&77.75&76.26&79.30&0.70 \tabularnewline
Distance Bogota&334.58&272.10&399.42&0 \tabularnewline
Gov. Transfers&0.02&0.03&0.01&0 \tabularnewline
Savings Capac.&35.90&36.34&35.41&0.61 \tabularnewline
\%Exp. in Investment&86.75&86.87&86.62&0.74 \tabularnewline
Fiscal Perf.&60.11&60.38&59.84&0.48 \tabularnewline
Overall Perf.&58.11&62.45&53.59&0 \tabularnewline
Mun. Develop.&50.36&51.04&49.66&0.35 \tabularnewline
Conflict 1901/30&0.20&0.17&0.25&0.13 \tabularnewline
Spanish Occup.&0.30&0.25&0.34&0.15 \tabularnewline
PC Expenditure&0.25&0.25&0.23&0.19 \tabularnewline
GINI&0.46&0.46&0.46&0.04 \tabularnewline
MDP&67.75&66.52&69.01&0.27 \tabularnewline
NBI&44.02&40.72&47.45&0.01 \tabularnewline
\# of Municipalities&216&146&70& \tabularnewline
\bottomrule \addlinespace[\belowrulesep]

\end{tabular} \end{adjustbox}
\\ \parbox{\linewidth}{\scriptsize \justifying \noindent \(Notes\): All the variables are measured in 2008 (the last period before the panel used in the main analysis) but for the last five variables that are only available for 2005 and 2016. Conf. 1901/1930 denotes whether the municipality experienced social conflict between 1901-1930. Distance variables are measured in kilometers. MDP is a multidimensional poverty index. Population in 1000s, expenditure per capita in 1.000.000s COP. The fourth column shows the p-value of the difference between surveyed and non-surveyed FARC municipalities.}
\end{table}


\renewcommand{\thefigure}{D\arabic{figure}}
\setcounter{figure}{0}
\renewcommand{\thetable}{D\arabic{table}}
\setcounter{table}{0}

\clearpage
\subsection{Additional Robustness Checks}
\label{app_addRobChecks}

In this Section, I present results of additional robustness checks to the baseline results. For brevity, I focus only on the results for the main (Anderson) Indices for violence, economic activity and state capacity. 

\begin{figure}[h!]
\begin{center}
\caption{Permutation Test -- Extensive Margin}
\label{perm_test}
\includegraphics[width=\textwidth]{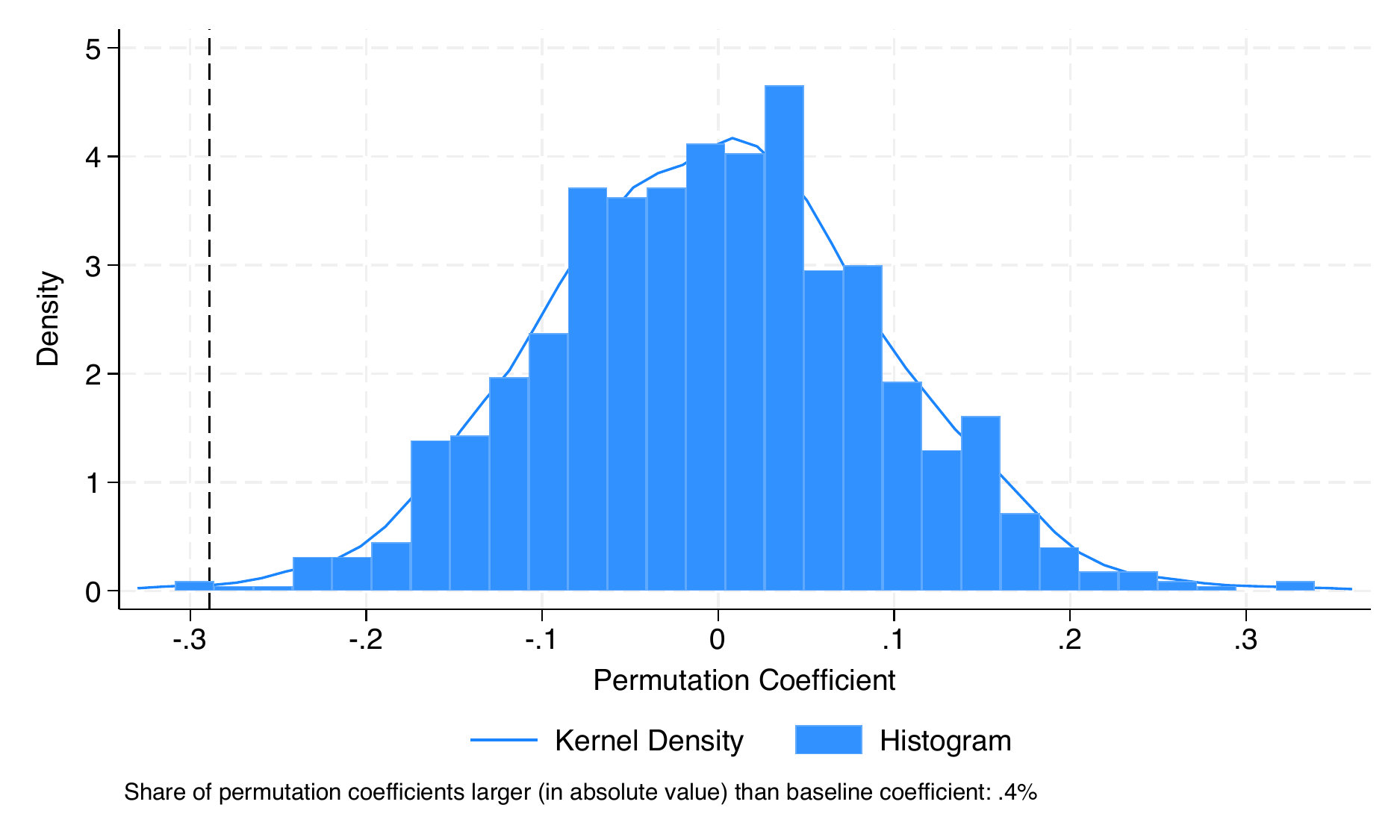}
\end{center}
\justifying 
\footnotesize{\textbf{Notes:} The Figure shows the distribution of estimates of Equation \eqref{eq_did} for the violence (Anderson) index, using 1000 randomly-created treatment and control groups (in the same proportion as in the original groups). The index is created following \citet{anderson2008multiple} and is based on the violence measures in Table \ref{DID_final_Vio_CEDE_Ext_p60}. The red line corresponds to the baseline results.}
\end{figure}

In Figure \ref{perm_test}, I assess how likely it is to observe the significant and sizeable negative violence results by chance, in a type of permutation test. \footnote{I focus only on the violence results because the others are insignificant.} More specifically, I randomly assign all municipalities in Colombia (but for those I classified as both FARC \textit{and} ELN, which I excluded from my original analysis) to a treatment and control group, each containing the same number of municipalities as the original treatment and control groups. I then estimate the usual diff-in-diff regression on the Anderson violence index. The Figure displays the distribution of coefficients from this exercise, with 1000 repetitions. Only in 0.4\% of repetitions do I get a coefficient larger in magnitude (in absolute value) than the original one, suggesting that this is not simply a chance event.

Table \ref{robCheck_CEDEExt_p60} shows the results of performing different robustness checks to the baseline results, focusing on the summary indices for brevity. Overall, the results align with the baseline ones: regardless of the specification, the effect on violence is significant, large, and negative, while the effects on economic activity and state capacity are insignificant and precisely-estimated. Thus, it is unlikely that this shows a problem with those results.

\begin{table}[tbp] \centering
\newcolumntype{R}{>{\raggedleft\arraybackslash}X}
\newcolumntype{L}{>{\raggedright\arraybackslash}X}
\newcolumntype{C}{>{\centering\arraybackslash}X}

\caption{Robustness Checks, Baseline Results}
\label{robCheck_CEDEExt_p60}
\begin{adjustbox}{max width=\linewidth, max height=0.6\textwidth}\begin{tabular}{lccc}

\toprule
& \multicolumn{3}{c}{\textbf{Index}} \tabularnewline & Violence & Economic & State Capacity \tabularnewline 
{}&{(1)}&{(2)}&{(3)} \tabularnewline
\midrule \addlinespace[\belowrulesep]
\midrule \textbf{Panel A. Alt. Inference}&&& \tabularnewline
Ceasefire \( \times \) FARC&--0.289&--0.004&0.067 \tabularnewline
Baseline&(0.104)***&(0.038)&(0.054) \tabularnewline
Conley Perc. 25&(0.076)***&(0.023)&(0.049) \tabularnewline
Conley Perc. 50&(0.082)***&(0.023)&(0.049) \tabularnewline
Conley Perc. 75&(0.085)***&(0.022)&(0.054) \tabularnewline
Wild Bootstrap p-value&[0.006]***&[0.919]&[0.219] \tabularnewline
&&& \tabularnewline
Observations&2,827&2,827&2,827 \tabularnewline
&&& \tabularnewline
\midrule \textbf{Panel B. Alt. (KLK) Index}&&& \tabularnewline
Ceasefire \( \times \) FARC&--0.293***&--0.027&0.008 \tabularnewline
&(0.043)&(0.021)&(0.053) \tabularnewline
&&& \tabularnewline
Observations&2,827&2,827&2,827 \tabularnewline
&&& \tabularnewline
\midrule \textbf{Panel C. Collapsing}&&& \tabularnewline
Ceasefire \( \times \) FARC&--0.289*&--0.004&0.067 \tabularnewline
&(0.147)&(0.053)&(0.076) \tabularnewline
&&& \tabularnewline
Observations&514&514&514 \tabularnewline
&&& \tabularnewline
\midrule \textbf{Panel D. Munic. Trends}&&& \tabularnewline
Ceasefire \( \times \) FARC&--0.366***&--0.057*&--0.043 \tabularnewline
&(0.064)&(0.034)&(0.031) \tabularnewline
&&& \tabularnewline
Observations&2,827&2,827&2,827 \tabularnewline
&&& \tabularnewline
\midrule \textbf{Panel E. No AUC Munic.}&&& \tabularnewline
Ceasefire \( \times \) FARC&--0.279**&--0.003&0.069 \tabularnewline
&(0.115)&(0.041)&(0.060) \tabularnewline
&&& \tabularnewline
Observations&2,563&2,563&2,563 \tabularnewline
&&& \tabularnewline
\midrule \textbf{Panel F. Alt. Definition}&&& \tabularnewline
Ceasefire \( \times \) FARC&--0.537***&0.021&0.074 \tabularnewline
&(0.115)&(0.037)&(0.046) \tabularnewline
&&& \tabularnewline
Observations&2,266&2,266&2,266 \tabularnewline
\bottomrule \addlinespace[\belowrulesep]

\end{tabular} \end{adjustbox}
\\ \parbox{\linewidth}{\scriptsize \justifying \noindent \(Notes \): Results from estimating Equation \eqref{eq_did}. For Panels B to G, standard errors clustered at the municipality level. Considers only years from 2009 and before 2020. Defines the post period as 2015. The first column uses the summary index based on the different violence variables. The second column uses the summary index based on weighted nighttime light intensity, value added per capita (from DANE), share of urban population, agricultural productivity, formal employment and firm entry measures. The third column uses the summary index based on the state capacity measures. Panel A uses alternative inference approaches, standard errors estimated following Conley (1999), allowing for spatial autocorrelation up to the 25th/50th/75th percentile of the distance to the department's capital, and the wild cluster bootstrap p-values (in squared brackets). Panel B creates the summary indices following the approach suggested by Kling, Liebman \& Katz (2008). Panel C follows Bertrand, Duflo \& Mullainathan (2004) and collapses all pre- and post-intervention periods together. Panel D includes municipality-specific time trends, in addition to the municipality fixed-effects. Panel E excludes municipalities that had had presence of the AUC, a paramilitary group. Panel F defines a municipality as FARC (ELN) if it is classified as FARC (ELN) using both the baseline intensive and extensive margins of presence. }
\end{table}

In Panel A, I estimate standard errors that take into account spatial correlation following \citet{conley1999gmm}\footnote{Up to the \nth{25}, \nth{50} and \nth{75} percentile of the distance of municipalities to their department's capital.} and the wild cluster bootstrap $t$-statistic method suggested by \citet{cameron2008bootstrap}, with the same results. In Panel B, I use the KLK index proposed by \citet{kling2007experimental} rather than the Anderson index. In Panel C, I follow \citet{bertrand2004much} and collapse all the pre- and post-ceasefire periods together to deal with serial correlation. Unsurprisingly, given that my panel is balanced for most of the variables, the coefficients are identical to the baseline ones, and while the SEs increase marginally, the significance of the coefficients remains unchanged. In Panel D, I add municipality-specific time trends in addition to the municipality fixed effects. Here, the coefficient on the economic index is marginally significant, although negative. In Panel E, I exclude municipalities that had the presence of a paramilitary group, the AUC (5 of the ELN and 19 of the FARC municipalities are excluded). This group was one of the main adversaries of the FARC and the ELN and demobilised mostly in 2005. The coefficients are basically identical but less precisely estimated due to the smaller sample size.  In Panel F, I classify municipalities as FARC/ELN municipalities if the FARC/ELN operated there for a long time \textit{and} were very violent. More specifically, I classify a municipality as FARC (ELN) if it is classified as FARC (ELN) using both the [baseline] intensive and extensive margin measures of presence (ignoring those that are classified as both FARC and ELN), finding qualitatively similar results, if anything there is an even larger decrease in violence. 

\begin{figure}[h!]
\caption{Robustness of Anderson Indices to Individual Component Variables}
\label{rob_indices}
\centering
\begin{subfigure}[t]{0.49\textwidth}
\includegraphics[width=\textwidth]{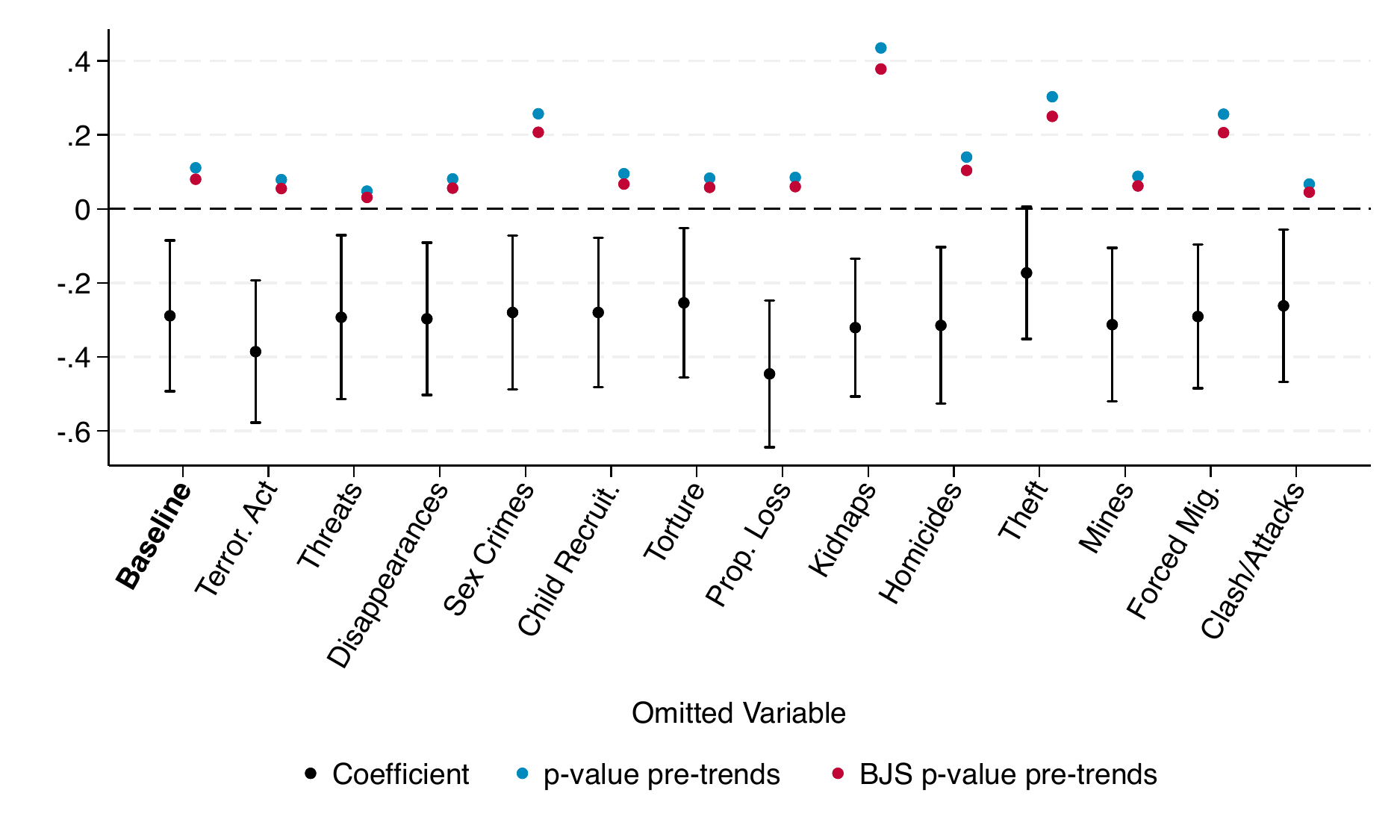}
\subcaption{Violence Outcomes}
\end{subfigure}
\begin{subfigure}[t]{0.49\textwidth}
\includegraphics[width=\textwidth]{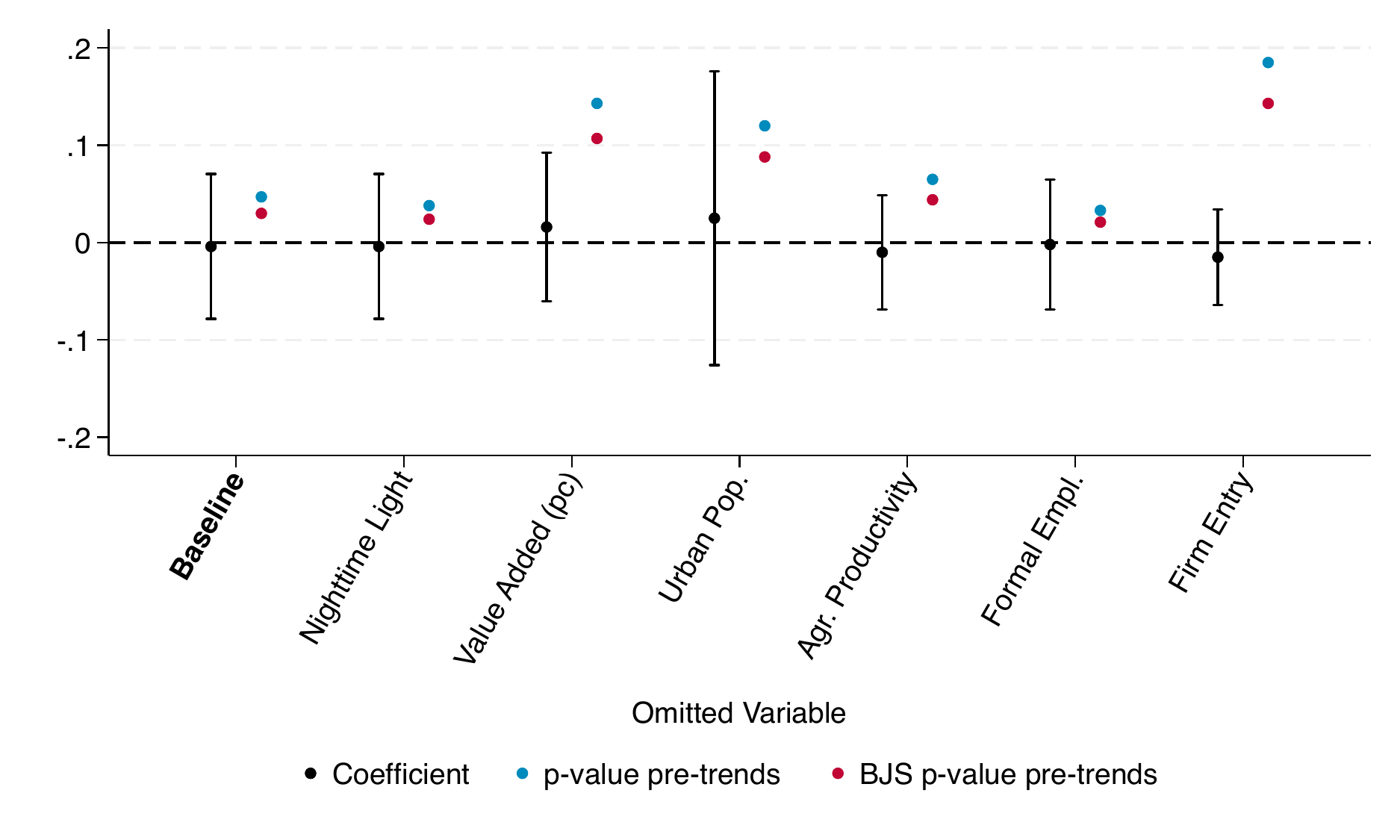}
\subcaption{Economic Outcomes}
\end{subfigure}
\begin{subfigure}[t]{0.49\textwidth}
\includegraphics[width=\textwidth]{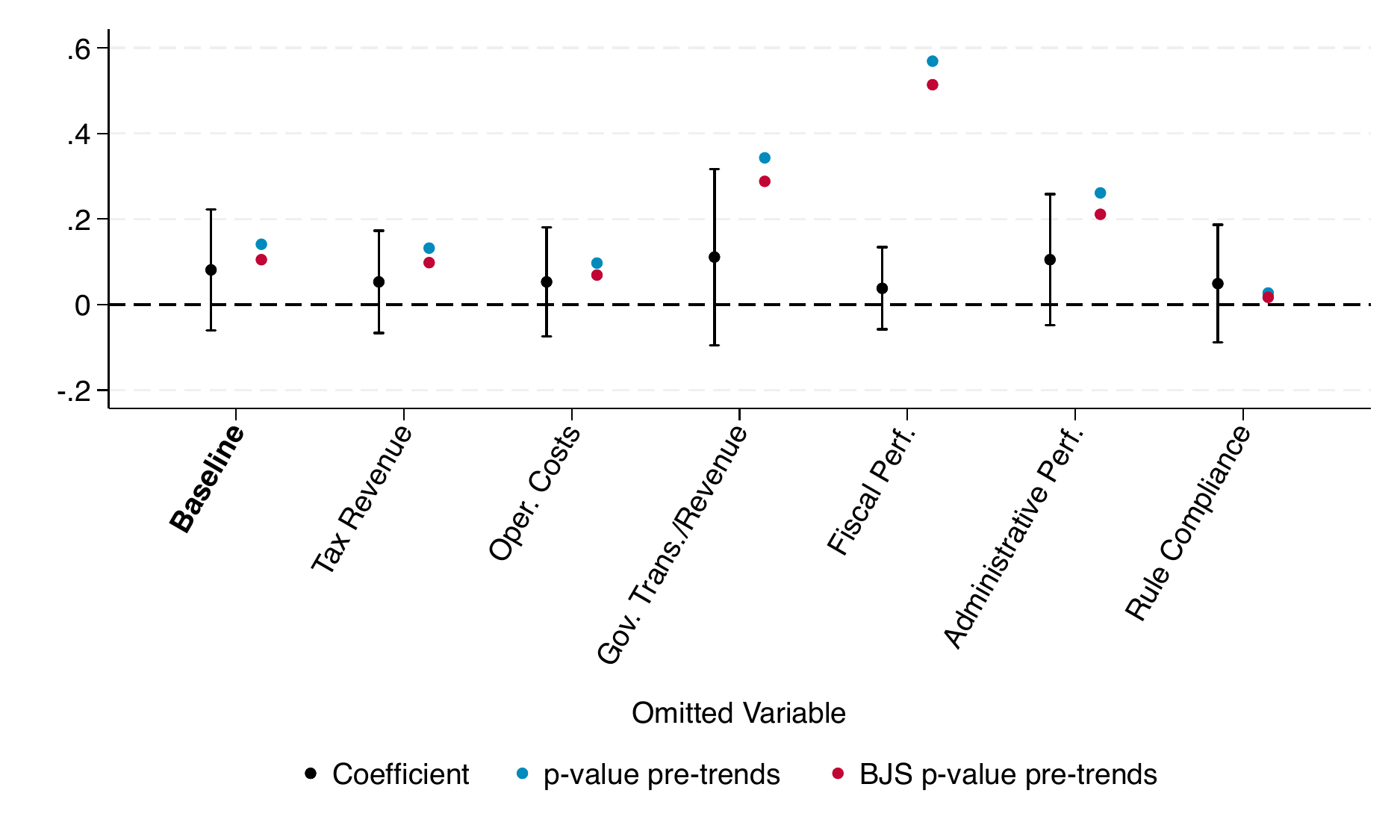}
\subcaption{State Capacity Outcomes}
\end{subfigure}
\justifying 

\footnotesize{\noindent \textbf{Notes:} Figures show in black estimates of Equation \eqref{eq_did} including 95\% confidence intervals (based on standard errors clustered at the municipality level), in blue $p$-values from a test of joint significance of all pre-treatment coefficients from estimates of Equation \eqref{eq_es} following \citet{freyaldenhoven2021visualization}, and in red $p$-values from a test of joint significance of all pre-treatment coefficients following \citet{borusyak2021revisiting}. All figures use indices created following \citet{anderson2008multiple}. Panel A uses the index based on the different violence variables. Panel B uses the index based on nighttime light intensity, value added per capita (from DANE), the share of urban population, agricultural productivity, firm entry and formal employment measures. Panel C uses the index based on state capacity measures. When estimating the results, the names in the $x$-axis correspond to the variable dropped from the corresponding index.}
\end{figure}

An additional concern could be that the results for the main indices are driven by the selection of variables that compose the indices. To allay these concerns, in Figure \ref{rob_indices}, I show the results of estimating the baseline set of results for each of the three main indices, eliminating each of the component variables individually. I present results for the estimates from Equation \eqref{eq_did} in black with 95\% CIs, as well as the $p$-values of the test of joint significance of pre-treatment coefficients i) suggested by \citet{freyaldenhoven2021visualization} in blue, and ii) suggested by \citet{borusyak2021revisiting} in red. The Figure also shows the baseline set of results for comparison. 

For the violence results in Panel A, the diff-in-diff estimates are always significant at the 95\% significance level and very close in magnitude to the baseline results, but when excluding terrorist attacks (slightly larger estimate in magnitude) and threats (marginally insignificant at this level). All the $p$-values for pre-trends are insignificant. The results for the economic index in Panel B are similar: the coefficients are very close in magnitude to the baseline results. While the $p$-value for the joint test is below 0.05, this is entirely driven by the firm entry variable, as shown in the last column. The state capacity index shows a similar pattern: it is always insignificant, close to 0, and precisely estimated. Overall, this indicates that the baseline results remain consistent when altering the composition of the index measures.

\begin{figure}[h!]
\caption{Time Series Economic Indicators}
\label{natLevel_timeSeries}
\centering
\begin{subfigure}[t]{0.48\textwidth}
\includegraphics[width=\textwidth]{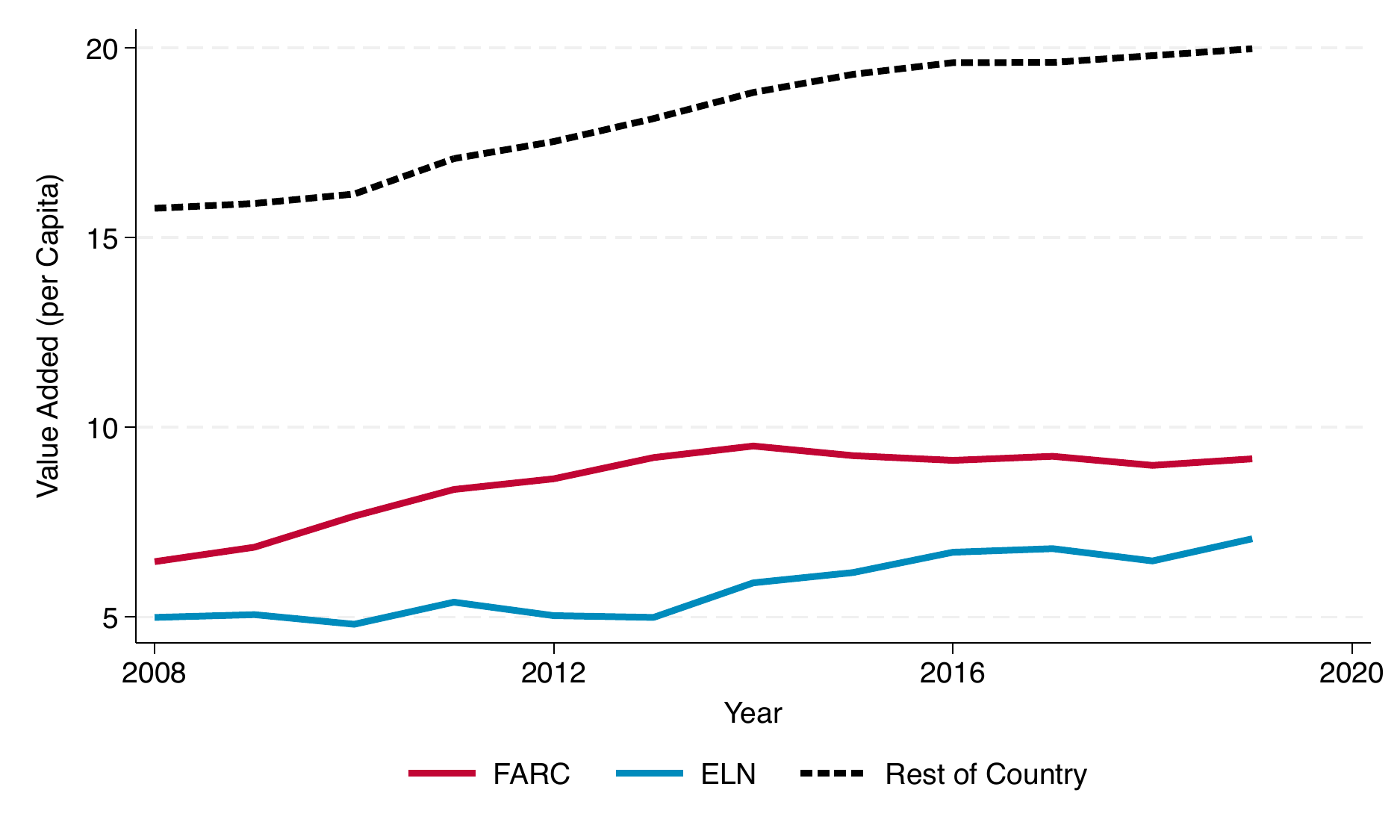}
\subcaption{Value Added per Capita (S\&E, 2012)}
\end{subfigure}
\begin{subfigure}[t]{0.48\textwidth}
\includegraphics[width=\textwidth]{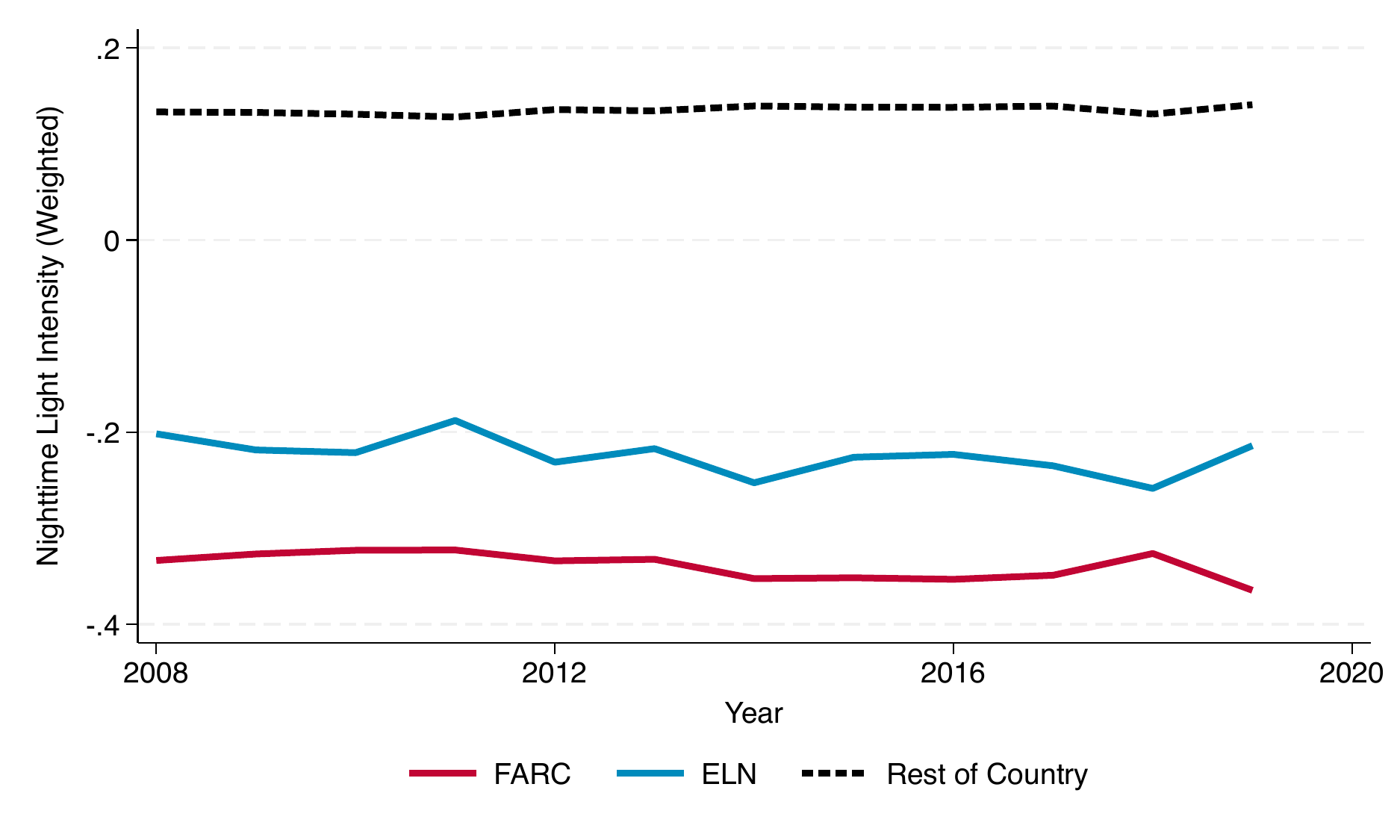}
\subcaption{Nighttime Light Intensity}
\end{subfigure}
\begin{subfigure}[t]{0.48\textwidth}
\includegraphics[width=\textwidth]{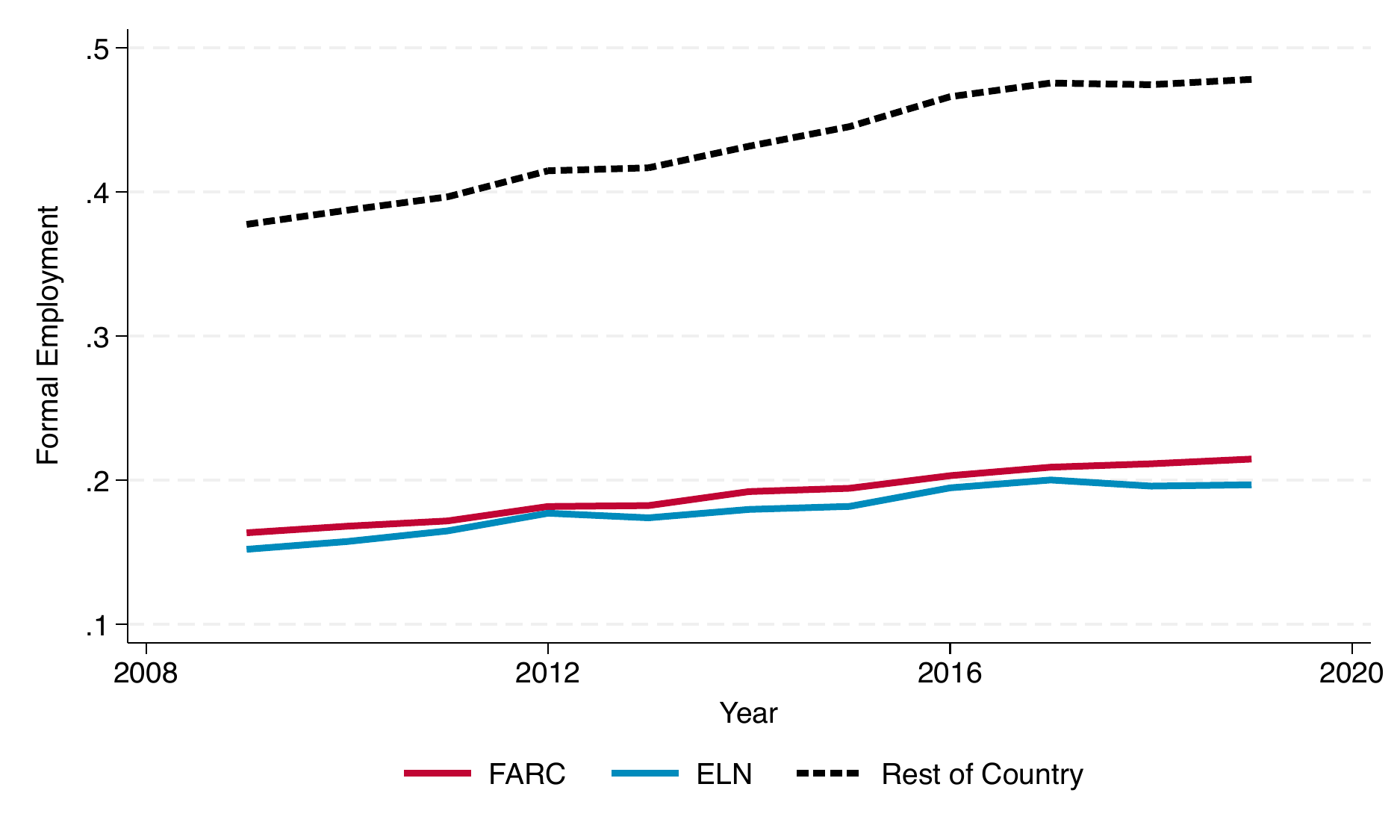}
\subcaption{Formal Employment}
\end{subfigure}
\begin{subfigure}[t]{0.48\textwidth}
\includegraphics[width=\textwidth]{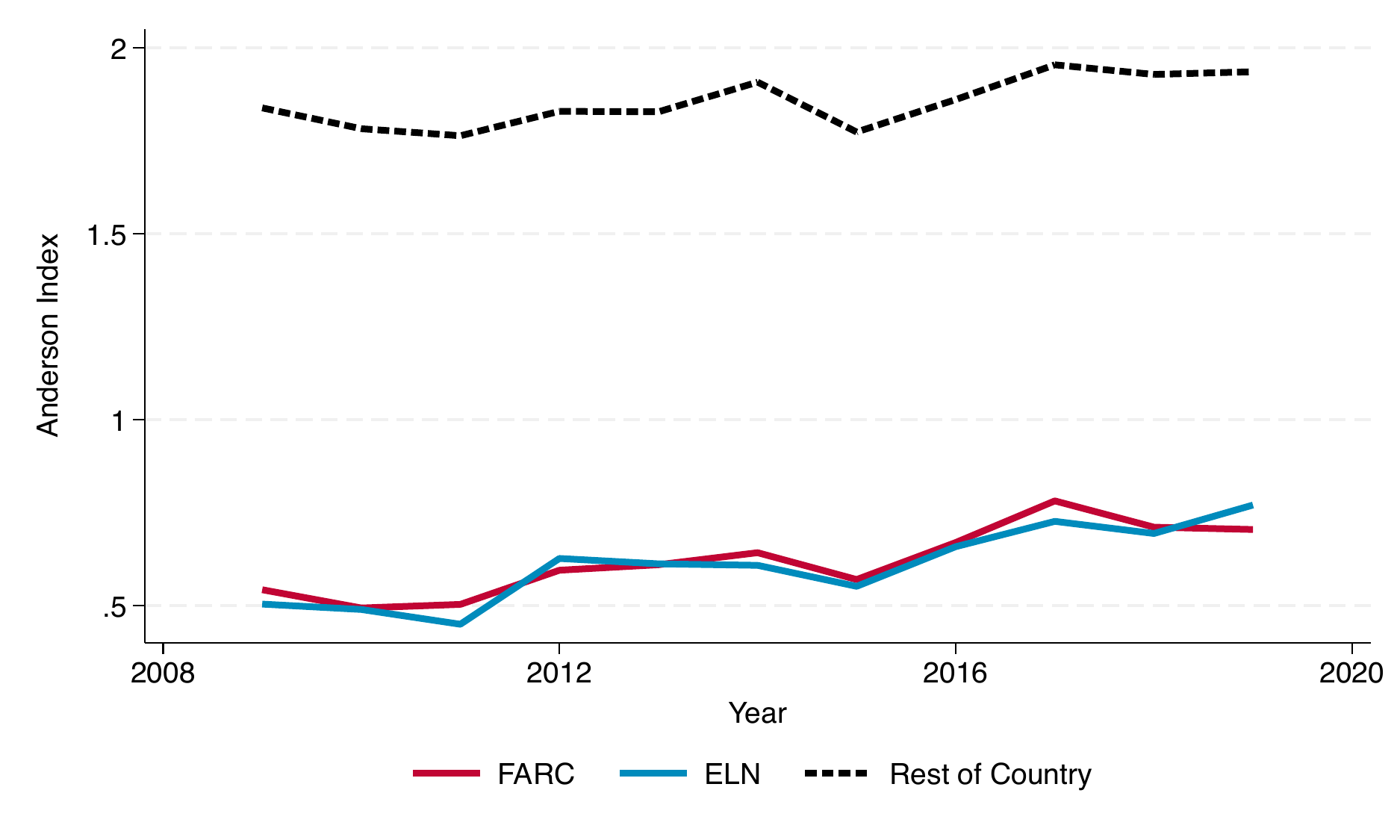}
\subcaption{Anderson Index}
\end{subfigure}
\justifying 
\footnotesize{\textbf{Notes:} Panel A shows the evolution of value added per capita \citep[following][]{sanchez2012urbanizacion}, Panel B of nighttime light intensity, Panel C of formal employment (using PILA data) and Panel D for the Anderson Index of the different economic indicators for FARC (red), ELN (blue) and the remaining municipalities (black), excluding those municipalities classified as both FARC and ELN.}
\end{figure}

Another reason for the lack of significant economic improvements in FARC municipalities could be that \textit{both} FARC and ELN municipalities were growing at the same rate and potentially faster than the rest of the country. Rather than FARC municipalities not experiencing economic improvements, these could be masked simply by virtue of the control group growing at a similar pace. However, Figure \ref{natLevel_timeSeries} shows this is not the case. Panel A plots the evolution of value added per capita \citep[estimated following][]{sanchez2012urbanizacion}, Panel B that of nighttime light intensity, Panel C of formal employment (using PILA data) and Panel D for the Anderson Index of the different economic indicators for FARC (red), ELN (blue) and the remaining municipalities (black).\footnote{The latter is standardised each year for comparison purposes.} The time series show that 1) FARC and ELN municipalities are much poorer than the rest of the country using all measures, and 2) they have not caught up with the rest of the country since 2008. Thus, the diff-in-diff estimates do not mask economic improvements in FARC and ELN municipalities relative to the rest of the country.

One concern of using the ELN as a counterfactual of the FARC is that, while they share a similar history and trajectory, the FARC and ELN operated in different areas of the country (see Figure \ref{maps}) and had different sources of income, with the FARC focusing on the drug trade while the ELN focused on smuggling and extorting oil companies. While this concern is partly alleviated by the inclusion of municipality and year fixed effects, Table \ref{robCheckControls_CEDEExt_p60} tests more rigorously that the results are robust to controlling for geographic characteristics and measures related to illegal activities by interacting several measures with the post-ceasefire dummy. Panel A replicates the baseline results. Panel B includes several geographic (area, distance to state capital) and demographic (baseline poverty, needs, population) characteristics. Panel C includes measures of a municipality's connectivity to other areas of the country (number of municipalities within 0-100kms, within 100-200kms, number of agricultural markets within 0-100kms, within 100-200kms, and the average distance to all other municipalities). Panel D includes several measures of the attractiveness of conducting illegal activities, including the farmgate price of coca leaves and paste, average price per kg of cocaine, international oil price, whether the municipality paid oil royalties, the coca suitability index from \citet{mejia2013bushes}, and a dummy for whether there are gold geochemical anomalies in the municipality from \citet{rozo2020unintended}.\footnote{Illegal gold mining has been a popular source of income for criminal groups in the past decades.} Finally, Panel E includes all these controls together. Overall, while the coefficient on violence reduces by around $\sim$ 15\%, it remains significant, while all the others remain precisely-estimated null results. 

\begin{table}[tbp] \centering
\newcolumntype{R}{>{\raggedleft\arraybackslash}X}
\newcolumntype{L}{>{\raggedright\arraybackslash}X}
\newcolumntype{C}{>{\centering\arraybackslash}X}

\caption{Robustness of Baseline Results to Addition of Controls}
\label{robCheckControls_CEDEExt_p60}
\begin{adjustbox}{max width=\linewidth, max height=0.6\textwidth}\begin{tabular}{lccc}

\toprule
& \multicolumn{3}{c}{\textbf{Index}} \tabularnewline & Violence & Economic & State Capacity \tabularnewline 
{}&{(1)}&{(2)}&{(3)} \tabularnewline
\midrule \addlinespace[\belowrulesep]
\midrule \textbf{Panel A. Baseline}&&& \tabularnewline
Ceasefire \( \times \) FARC&--0.289***&--0.004&0.067 \tabularnewline
Baseline&(0.104)&(0.038)&(0.054) \tabularnewline
Observations&2,827&2,827&2,827 \tabularnewline
&&& \tabularnewline
\midrule \textbf{Panel B. Characteristics}&&& \tabularnewline
Ceasefire \( \times \) FARC&--0.243**&--0.012&0.030 \tabularnewline
&(0.107)&(0.035)&(0.052) \tabularnewline
Observations&2,827&2,827&2,827 \tabularnewline
&&& \tabularnewline
\midrule \textbf{Panel C. Connectivity}&&& \tabularnewline
Ceasefire \( \times \) FARC&--0.266**&0.025&0.063 \tabularnewline
&(0.119)&(0.040)&(0.054) \tabularnewline
Observations&2,827&2,827&2,827 \tabularnewline
&&& \tabularnewline
\midrule \textbf{Panel D. Illegal Activity}&&& \tabularnewline
Ceasefire \( \times \) FARC&--0.274**&0.012&0.035 \tabularnewline
&(0.106)&(0.038)&(0.050) \tabularnewline
Observations&2,805&2,805&2,805 \tabularnewline
&&& \tabularnewline
\midrule \textbf{Panel E. All}&&& \tabularnewline
Ceasefire \( \times \) FARC&--0.241**&0.009&0.002 \tabularnewline
&(0.117)&(0.038)&(0.052) \tabularnewline
Observations&2,805&2,805&2,805 \tabularnewline
&&& \tabularnewline
\bottomrule \addlinespace[\belowrulesep]

\end{tabular} \end{adjustbox}
\\ \parbox{\linewidth}{\scriptsize \justifying \noindent \(Notes \): Results from estimating Equation \eqref{eq_did}. Standard errors clustered at the municipality level. Considers only years from 2009 and before 2020. Defines the post period as 2015. The first column uses the summary index based on the different violence variables. The second column uses the summary index based on weighted nighttime light intensity, value added per capita (from DANE), share of urban population, agricultural productivity, formal employment and firm entry measures. The third column uses the summary index based on the state capacity measures. Panel A shows the baseline results. Each of the following panels adds a different set of control variables interacted with the ceasefire dummy. Panel B includes municipality's baseline characteristics (area in \(km^2\), measures of poverty and basic needs, (log) population, and distance to the state capital). Panel C includes measures of connectivity to other areas of the country (the number of municipalities between 0-100kms, and 100-200kms, the number of agricultural markets between 0-100kms, and 100-200kms [based on 2014's driving distances from G\'{a}faro \& Pellegrina, 2022], and the average (birdseye) distance to all other municipalities in km). Panel D includes controls for some of the main sources of income of criminal groups (farmgate price of coca leaves and paste, average price per kg of cocaine in the country, European Brent Spot Market price, values of regal\'{i}as paid by municipality, a dummy for whether regal\'{i}as were paid, a dummy for whether there are gold geochemical anomalies in the municipality, and a coca suitability index). Panel E controls for all these variables simultaneously.}
\end{table}

Lastly, the measure of FARC presence is dichotomous: It is one if the municipality is categorised as having historically been under the control of the FARC, and 0 if it is classified as having historically been under the control of the ELN. For the extensive margin measure of presence, to be categorised as either, a municipality must have a criminal event by a given criminal group in 60\% of the years between 1996 and 2008. This means that some of the FARC (ELN) municipalities experienced FARC (ELN) events in 8 years in that timeframe, while others experienced up to 12. I use this variation to create two alternative, continuous measures of FARC presence and re-do the baseline results using these continuous measures interacted with the post-ceasefire dummy. Results are shown in Table \ref{contMeasAlt_CEDEExt_p60}. In Panel A, I focus only on FARC municipalities, and the FARC measure takes values between 0 (8 years of FARC's events, the minimum to be categorised as a FARC municipality) and 1 (12 years of FARC's events, the maximum in this timeframe). In Panel B, I use this same measure but code ELN municipalities with values between -1 (12 years of ELN's events, the maximum in this timeframe) and 0 (8 years of ELN's events, the minimum in this timeframe). This way, moving from -1 to 1 means moving from a municipality with a very strong ELN presence to one with a very strong FARC presence. Using both these measures, I find results in line with the baseline results: Municipalities with stronger FARC presence experienced a larger reduction in violence, yet there are no significant changes in economic or state capacity indicators. 

\begin{table}[h!] \centering
\newcolumntype{R}{>{\raggedleft\arraybackslash}X}
\newcolumntype{L}{>{\raggedright\arraybackslash}X}
\newcolumntype{C}{>{\centering\arraybackslash}X}

\caption{Continuous Measure of Presence}
\label{contMeasAlt_CEDEExt_p60}
\begin{adjustbox}{max width=\linewidth, max height=\textwidth}\begin{tabular}{lccc}

\toprule
& Violence & Economic & State Capacity \tabularnewline 
{}&{(1)}&{(2)}&{(3)} \tabularnewline
\midrule \addlinespace[\belowrulesep]
\midrule \textbf{Panel A. Only FARC}&&& \tabularnewline
Ceasefire \( \times \) FARC (Continuous)&--0.486***&0.022&--0.021 \tabularnewline
&(0.123)&(0.046)&(0.071) \tabularnewline
&&& \tabularnewline
Observations&2,376&2,376&2,376 \tabularnewline
\(R^2\)&0.439&0.951&0.512 \tabularnewline
Mean Dep. Var.&0.105&--0.101&0.017 \tabularnewline
&&& \tabularnewline
\midrule \textbf{Panel B. Both FARC \& ELN}&&& \tabularnewline
Ceasefire \( \times \) FARC (Continuous)&--0.415***&0.031&--0.023 \tabularnewline
&(0.116)&(0.043)&(0.065) \tabularnewline
&&& \tabularnewline
Observations&2,827&2,827&2,827 \tabularnewline
\(R^2\)&0.434&0.952&0.594 \tabularnewline
Mean Dep. Var.&0.079&--0.097&0.014 \tabularnewline
\bottomrule \addlinespace[\belowrulesep]

\end{tabular} \end{adjustbox}
\\ \parbox{\linewidth}{\scriptsize \justifying \noindent \(Notes \): Results from estimating Equation \eqref{eq_did}. Standard errors clustered at the municipality level. Considers only years from 2009 and before 2020. Defines the post period as 2015. The first column uses the summary index based on the different violence variables. The second column uses the summary index based on weighted nighttime light intensity, value added per capita (from DANE), share of urban population, agricultural productivity, formal employment and firm entry measures. The third column uses the summary index based on the state capacity measures. Panel A uses only FARC municipalities, and the presence variable varies between 0 (minimum years with FARC presence among those catalogued as FARC) and 1 (maximum years with FARC presence among those catalogued as FARC). Panel B uses both FARC and ELN municipalities, with ELN ones taking values between -1 (maximum years under ELN's presence) and 0 (minimum years under ELN's presence) and FARC ones taking values between 0 (minimum years under FARC's presence) and 1 (maximum years under FARC's presence).}
\end{table}

\begin{figure}[h!]
\begin{center}
\caption{Placebo Treatment -- Extensive Margin}
\label{placebo_treat}
\includegraphics[width=\textwidth]{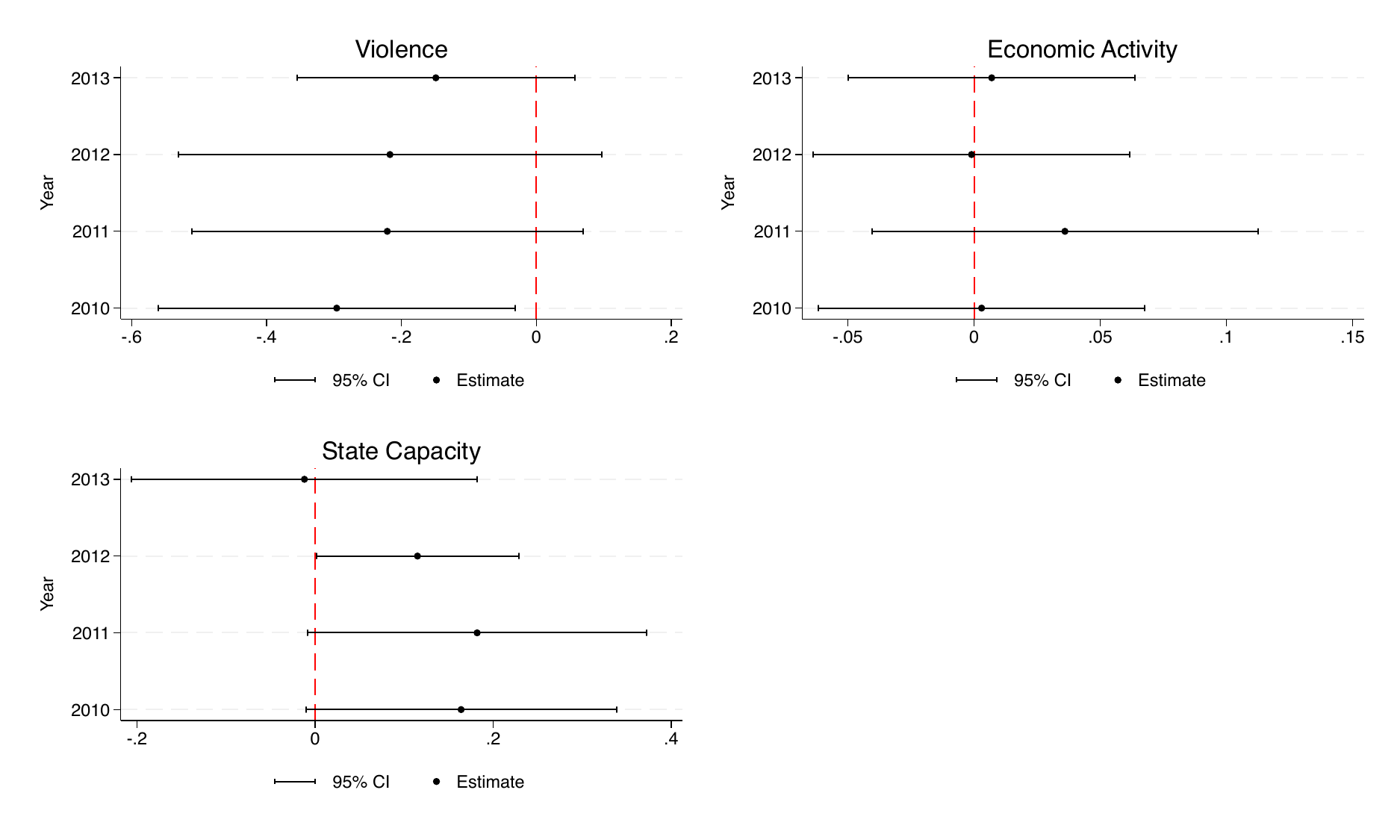}
\end{center}
\justifying 
\footnotesize{\textbf{Notes:} Estimates of Equation \eqref{eq_did}, including 95\% confidence intervals (based on standard errors clustered at the municipality level), using data only from the pre-treatment period (2009-2014). The number in the $y$-axis corresponds to the ``placebo'' treatment year. All figures use indices created following \citet{anderson2008multiple}. The top left Figure uses the index based on the different violence variables. The top right Figure uses the index based on the different economic variables. The bottom left Figure uses the index based on the state capacity variables.}
\end{figure}

Figure \ref{placebo_treat} shows the results for the indices of a placebo exercise. In the exercise, I use data only from the pre-ceasefire period, 2009-2014, and run the usual diff-in-diff regression, assigning the treatment to be 2011, 2012, or 2013 (so that there are always at least two pre-treatment and post-treatment periods). The coefficients on the placebo treatments are mostly insignificant for the violence and economic indices, while for the state capacity index, those of the earlier years are marginally significant. This suggests that anticipation problems are unlikely.

\renewcommand{\thefigure}{E\arabic{figure}}
\setcounter{figure}{0}
\renewcommand{\thetable}{E\arabic{table}}
\setcounter{table}{0}

\clearpage
\subsection{Additional Figures}
\label{app_figures}

\subsubsection{Violence Results by Likelihood of Gold and Coca Suitability}

\begin{figure}[h!]
\begin{center}
\caption{Violence by Type of Municipality}
\label{viol_cocaGold}
\includegraphics[width=0.8\textwidth]{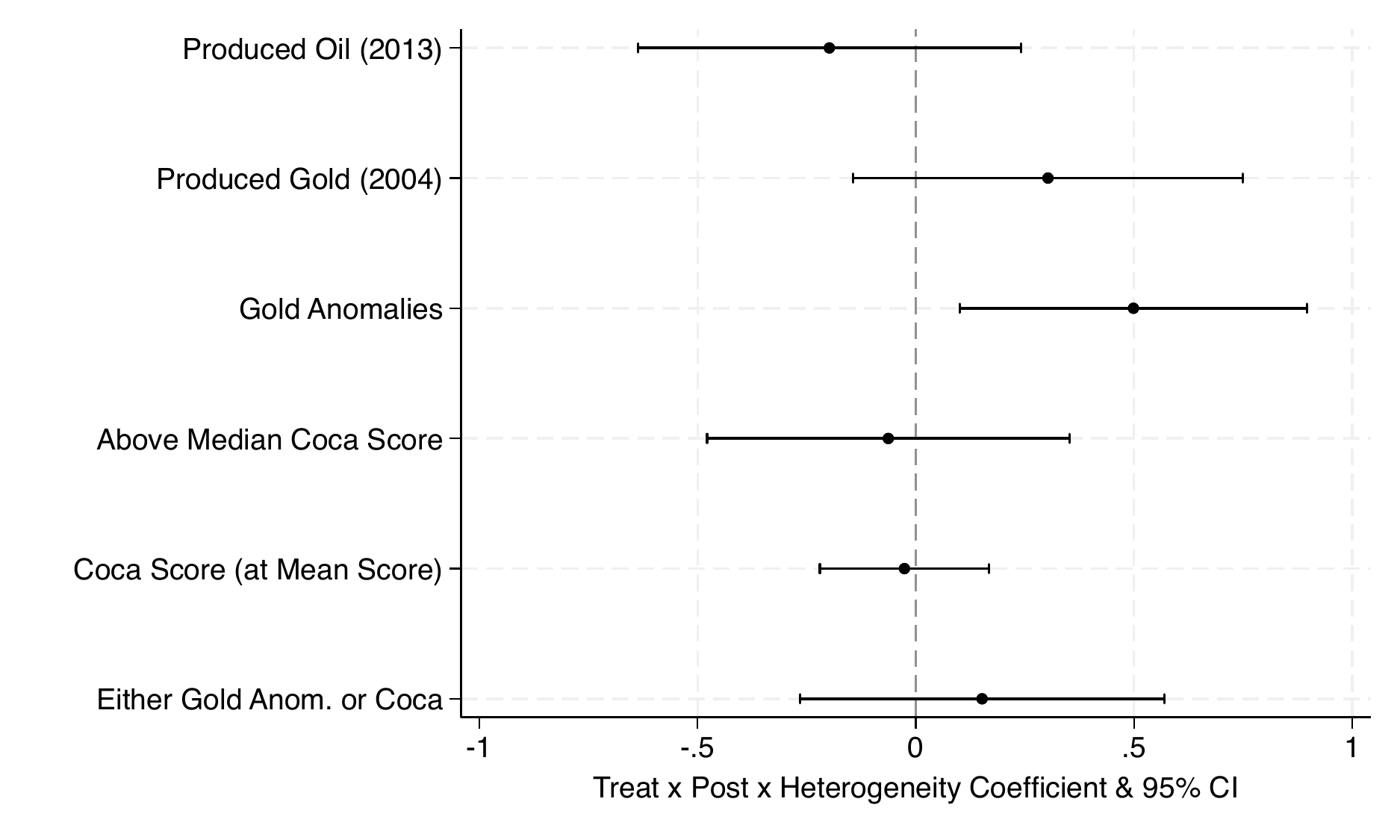}
\end{center}
\justifying 
\footnotesize{\textbf{Notes:}  Estimates of $\gamma$ and their 95\% confidence intervals from estimating Equation \eqref{eq_het}. The first row uses a dummy for whether the municipality produced oil in 2013. The second row uses a dummy for whether the municipality produced gold in 2004. The third row uses a dummy for whether the municipality has geochemical anomalies associated with the presence of gold. The fourth and fifth rows use the coca suitability index created by \citet{mejia2013bushes}. The fourth row uses a dummy for whether the municipality is above the median coca suitability score among FARC and ELN municipalities, while the fifth one uses the continuous score (the coefficient is then evaluated at the mean score). The last row uses a dummy for whether the municipality has gold geochemical anomalies or is above the median regarding coca suitability. The dependent variable is the Anderson Index of the violence measures.}
\end{figure}

\subsubsection{Violence Event Studies}

\begin{landscape}
\begin{figure}[htp]
\caption{Dynamic Estimation, Extensive Margin, Events in Over 60\% of Years, Victims' Unit Measures}
\label{dyn_vioUARIV_ext}
\centering
\begin{subfigure}[b]{0.4\textwidth}
\includegraphics[width=\textwidth]{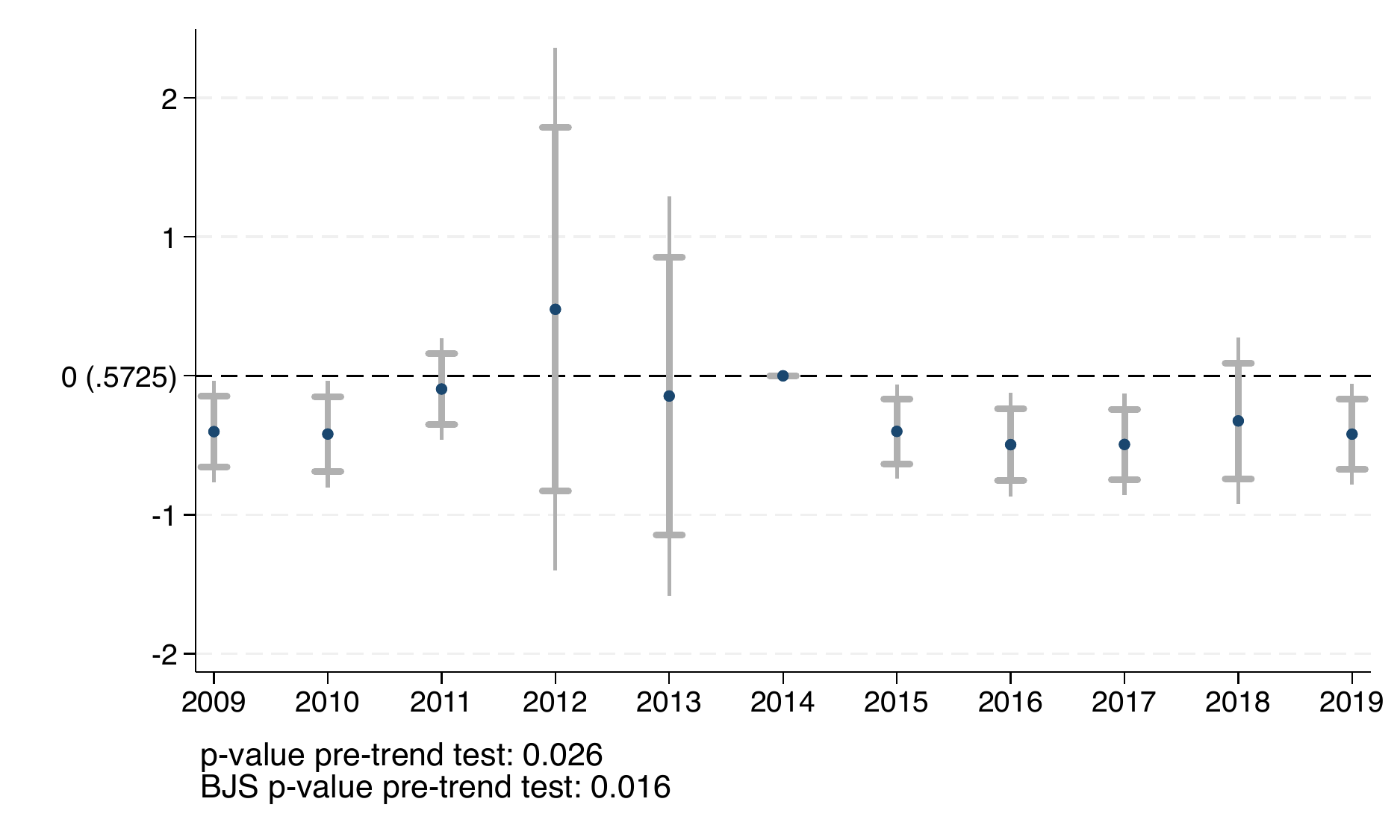}
\subcaption{Terrorist Acts/Armed Fights}
\end{subfigure}
\begin{subfigure}[b]{0.4\textwidth}
\includegraphics[width=\textwidth]{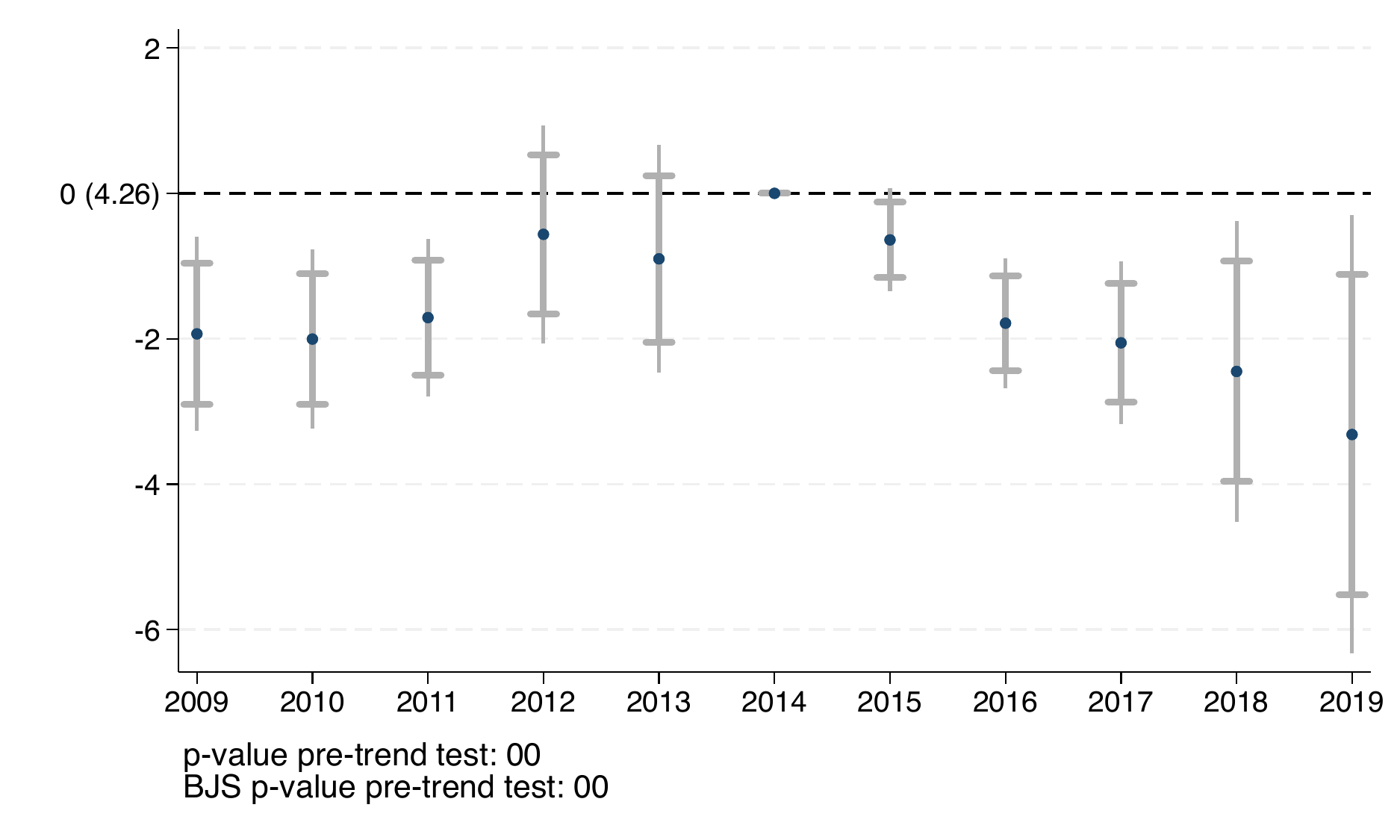}
\subcaption{Threats}
\end{subfigure}
\begin{subfigure}[b]{0.4\textwidth}
\includegraphics[width=\textwidth]{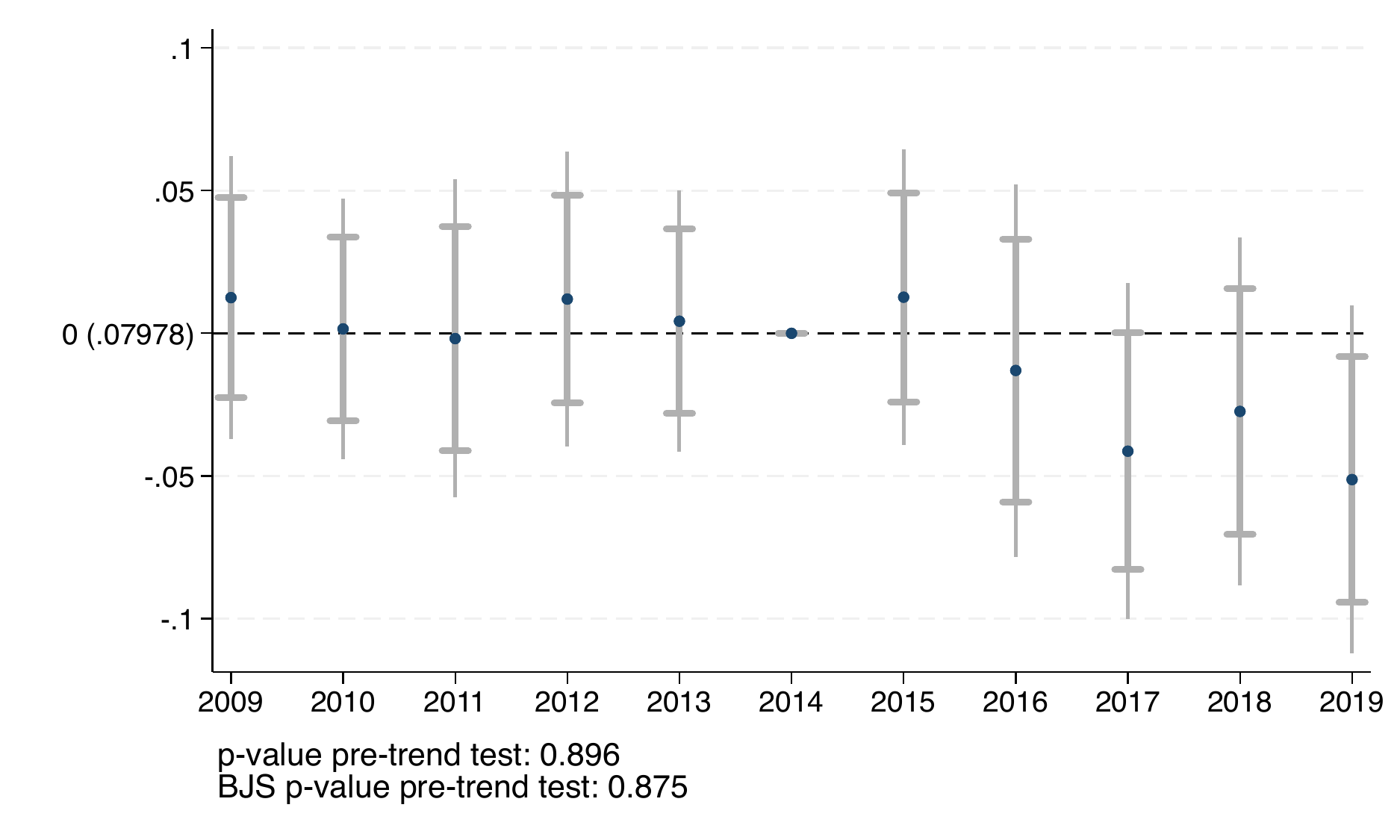}
\subcaption{Sex Crimes}
\end{subfigure}
\begin{subfigure}[b]{0.4\textwidth}
\includegraphics[width=\textwidth]{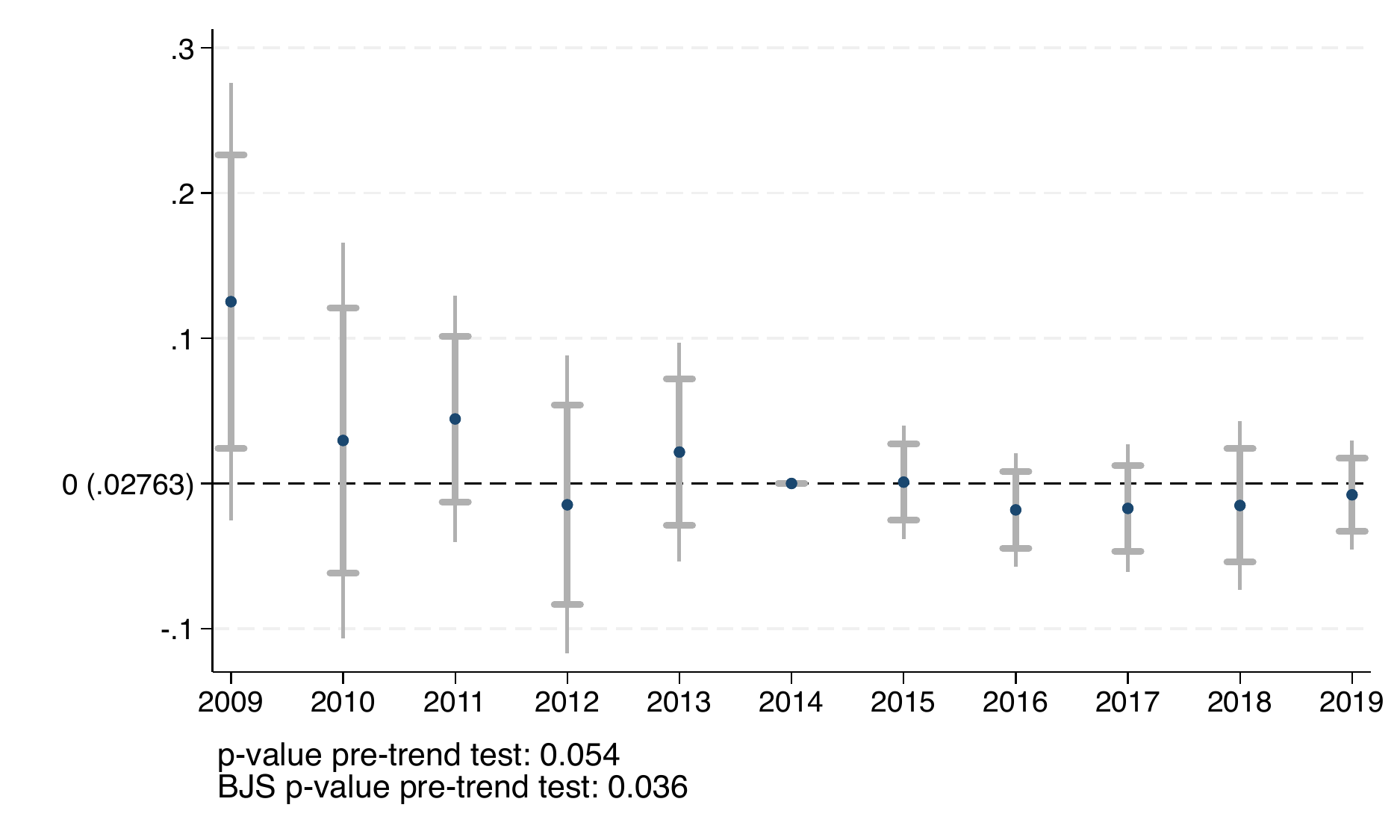}
\subcaption{Forced Disappearences}
\end{subfigure}
\begin{subfigure}[b]{0.4\textwidth}
\includegraphics[width=\textwidth]{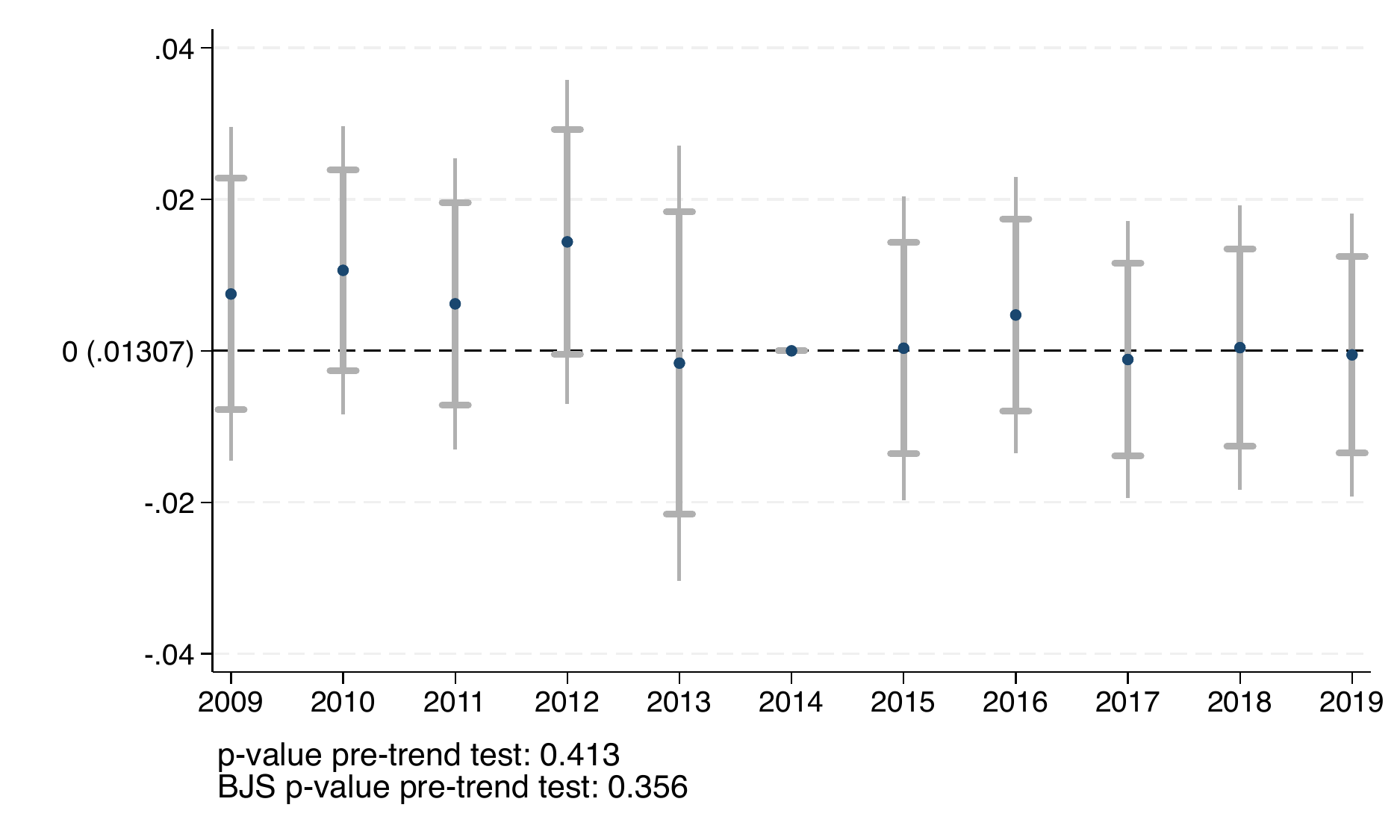}
\subcaption{Torture}
\end{subfigure}
\begin{subfigure}[b]{0.4\textwidth}
\includegraphics[width=\textwidth]{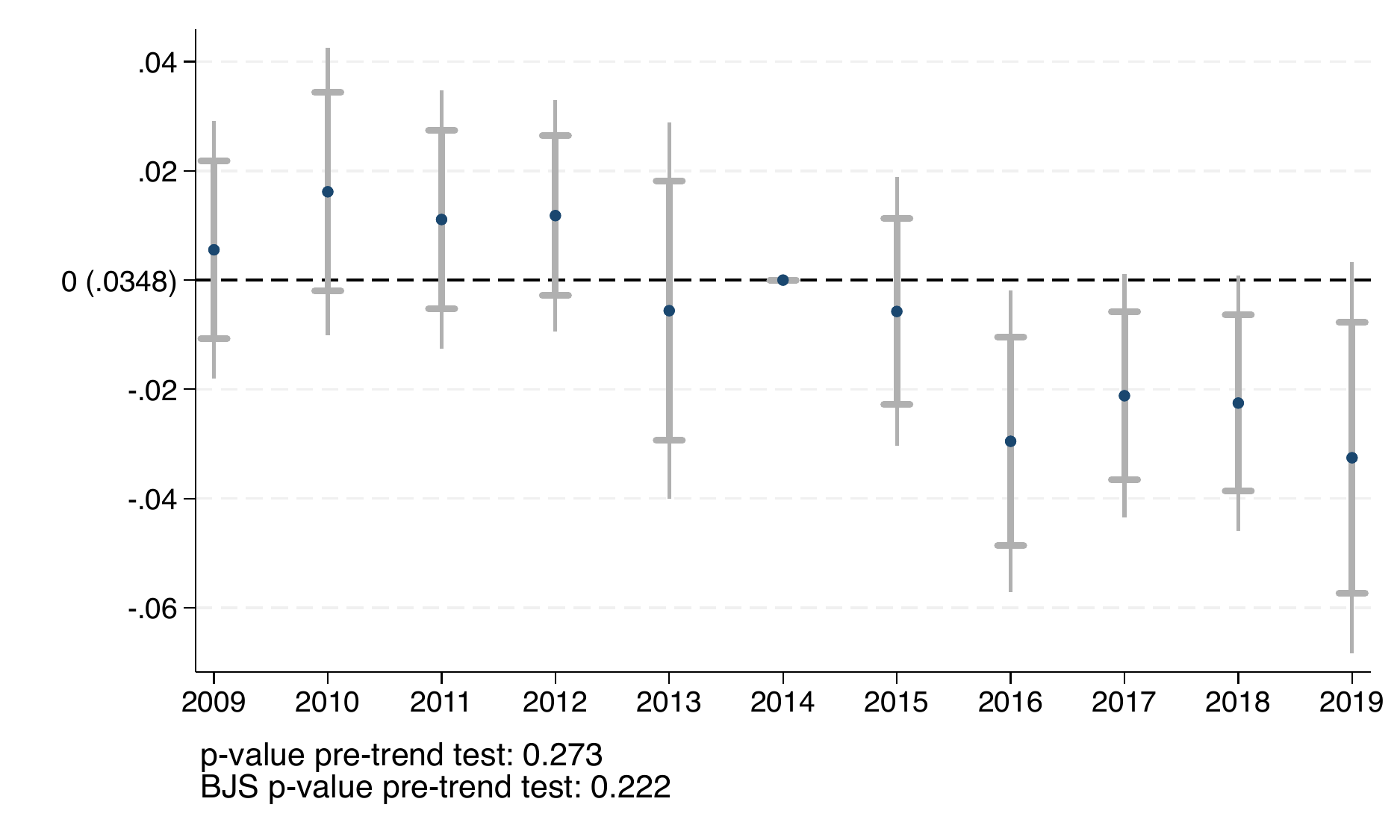}
\subcaption{Child Recruitment}
\end{subfigure}
\\
\justifying
\footnotesize{\noindent \textbf{Notes:}  Event study plots from estimating Equation \eqref{eq_es} for different violence measures from the Victims' Unit, including including 95\% confidence intervals (based on standard errors clustered at the municipality level).}
\end{figure}
\end{landscape}

\begin{landscape}
\begin{figure}[htp]
\caption{Dynamic Estimation, Extensive Margin, Events in Over 60\% of Years, Different Sources}
\label{dyn_vioCEDE_ext}
\centering
\begin{subfigure}[b]{0.4\textwidth}
\includegraphics[width=\textwidth]{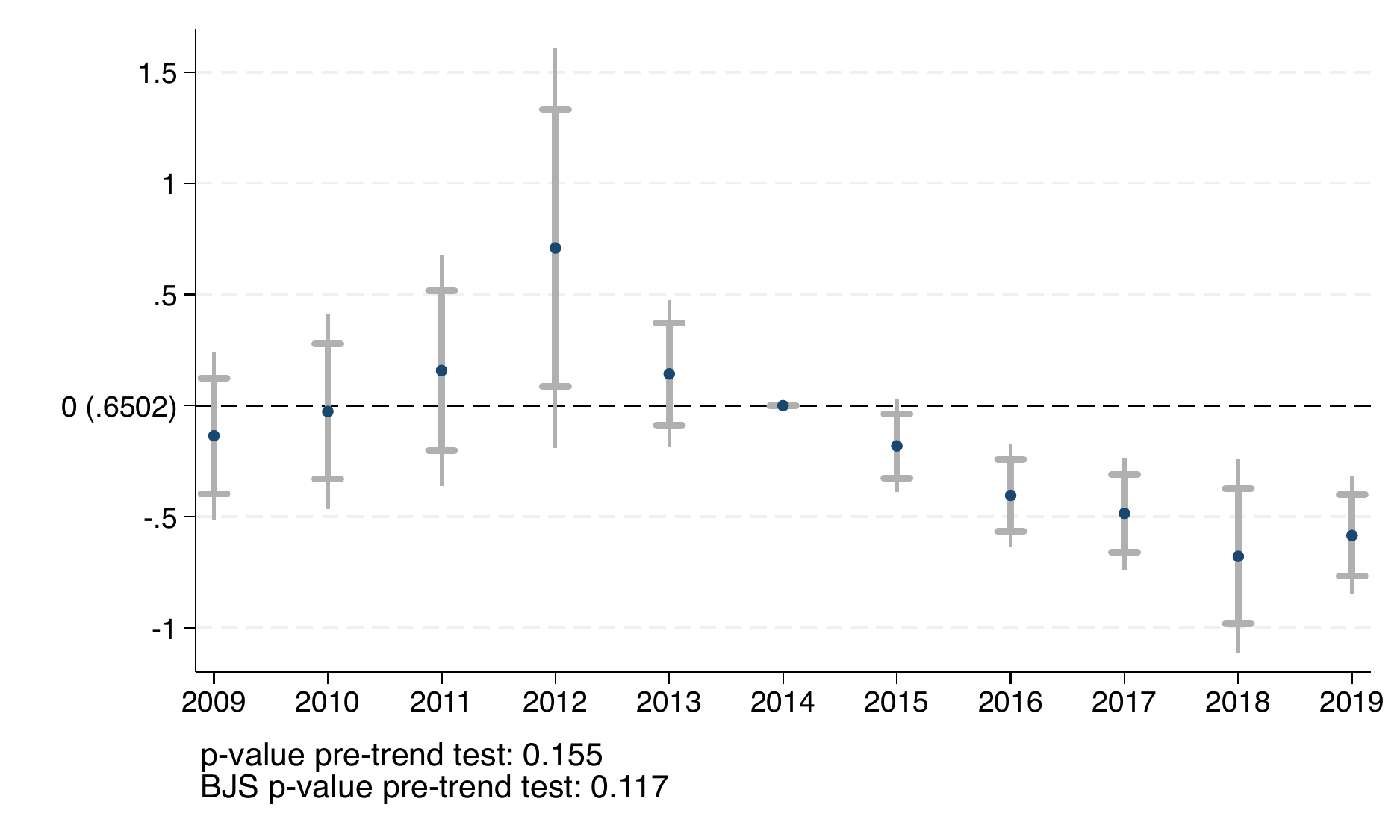}
\subcaption{Property Losses (Victims' Unit)}
\end{subfigure}
\begin{subfigure}[b]{0.45\textwidth}
\includegraphics[width=\textwidth]{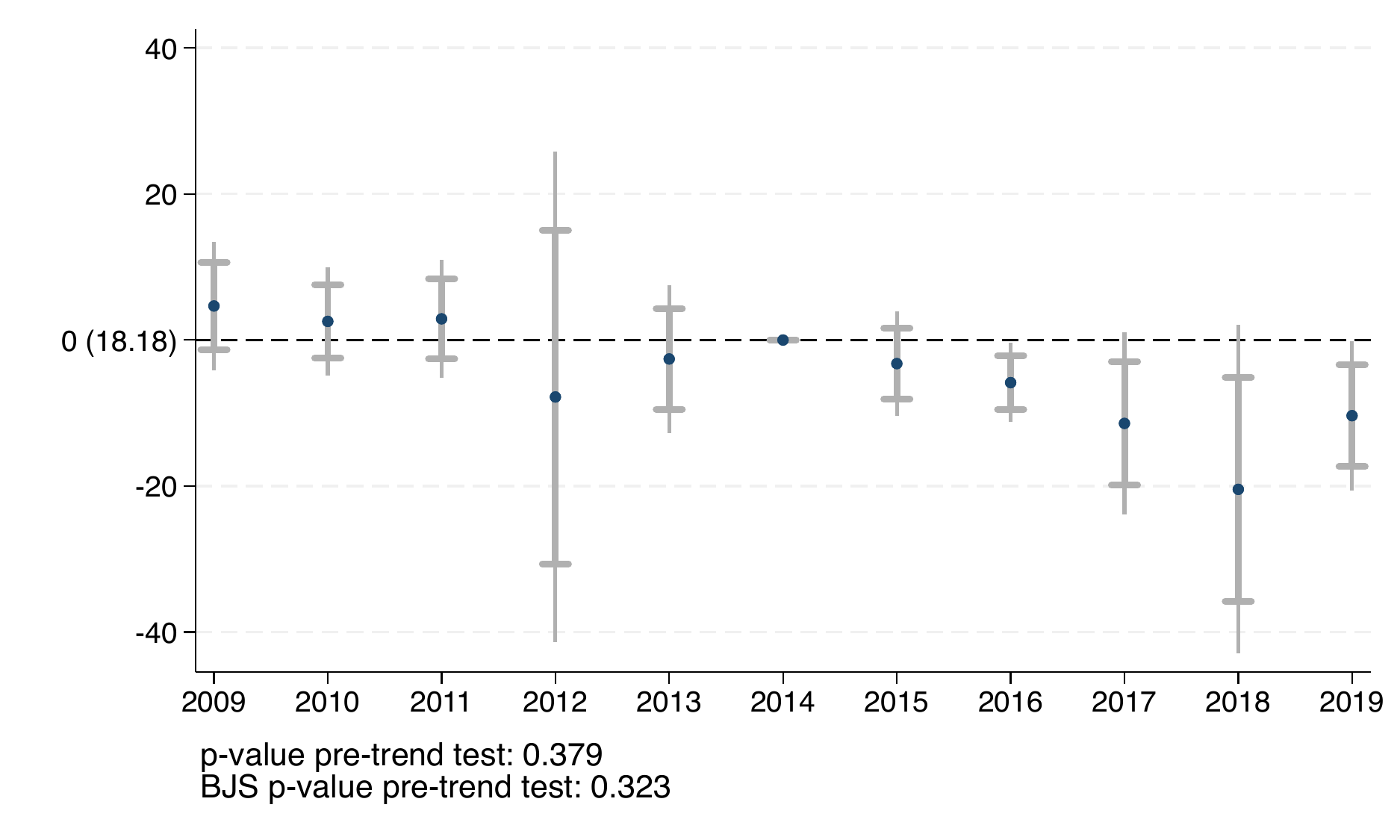}
\subcaption{Forced Migration (Victims' Unit)}
\end{subfigure}
\begin{subfigure}[b]{0.45\textwidth}
\includegraphics[width=\textwidth]{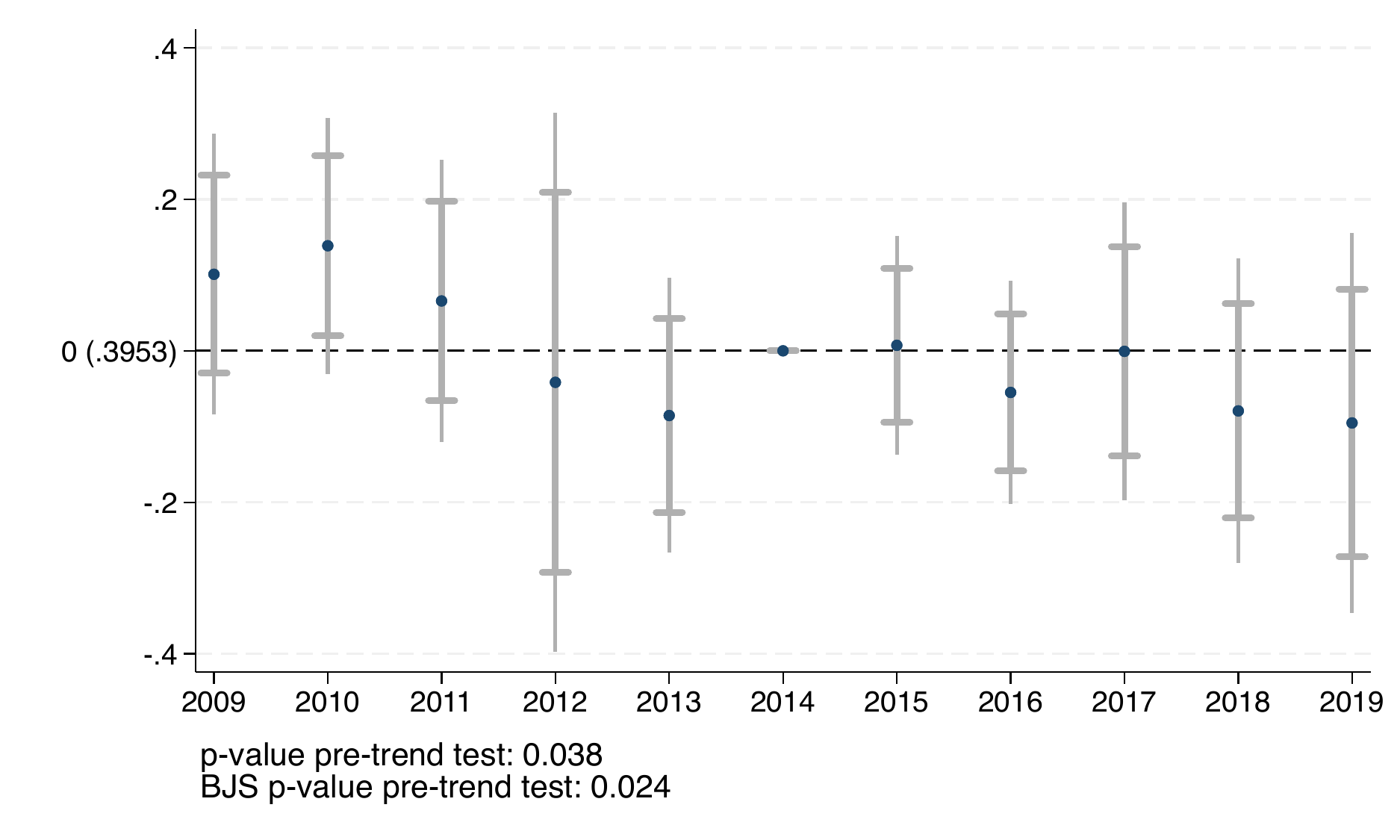}
\subcaption{Homicides (Ministry of Defense)}
\end{subfigure}
\begin{subfigure}[b]{0.45\textwidth}
\includegraphics[width=\textwidth]{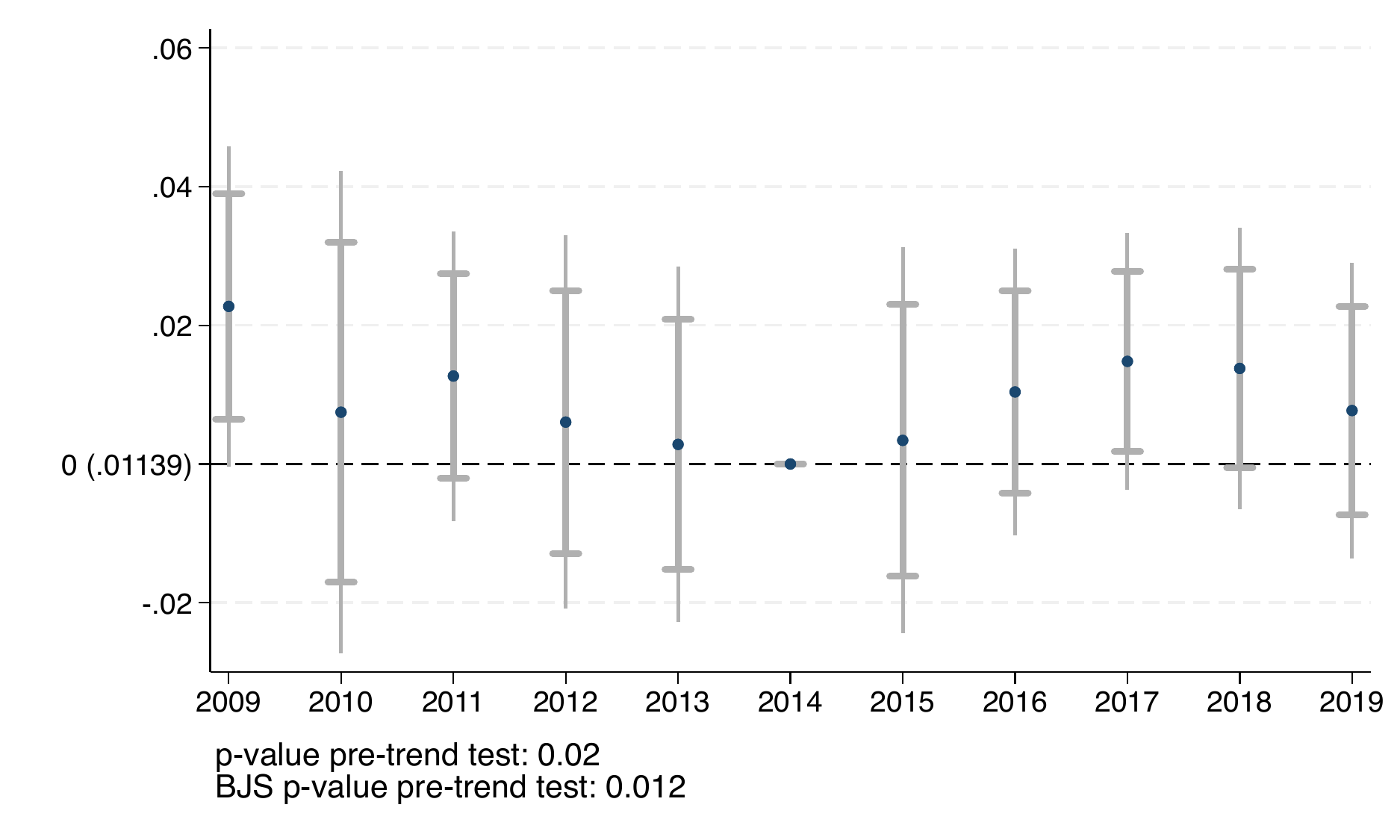}
\subcaption{Kidnappings (Ministry of Defense)}
\end{subfigure}
\begin{subfigure}[b]{0.45\textwidth}
\includegraphics[width=\textwidth]{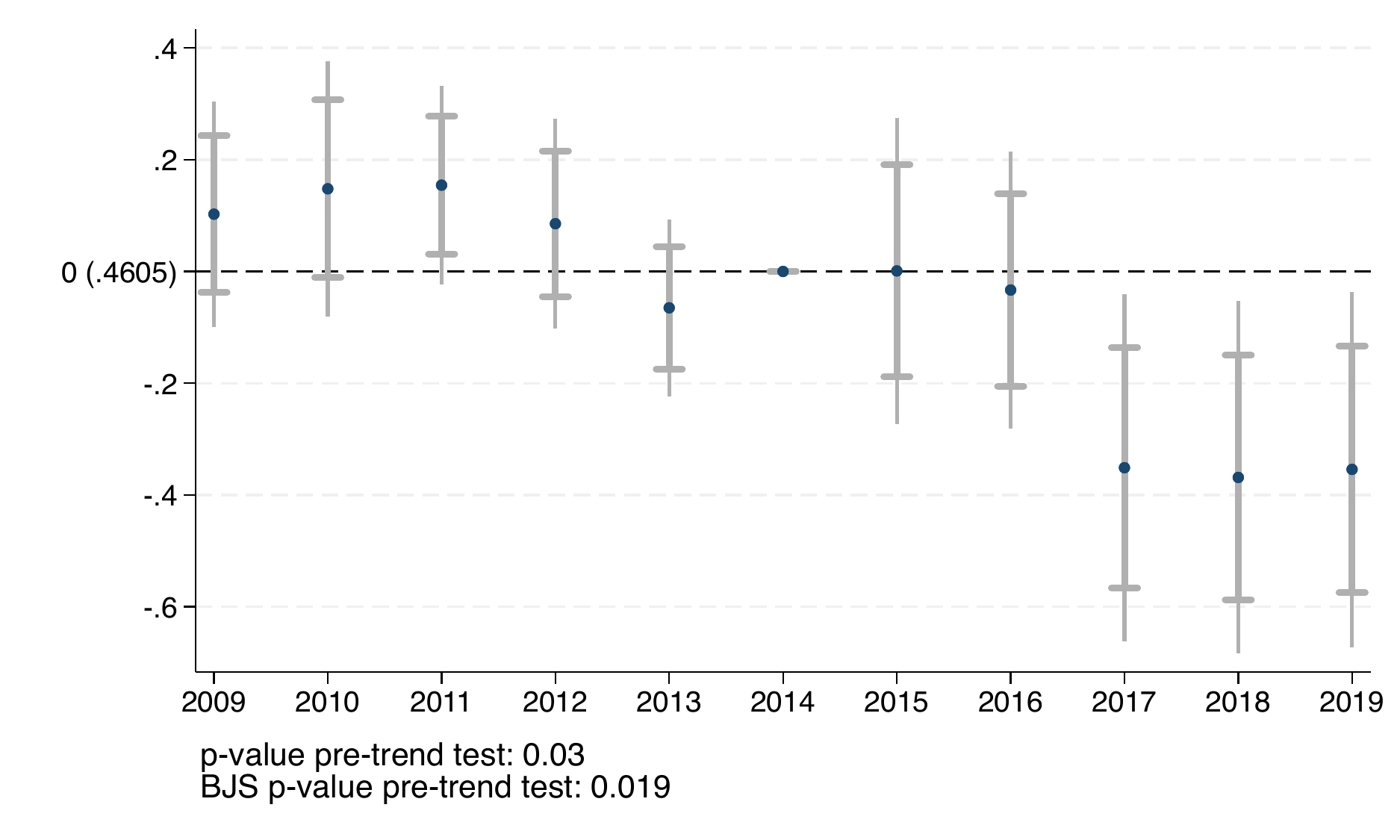}
\subcaption{Mines (DAIMA)}
\end{subfigure}
\begin{subfigure}[b]{0.45\textwidth}
\includegraphics[width=\textwidth]{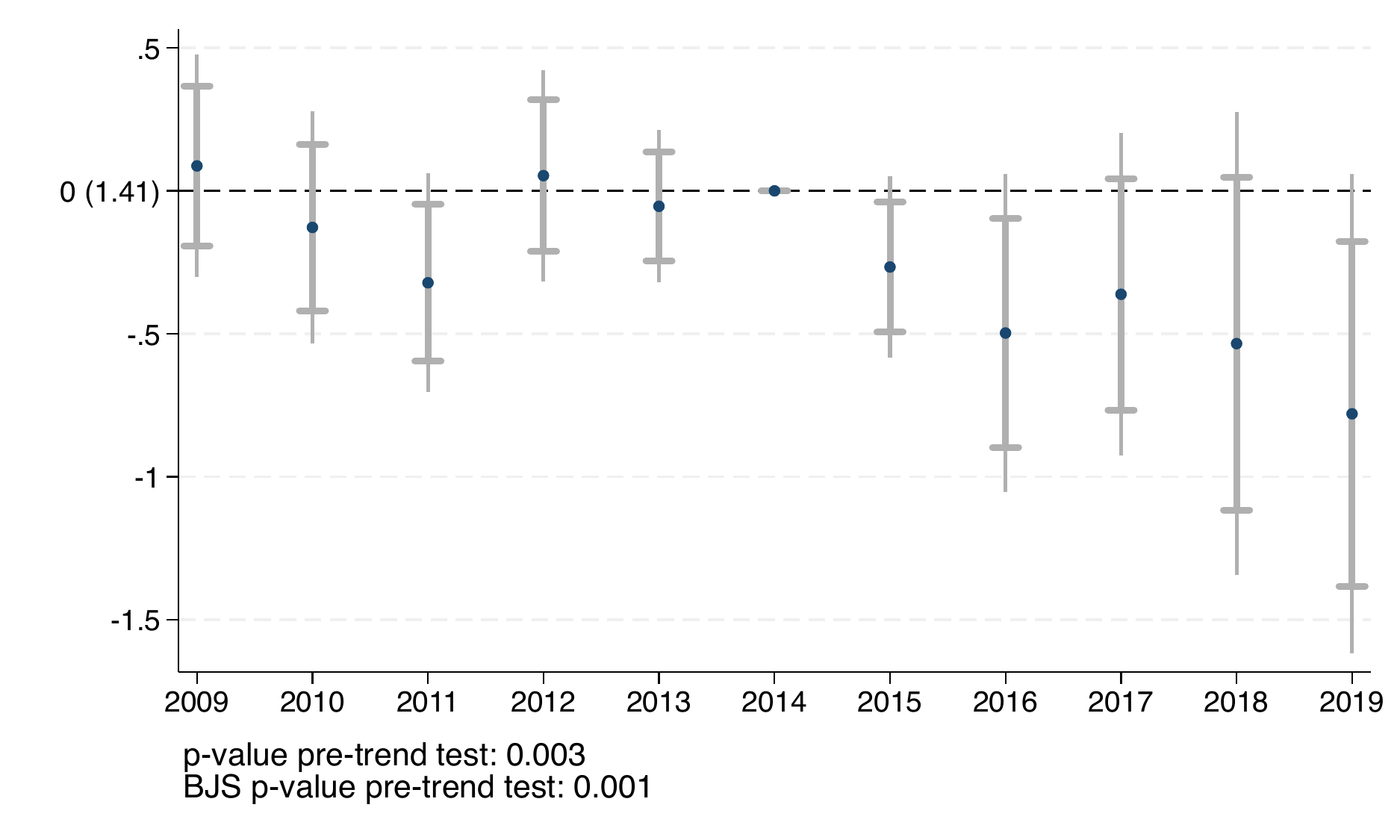}
\subcaption{Theft (Ministry of Defense)}
\end{subfigure}
\\
\justifying
\footnotesize{\noindent \textbf{Notes:}  Event study plots from estimating Equation \eqref{eq_es} for different violence measures from several sources, including 95\% confidence intervals (based on standard errors clustered at the municipality level).}
\end{figure}
\end{landscape}

The following Figures show the event-study estimates for each violence measure presented in Section \ref{sec_viol}. Figure \ref{dyn_vioUARIV_ext} shows the results for the measures from the Victims' Unit, while Figure \ref{dyn_vioCEDE_ext} shows the results for the other measures of conflict from different sources. The pre-treatment coefficients are jointly and individually significant for most of the measures. For those with significant joint tests, these are driven by a single significant year (expected due to the number of coefficients estimated), no trend is apparent, and the sup-t confidence bands cover the zero. 

\clearpage
\subsubsection{Alternative Measures of Insurgent Groups' Presence}

\begin{figure}[h!]
\begin{center}
\caption{Violence in FARC Municipalities vs. ELN Municipalities -- Intensive Margin, Top 20\% Most Violent}
\label{viol_es_int}
\includegraphics[width=0.8\textwidth]{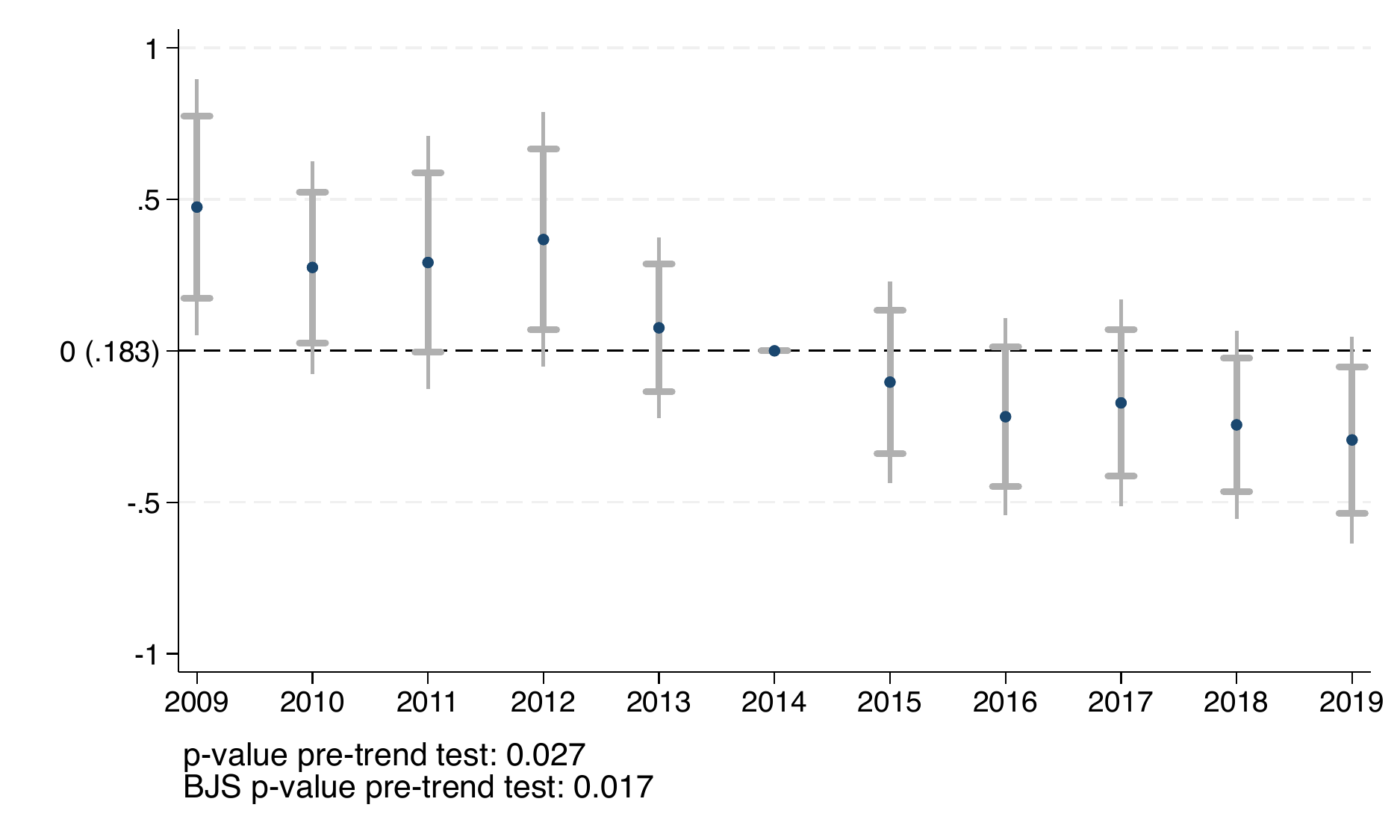}
\end{center}
\justifying 
\footnotesize{\textbf{Notes:}  Event study plots from estimating Equation \eqref{eq_es}, including including 95\% confidence intervals (based on standard errors clustered at the municipality level). The index is created following \citet{anderson2008multiple} and is based on the violence measures in Table \ref{DID_final_Vio_CEDE_Ext_p60}.}
\end{figure}

\clearpage
\subsubsection{Heterogeneity in Economic Activity}

\begin{figure}[h!]
\caption{Economic Heterogeneity by Time-Invariant Characteristics -- Triple Interactions}
\label{econHeterogeneity}
\begin{center}
\begin{subfigure}[b]{0.65\textwidth}
	\includegraphics[width=\textwidth]{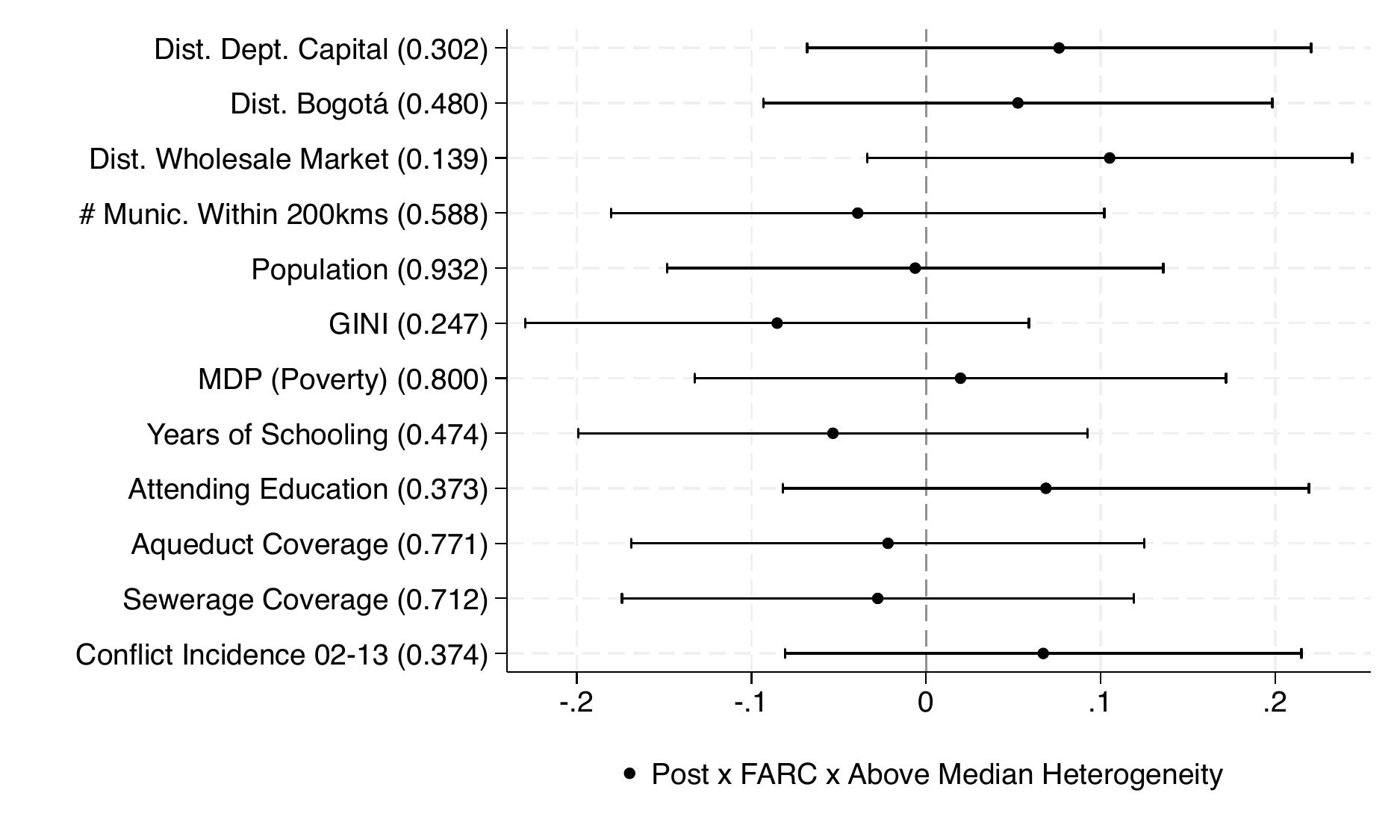}
	\subcaption{Municipality's Characteristics}
\end{subfigure}
\begin{subfigure}[b]{0.65\textwidth}
	\includegraphics[width=\textwidth]{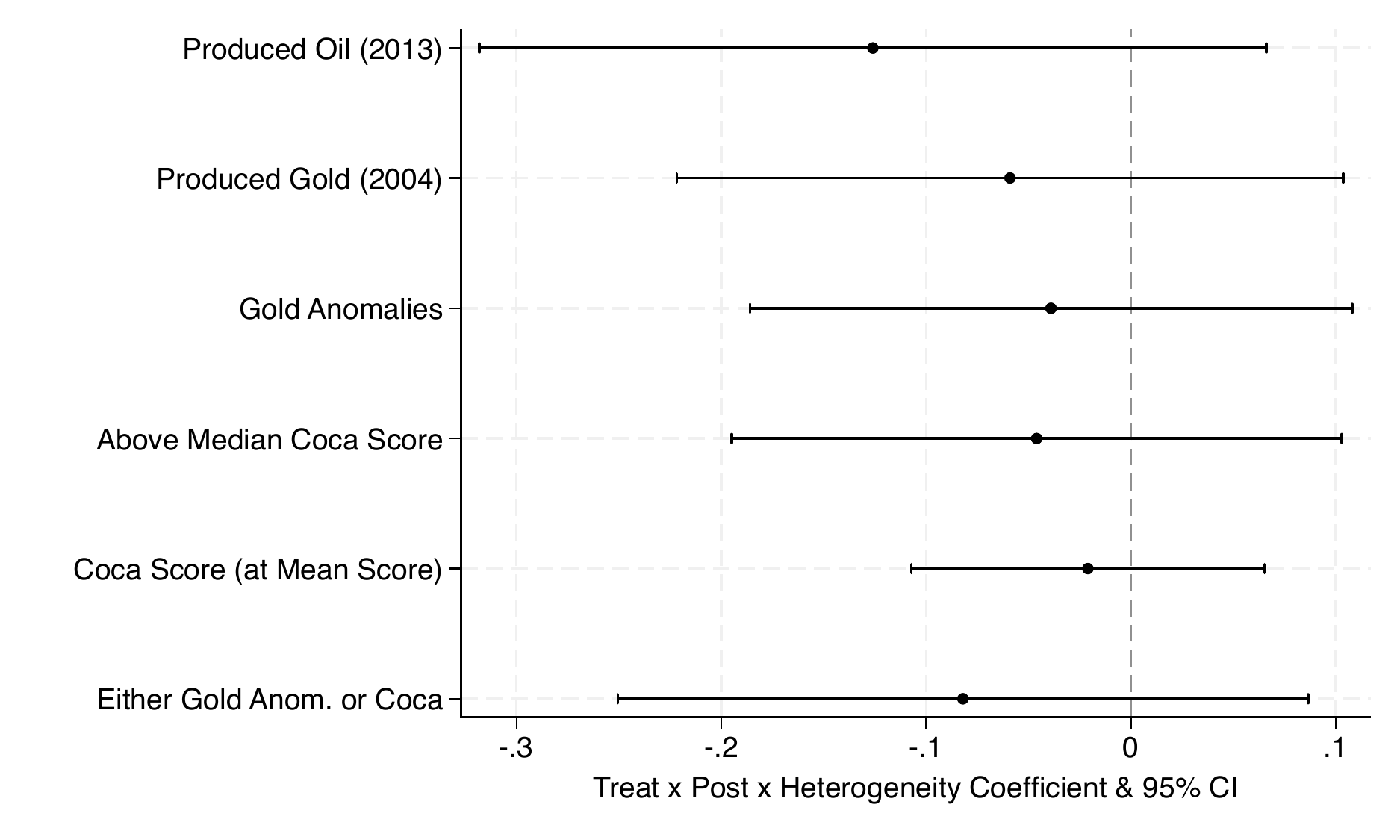}
	\subcaption{Presence of Resources}
\end{subfigure}
\end{center}
\justifying \footnotesize{\textbf{Notes:} Coefficients and 95\% confidence intervals from estimating Equation \eqref{eq_het}, plotting the triple interaction term between the post-ceasefire dummy, the FARC dummy, and a dummy for being above the median value of the heterogeneity variable. The dependent variable is the Anderson Index composed of economic indicators. Each row corresponds to a different variable used to divide treated municipalities, based on whether the municipality is above or below the median of said variable among treated municipalities. In Panel A, the first three rows show birdseye distance in kms to the department capital, Bogotá, and the closest food wholesale market, respectively. The next rows correspond to: number of municipalities within 200 kms (birdseye), population, GINI coefficient, poverty incidence, average years of schooling, average school attendance of people aged 5-24, and the aqueduct and sewerage coverage. All figures are from 2005, based on the 2005 Census. $p$-value of the triple interaction coefficient shown in parentheses. Heterogeneity variables have been recoded so that above the median corresponds to characteristics more conductive to economic activity (recoded variables are the ones related to distances, the GINI coefficient, and the poverty incidence). Panel B uses indicators for whether the municipality produced oil in 2013, produced gold in 2004, has gold geochemical anomalies, is above the median coca suitability score, a continuous measure of coca suitability (evaluated at the mean suitability score), and whether it has gold geochemical anomalies, or is above the median coca suitability score.}
\end{figure}

\clearpage
\subsubsection{Robustness to Alternative Measure of Presence}

\begin{figure}[htp]
\caption{Economic Activity in FARC vs. ELN Municipalities -- Intensive Margin, Top 20\% Most Violent}
\label{dyn_econ_int}
\begin{center}
\begin{subfigure}[b]{0.44\textwidth}
\includegraphics[width=\textwidth]{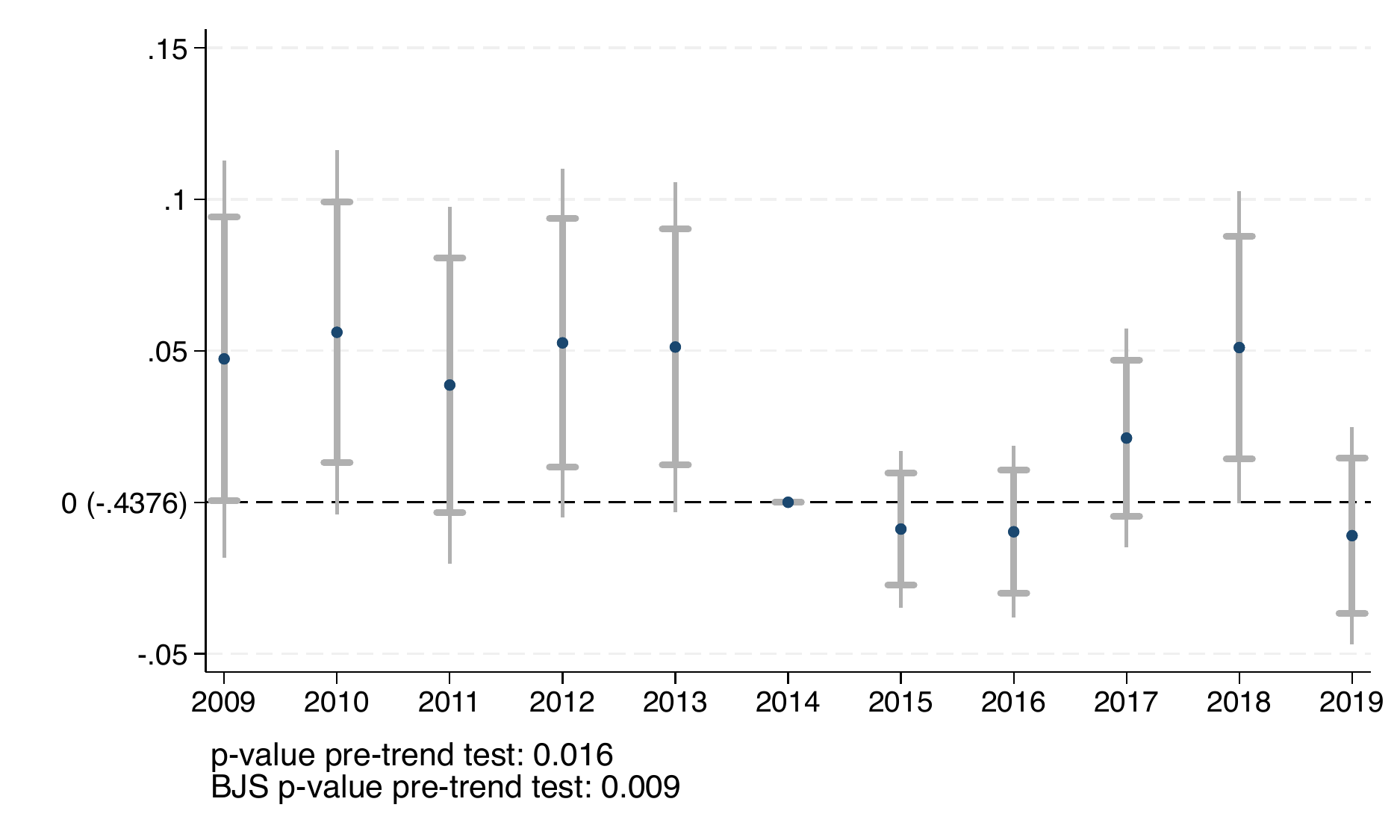}
\subcaption{Nighttime Light Weighted}
\end{subfigure}
\begin{subfigure}[b]{0.44\textwidth}
\includegraphics[width=\textwidth]{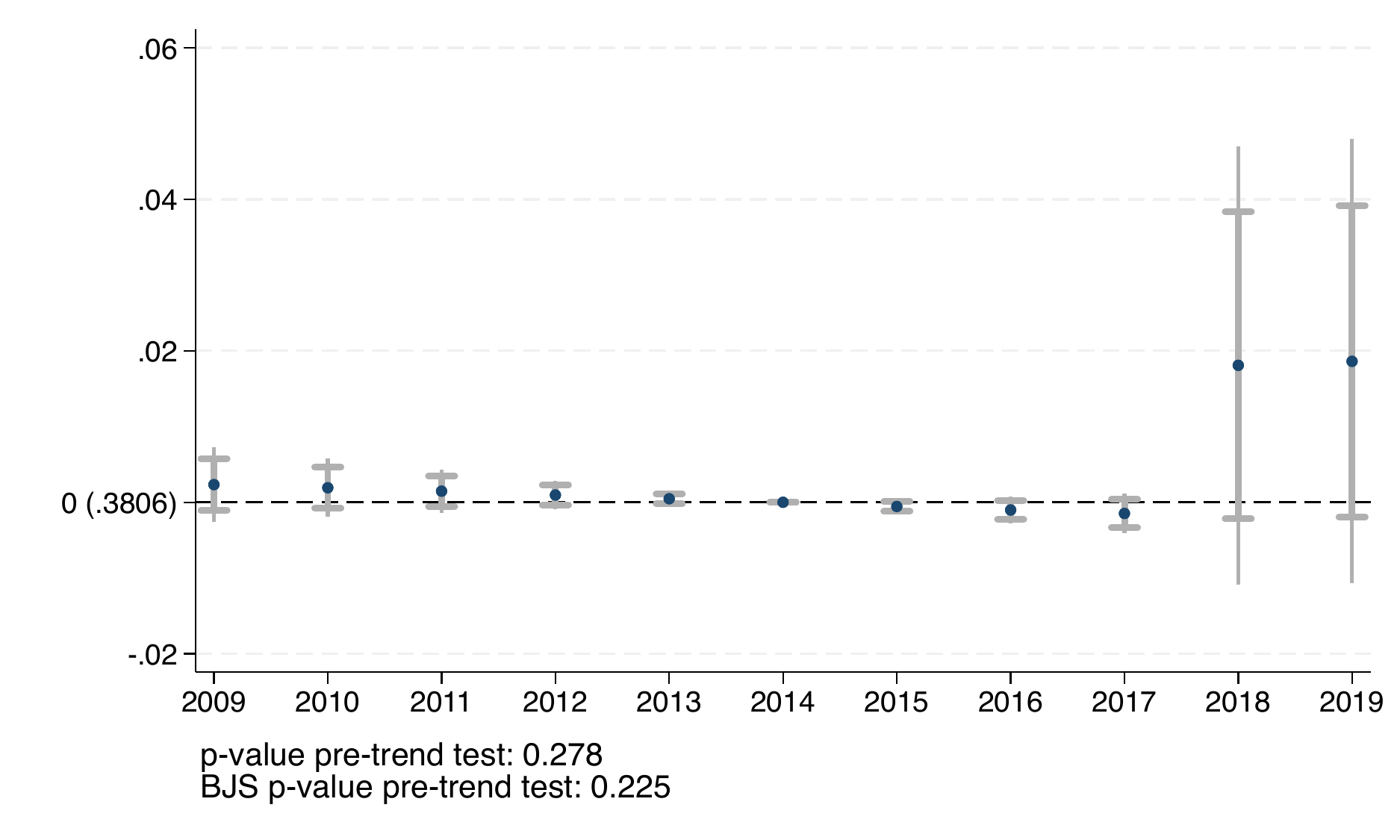}
\subcaption{Share Urban Population}
\end{subfigure}
\begin{subfigure}[b]{0.44\textwidth}
\includegraphics[width=\textwidth]{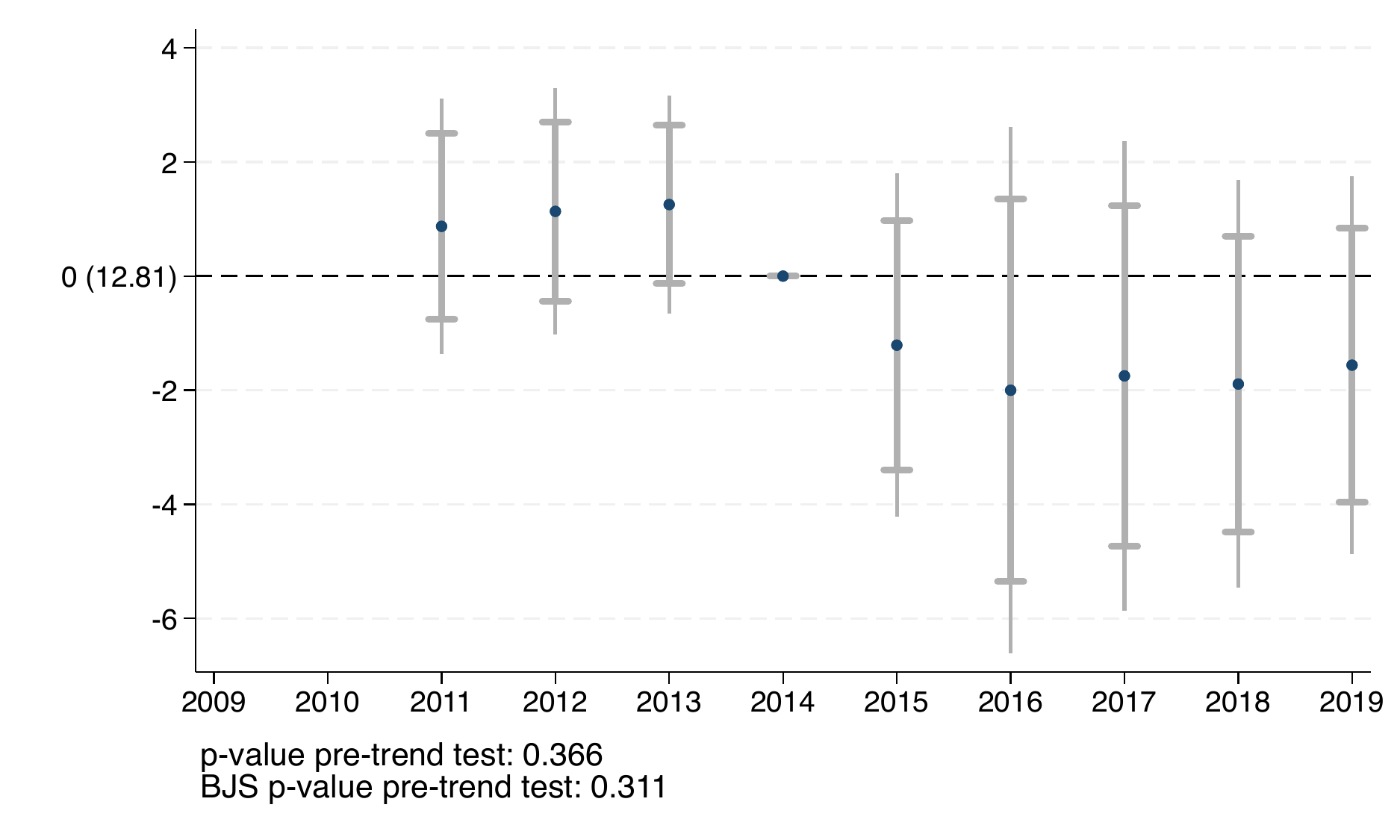}
\subcaption{Value Added (pc, DANE)}
\end{subfigure}
\begin{subfigure}[b]{0.44\textwidth}
\includegraphics[width=\textwidth]{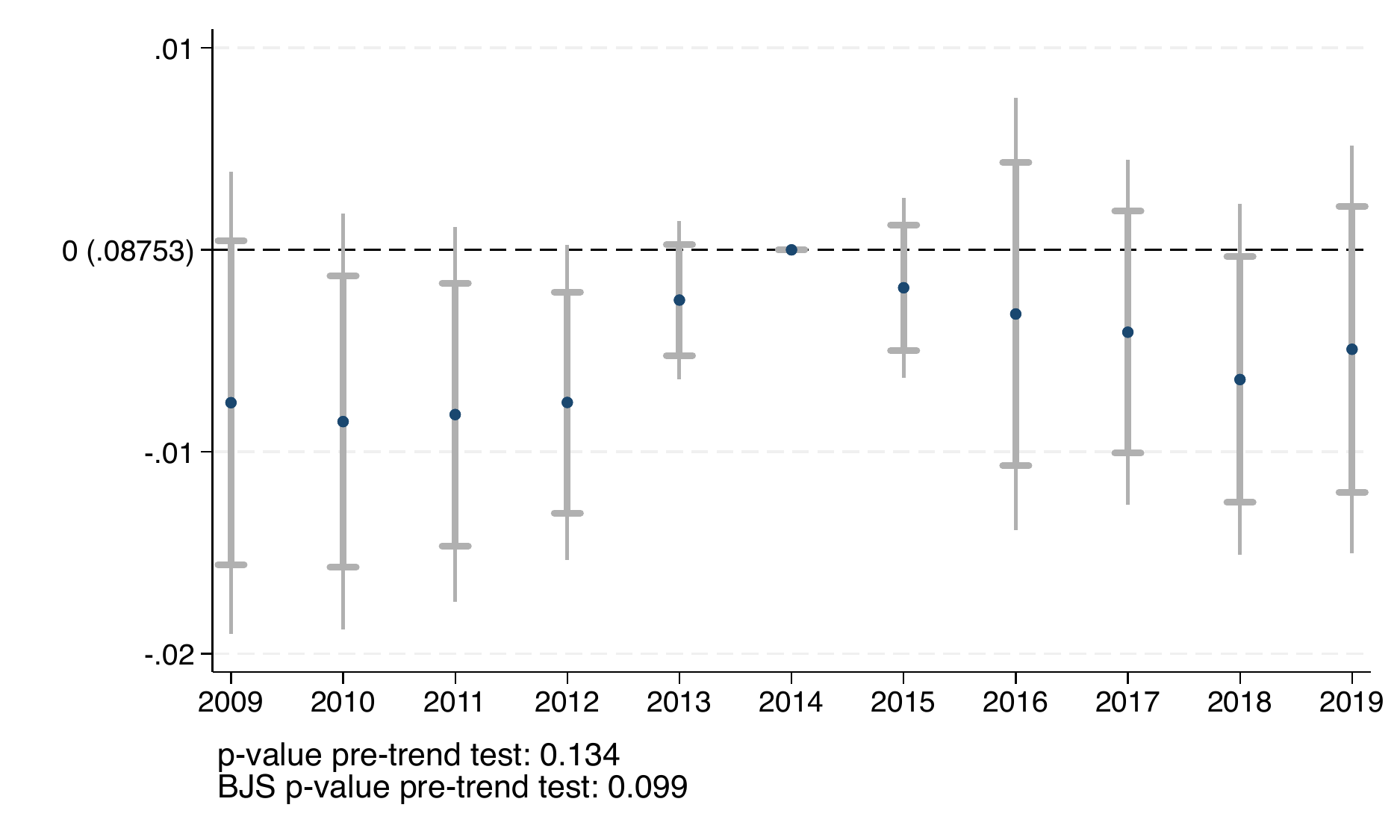}
\subcaption{Formal Employment}
\end{subfigure}
\begin{subfigure}[b]{0.44\textwidth}
\includegraphics[width=\textwidth]{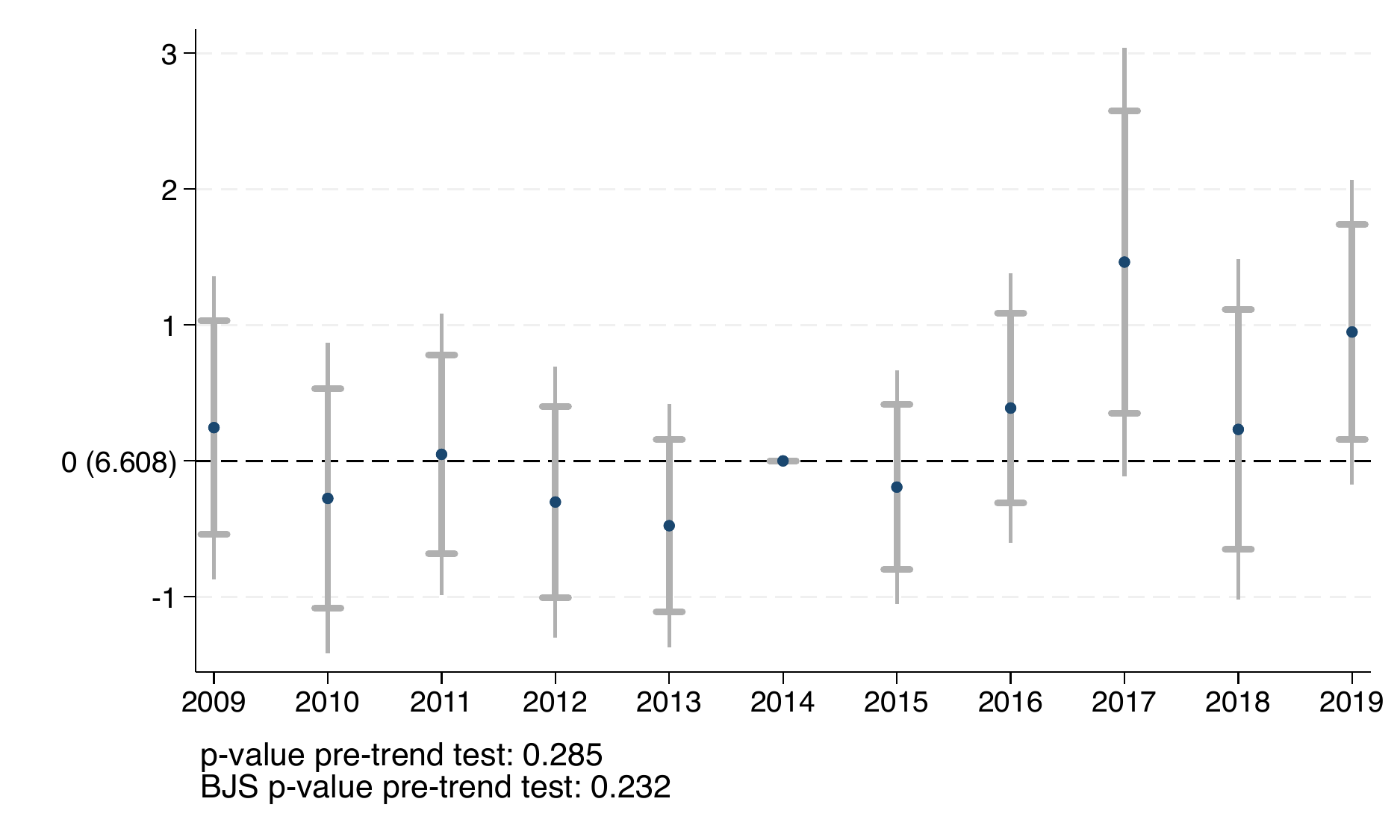}
\subcaption{Firm Entry}
\end{subfigure}
\begin{subfigure}[b]{0.44\textwidth}
\includegraphics[width=\textwidth]{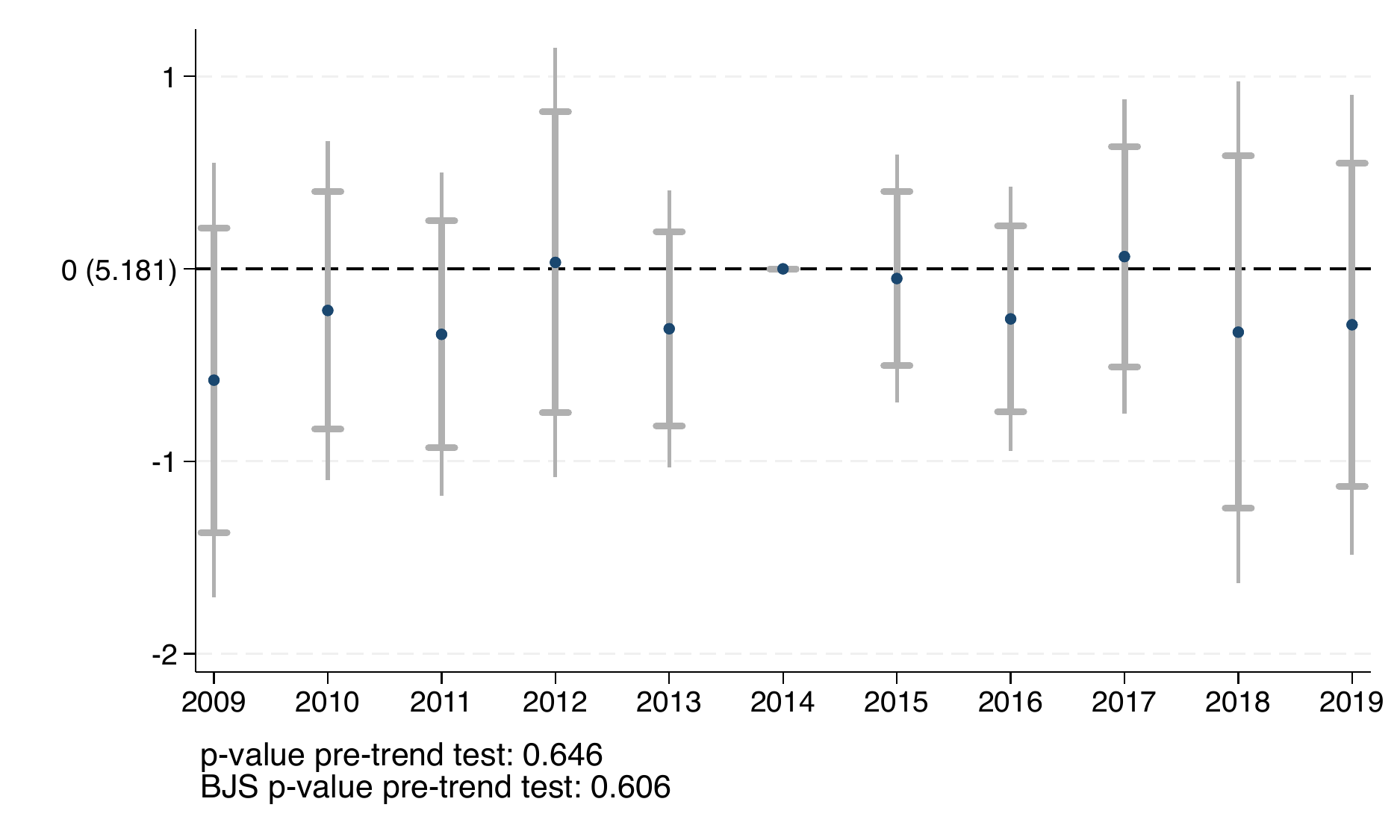}
\subcaption{Agricultural Productivity}
\end{subfigure}
\begin{subfigure}[b]{0.44\textwidth}
\includegraphics[width=\textwidth]{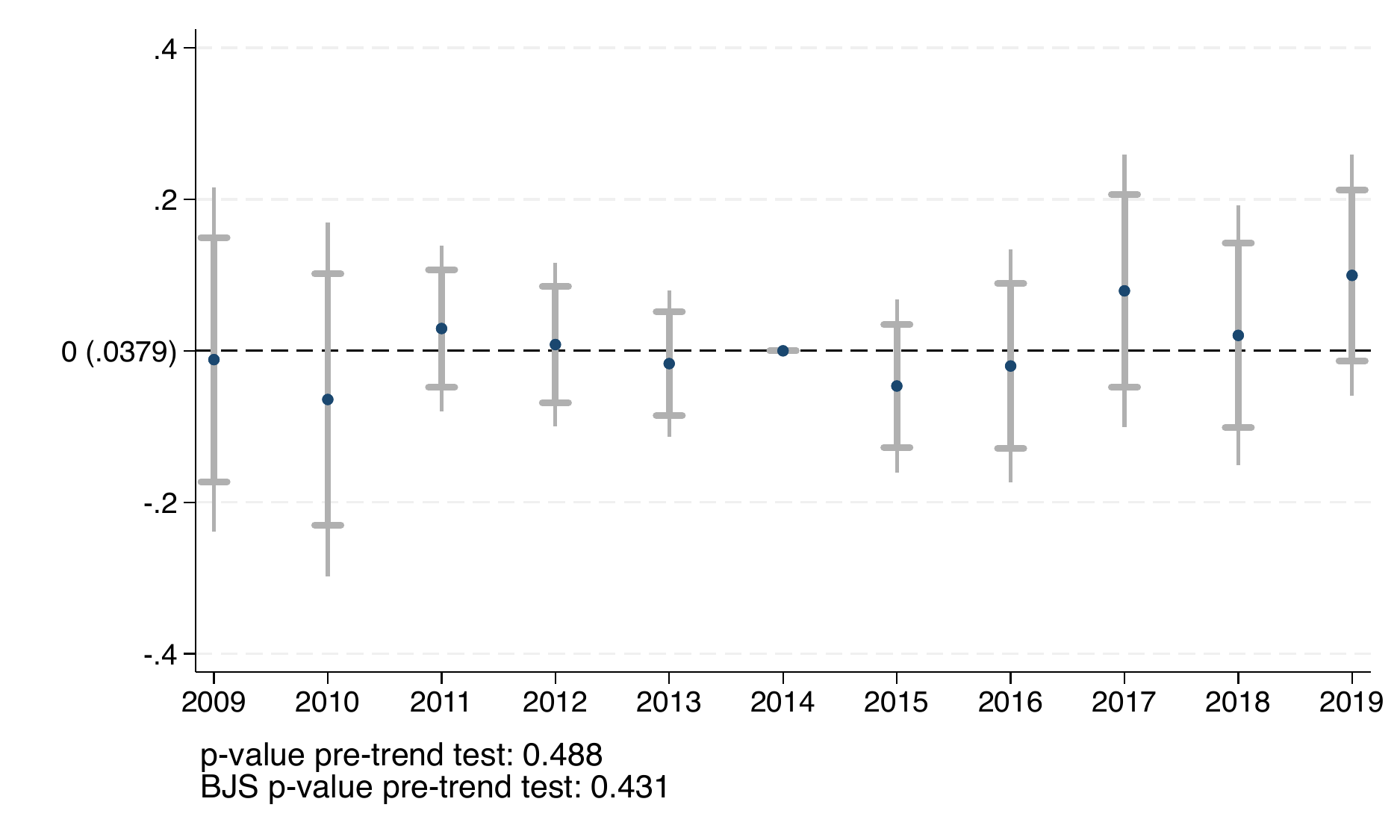}
\subcaption{Anderson Index}
\end{subfigure}
\end{center}
\justifying 
\footnotesize{\textbf{Notes:}  Event study plots from estimating Equation \eqref{eq_es}, including including 95\% confidence intervals (based on standard errors clustered at the municipality level). The index is created following \citet{anderson2008multiple} and is based on weighted nighttime light intensity, value added per capita (from DANE), share of urban population, agricultural productivity, firm creation and formal employment.}
\end{figure}

\begin{figure}[h!]
\caption{State Capacity Outcomes in FARC vs. ELN Municipalities -- Intensive Margin, Top 20\% Most Violent}
\label{dyn_stateCap_int}
\centering
\begin{subfigure}[b]{0.48\textwidth}
\includegraphics[width=\textwidth]{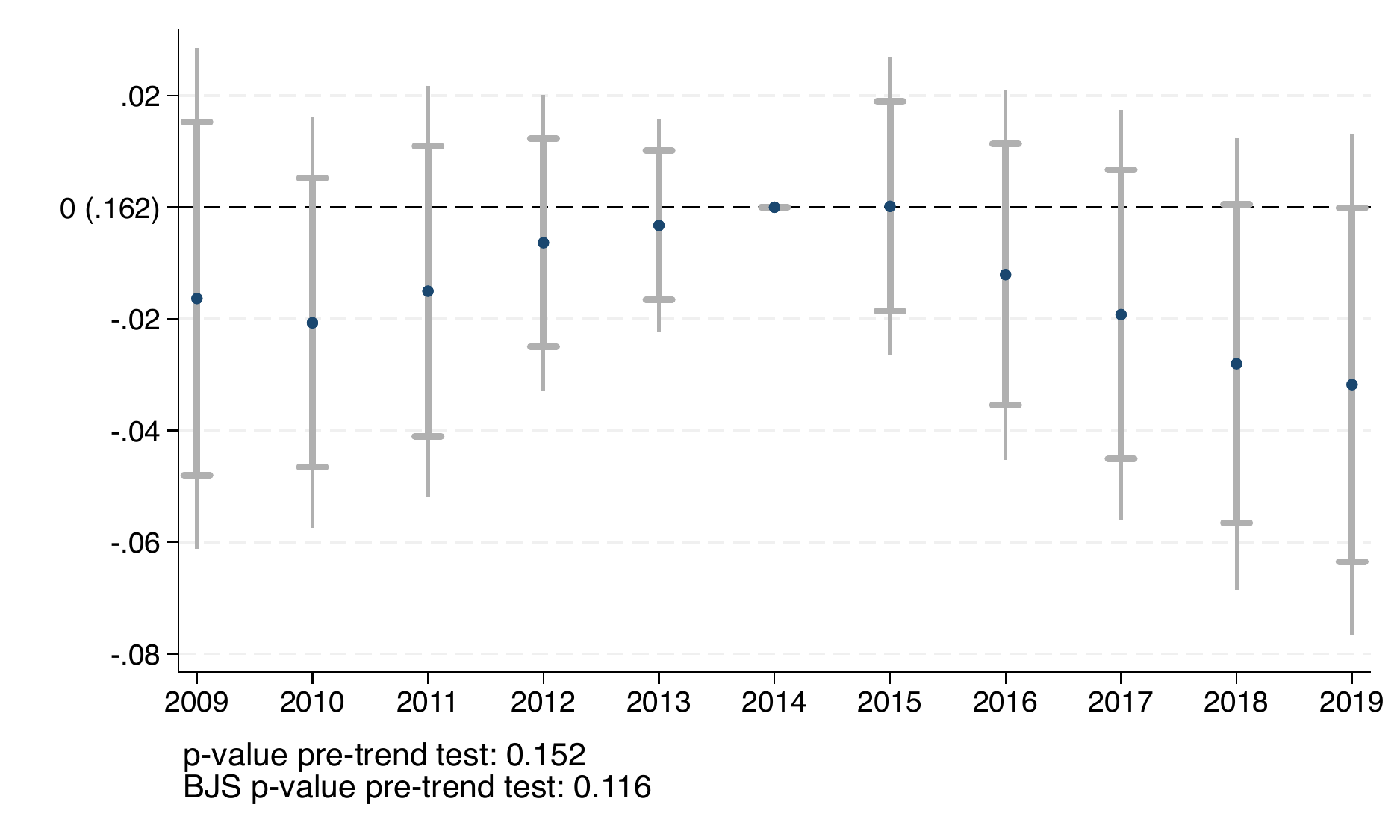}
\subcaption{Tax Revenue per Capita}
\end{subfigure}
\begin{subfigure}[b]{0.48\textwidth}
\includegraphics[width=\textwidth]{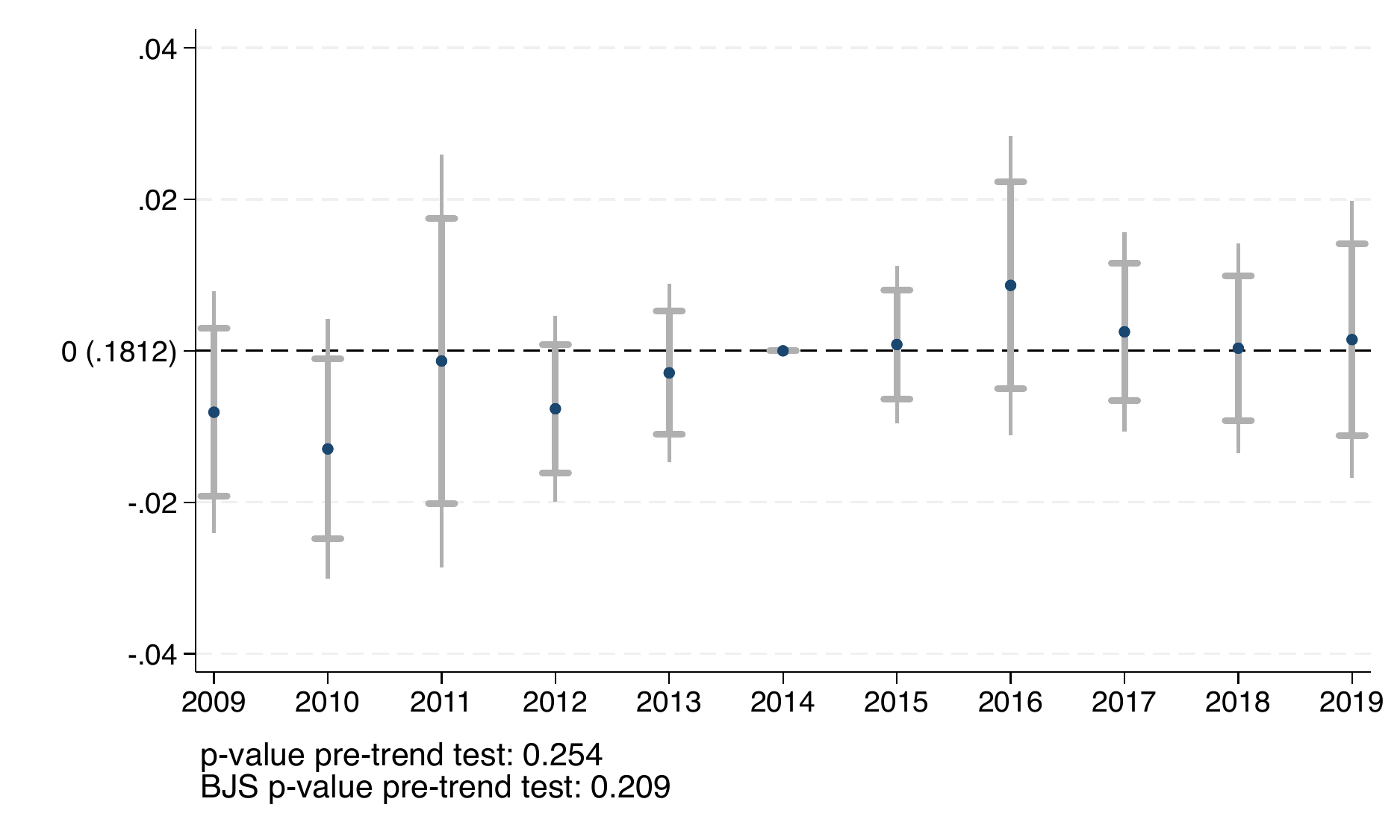}
\subcaption{Operational Costs (pc)}
\end{subfigure}
\begin{subfigure}[b]{0.48\textwidth}
\includegraphics[width=\textwidth]{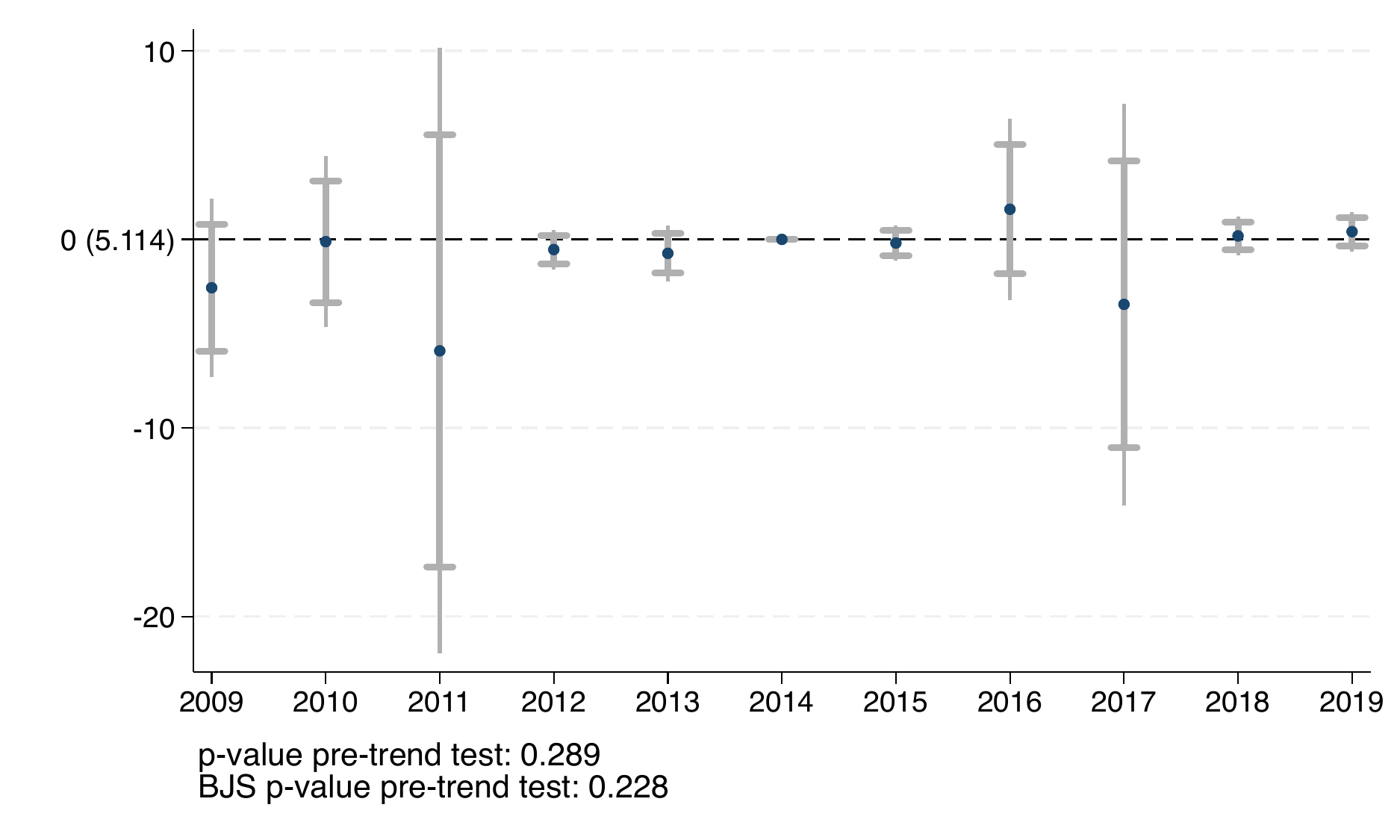}
\subcaption{Ratio Gov. Trans. to Revenue}
\end{subfigure}
\begin{subfigure}[b]{0.48\textwidth}
\includegraphics[width=\textwidth]{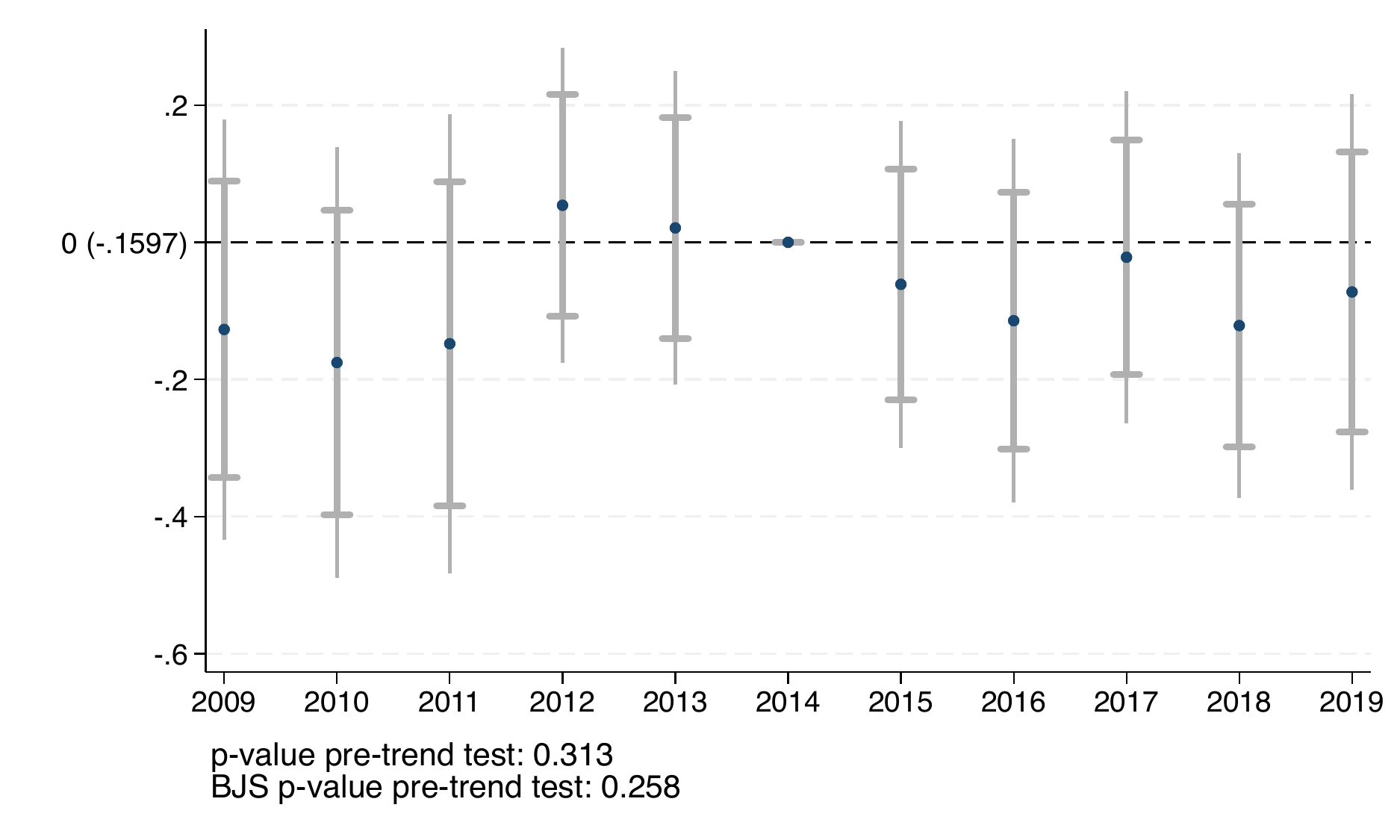}
\subcaption{Fiscal Performance}
\end{subfigure}
\begin{subfigure}[b]{0.48\textwidth}
\includegraphics[width=\textwidth]{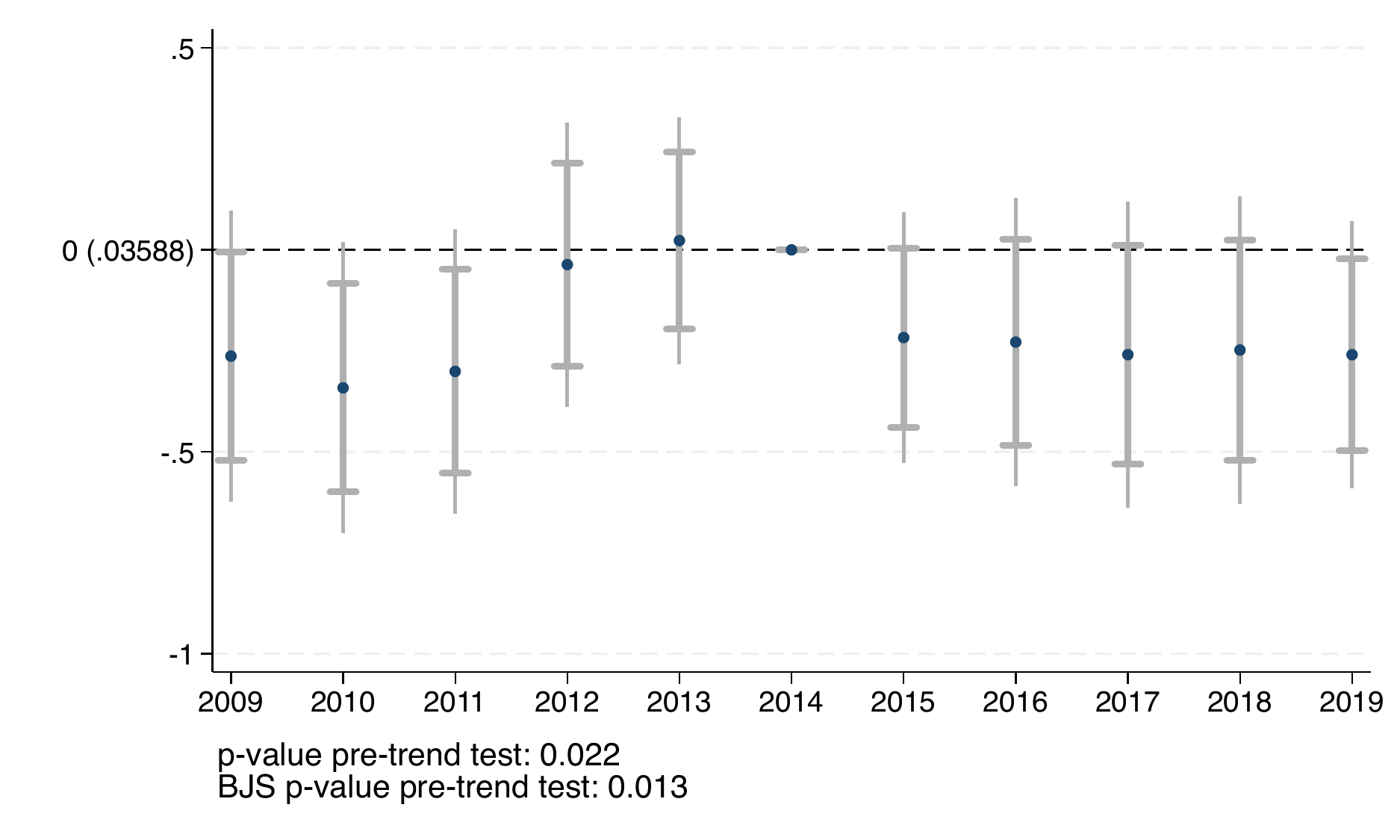}
\subcaption{Administrative Performance}
\end{subfigure}
\begin{subfigure}[b]{0.48\textwidth}
\includegraphics[width=\textwidth]{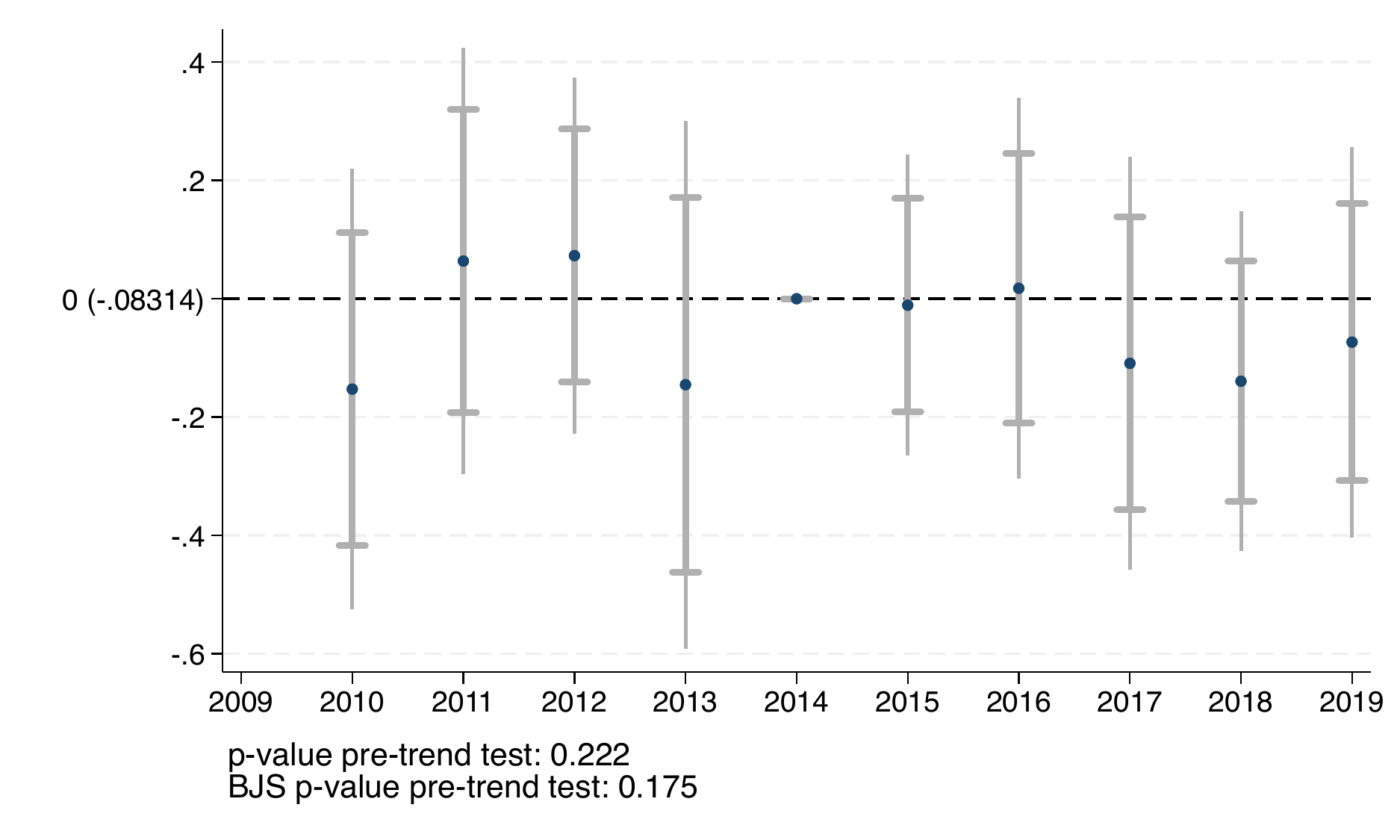}
\subcaption{Information Openness}
\end{subfigure}
\begin{subfigure}[b]{0.48\textwidth}
\includegraphics[width=\textwidth]{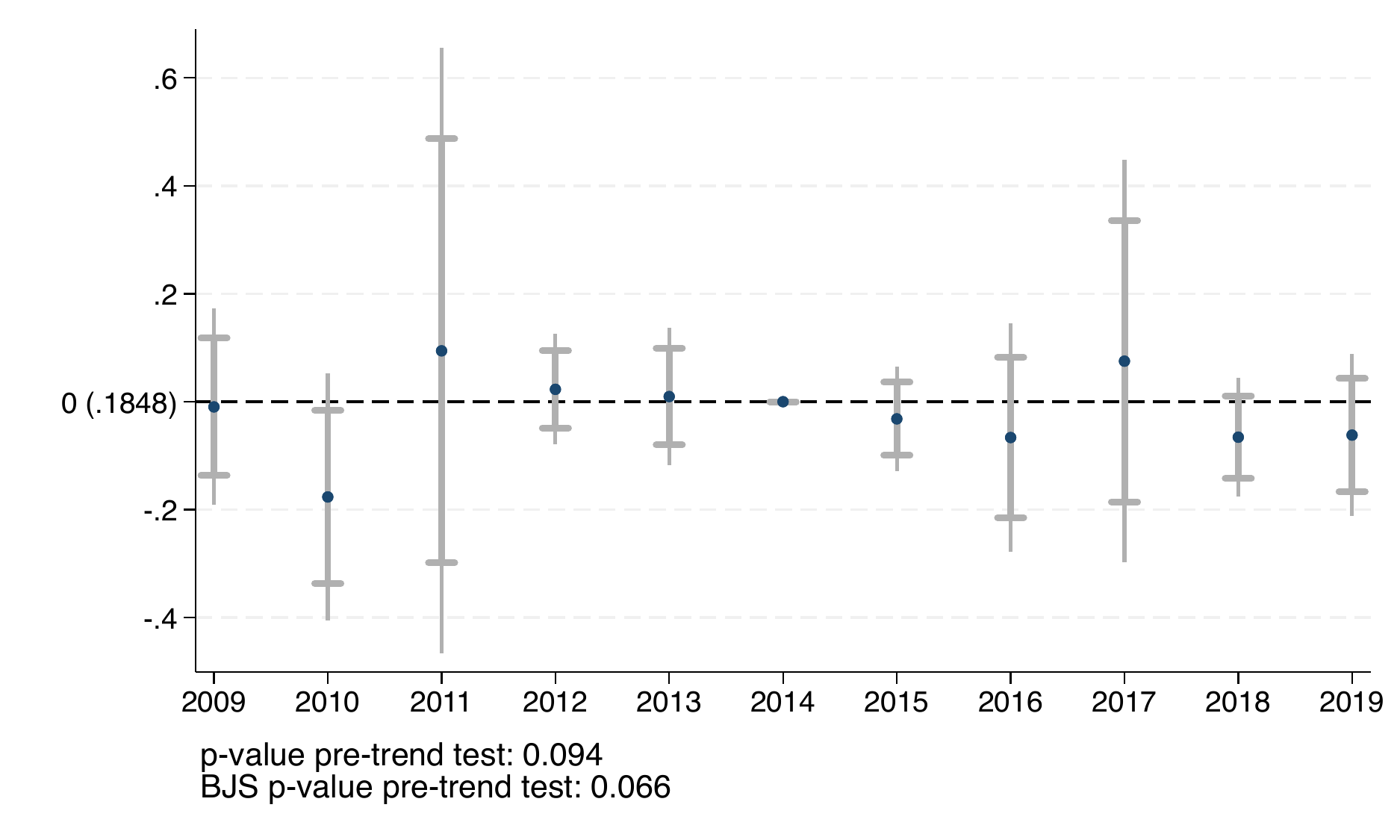}
\subcaption{Anderson Index}
\end{subfigure}
\justifying 

\footnotesize{\noindent \textbf{Notes:}  Event study plots from estimating Equation \eqref{eq_es} for the different state capacity measures, including including 95\% confidence intervals (based on standard errors clustered at the municipality level). The panels show results using tax revenue, operational costs, the ratio of transfers from the central government to total municipality revenue (excluding transfers from the central government), an indicator of fiscal performance created by the National Department of Planning (DNP), an indicator of the municipality's overall administrative performance created by the National Department of Planning (DNP), and an indicator of information openness created by the Office of the Inspector General, respectively. The last panel shows the results using a summary index of all these variables following \citet{anderson2008multiple}.}

\end{figure}

\begin{figure}[htp]
\caption{Robustness to Alt. Thresholds of Presence Measures -- Intensive Margin}
\label{rob_thresholds_int}
\centering
\begin{subfigure}[b]{0.63\textwidth}
\includegraphics[width=\textwidth]{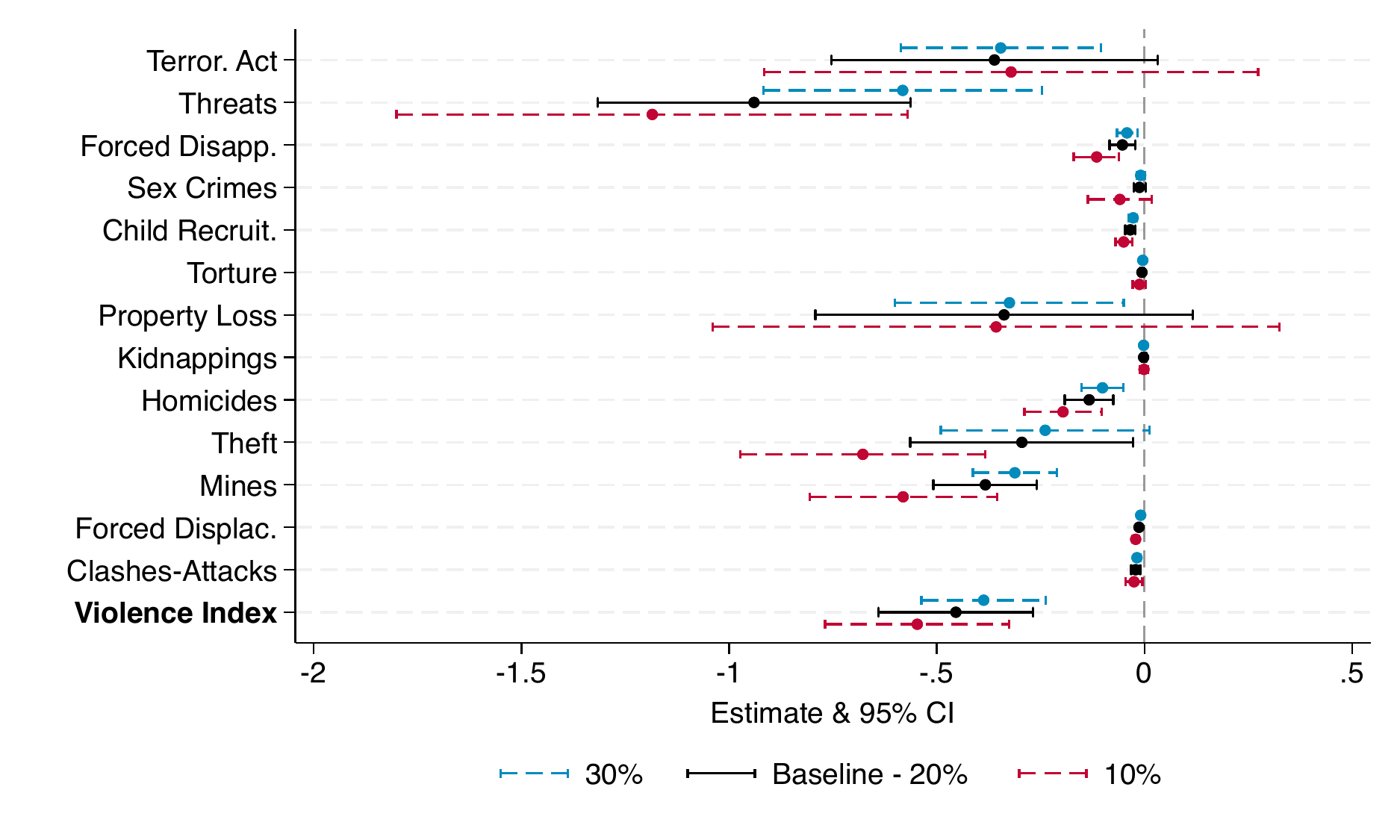}
\subcaption{Violence Measures}
\end{subfigure}
\begin{subfigure}[b]{0.63\textwidth}
\includegraphics[width=\textwidth]{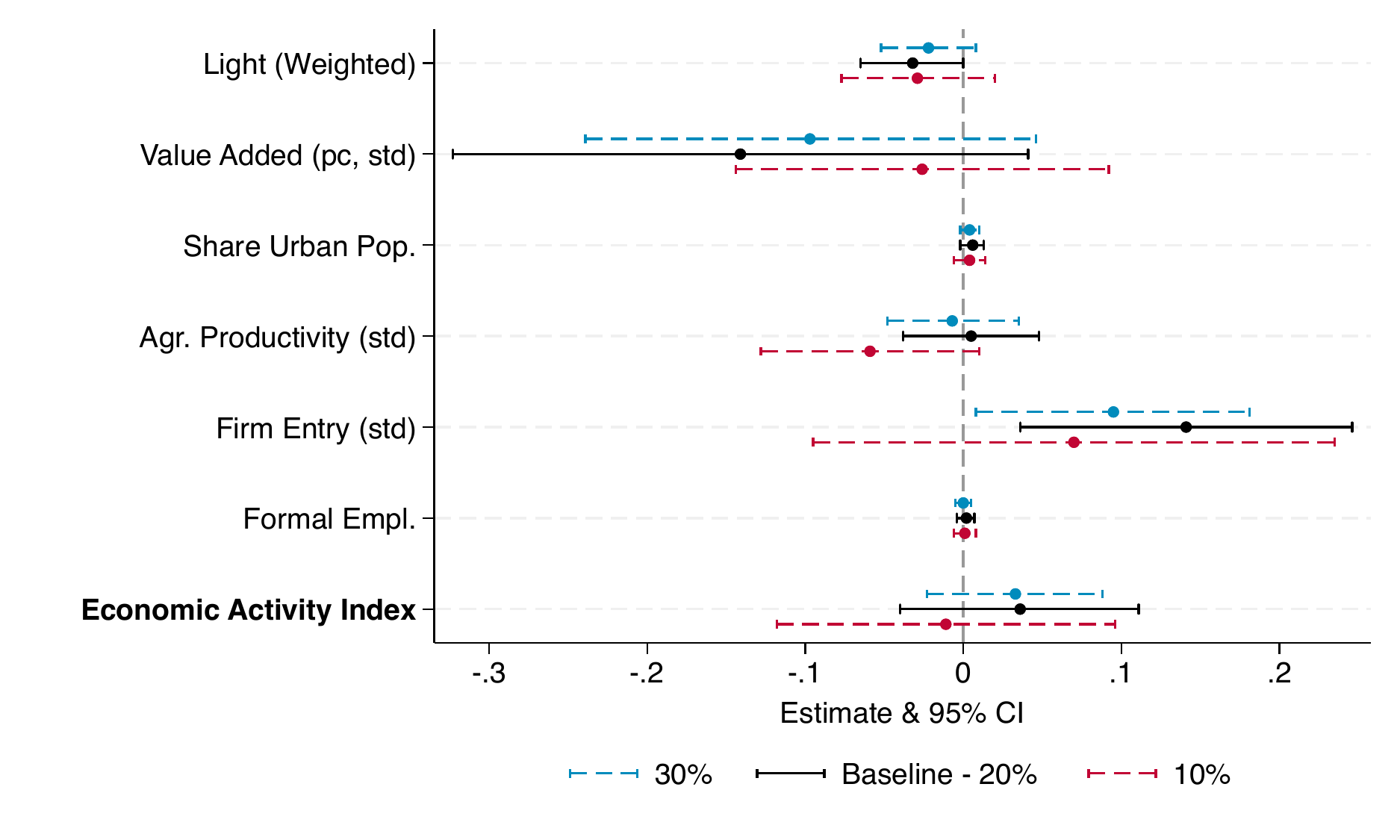}
\subcaption{Economic Indicators}
\end{subfigure}
\begin{subfigure}[b]{0.63\textwidth}
\includegraphics[width=\textwidth]{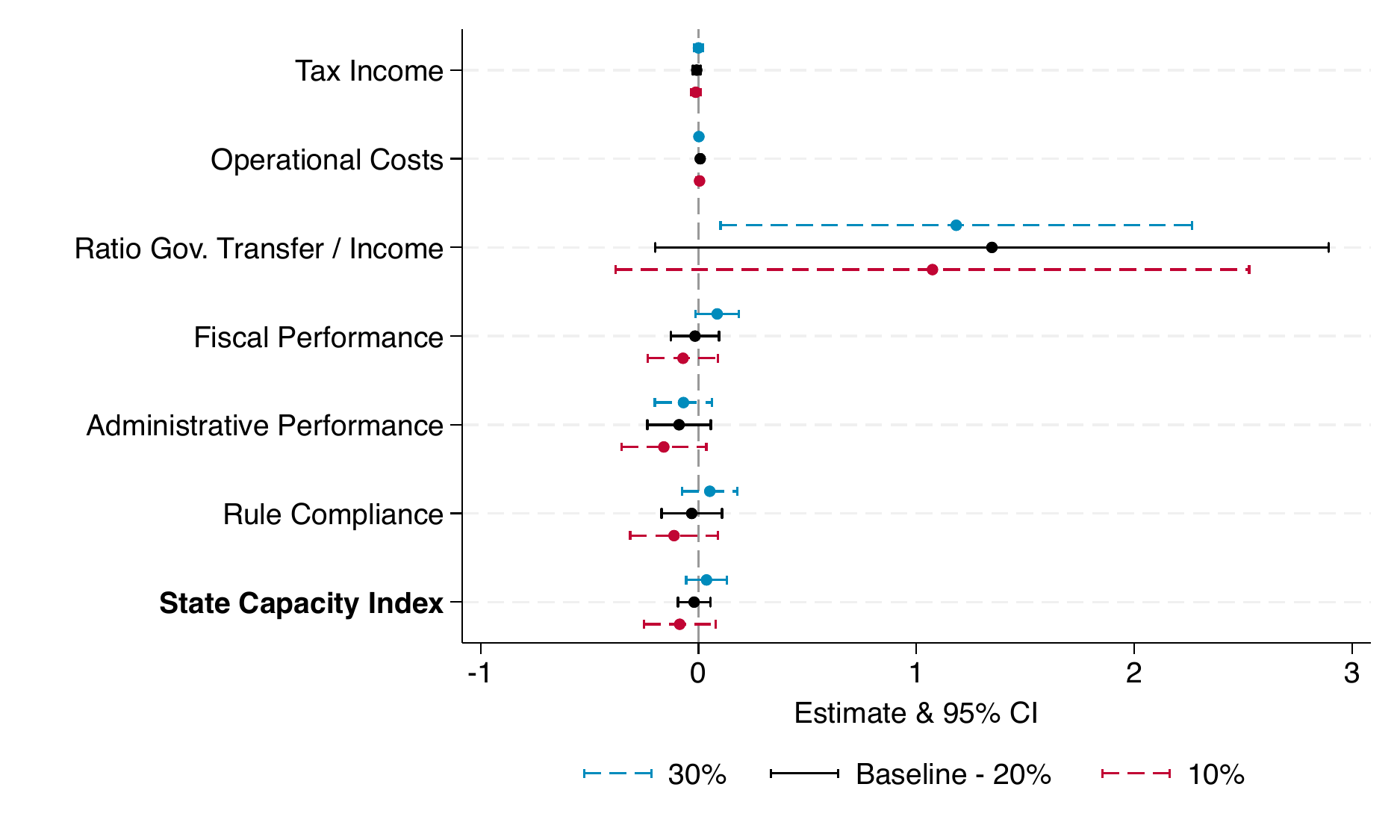}
\subcaption{State Capacity Measures}
\end{subfigure}
\\
\justifying
\noindent\footnotesize{\textbf{Notes:} In Panel A, all variables are in 1000s of inhabitants, except for the migration ones (forced displaced and forced migration) which are measured in per capita terms. In Panel B, the measures of value added per capita have been standardised for comparability.}
\end{figure}

\clearpage
\subsubsection{Synthetic Difference-in-Difference Results}

\begin{figure}[h!]
\caption{Synthetic Difference-in-Difference -- Violence Index}
\label{synt_Viol}
\centering

\bigskip
\textit{Extensive Margin, Events in Over 60\% of Years}
\medskip

\begin{subfigure}[b]{0.49\textwidth}
\includegraphics[width=\textwidth]{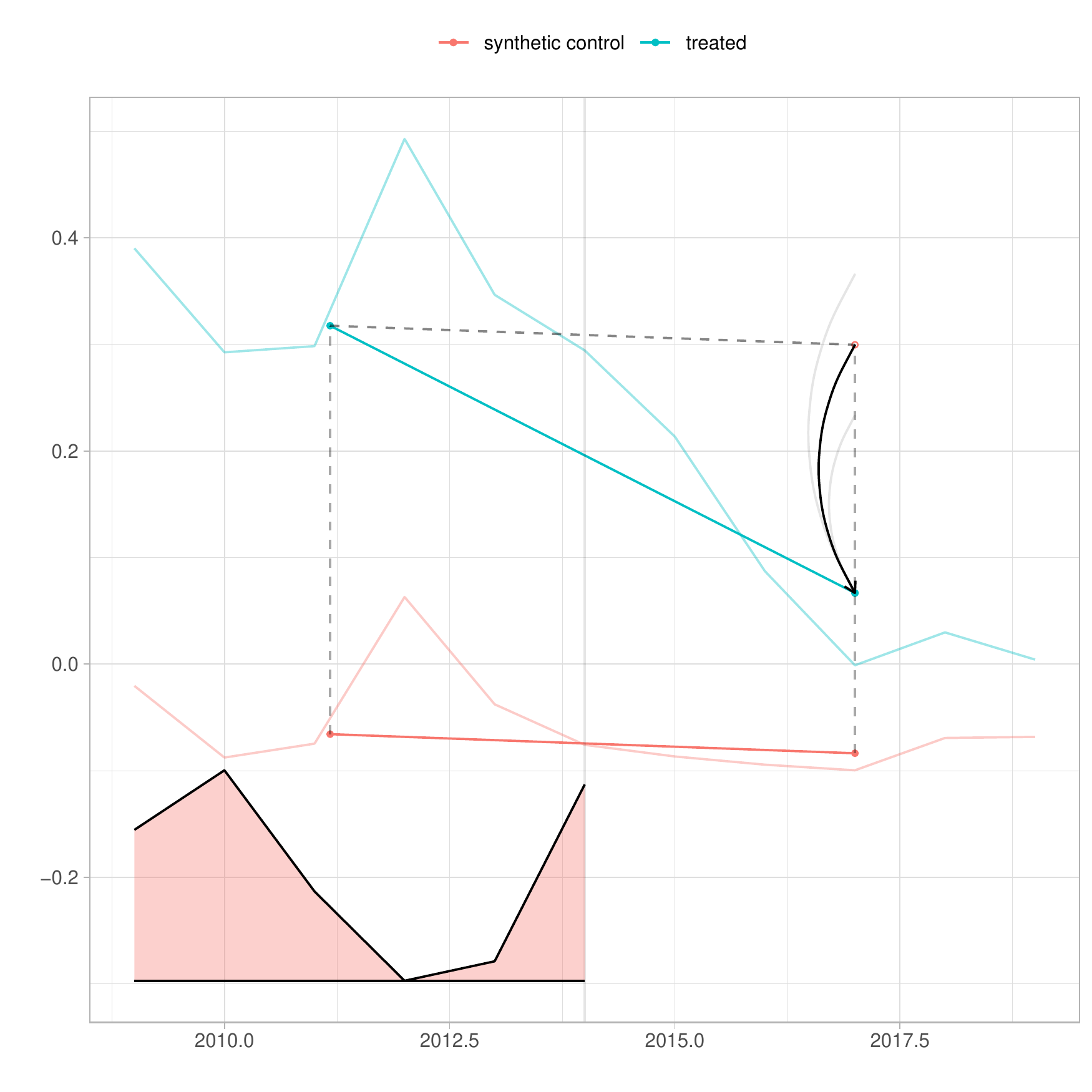}
\subcaption{Extensive Margin -- KLK Index}
\end{subfigure}
\begin{subfigure}[b]{0.49\textwidth}
\includegraphics[width=\textwidth]{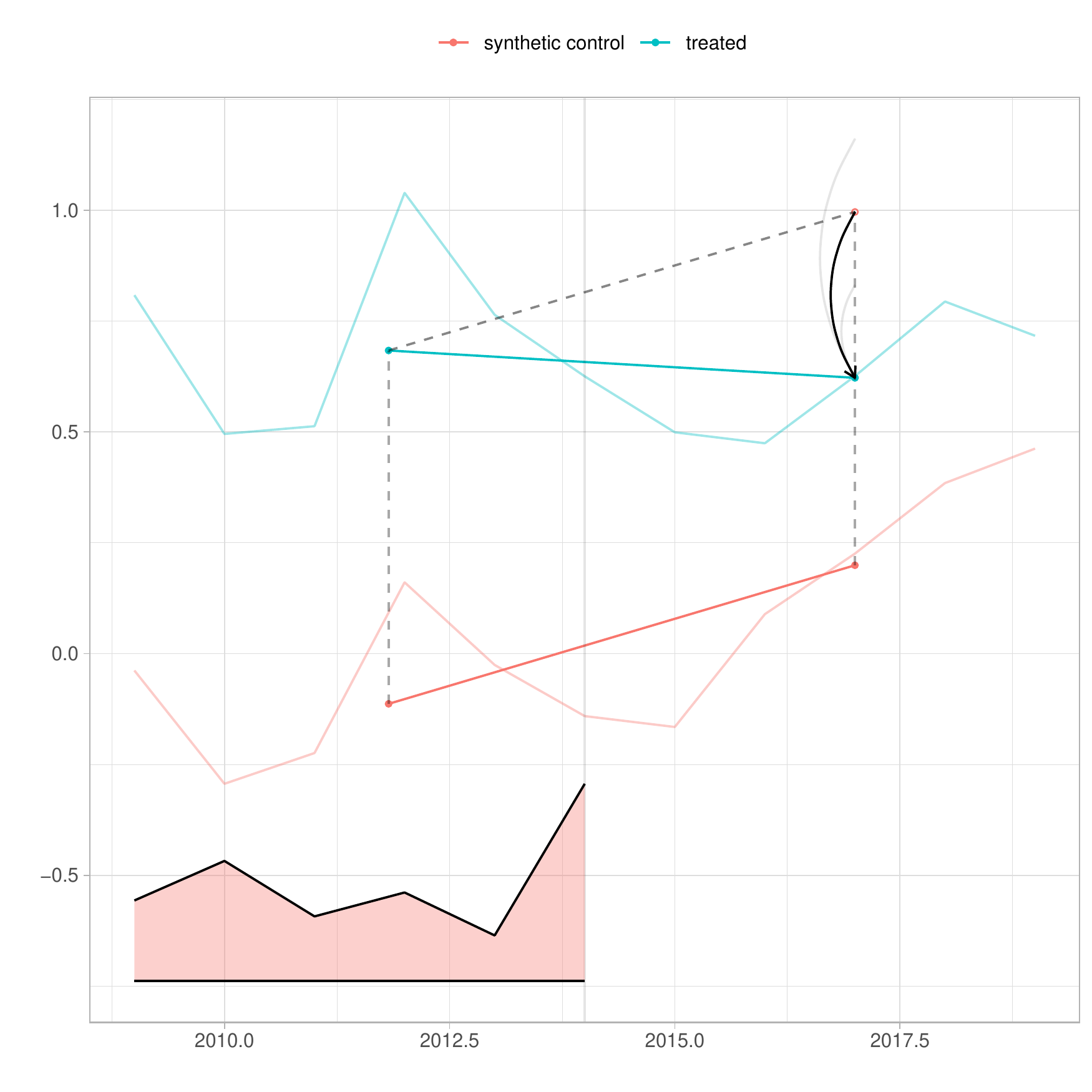}
\subcaption{Extensive Margin -- Anderson Index}
\end{subfigure}

\bigskip
\textit{Intensive Margin, Top 20\% Most Violent}
\medskip

\begin{subfigure}[b]{0.49\textwidth}
\includegraphics[width=\textwidth]{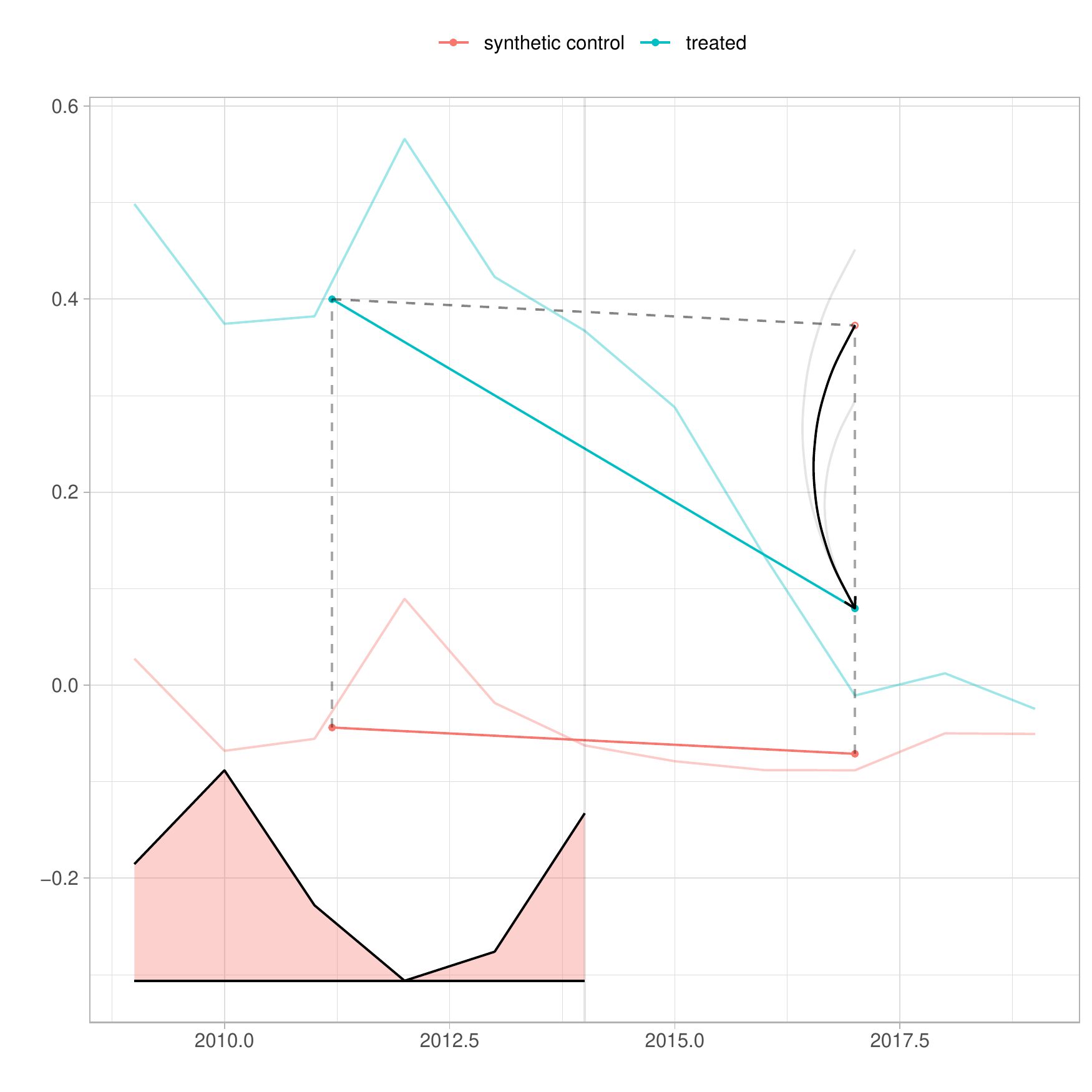}
\subcaption{KLK Index}
\end{subfigure}
\begin{subfigure}[b]{0.49\textwidth}
\includegraphics[width=\textwidth]{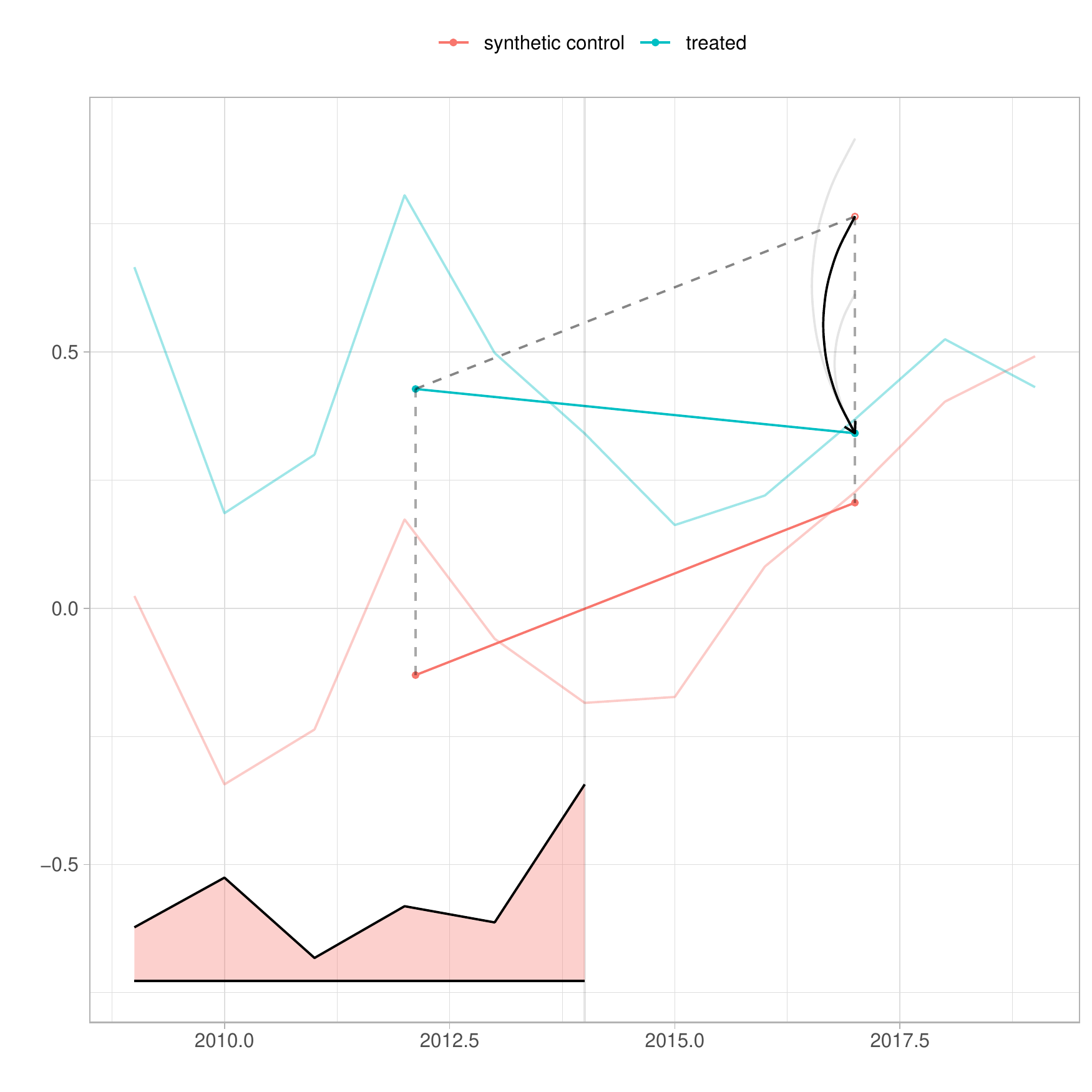}
\subcaption{Anderson Index}
\end{subfigure}
\justifying 
\footnotesize{\textbf{Notes:} Event study plots from the synthetic difference-in-difference estimator developed by \citet{arkhangelsky2021synthetic}. KLK is a summary index created following \citet{kling2007experimental}, while Anderson is a summary index created following \citet{anderson2008multiple}. The index is based on the violence measures in Table \ref{DID_final_Vio_CEDE_Ext_p60}.}
\end{figure}

\begin{figure}[h!]
\caption{Synthetic Difference-in-Difference -- Economic Activity Index}
\label{synt_Econ}
\centering

\bigskip
\textit{Extensive Margin, Events in Over 60\% of Years}
\medskip

\begin{subfigure}[b]{0.49\textwidth}
	\includegraphics[width=\textwidth]{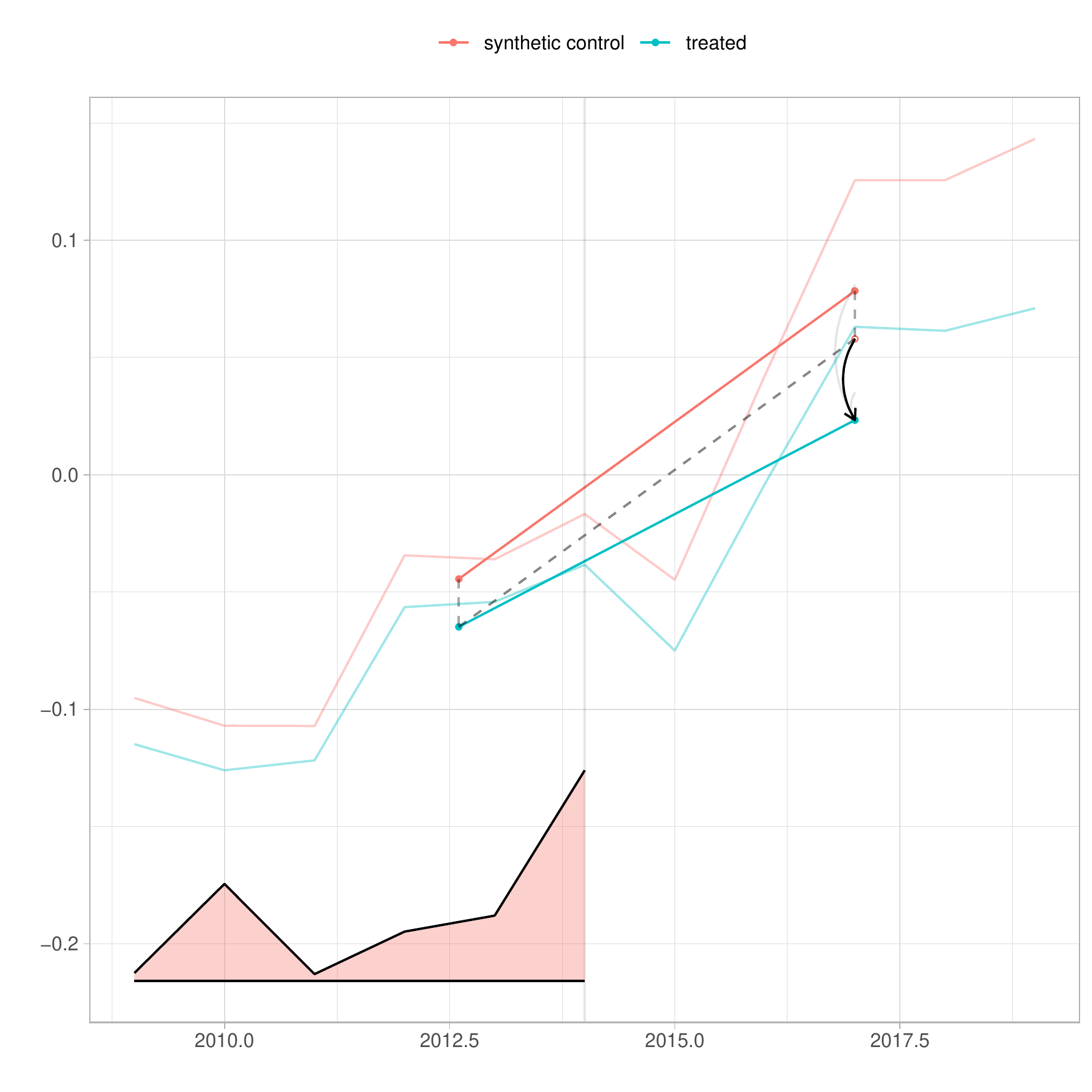}
	\subcaption{Extensive Margin -- KLK Index}
\end{subfigure}
\begin{subfigure}[b]{0.49\textwidth}
	\includegraphics[width=\textwidth]{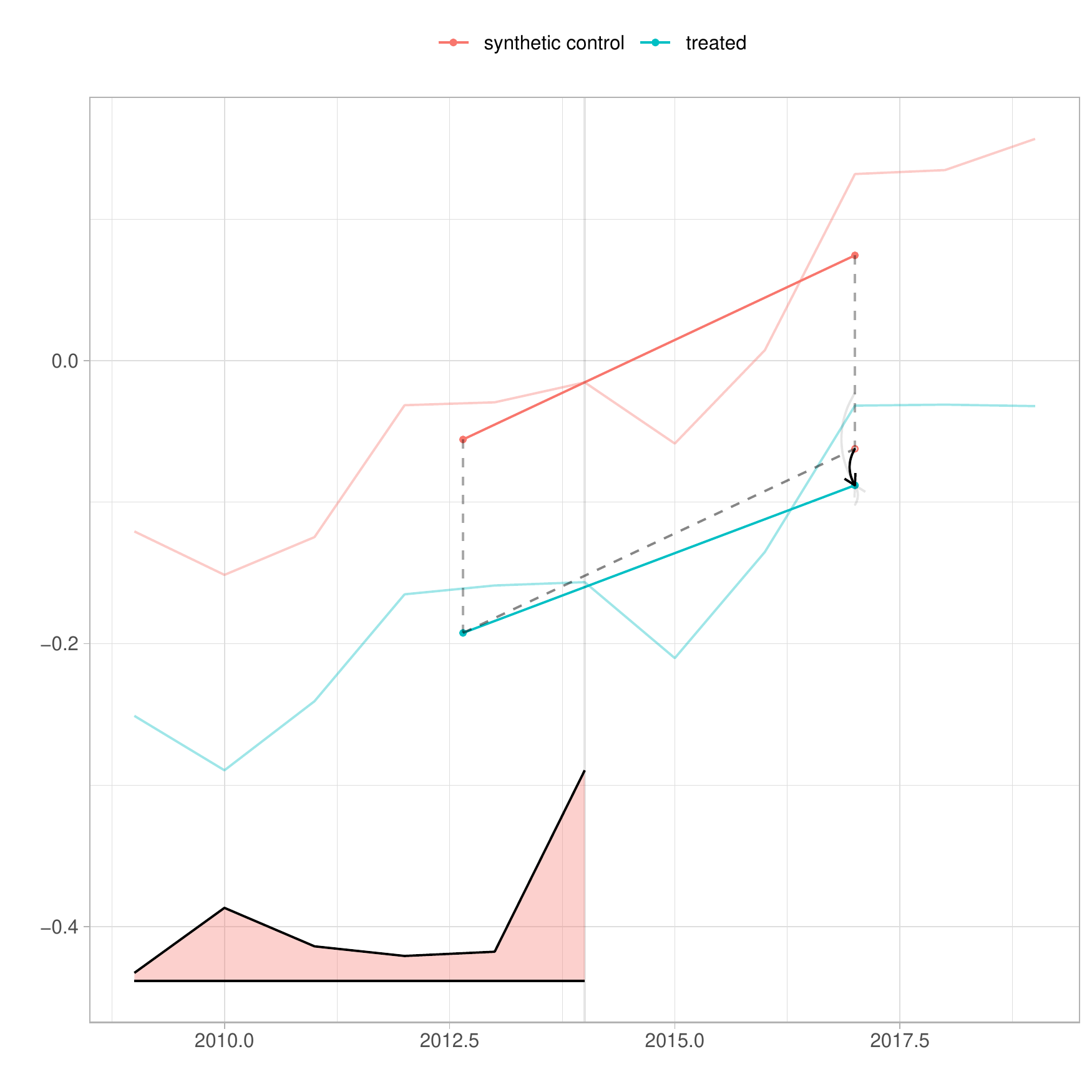}
	\subcaption{Extensive Margin -- Anderson Index}
\end{subfigure}

\bigskip
\textit{Intensive Margin, Top 20\% Most Violent}
\medskip

\begin{subfigure}[b]{0.49\textwidth}
	\includegraphics[width=\textwidth]{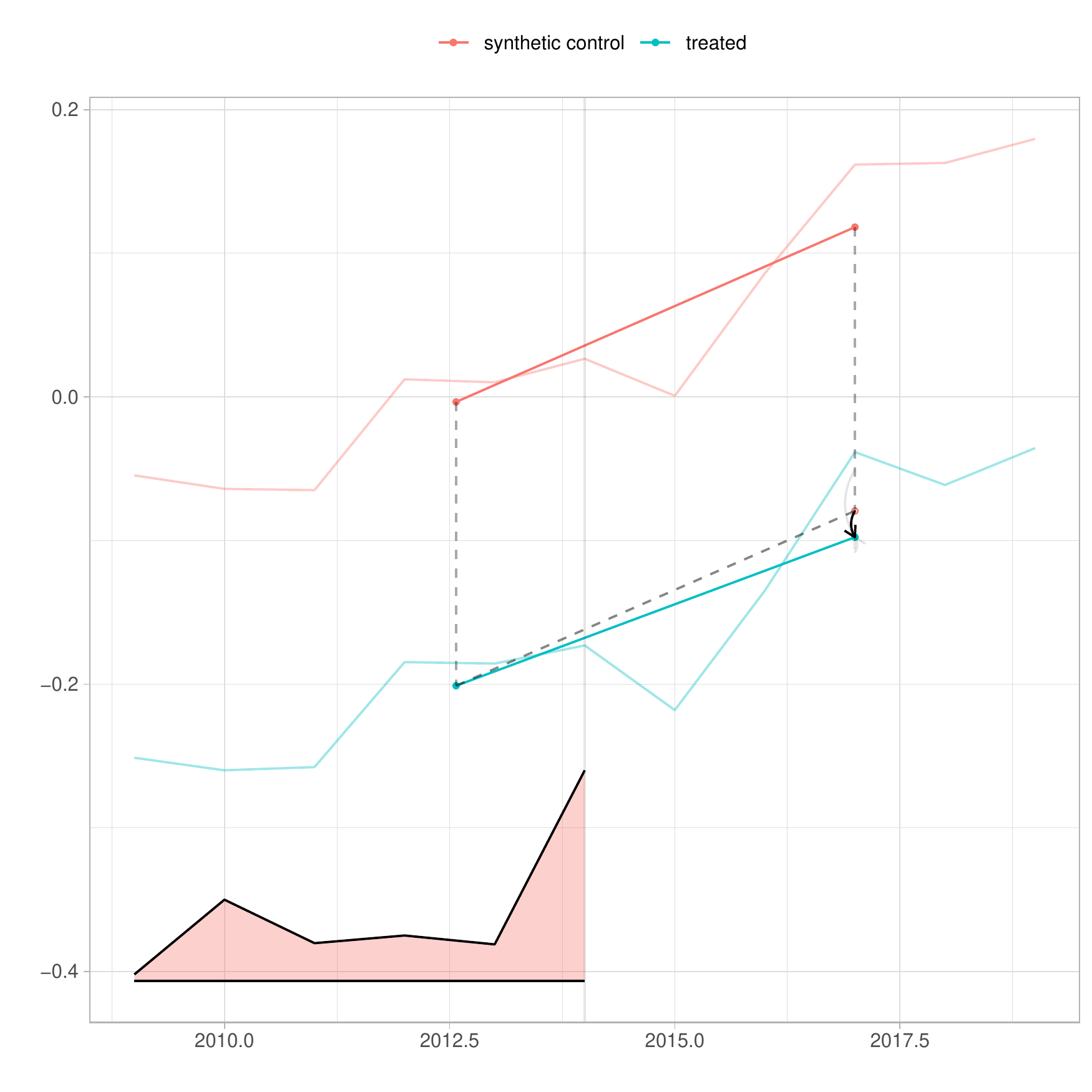}
	\subcaption{Intensive Margin -- KLK Index}
\end{subfigure}
\begin{subfigure}[b]{0.49\textwidth}
	\includegraphics[width=\textwidth]{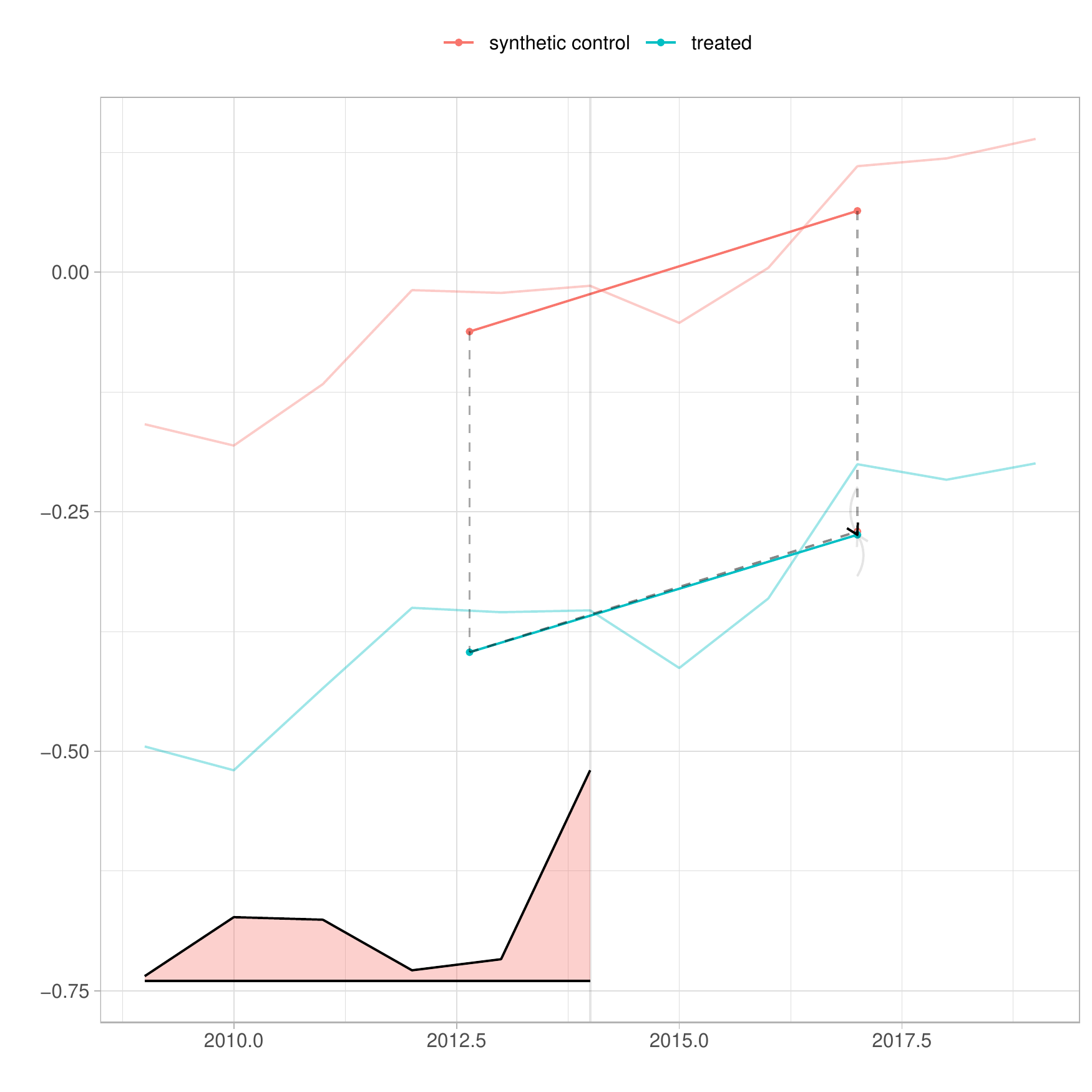}
	\subcaption{Intensive Margin -- Anderson Index}
\end{subfigure}
\justifying 
\footnotesize{\textbf{Notes:} Event study plots from the synthetic difference-in-difference estimator developed by \citet{arkhangelsky2021synthetic}. KLK is a summary index created following \citet{kling2007experimental}, while Anderson is a summary index created following \citet{anderson2008multiple}. The index is based on weighted nighttime light intensity, value added per capita (from DANE), share of urban population, agricultural productivity, firm creation and formal employment.}
\end{figure}

\begin{figure}[h!]
\caption{Synthetic Difference-in-Difference -- State Capacity Index}
\label{synt_SC}
\centering

\bigskip
\textit{Extensive Margin, Events in Over 60\% of Years}
\medskip

\begin{subfigure}[t]{0.49\textwidth}
	\includegraphics[width=\textwidth]{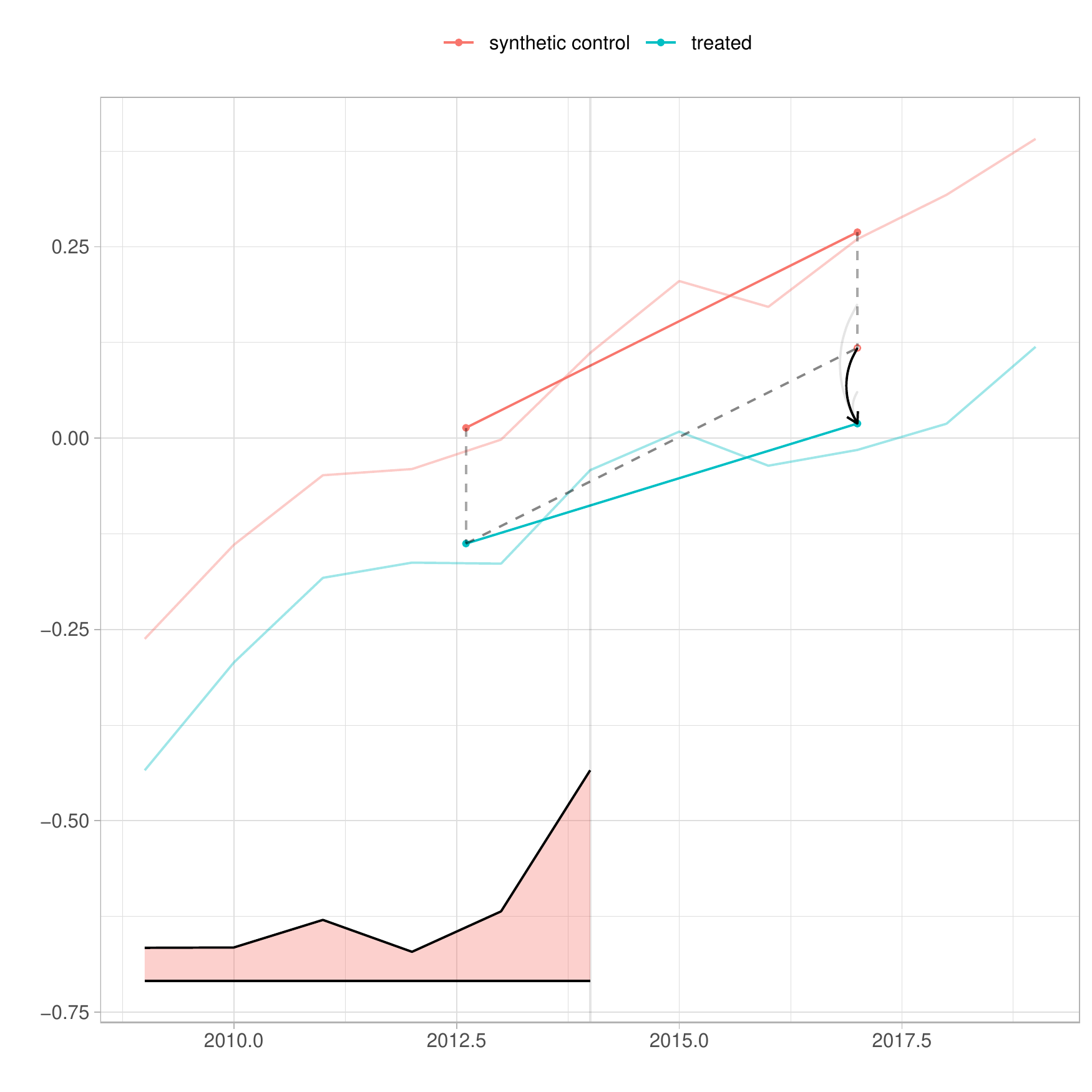}
	\subcaption{Extensive Margin -- KLK Index}
\end{subfigure}
\begin{subfigure}[t]{0.49\textwidth}
	\includegraphics[width=\textwidth]{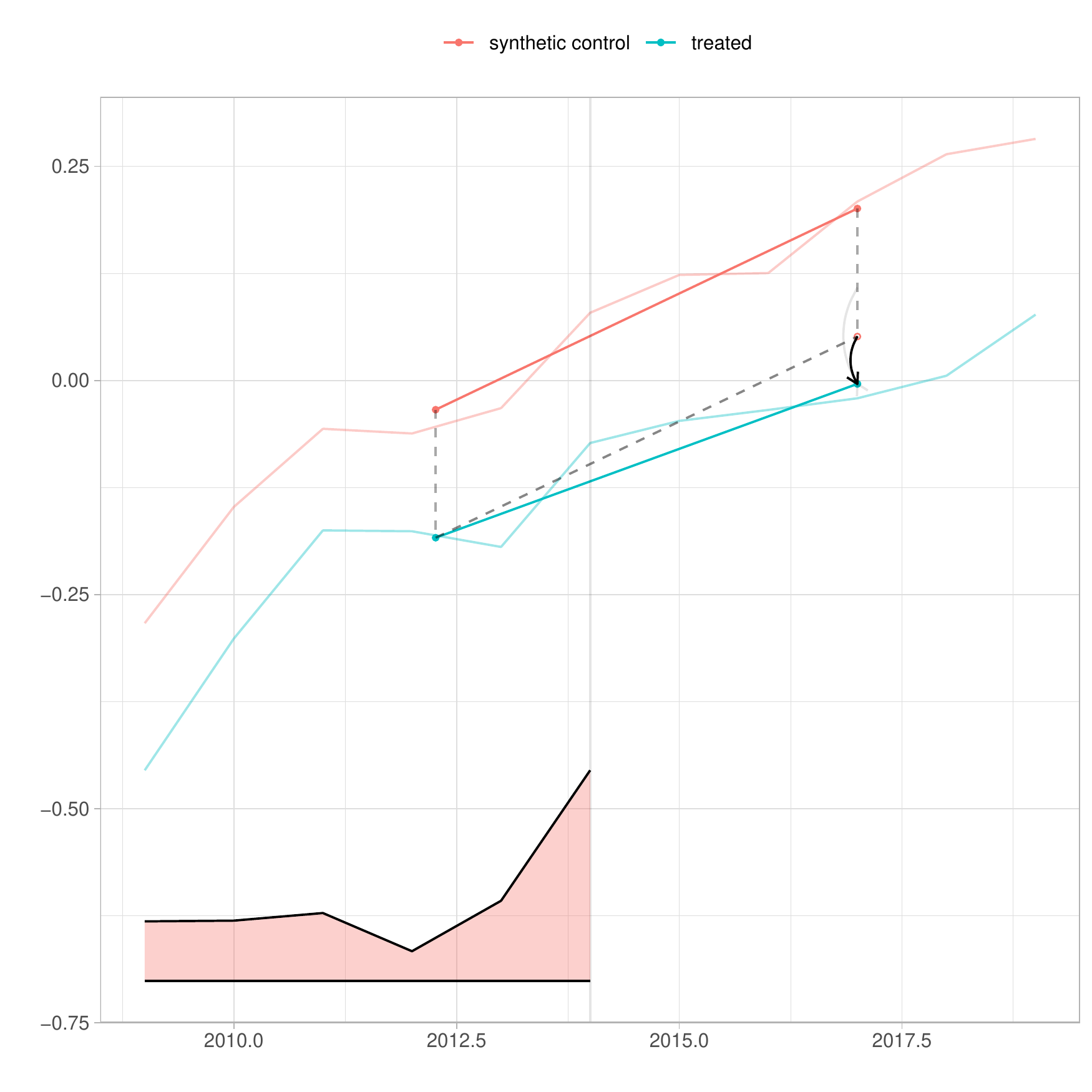}
	\subcaption{Extensive Margin -- Anderson Index}
\end{subfigure}

\bigskip
\textit{Intensive Margin, Top 20\% Most Violent}
\medskip

\begin{subfigure}[c]{0.49\textwidth}
	\includegraphics[width=\textwidth]{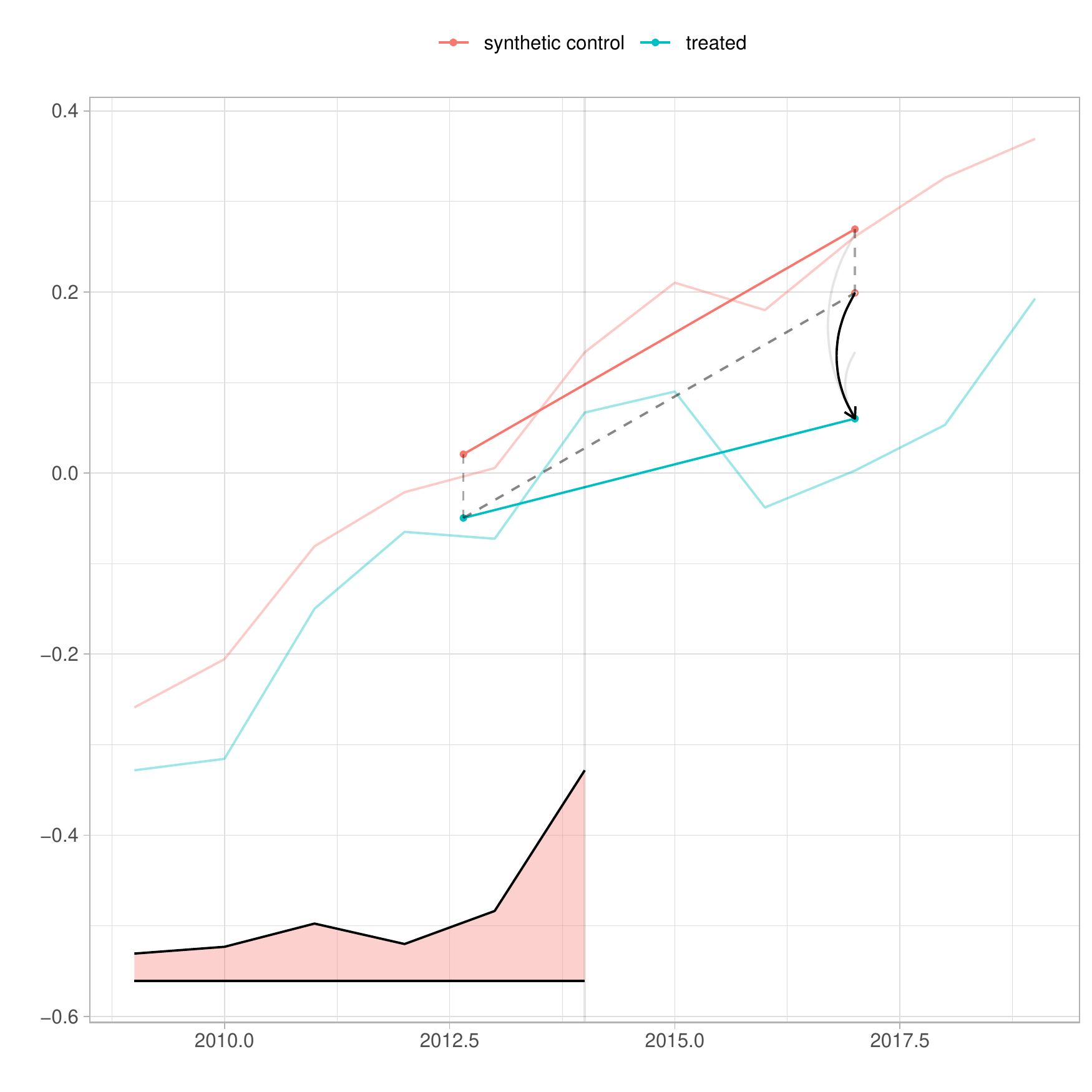}
	\subcaption{Intensive Margin -- KLK Index}
\end{subfigure}
\begin{subfigure}[c]{0.49\textwidth}
	\includegraphics[width=\textwidth]{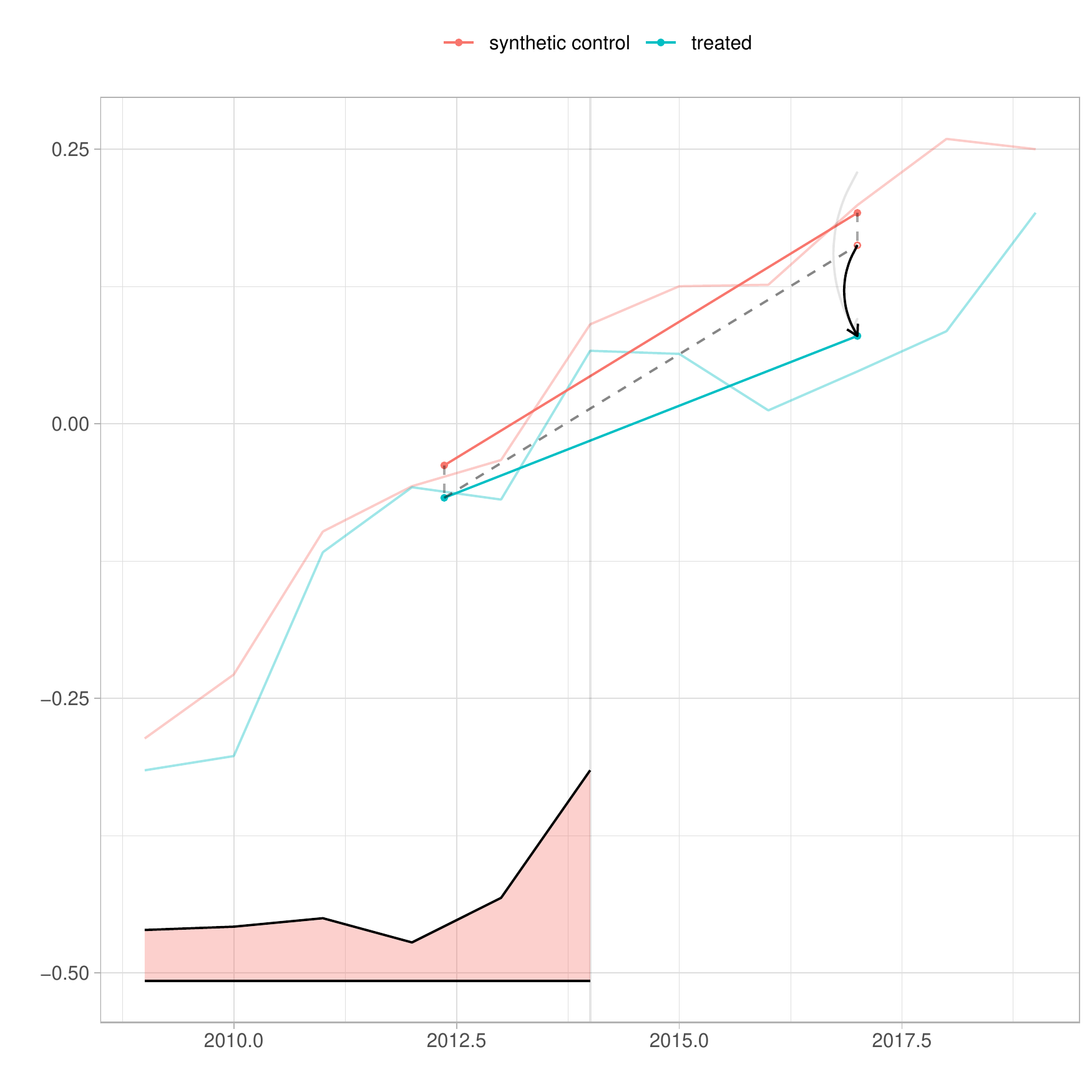}
	\subcaption{Intensive Margin -- Anderson Index}
\end{subfigure}
\justifying 
\footnotesize{\textbf{Notes:} Event study plots from the synthetic difference-in-difference estimator developed by \citet{arkhangelsky2021synthetic}. KLK is a summary index created following \citet{kling2007experimental}, while Anderson is a summary index created following \citet{anderson2008multiple}. The indices are composed of the following variables: tax revenue (per capita), operational costs (per capita), the ratio of government transfers to total municipality revenue (excluding transfers from the government), a measure of financial performance, a measure of administrative performance, and a measure of compliance with the rules set out by the national government.}
\end{figure}

\clearpage
\subsubsection{Violations of Parallel Trend Assumption}

\begin{figure}[h!]
\caption{Linear Violation of Parallel Trends Assumption at 80\% Power}
\label{roth_pretrend_fig}
\centering
\begin{subfigure}[t]{0.49\textwidth}
	\includegraphics[width=\textwidth]{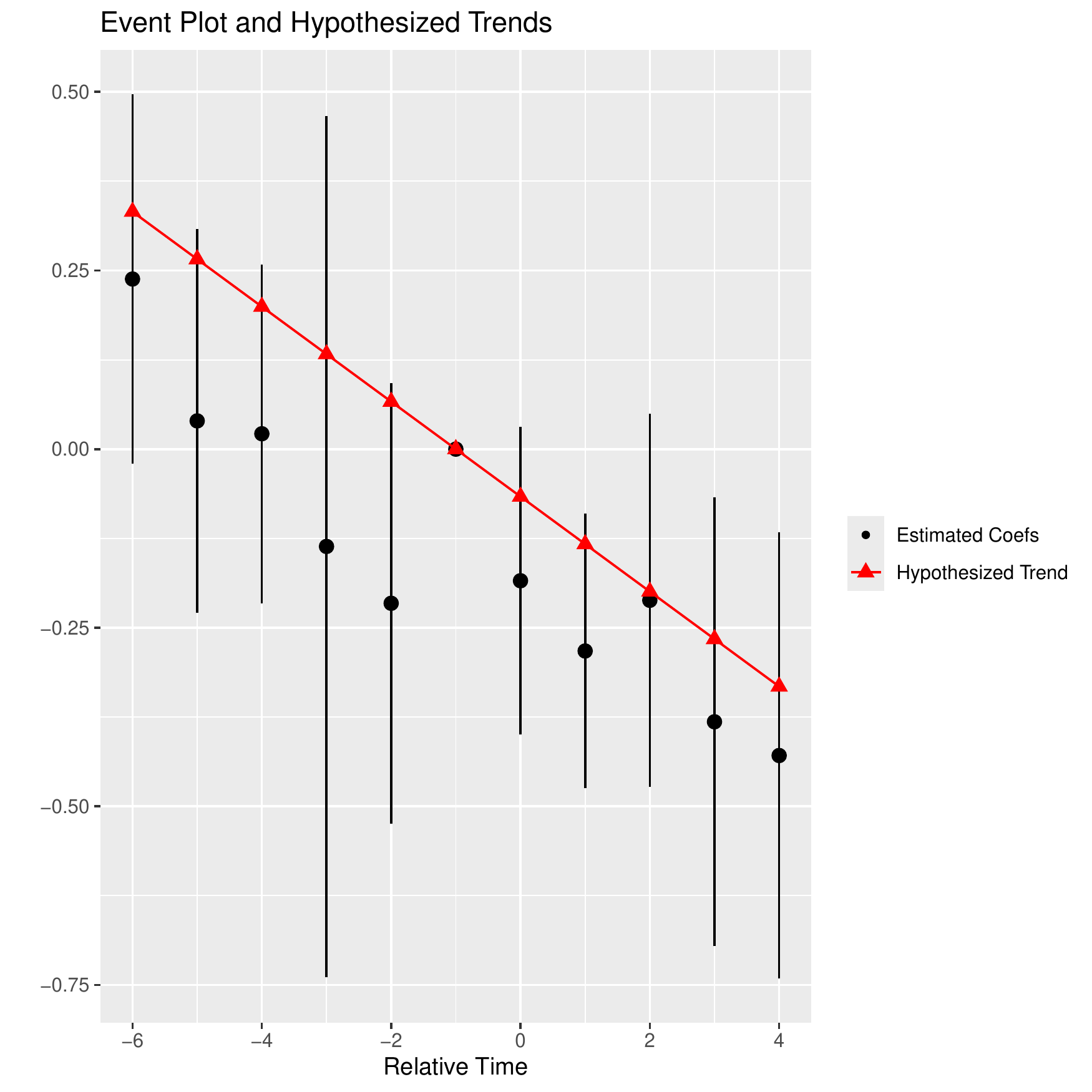}
	\subcaption{Extensive Margin, Events in Over 60\% of Years}
\end{subfigure}
\begin{subfigure}[t]{0.49\textwidth}
	\includegraphics[width=\textwidth]{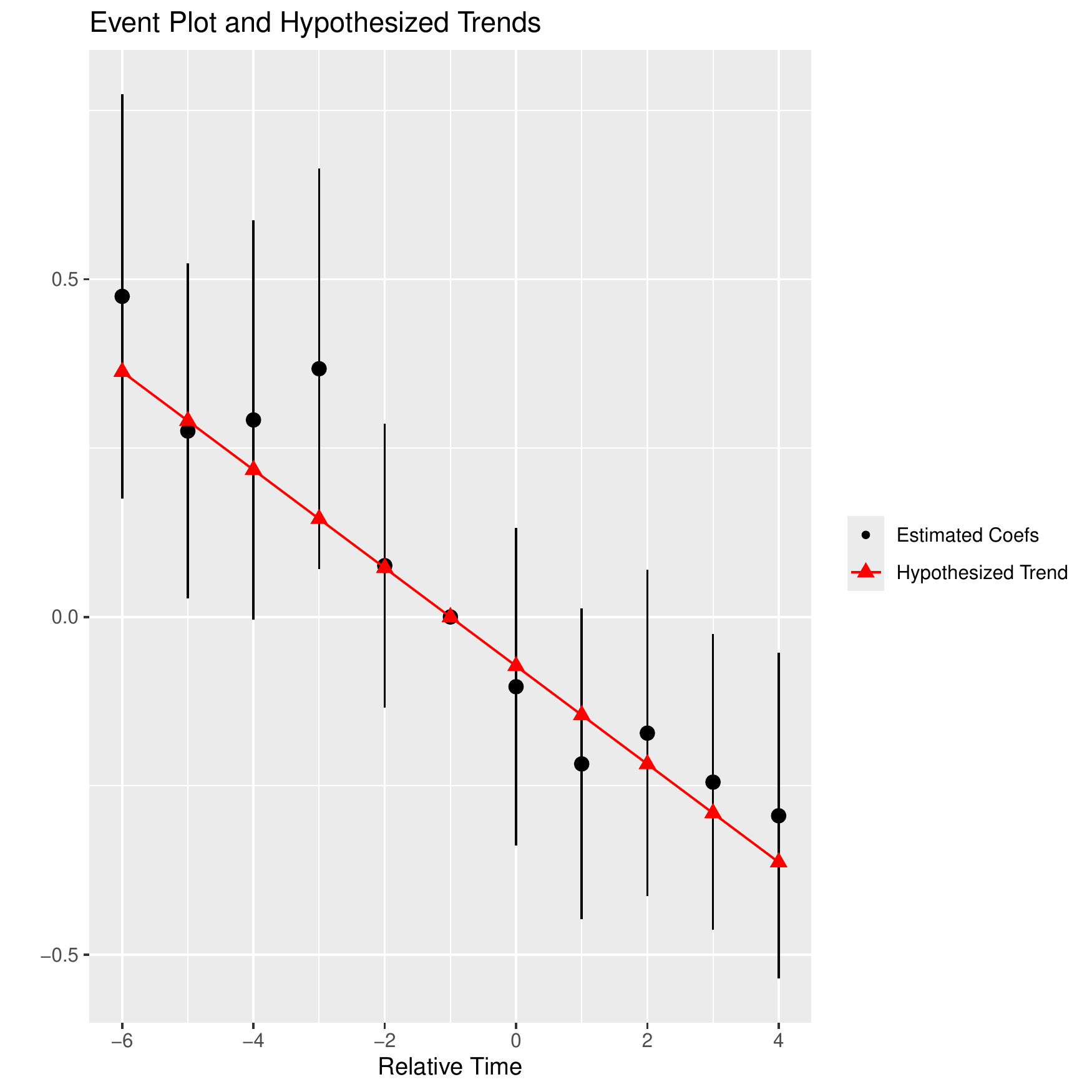}
	\subcaption{Intensive Margin, Top 20\% Most Violent}
\end{subfigure}
\justifying 
\footnotesize{\textbf{Notes:}  Event study plots from estimating Equation \eqref{eq_es} for the violence (Anderson) index, including including 95\% confidence intervals (based on standard errors clustered at the municipality level). The red line is the hypothetical linear violation of the parallel trends assumption that would be detected with 80\% power, estimated following \citet{rothpre}.}
\end{figure}

\clearpage
\subsubsection{International Tourism}

\begin{figure}[h!]
\caption{International Tourism in Colombia -- Extensive Margin, Events in Over 60\% ofYears}
\label{internationalTourists}
\centering
\begin{subfigure}[b]{0.48\textwidth}
	\includegraphics[width=\textwidth]{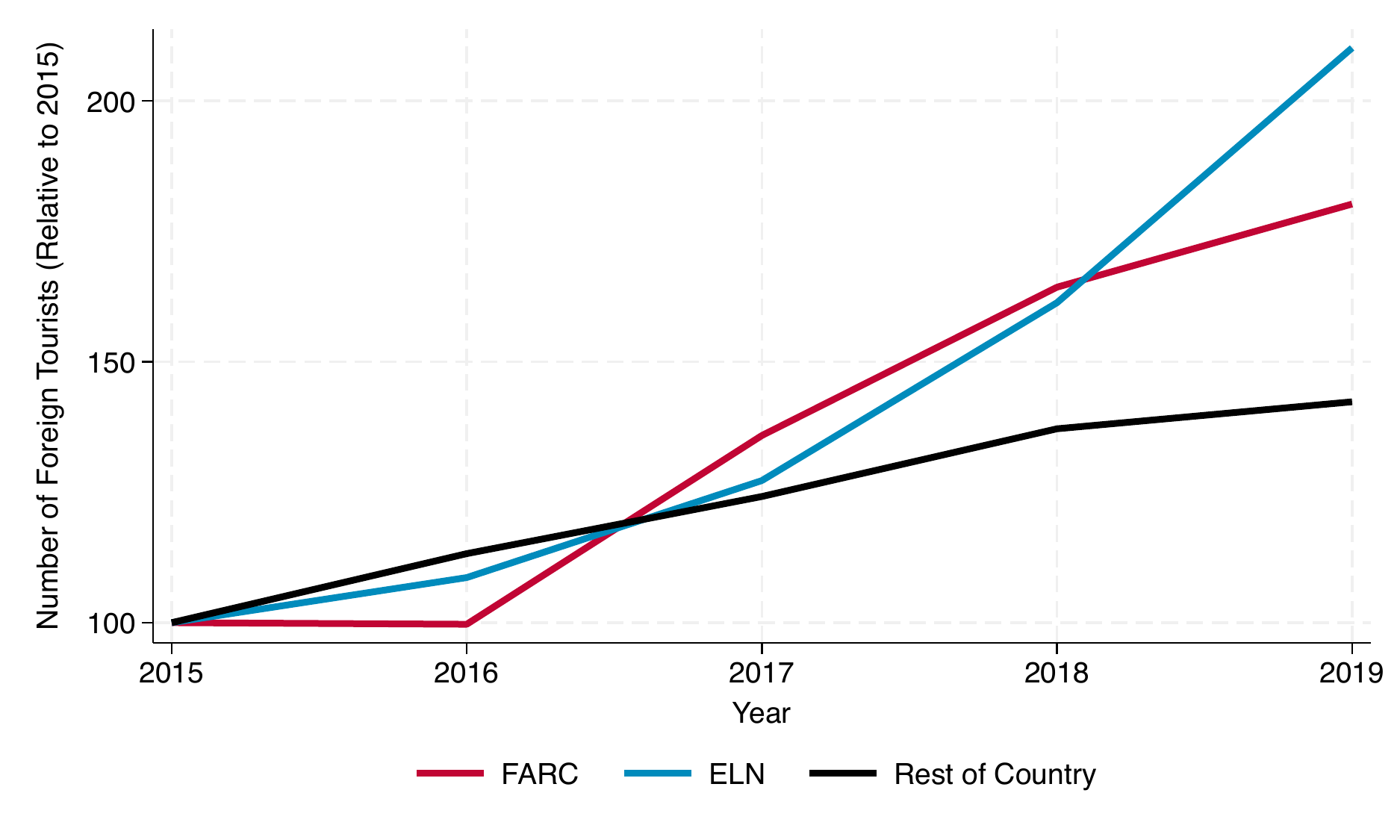}
	\subcaption{Unweighted}
\end{subfigure}
\begin{subfigure}[b]{0.48\textwidth}
	\includegraphics[width=\textwidth]{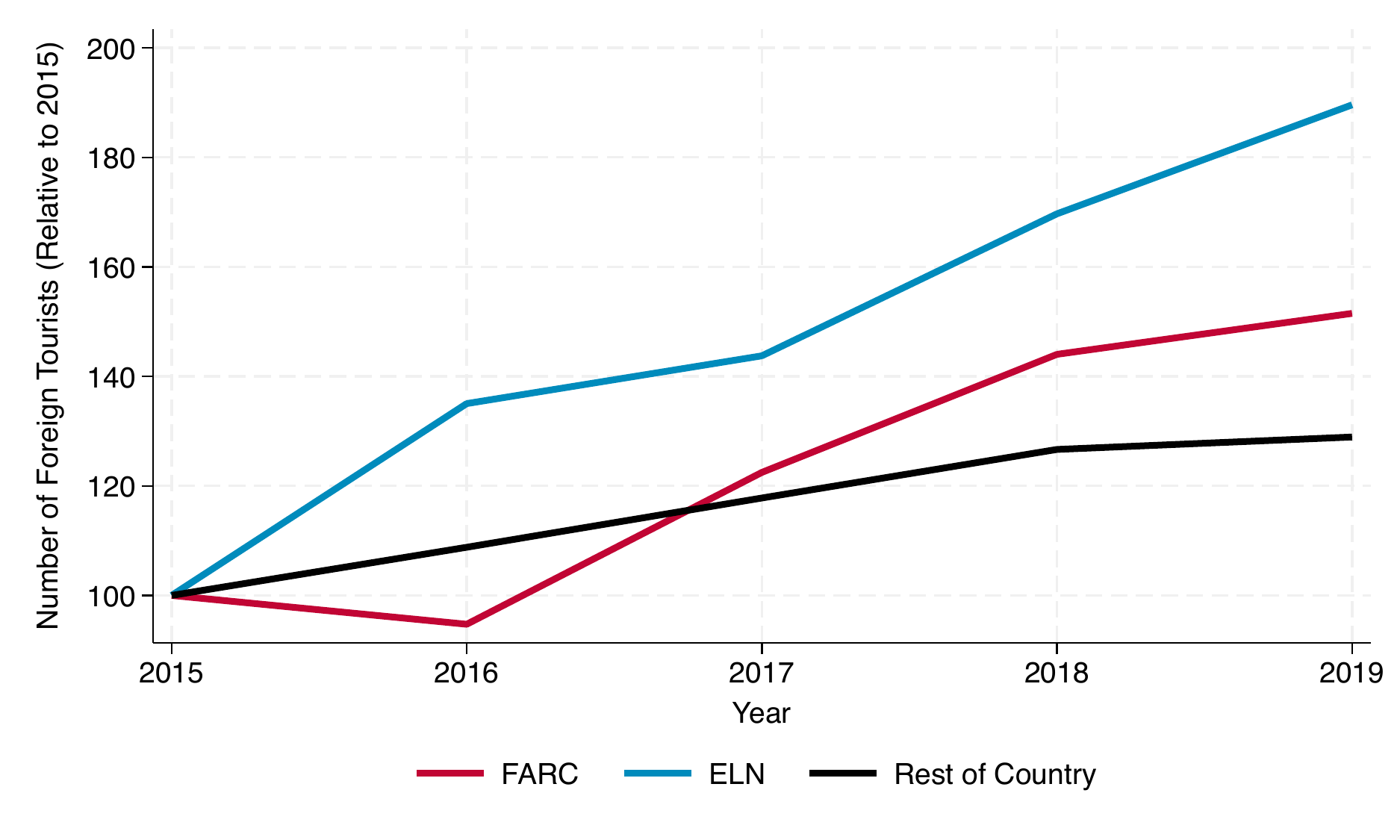}
	\subcaption{Weighted}
\end{subfigure}
\justifying
\footnotesize{\noindent \textbf{Notes:} Evolution of international tourists reporting visiting a given municipality, relative to 2015. Data come from the Ministry of Commerce, Industry and Tourism. Panel A does not apply any weighting, while Panel B weights by the municipality's population.}
\end{figure}

\clearpage
\subsubsection{Alternative Definitions of Control Group -- Non-FARC Municipalities}

\begin{figure}[h!]
\caption{Robustness of Results to Using Non-FARC Municipalities as Control Group}
\label{robNonFARCControl}
\centering
\begin{subfigure}[b]{0.81\textwidth}
	\includegraphics[width=\textwidth]{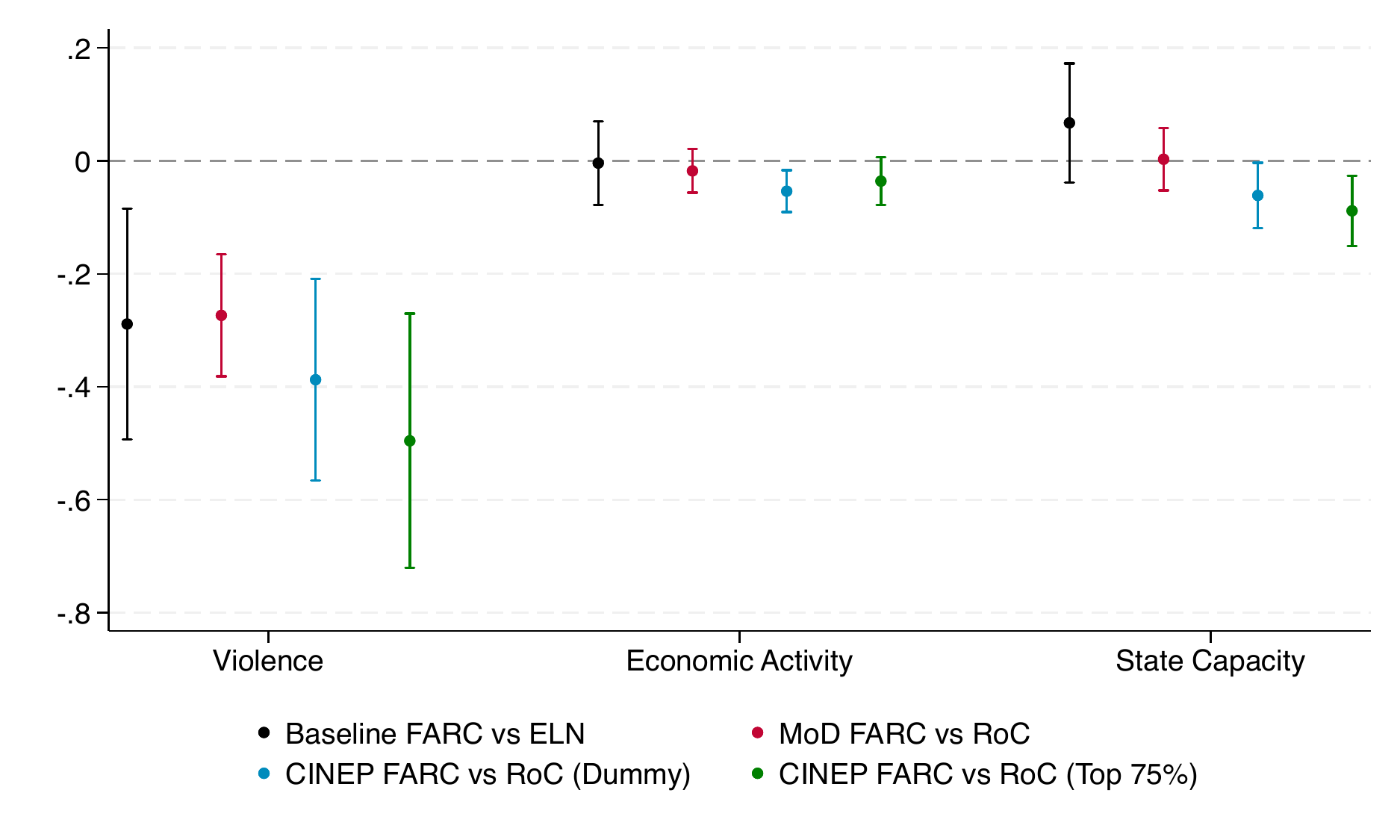}
	\subcaption{Extensive Margin -- Over 60\% of Years}
\end{subfigure}
\begin{subfigure}[b]{0.81\textwidth}
	\includegraphics[width=\textwidth]{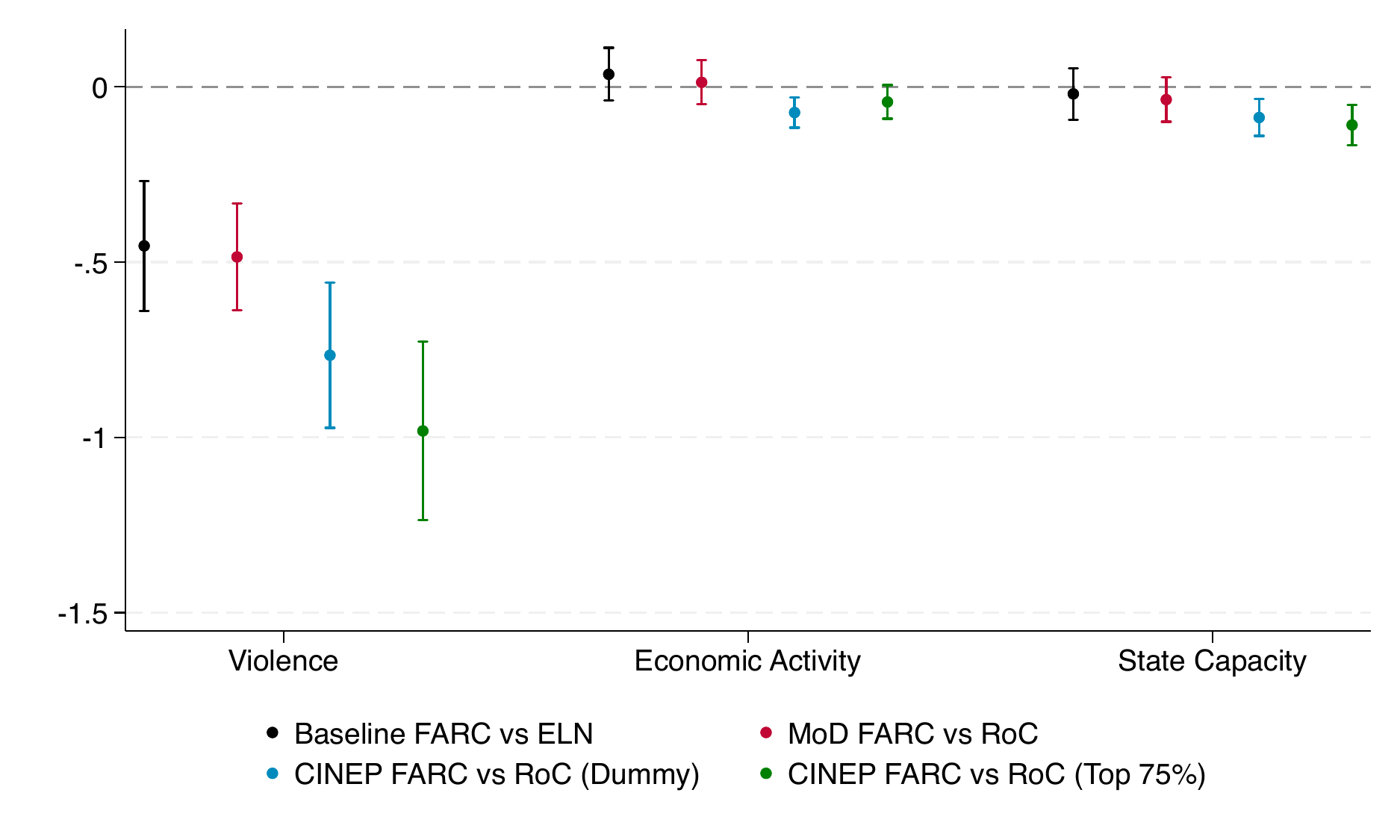}
	\subcaption{Intensive Margin -- Top 80\% Most Violent}
\end{subfigure}

\justifying
\footnotesize{\noindent \textbf{Notes:} Each coefficient corresponds to the estimation of the DiD Equation \eqref{eq_did} using one of the three summary indices. The coefficient in black corresponds to the baseline results, which uses data from the Ministry of Defence between 1996 and 2008 to identify FARC (treatment) and ELN (control) municipalities. The coefficient in red uses the same data and keeps the same treatment group, but for the control group uses all non-FARC municipalities rather than only ELN ones. The coefficients in blue and green use data from CINEP between 2011 and 2014. In blue, FARC municipalities are those with at least one FARC event in that timeframe, and the control is the rest of the country (RoC). In green, FARC municipalities are those in the top 75\% of FARC events per capita in that timeframe, and the control is the RoC.}

\end{figure}

\clearpage
\subsubsection{Event Studies Main Indices Until 2021}

\begin{figure}[h!]
\caption{Event Studies Extended Timeframe}
\label{timeframes_CEDEExt_p60}
\centering
\begin{subfigure}[b]{0.48\textwidth}
	\includegraphics[width=\textwidth]{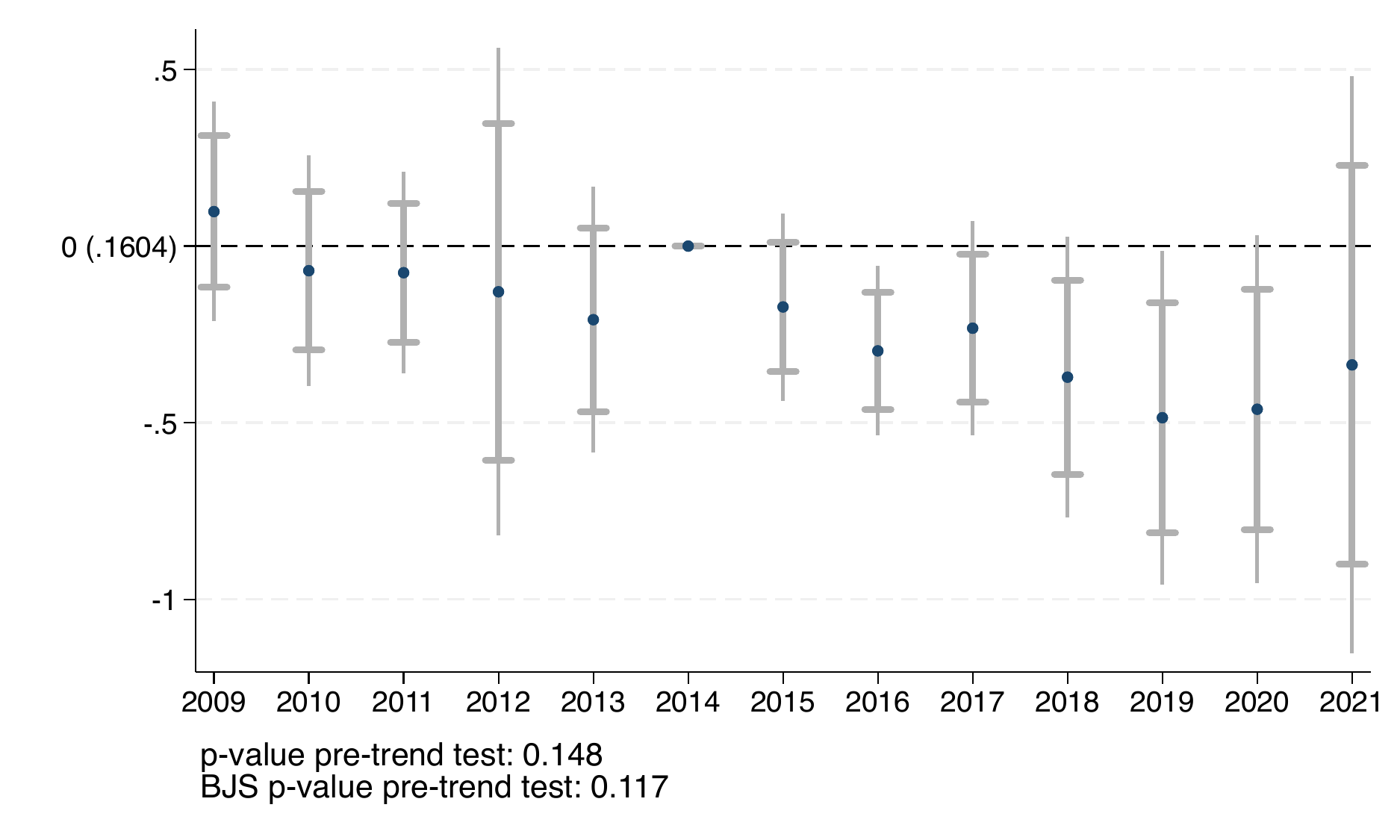}
	\subcaption{Violence Index}
\end{subfigure}
\begin{subfigure}[b]{0.48\textwidth}
	\includegraphics[width=\textwidth]{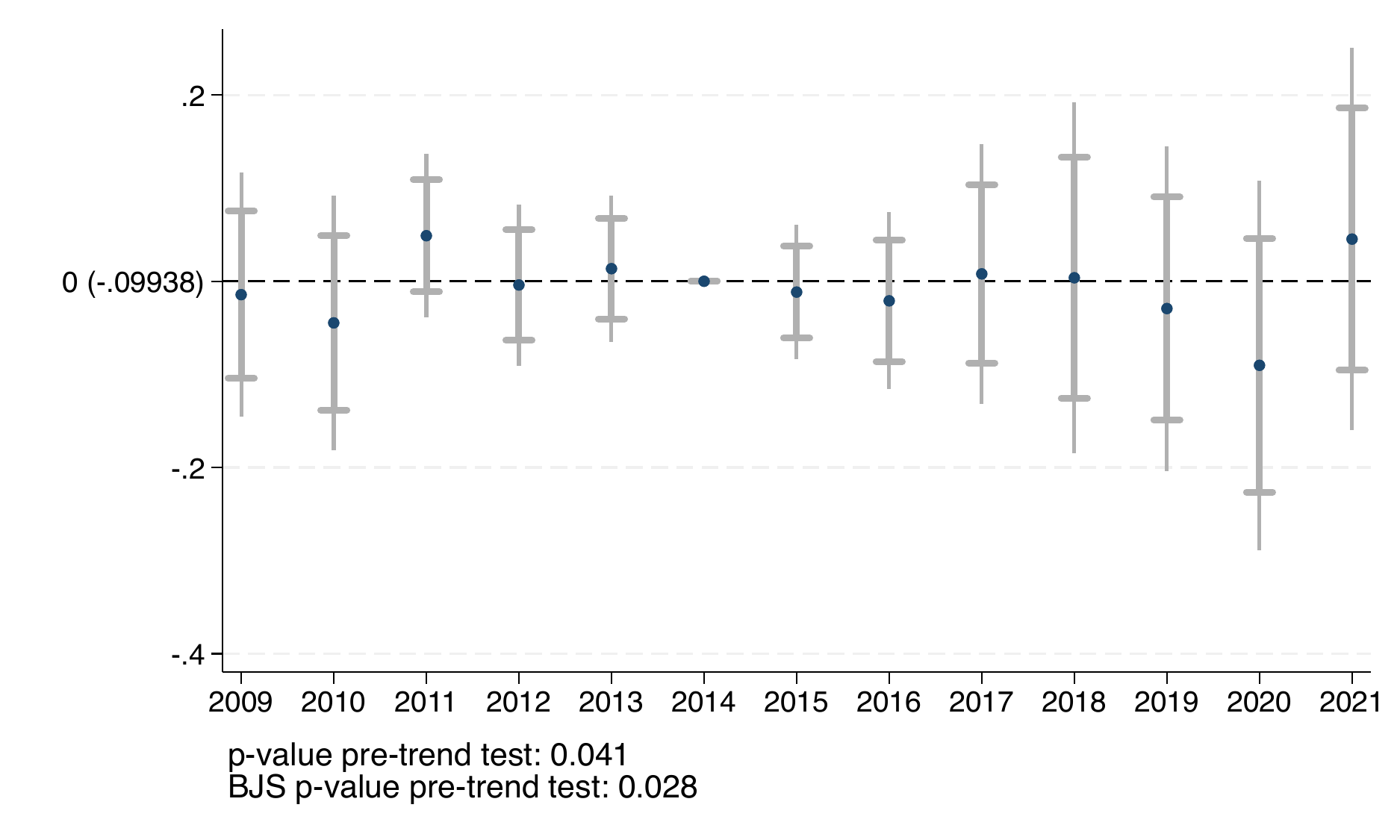}
	\subcaption{Economic Index}
\end{subfigure}
\begin{subfigure}[b]{0.48\textwidth}
	\includegraphics[width=\textwidth]{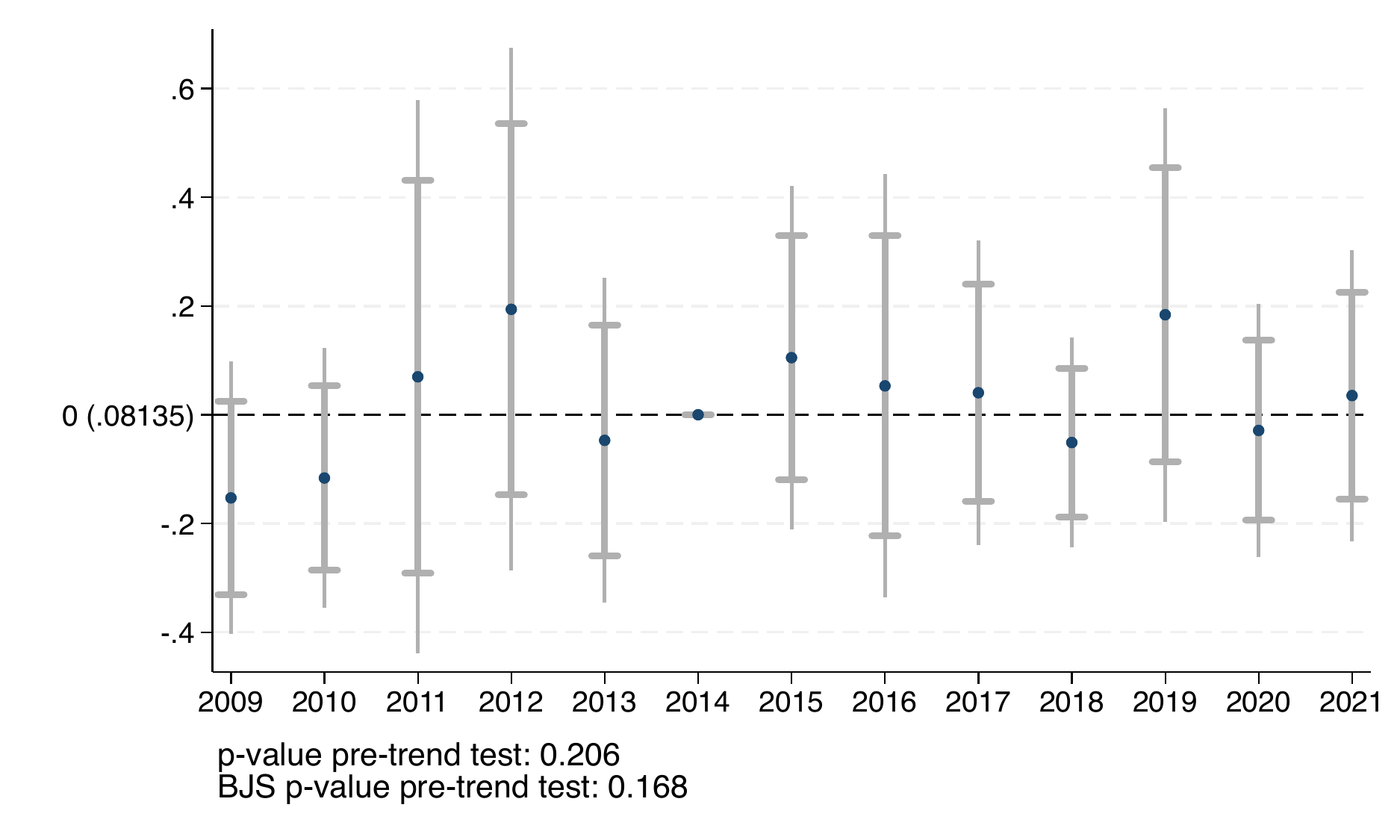}
	\subcaption{State Capacity Index}
\end{subfigure}

\justifying \noindent
\footnotesize{\textbf{Notes:}  Event study plots from estimating Equation \eqref{eq_es}, including including 95\% confidence intervals (based on standard errors clustered at the municipality level). Timeframe considered: 2009-2021.}
\end{figure}


\renewcommand{\thefigure}{F\arabic{figure}}
\setcounter{figure}{0}
\renewcommand{\thetable}{F\arabic{table}}
\setcounter{table}{0}

\clearpage
\subsection{Additional Tables}
\label{app_tables}

\subsubsection{Location of Demobilized FARC/ELN Members}

\begin{table}[h!] \centering
\newcolumntype{C}{>{\centering\arraybackslash}X}

\caption{Demobilizations by Criminal Organization Over Time -- Extensive Margin, Events in Over 60\% of Years}
\label{demob_CEDEExt_p60}
\begin{adjustbox}{max width=\linewidth, max height=\textwidth}\begin{tabular}{ccccccccc}

\toprule
& \multicolumn{4}{c}{\textbf{Demobilizations FARC Members}} & \multicolumn{4}{c}{\textbf{Demobilizations ELN Members}} \tabularnewline & Rest Country & Treatment & Control & Treat-Control & Rest Country & Treatment & Control & Treat-Control \tabularnewline
{Year}&{(1)}&{(2)}&{(3)}&{(4)}&{(5)}&{(6)}&{(7)}&{(8)} \tabularnewline
\midrule \addlinespace[\belowrulesep]
\midrule 2005&0.12&0.3&0.03&&0.02&0.04&0.05& \tabularnewline
2006&0.18&0.61&0.22&&0.04&0.06&0.15& \tabularnewline
2007&0.2&0.73&0.13&&0.06&0.06&0.12& \tabularnewline
2008&0.26&1.12&0.12&&0.02&0.04&0.13& \tabularnewline
2009&0.24&0.95&0.28&&0.04&0.06&0.04& \tabularnewline
2010&0.23&0.87&0.24&&0.04&0.06&0.07& \tabularnewline
2011&0.16&0.56&0.05&&0.02&0.04&0.04& \tabularnewline
2012&0.09&0.41&0.24&&0.01&0.01&0.03& \tabularnewline
2013&0.11&0.5&0.1&&0.02&0.05&0.07& \tabularnewline
2014&0.12&0.56&0.13&&0.03&0.03&0.07& \tabularnewline
2015&0.09&0.45&0.08&&0.03&0.05&0.07& \tabularnewline
2016&0.05&0.4&0.05&&0.04&0.04&0.2& \tabularnewline
2017&0.07&0.45&0&&0.05&0.07&0.25& \tabularnewline
2018&0.01&0.08&0&&0.05&0.05&0.21& \tabularnewline
2019&0&0.01&0&&0.03&0.04&0.06& \tabularnewline
&&&&&&&& \tabularnewline
\textbf{Pre-Cease}&0.17&0.66&0.15&0&0.03&0.05&0.08&0.017 \tabularnewline
\textbf{Post-Cease}&0.05&0.28&0.03&0&0.04&0.05&0.16&0.011 \tabularnewline
\bottomrule \addlinespace[\belowrulesep]

\end{tabular} \end{adjustbox}
\\ \parbox{\linewidth}{\scriptsize \justifying \noindent \(Notes\): Demobilization numbers from President's Office. All measures per 10.000 inhabitants. Columns 4 and 8 show p-values between the treatment and control means.}
\end{table}

\subsubsection{Activity by Criminal Group}

\begin{table}[h!] \centering
\newcolumntype{C}{>{\centering\arraybackslash}X}

\caption{DiD Analysis -- Violence Spillovers: Extensive Margin, Events in Over 60\% of Years}
\label{DID_violSpill_CEDE_Ext_p60}
\begin{adjustbox}{max width=\linewidth, max height=\textwidth}\begin{tabular}{l|cccccccc}

\toprule
& Gov. & FARC & ELN & Crim. Bands & FARC Diss. & Paramilitaries & Other & Total \tabularnewline
{}&{(1)}&{(2)}&{(3)}&{(4)}&{(5)}&{(6)}&{(7)}&{(8)} \tabularnewline
\midrule \addlinespace[\belowrulesep]
\midrule Ceasefire \( \times \) FARC&--0.154&--0.162***&0.003&0.003&0.031**&0.029&--0.147***&--0.397* \tabularnewline
&(0.094)&(0.035)&(0.033)&(0.031)&(0.012)&(0.075)&(0.045)&(0.202) \tabularnewline
&&&&&&&& \tabularnewline
Mean Sample (Pre)&0.551&0.263&0.036&0.114&0.000&0.230&0.124&1.319 \tabularnewline
Mean Sample (Post)&0.373&0.113&0.109&0.114&0.043&0.264&0.088&1.104 \tabularnewline
Observations&2,827&2,827&2,827&2,827&2,827&2,827&2,827&2,827 \tabularnewline
R Squared&0.361&0.373&0.251&0.209&0.178&0.284&0.264&0.411 \tabularnewline
\bottomrule \addlinespace[\belowrulesep]

\end{tabular} \end{adjustbox}
\\ \parbox{\linewidth}{\scriptsize \justifying \(Notes \):  Standard errors clustered at the municipality level. Includes municipality and year fixed effects. Considers only years from 2009 and before 2020. Defines the post period as 2015. Mean Sample (Pre/Post): Mean of the dependent variable in the pre/post-intervention period for municipalities in the sample (FARC and ELN). In each column, the dependent variable is the number of human right violations committed by a given criminal organization per 10.000 inhabitants. Data from Osorio et al. (2019). The total in the last column is the sum across all columns.}
\end{table}

\clearpage

\subsubsection{Additional Results on Economic Outcomes}

\begin{table}[h!] \centering
\newcolumntype{R}{>{\raggedleft\arraybackslash}X}
\newcolumntype{L}{>{\raggedright\arraybackslash}X}
\newcolumntype{C}{>{\centering\arraybackslash}X}

\caption{DiD Additional Economic Outcomes: Extensive Margin, Events in Over 60\% of Years}
\label{DID_Other_CEDE_Ext_p60}
\begin{adjustbox}{max width=\linewidth, max height=\textwidth}\begin{tabular}{lccccccc}

\toprule
{}&{(1)}&{(2)}&{(3)}&{(4)}&{(5)}&{(6)}&{(7)} \tabularnewline
\midrule \addlinespace[\belowrulesep]
\textbf{Panel A. Robustness}&Nighttime Light&GDP per Capita&Agr. Production&Agr. Productivity&Urban Built Up&Human Dev. Index& \tabularnewline
\midrule Ceasefire \( \times \) FARC&--0.008&--0.346&0.222&0.301&--0.010***&--0.001& \tabularnewline
&(0.021)&(0.402)&(0.138)&(0.493)&(0.002)&(0.001)& \tabularnewline
&&&&&&& \tabularnewline
Mean Dep. Var.&--0.314&6.364&1.693&6.361&0.025&0.716& \tabularnewline
Observations&2,827&2,822&2,827&2,827&2,827&2,056& \tabularnewline
&&&&&&& \tabularnewline
\midrule \textbf{Panel B. Agricultural Trade}&\# Products Supplied&\# Markets Supplied&\# Deliveries&Avr. Price&Total Value&Total Quantity& \tabularnewline
\midrule Ceasefire \( \times \) FARC&--0.510&0.452*&--11.841&0.026&675.025&3.488& \tabularnewline
&(0.782)&(0.252)&(15.600)&(0.167)&(653.091)&(2.241)& \tabularnewline
&&&&&&& \tabularnewline
Mean Dep. Var.&14.420&2.907&86.224&1.564&3887.102&6.771& \tabularnewline
Observations&1,799&1,799&1,799&1,605&1,799&1,799& \tabularnewline
&&&&&&& \tabularnewline
\midrule \textbf{Panel C. Tourism \& Migration}&\# Beds&Acc. Employees (pc)&Tourists (Total)&Tourists (pc)&Travel Expenses&Immigrated&Immigrated (20+) \tabularnewline
\midrule Ceasefire \( \times \) FARC&--3.882&1.643&1.853***&1.112***&--1.598&7.502&--12.977 \tabularnewline
&(31.143)&(1.413)&(0.497)&(0.372)&(33.771)&(203.868)&(139.428) \tabularnewline
&&&&&&& \tabularnewline
Mean Dep. Var.&129.053&4.990&4.319&1.791&142.614&460.222&251.603 \tabularnewline
Observations&2,827&2,827&1,028&1,028&1,028&514&514 \tabularnewline
&&&&&&& \tabularnewline
\midrule \textbf{Panel D. Health \& Education}&Birth Weight&Birth Size&Apgar Score (5 Min.)&Child Mortality&\# Off. Schools&\# Teachers&\# Students \tabularnewline
\midrule Ceasefire \( \times \) FARC&2.447&--0.178&--0.027&--0.001&--0.075*&--0.337**&--8.836* \tabularnewline
&(5.833)&(0.113)&(0.021)&(0.007)&(0.044)&(0.158)&(5.192) \tabularnewline
&&&&&&& \tabularnewline
Mean Dep. Var.&3193.500&50.772&9.588&0.050&3.212&11.345&231.496 \tabularnewline
Observations&2,827&2,827&2,827&2,827&2,827&2,827&2,827 \tabularnewline
&&&&&&& \tabularnewline
\midrule \textbf{Panel E. GEIH Survey Data}&Min. House Price&Asset Index&Weeks Job Search&Months Current Job&Months Emp. Past Year&Weekly Work Hours&Monthly Salary \tabularnewline
\midrule Ceasefire \( \times \) FARC&--3.316&0.025&--1.294&--3.173&--0.071&--0.360&0.218 \tabularnewline
&(4.624)&(0.025)&(1.080)&(2.451)&(0.057)&(0.469)&(0.215) \tabularnewline
&&&&&&& \tabularnewline
Mean Dep. Var.&32.994&--0.521&17.194&81.379&11.076&46.138&7.561 \tabularnewline
Observations&112,268&295,070&50,642&382,977&382,977&382,974&158,039 \tabularnewline
\bottomrule \addlinespace[\belowrulesep]

\end{tabular} \end{adjustbox}
\\ \parbox{\linewidth}{\scriptsize \justifying \noindent \(Notes \): Results from estimating Equation \eqref{eq_did}. Standard errors clustered at the municipality level. Includes municipality and year fixed effects. For most outcomes, considers only years from 2009 and before 2020. Mean Dep. Var.: Mean of the dependent variable in the pre-intervention period for municipalities in the sample (FARC and ELN). Defines the post period as 2015. In Panel A, the outcome variables are nighttime light intensity (unweighted by the share of the municipality in a given grid), an estimate of municipal GDP per capita based on calculations following Sanchez \& Espana (2012, in millions of COP), agricultural production (total tonnes produced of 271 agricultural crops per capita), agricultural productivity based only on major crops (total tonnes produced divided by total area cultivated in hectares, excluding those crops that are planted in less than 2\% of all municipality-year pairs), a measure of urban built up based on satellite images that measures the average amount of pixels within a municipality that are classified as urban built up based on the Band Ratio for Built-Up Area (BRBA) index developed by Waqar et al. (2012), and an index of Human Development based on satellite data and machine learning (only since 2012, from Sherman et al., 2023). Panel B uses data from the SIPSA on agricultural deliveries to each of the main markets in the country and prices the Ministry of Agriculture. The outcome variables are the number of agricultural products supplied from a municipality to any market, the number of markets that receive at least one delivery for each municipality, the number of yearly deliveries from a municipality to all markets, the average price of products from a municipality (weighted by the total amount of each product delivered), the total yearly value of all deliveries to all markets (in 1000 COP, per capita), and the total yearly quantity delivered to all markets from a municipality (in tonnes, per capita). For this Panel, data start in 2013. In Panel C, the outcome variables are the number of beds for touristic purposes, number of employees working in touristic accommodations (per capita, both of these from the National Tourism Registry, or Registro Nacional de Turismo in Spanish), the total and per capita number of national tourists, the total travel expenses of national tourists (these last three from the Survey of Internal Tourism Expenditures, or Encuesta de Gasto Interno en Turismo in Spanish, for which data are available only for 2012, 2014, 2015 and 2019), and the number of people who immigrated to a given municipality in the past year (all, and above 20 years, based on the census data from 2005 and 2018). In Panel D, the outcome variables are birth weight (in grams), birth size, the Apgar score (after 5 minutes), child mortality (number of deaths of kids under 1 relative to newborn kids), number of official schools, teachers, and students (all per 10.000 inhabitants). Panel E uses data from the Gran Encuesta Integrada de Hogares (Large Integrated Household Survey, GEIH), which surveys repeated cross sections every month across different municipalities. It controls for municipality fixed effects and month-year fixed effects. The outcome variables are the minimum hypothetical price the household would be willing to sell the house they currently reside in (in COP millions), an index based on ownership of assets using weights from Colombia's 2015 DHS survey, the number of weeks searching for a job (for those unemployed), the number of months working at the main employer, the number of months worked in the past year, the number of hours regularly worked in a week, and the monthly salary (in COP hundreds of thousands). These last four variables, for those employed at the time of the survey.}
\end{table}

\begin{table}[h!] \centering
\newcolumntype{R}{>{\raggedleft\arraybackslash}X}
\newcolumntype{L}{>{\raggedright\arraybackslash}X}
\newcolumntype{C}{>{\centering\arraybackslash}X}

\caption{DiD Additional Economic Outcomes: Intensive Margin, Top 20\% Most Violent}
\label{DID_Other_CEDE_Int_p80}
\begin{adjustbox}{max width=\linewidth, max height=\textwidth}\begin{tabular}{lccccccc}

\toprule
{}&{(1)}&{(2)}&{(3)}&{(4)}&{(5)}&{(6)}&{(7)} \tabularnewline
\midrule \addlinespace[\belowrulesep]
\textbf{Panel A. Robustness}&Nighttime Light&GDP per Capita&Agr. Production&Agr. Productivity&Urban Built Up&Human Dev. Index& \tabularnewline
\midrule Ceasefire \( \times \) FARC&--0.033*&--0.310&0.243*&--0.054&--0.006**&0.000& \tabularnewline
&(0.017)&(0.504)&(0.132)&(0.263)&(0.003)&(0.001)& \tabularnewline
&&&&&&& \tabularnewline
Mean Dep. Var.&--0.365&6.416&1.480&4.713&0.024&0.714& \tabularnewline
Observations&3,366&3,362&3,366&3,363&3,365&2,448& \tabularnewline
&&&&&&& \tabularnewline
\midrule \textbf{Panel B. Agricultural Trade}&\# Products Supplied&\# Markets Supplied&\# Deliveries&Avr. Price&Total Value&Total Quantity& \tabularnewline
\midrule Ceasefire \( \times \) FARC&--0.999**&0.084&--8.224&0.256**&1034.748&1.680& \tabularnewline
&(0.506)&(0.155)&(4.996)&(0.122)&(996.788)&(1.277)& \tabularnewline
&&&&&&& \tabularnewline
Mean Dep. Var.&11.172&2.291&58.369&1.492&4955.734&7.456& \tabularnewline
Observations&2,142&2,142&2,142&1,853&2,142&2,142& \tabularnewline
&&&&&&& \tabularnewline
\midrule \textbf{Panel C. Tourism \& Migration}&\# Beds&Acc. Employees (pc)&Tourists (Total)&Tourists (pc)&Travel Expenses&Immigrated&Immigrated (20+) \tabularnewline
\midrule Ceasefire \( \times \) FARC&--25.944&--0.826&--0.284&0.313&--15.147&--70.772&--49.249 \tabularnewline
&(20.706)&(1.070)&(0.430)&(0.329)&(28.282)&(81.989)&(51.092) \tabularnewline
&&&&&&& \tabularnewline
Mean Dep. Var.&80.756&4.217&2.794&1.742&129.120&359.833&196.348 \tabularnewline
Observations&3,366&3,366&1,224&1,224&1,224&610&610 \tabularnewline
&&&&&&& \tabularnewline
\midrule \textbf{Panel D. Health \& Education}&Birth Weight&Birth Size&Apgar Score (5 Min.)&Child Mortality&\# Off. Schools&\# Teachers&\# Students \tabularnewline
\midrule Ceasefire \( \times \) FARC&1.340&--0.126&--0.063***&--0.006&--0.164***&--0.501***&--20.027*** \tabularnewline
&(5.587)&(0.077)&(0.017)&(0.005)&(0.038)&(0.125)&(3.774) \tabularnewline
&&&&&&& \tabularnewline
Mean Dep. Var.&3182.402&50.800&9.568&0.043&3.611&11.288&221.519 \tabularnewline
Observations&3,366&3,366&3,366&3,366&3,366&3,366&3,366 \tabularnewline
&&&&&&& \tabularnewline
\midrule \textbf{Panel E. GEIH Survey Data}&Min. House Price&Asset Index&Weeks Job Search&Months Current Job&Months Emp. Past Year&Weekly Work Hours&Monthly Salary \tabularnewline
\midrule Ceasefire \( \times \) FARC&--1.344&0.011&0.983&--3.590**&--0.235**&0.714&--0.430* \tabularnewline
&(5.345)&(0.027)&(1.035)&(1.567)&(0.092)&(0.685)&(0.251) \tabularnewline
&&&&&&& \tabularnewline
Mean Dep. Var.&31.461&--0.602&19.723&85.315&10.908&46.283&7.998 \tabularnewline
Observations&93,198&261,619&47,591&307,796&307,796&307,792&116,927 \tabularnewline
\bottomrule \addlinespace[\belowrulesep]

\end{tabular} \end{adjustbox}
\\ \parbox{\linewidth}{\scriptsize \justifying \noindent \(Notes \): Results from estimating Equation \eqref{eq_did}. Standard errors clustered at the municipality level. Includes municipality and year fixed effects. For most outcomes, considers only years from 2009 and before 2020. Mean Dep. Var.: Mean of the dependent variable in the pre-intervention period for municipalities in the sample (FARC and ELN). Defines the post period as 2015. In Panel A, the outcome variables are nighttime light intensity (unweighted by the share of the municipality in a given grid), an estimate of municipal GDP per capita based on calculations following Sanchez \& Espana (2012, in millions of COP), agricultural production (total tonnes produced of 271 agricultural crops per capita), agricultural productivity based only on major crops (total tonnes produced divided by total area cultivated in hectares, excluding those crops that are planted in less than 2\% of all municipality-year pairs), a measure of urban built up based on satellite images that measures the average amount of pixels within a municipality that are classified as urban built up based on the Band Ratio for Built-Up Area (BRBA) index developed by Waqar et al. (2012), and an index of Human Development based on satellite data and machine learning (only since 2012, from Sherman et al., 2023). Panel B uses data from the SIPSA on agricultural deliveries to each of the main markets in the country and prices the Ministry of Agriculture. The outcome variables are the number of agricultural products supplied from a municipality to any market, the number of markets that receive at least one delivery for each municipality, the number of yearly deliveries from a municipality to all markets, the average price of products from a municipality (weighted by the total amount of each product delivered), the total yearly value of all deliveries to all markets (in 1000 COP, per capita), and the total yearly quantity delivered to all markets from a municipality (in tonnes, per capita). For this Panel, data start in 2013. In Panel C, the outcome variables are the number of beds for touristic purposes, number of employees working in touristic accommodations (per capita, both of these from the National Tourism Registry, or Registro Nacional de Turismo in Spanish), the total and per capita number of national tourists, the total travel expenses of national tourists (these last three from the Survey of Internal Tourism Expenditures, or Encuesta de Gasto Interno en Turismo in Spanish, for which data are available only for 2012, 2014, 2015 and 2019), and the number of people who immigrated to a given municipality in the past year (all, and above 20 years, based on the census data from 2005 and 2018). In Panel D, the outcome variables are birth weight (in grams), birth size, the Apgar score (after 5 minutes), child mortality (number of deaths of kids under 1 relative to newborn kids), number of official schools, teachers, and students (all per 10.000 inhabitants). Panel E uses data from the Gran Encuesta Integrada de Hogares (Large Integrated Household Survey, GEIH), which surveys repeated cross sections every month across different municipalities. It controls for municipality fixed effects and month-year fixed effects. The outcome variables are the minimum hypothetical price the household would be willing to sell the house they currently reside in (in COP millions), an index based on ownership of assets using weights from Colombia's 2015 DHS survey, the number of weeks searching for a job (for those unemployed), the number of months working at the main employer, the number of months worked in the past year, the number of hours regularly worked in a week, and the monthly salary (in COP hundreds of thousands). These last four variables, for those employed at the time of the survey.}
\end{table}

\begin{table}[h!] \centering
\newcolumntype{R}{>{\raggedleft\arraybackslash}X}
\newcolumntype{L}{>{\raggedright\arraybackslash}X}
\newcolumntype{C}{>{\centering\arraybackslash}X}

\caption{IV-DiD Results -- (Instrumented) Effect of Violence on Economic Activity: Extensive Margin, Events in Over 60\% of Years}
\label{DIDIV_Econ_CEDE_Ext_p60}
\begin{adjustbox}{max width=\linewidth, max height=\textwidth}\begin{tabular}{lccccccc}

\toprule
& Nighttime & Value Added (pc) & Agricultural & Share Urban & Firm  & Formal & \textbf{Anderson} \tabularnewline & Light & DANE & Productivity & Population & Entry & Empl. & \textbf{Index} \tabularnewline 
{}&{(1)}&{(2)}&{(3)}&{(4)}&{(5)}&{(6)}&{(7)} \tabularnewline
\midrule \addlinespace[\belowrulesep]
\midrule Violence (Anderson) Index&0.025&10.914&--0.914&0.003&--0.612&--0.007&0.014 \tabularnewline
&(0.071)&(8.594)&(1.763)&(0.015)&(1.157)&(0.014)&(0.131) \tabularnewline
&&&&&&& \tabularnewline
Observations&2,827&2,313&2,827&2,827&2,827&2,827&2,827 \tabularnewline
\(R^2\)&--0.003&--0.905&--0.127&--0.019&--0.053&--0.119&--0.003 \tabularnewline
Treated Munic.&216&216&216&216&216&216&216 \tabularnewline
Control Munic.&41&41&41&41&41&41&41 \tabularnewline
Mean Dep. Var.&--0.314&12.007&6.654&0.430&6.078&0.107&--0.097 \tabularnewline
1st Stage CD Wald F Stat.&13.773&7&13.773&13.773&13.773&13.773&13.773 \tabularnewline
\bottomrule \addlinespace[\belowrulesep]

\end{tabular} \end{adjustbox}
\\ \parbox{\linewidth}{\scriptsize \justifying \noindent \(Notes \): Results from regression the different economic outcomes on the Anderson Index for violence as the independent variable, with this violence index instrumented by the post ceasefire - FARC interaction term, as in Equation \eqref{eq_did}. Standard errors clustered at the municipality level. Includes municipality and year fixed effects. Considers only years from 2009 and before 2020. Mean Dep. Var.: Mean of the dependent variable in the pre-intervention period for municipalities in the sample (FARC and ELN). Defines the post period as 2015. Nighttime light intensity defined so that grid cells in the border of multiple municipalities are assigned in proportion to the share of the grid cell in each municipality (weighted). Value added per capita comes DANE (National Department of Statistics), in millions of COP. Agricultural productivity is defined as total tonnes produced of 271 agricultural crops divided by total area cultivated in hectares, based on data from the Ministry of Agriculture. Firm entry comes from RUES and is measured per 1000 inhabitants. Formal employment is measured as the average number of individuals paying contributions to healthcare, pension funds and workers' compensations across the year in the municipality per 18-60 years old, from PILA. Anderson Index is a summary measure created following Anderson (2008) that summarizes the measures in columns 1-6. It is the weighted average of the standardized outcomes, weighted by their inverted covariance matrix. First-stage Cragg-Donald Wald F-statistic shown. }
\end{table}

\begin{table}[h!] \centering
\newcolumntype{R}{>{\raggedleft\arraybackslash}X}
\newcolumntype{L}{>{\raggedright\arraybackslash}X}
\newcolumntype{C}{>{\centering\arraybackslash}X}

\caption{Summary Statistics: Extensive Margin, Events in Over 60\% of Years}
\label{SumStatsGEIH_CEDEExt_p60}
\begin{adjustbox}{max width=\linewidth, max height=\textwidth}\begin{tabular}{l|cccc|cccc}

\toprule
& \multicolumn{4}{c|}{\textbf{ELN Municipalities}} & \multicolumn{4}{c}{\textbf{FARC Municipalities}} \tabularnewline & All & In GEIH & Not In GEIH & \(p\)-value & All & In GEIH & Not In GEIH & \(p\)-value \tabularnewline 
{Variable}&{(1)}&{(2)}&{(3)}&{(4)}&{(5)}&{(6)}&{(7)}&{(8)} \tabularnewline
\midrule \addlinespace[\belowrulesep]
\midrule Population&27.36&41.47&17.36&0&25.15&36.83&15.81&0 \tabularnewline
Area&711.46&890.76&584.46&0.05&1842.32&1502.64&2114.08&0.21 \tabularnewline
Distance Capital&101.34&100.01&102.29&0.92&79.73&84.33&76.05&0.25 \tabularnewline
Distance Bogot\'a&345.42&382.63&319.07&0.14&303.70&310.19&298.51&0.59 \tabularnewline
Gov. Transfers&44.58&32.68&53&0.02&56.09&42.43&67.01&0 \tabularnewline
Savings Capacity&32.15&33.75&31.06&0.53&29.26&29.05&29.43&0.87 \tabularnewline
\%Exp. in Investment&84.09&82.44&85.20&0.12&84.22&84.23&84.20&0.98 \tabularnewline
Fiscal Performance&61.29&61.88&60.88&0.68&60.61&61.16&60.16&0.36 \tabularnewline
Overall Perf.&59.41&59.63&59.25&0.92&55.29&57.25&53.72&0.10 \tabularnewline
Municipal Develop.&65.31&66.62&64.41&0.56&63.84&64.12&63.63&0.76 \tabularnewline
Conflict in 1901/30&0.11&0.05&0.17&0.31&0.07&0.12&0.02&0 \tabularnewline
Spanish Occup.&0.07&0.05&0.07&0.77&0.18&0.23&0.14&0.09 \tabularnewline
Aqueduct&57.18&65.51&50.50&0.14&56.58&58.54&55.02&0.46 \tabularnewline
Garbage Collection&43.04&52.88&35.16&0.07&47.33&51.47&44.02&0.10 \tabularnewline
Sewage&41.58&47.65&36.72&0.21&44.70&47.97&42.11&0.20 \tabularnewline
GINI&0.46&0.46&0.46&0.95&0.46&0.46&0.46&0.25 \tabularnewline
MDP&72.80&71.55&73.69&0.58&73.41&71.95&74.56&0.16 \tabularnewline
NBI&45.41&45.86&45.09&0.89&49.06&49.34&48.83&0.85 \tabularnewline
\bottomrule \addlinespace[\belowrulesep]

\end{tabular} \end{adjustbox}
\\ \parbox{\linewidth}{\scriptsize \justifying \noindent \(Notes\): All the variables are measured in the last pre-treatment year for which data are available (either 2008 or 2005). Conf. 1901/1930 denotes whether the municipality experienced social conflict between 1901-1930. Distance variables are measured in kilometers. MDP is a multidimensional poverty index. Population in 1000s. Last two rows correspond to the share of votes in favor of the peace agreement in 2016 and the share of registered voters voting in the plebiscite in 2016. For a given measure of presence, columns 1 and 5 show the mean of the variable for all ELN and FARC municipalities, respectively. Columns 2 and 6 show the mean of the variable for ELN and FARC municipalities for which GEIH data are available, while columns 3 and 7 show the mean of the variable for ELN and FARC municipalities in which no GEIH surveys were conducted. Columns 4 and 8 show the \(p\)-value of the difference between municipalities in which GEIH was and was not collected, for ELN and FARC municipalities, respectively.}
\end{table}




\clearpage

\subsubsection{Additional Results on State Capacity Outcomes}

\begin{table}[h!] \centering
\newcolumntype{R}{>{\raggedleft\arraybackslash}X}
\newcolumntype{L}{>{\raggedright\arraybackslash}X}
\newcolumntype{C}{>{\centering\arraybackslash}X}

\caption{Robustness DiD State Capacity Analysis: Extensive Margin, Events in Over 60\% of Years}
\label{rob_DID_SC_CEDE_Ext_p60}
\begin{adjustbox}{max width=\linewidth, max height=\textwidth}\begin{tabular}{lccccccc}

\toprule
& \multicolumn{2}{c}{Tax Revenue per GDP} & \multicolumn{2}{c}{Tax Revenue per Nighttime Light} & \multicolumn{3}{c}{No Windsorizing} \tabularnewline & Tax Rev. to & Anderson & Tax Rev. to & Anderson & Operational  & Ratio Gov. Trans. & Anderson \tabularnewline & GDP (DANE) & Index & Nighttime Light & Index & Costs (pc) & to Revenue & Anderson \tabularnewline 
{}&{(1)}&{(2)}&{(3)}&{(4)}&{(5)}&{(6)}&{(7)} \tabularnewline
\midrule \addlinespace[\belowrulesep]
\midrule Ceasefire \( \times \) FARC&--0.000&0.040&5.201&0.041&--0.002&--0.050&0.075 \tabularnewline
&(0.001)&(0.045)&(4.541)&(0.044)&(0.009)&(0.539)&(0.102) \tabularnewline
&&&&&&& \tabularnewline
Treated Municip.&216&216&216&216&216&216&216 \tabularnewline
Control Municip.&41&41&41&41&41&41&41 \tabularnewline
Mean Dependent Var.&0.012&0.001&--0.091&0.003&0.118&7.149&--0.016 \tabularnewline
Observations&2,308&2,827&2,822&2,827&2,822&2,815&2,827 \tabularnewline
\(R^2\)&0.659&0.662&0.096&0.668&0.219&0.676&0.507 \tabularnewline
\bottomrule \addlinespace[\belowrulesep]

\end{tabular} \end{adjustbox}
\\ \parbox{\linewidth}{\scriptsize \justifying \noindent \(Notes \): Results from estimating Equation \eqref{eq_did}. Standard errors clustered at the municipality level. Includes municipality and year fixed effects. Considers only years from 2009 and before 2020. Mean Dep. Var.: Mean of the dependent variable in the pre-intervention period for municipalities in the sample (FARC and ELN). Defines the post period as 2015. Column 1 uses the ratio of tax revenue to the estimate of GDP from DANE. Column 2 creates the State Capacity Index following Anderson (2008), using the measure in column 1 instead of tax revenue per capita. Column 3 uses the ratio of tax revenue to nighttime light intensity. Column 4 creates the State Capacity Index following Anderson (2008), using the measure in column 3 instead of tax revenue per capita. Columns 5 and 6 use the measure of operational costs (per capita) and the ratio of government transfers to total municipality revenue, without windsorizing. Column 7 creates the State Capacity Index following Anderson (2008), using the non-windsorized measures in columns 5 and 6.}
\end{table}

\begin{table}[h!] \centering
\newcolumntype{R}{>{\raggedleft\arraybackslash}X}
\newcolumntype{L}{>{\raggedright\arraybackslash}X}
\newcolumntype{C}{>{\centering\arraybackslash}X}

\caption{PDET Projects' Descriptives}
\label{pdetDescriptives}
\begin{adjustbox}{max width=\linewidth, max height=\textwidth}\begin{tabular}{lccccccc}

\toprule
& \# Projects & \% Projects & \# Finished & \% Finished & \# Municipalities & Value Projects & Value Projects \tabularnewline &  &  & Projects & Projects &  & Mean & Median \tabularnewline 
{Sector}&{(1)}&{(2)}&{(3)}&{(4)}&{(5)}&{(6)}&{(7)} \tabularnewline
\midrule \addlinespace[\belowrulesep]
\midrule All Projects&5393&100&1161&100&1.17&1507.92&155.62 \tabularnewline
All Projects (Until 2019)&1773&32.88&1161&100&1.26&994.12&155.12 \tabularnewline
&&&&&&& \tabularnewline
\textbf{Non-Economic Projects}&&&&&&& \tabularnewline
Culture&17&0.32&9&0.78&1&159.31&113.63 \tabularnewline
Education&798&14.8&215&18.52&1.26&421.28&137.79 \tabularnewline
Environment&58&1.08&1&0.09&1.29&3268.6&1769.74 \tabularnewline
Health, Social Protection&61&1.13&6&0.52&1.28&566.78&124.72 \tabularnewline
IT&18&0.33&0&0&1.11&8864.52&6904.68 \tabularnewline
Justice&86&1.59&5&0.43&1.1&55.5&42.88 \tabularnewline
Mixed&620&11.5&18&1.55&1.02&138.66&52.03 \tabularnewline
Planning&30&0.56&0&0&1.5&1773.04&2.63 \tabularnewline
Sport&223&4.13&92&7.92&1&244.45&129.52 \tabularnewline
Statistical Information&7&0.13&0&0&1&3026.6&3062.62 \tabularnewline
\textit{Total}&1887&34.99&346&29.8&1.15&488.94&67.47 \tabularnewline
&&&&&&& \tabularnewline
\textbf{Economic Projects}&&&&&&& \tabularnewline
Agriculture, Rural Development&618&11.46&78&6.72&1.42&1930.6&598.15 \tabularnewline
Commerce, Industry, Tourism&55&1.02&1&0.09&1.2&389.25&83.33 \tabularnewline
Housing&242&4.49&34&2.93&1.05&4873.59&2143.87 \tabularnewline
Mines, Energy&210&3.89&49&4.22&1.11&6441.84&3386.74 \tabularnewline
Social Inclusion, Reconciliation&375&6.95&25&2.15&1.55&115.57&55.77 \tabularnewline
Territorial Government&704&13.05&173&14.9&1.07&115.88&56.14 \tabularnewline
Transport&1269&23.53&454&39.1&1.07&2607.67&243.66 \tabularnewline
Work&2&0.04&1&0.09&1&4607.28&4607.28 \tabularnewline
\textit{Total}&3506&65.01&815&70.2&1.19&2056.36&213.4 \tabularnewline
\bottomrule \addlinespace[\belowrulesep]

\end{tabular} \end{adjustbox}
\\ \parbox{\linewidth}{\scriptsize \justifying \noindent \(Notes \): First column shows the total number of PDET projects until 2023 (all years in the first row, until the end of 2019 in the second row, and for all years by sector in following rows). Second column shows the share of projects relative to the overall number of projects until 2023. Third column shows the total number of finished PDET projects until 2019. Fourth column shows the share of projects relative to the overall number of finished projects until 2019. Fifth column shows the average number of municipalities that benefitted from the project. Sixth column shows the average value of a PDET project (for those projects that cover multiple municipalities, the value is divided evenly among beneficiary municipalities. In millions COP). Seventh column shows the median value of a PDET project (for those projects that cover multiple municipalities, the value is divided evenly among beneficiary municipalities. In millions COP).}
\end{table}

\clearpage

\subsubsection{Alternative Measures of Insurgent Groups' Presence}

\begin{table}[h!] \centering
\newcolumntype{R}{>{\raggedleft\arraybackslash}X}
\newcolumntype{L}{>{\raggedright\arraybackslash}X}
\newcolumntype{C}{>{\centering\arraybackslash}X}

\caption{DiD Violence Analysis: Intensive Margin, Top 20\% Most Violent}
\label{DID_final_Vio_CEDE_Int_p80}
\begin{adjustbox}{max width=\linewidth, max height=\textwidth}\begin{tabular}{lccccccc}

\toprule
{}&{(1)}&{(2)}&{(3)}&{(4)}&{(5)}&{(6)}&{(7)} \tabularnewline
\midrule \addlinespace[\belowrulesep]
\midrule \textbf{Panel A. UARIV}&Terror. Act&Threats&Disapp.&Sex Crimes&Child Recruit.&Torture&Prop. Loss \tabularnewline
\midrule Ceasefire \( \times \) FARC&--0.361*&--0.940***&--0.053***&--0.012&--0.034***&--0.006**&--0.338 \tabularnewline
&(0.200)&(0.192)&(0.016)&(0.008)&(0.006)&(0.002)&(0.231) \tabularnewline
&&&&&&& \tabularnewline
Treated Munic.&153&153&153&153&153&153&153 \tabularnewline
Control Munic.&153&153&153&153&153&153&153 \tabularnewline
Mean Dep. Var.&0.555&2.124&0.087&0.059&0.031&0.012&0.669 \tabularnewline
Observations&3,366&3,366&3,366&3,366&3,366&3,366&3,366 \tabularnewline
\(R^2\)&0.273&0.549&0.277&0.547&0.453&0.196&0.362 \tabularnewline
&&&&&&& \tabularnewline
\midrule \textbf{Panel B. Other}&Kidnap.&Homicides&Theft&Mines&Forced Mig.&Clash/Att.&\textbf{\textbf{And. Index}} \tabularnewline
\midrule Ceasefire \( \times \) FARC&--0.002&--0.133***&--0.295**&--0.383***&--12.996***&--0.021***&--0.454*** \tabularnewline
&(0.003)&(0.030)&(0.137)&(0.063)&(2.117)&(0.006)&(0.095) \tabularnewline
&&&&&&& \tabularnewline
Treated Munic.&153&153&153&153&153&153&153 \tabularnewline
Control Munic.&153&153&153&153&153&153&153 \tabularnewline
Mean Dep. Var.&0.013&0.376&0.817&0.370&17.196&0.031&0.161 \tabularnewline
Observations&3,366&3,366&3,366&3,366&3,366&3,060&3,366 \tabularnewline
\(R^2\)&0.156&0.537&0.722&0.625&0.482&0.329&0.483 \tabularnewline
\bottomrule \addlinespace[\belowrulesep]

\end{tabular} \end{adjustbox}
\\ \parbox{\linewidth}{\scriptsize \justifying \noindent \(Notes \): Results from estimating Equation \eqref{eq_did}. Standard errors clustered at the municipality level. Includes municipality and year fixed effects. Considers only years from 2009 and before 2020. Mean Dep. Var.: Mean of the dependent variable in the pre-intervention period for municipalities in the sample (FARC and ELN). Defines the post period as 2015. All variables are measures in 1000's of inhabitants. Prop. Loss: Property Loss. Clash/Att.: Number of clashes and attacks between government and paramilitary and guerrilla groups. Panel B shows variables from other data sources, with the first three columns coming from the Ministry of Defense, the fourth from the agency against anti-personnel mines (DAIMA), the fifth from the Victims' Unit, and the sixth from Juan Vargas. And. Index is a summary measure created following Anderson (2008) that summarizes all the different outcomes variables. It is the weighted average of the standardized outcomes, weighted by their inverted covariance matrix.}
\end{table}

\begin{table}[h!] \centering
\newcolumntype{R}{>{\raggedleft\arraybackslash}X}
\newcolumntype{L}{>{\raggedright\arraybackslash}X}
\newcolumntype{C}{>{\centering\arraybackslash}X}

\caption{DiD Economic Outcomes: Intensive Margin, Top 20\% Most Violent}
\label{DID_Econ_CEDE_Int_p80}
\begin{adjustbox}{max width=\linewidth, max height=\textwidth}\begin{tabular}{lccccccc}

\toprule
& Nighttime & Value Added (pc) & Agricultural & Share Urban & Firm  & Formal & \textbf{Anderson} \tabularnewline & Light & DANE & Productivity & Population & Entry & Empl. & \textbf{Index} \tabularnewline 
{}&{(1)}&{(2)}&{(3)}&{(4)}&{(5)}&{(6)}&{(7)} \tabularnewline
\midrule \addlinespace[\belowrulesep]
\midrule Ceasefire \( \times \) FARC&--0.032*&--2.498&0.062&0.006&0.695***&0.002&0.036 \tabularnewline
&(0.017)&(1.639)&(0.271)&(0.004)&(0.263)&(0.003)&(0.038) \tabularnewline
&&&&&&& \tabularnewline
Treated Munic.&153&153&153&153&153&153&153 \tabularnewline
Control Munic.&153&153&153&153&153&153&153 \tabularnewline
Mean Dep. Var.&--0.365&11.673&4.736&0.395&5.431&0.082&--0.071 \tabularnewline
Observations&3,366&2,754&3,365&3,366&3,366&3,366&3,366 \tabularnewline
\(R^2\)&0.917&0.859&0.904&0.974&0.696&0.956&0.871 \tabularnewline
\bottomrule \addlinespace[\belowrulesep]

\end{tabular} \end{adjustbox}
\\ \parbox{\linewidth}{\scriptsize \justifying \noindent \(Notes \): Results from estimating Equation \eqref{eq_did}. Standard errors clustered at the municipality level. Includes municipality and year fixed effects. Considers only years from 2009 and before 2020. Mean Dep. Var.: Mean of the dependent variable in the pre-intervention period for municipalities in the sample (FARC and ELN). Defines the post period as 2015. Nighttime light intensity defined so that grid cells in the border of multiple municipalities are assigned in proportion to the share of the grid cell in each municipality (weighted). Value added per capita comes DANE (National Department of Statistics), in millions of COP. Agricultural productivity is defined as total tonnes produced of 271 agricultural crops divided by total area cultivated in hectares, based on data from the Ministry of Agriculture. Firm entry comes from RUES and is measured per 1000 inhabitants. Formal employment is measured as the average number of individuals paying contributions to healthcare, pension funds and workers' compensations across the year in the municipality per 18-60 years old, from PILA. Anderson Index is a summary measure created following Anderson (2008) that summarizes the measures in columns 1-6. It is the weighted average of the standardized outcomes, weighted by their inverted covariance matrix.}
\end{table}

\begin{table}[h!] \centering
\newcolumntype{R}{>{\raggedleft\arraybackslash}X}
\newcolumntype{L}{>{\raggedright\arraybackslash}X}
\newcolumntype{C}{>{\centering\arraybackslash}X}

\caption{Baseline State Capacity Summary Statistics: Intensive Margin, Top 20\% Most Violent}
\label{SC_SumStats_CEDEInt_p80}
\begin{adjustbox}{max width=\linewidth, max height=\textwidth}\begin{tabular}{l|ccc|ccc}

\toprule
& \multicolumn{3}{c|}{\textbf{Mean}} & \multicolumn{3}{c}{\textbf{p-value of Difference}} \tabularnewline & Rest Country & FARC & ELN & Rest-FARC & Rest-ELN & FARC-ELN \tabularnewline 
{Variable}&{(1)}&{(2)}&{(3)}&{(4)}&{(5)}&{(6)} \tabularnewline
\midrule \addlinespace[\belowrulesep]
\midrule 1. Total Revenue&747.47&821.5&750.59&0.05&0.93&0.15 \tabularnewline
2. Tax Revenue&104.33&72.18&72.95&0&0&0.92 \tabularnewline
3. Gov. Transfers&77.02&94.16&78.43&0.04&0.86&0.04 \tabularnewline
4. Operational Expenditures&111.66&118.83&98.80&0.28&0.05&0 \tabularnewline
5. Tax Rev. / Nighttime Light&5732.24&50.04&--1040.38&0.64&0.58&0.10 \tabularnewline
6. Savings Capacity&0.07&--0.25&--0.07&0&0.06&0.09 \tabularnewline
7. Fiscal Perf.&0.07&--0.21&--0.11&0&0.02&0.30 \tabularnewline
8. Administrative Perf.&0&--0.05&0&0.54&0.96&0.55 \tabularnewline
9. Rule Compliance&0.05&--0.21&--0.05&0&0.20&0.12 \tabularnewline
10. Aqueduct Coverage&60.54&53.36&59.95&0.01&0.84&0.10 \tabularnewline
11. Garbage Collection&45.52&42.99&46.43&0.39&0.76&0.36 \tabularnewline
\bottomrule \addlinespace[\belowrulesep]

\end{tabular} \end{adjustbox}
\\ \parbox{\linewidth}{\scriptsize \justifying \noindent \(Notes\): All the variables are measured in 2008 (the last period before the panel used in the main analysis) but for the rule compliance variable that is only available from 2010. The financial measures are in per capita terms (in thousand COP), and the performance measures are standardised (by year).}
\end{table}

\begin{table}[h!] \centering
\newcolumntype{R}{>{\raggedleft\arraybackslash}X}
\newcolumntype{L}{>{\raggedright\arraybackslash}X}
\newcolumntype{C}{>{\centering\arraybackslash}X}

\caption{DiD State Capacity Analysis: Intensive Margin, Top 20\% Most Violent}
\label{DID_SC_CEDE_Int_p80}
\begin{adjustbox}{max width=\linewidth, max height=\textwidth}\begin{tabular}{lccccccc}

\toprule
& Tax & Operational  & Ratio Gov. Trans. & Fiscal & Administrative & Rule  & \textbf{Anderson} \tabularnewline & Revenue (pc) & Costs (pc) & to Revenue & Performance  & Performance & Compliance & \textbf{Index} \tabularnewline 
{}&{(1)}&{(2)}&{(3)}&{(4)}&{(5)}&{(6)}&{(7)} \tabularnewline
\midrule \addlinespace[\belowrulesep]
\midrule Ceasefire \( \times \) FARC&--0.008&0.008*&1.347*&--0.016&--0.089&--0.031&--0.020 \tabularnewline
&(0.009)&(0.005)&(0.786)&(0.056)&(0.074)&(0.070)&(0.038) \tabularnewline
&&&&&&& \tabularnewline
Treated Municip.&153&153&153&153&153&153&153 \tabularnewline
Control Municip.&153&153&153&153&153&153&153 \tabularnewline
Mean Dependent Var.&0.106&0.134&7.904&--0.187&--0.061&--0.087&--0.023 \tabularnewline
Observations&3,362&3,362&3,353&3,358&3,366&3,055&3,366 \tabularnewline
\(R^2\)&0.817&0.843&0.318&0.547&0.412&0.406&0.453 \tabularnewline
\bottomrule \addlinespace[\belowrulesep]

\end{tabular} \end{adjustbox}
\\ \parbox{\linewidth}{\scriptsize \justifying \noindent \(Notes \): Results from estimating Equation \eqref{eq_did}. Standard errors clustered at the municipality level. Includes municipality and year fixed effects. Considers only years from 2009 and before 2020. Mean Dep. Var.: Mean of the dependent variable in the pre-intervention period for municipalities in the sample (FARC and ELN). Defines the post period as 2015. Column 1 uses tax revenue. Column 2 uses the municipality's operational costs. Both these measures are in millions of COP per capita. Column 3 uses the ratio of transfers from the central government to total municipality revenue (excluding transfers from the central government). These three measures have been winsorized for observations ten SDs away from the mean for the whole country. Column 4 uses an indicator of fiscal performance created by the National Department of Planning and measures how well the municipality spends its resources. Column 5 uses an indicator of the municipality's overall administrative performance created by the National Department of Planning. It takes into account the local administration's efficiency and efficacy, compliance with rules, and the municipality's administrative capacity. Column 6 combines two indicators created by the Office of the Inspector General, which measure how well municipalities report information, implement internal monitoring tools, and the transparency of contracting practices. These last three are standardised. Column 7 summarizes these variables into a single index following Anderson (2008). It is the weighted average of the standardized outcomes, weighted by their inverted covariance matrix.}
\end{table}

\begin{table}[h!] \centering
\newcolumntype{R}{>{\raggedleft\arraybackslash}X}
\newcolumntype{L}{>{\raggedright\arraybackslash}X}
\newcolumntype{C}{>{\centering\arraybackslash}X}

\caption{Robustness DiD State Capacity Analysis: Intensive Margin, Top 20\% Most Violent}
\label{rob_DID_SC_CEDE_Int_p80}
\begin{adjustbox}{max width=\linewidth, max height=\textwidth}\begin{tabular}{lccccccc}

\toprule
& \multicolumn{2}{c}{Tax Revenue per GDP} & \multicolumn{2}{c}{Tax Revenue per Nighttime Light} & \multicolumn{3}{c}{No Windsorizing} \tabularnewline & Tax Rev. to & Anderson & Tax Rev. to & Anderson & Operational  & Ratio Gov. Trans. & Anderson \tabularnewline & GDP (DANE) & Index & Nighttime Light & Index & Costs (pc) & to Revenue & Anderson \tabularnewline 
{}&{(1)}&{(2)}&{(3)}&{(4)}&{(5)}&{(6)}&{(7)} \tabularnewline
\midrule \addlinespace[\belowrulesep]
\midrule Ceasefire \( \times \) FARC&--0.001&--0.037&6.286&--0.028&--0.001&495.730&--0.045 \tabularnewline
&(0.001)&(0.035)&(6.239)&(0.034)&(0.012)&(495.220)&(0.055) \tabularnewline
&&&&&&& \tabularnewline
Treated Municip.&153&153&153&153&153&153&153 \tabularnewline
Control Municip.&153&153&153&153&153&153&153 \tabularnewline
Mean Dependent Var.&0.013&--0.039&--0.116&--0.039&0.138&211.729&--0.008 \tabularnewline
Observations&2,750&3,366&3,362&3,366&3,362&3,353&3,366 \tabularnewline
\(R^2\)&0.630&0.447&0.097&0.452&0.278&0.115&0.177 \tabularnewline
\bottomrule \addlinespace[\belowrulesep]

\end{tabular} \end{adjustbox}
\\ \parbox{\linewidth}{\scriptsize \justifying \noindent \(Notes \): Results from estimating Equation \eqref{eq_did}. Standard errors clustered at the municipality level. Includes municipality and year fixed effects. Considers only years from 2009 and before 2020. Mean Dep. Var.: Mean of the dependent variable in the pre-intervention period for municipalities in the sample (FARC and ELN). Defines the post period as 2015. Column 1 uses the ratio of tax revenue to the estimate of GDP from DANE. Column 2 creates the State Capacity Index following Anderson (2008), using the measure in column 1 instead of tax revenue per capita. Column 3 uses the ratio of tax revenue to nighttime light intensity. Column 4 creates the State Capacity Index following Anderson (2008), using the measure in column 3 instead of tax revenue per capita. Columns 5 and 6 use the measure of operational costs (per capita) and the ratio of government transfers to total municipality revenue, without windsorizing. Column 7 creates the State Capacity Index following Anderson (2008), using the non-windsorized measures in columns 5 and 6.}
\end{table}

\clearpage

\subsubsection{Synthetic Difference-in-Difference}

\begin{landscape}
\begin{table}[h!] \centering
\newcolumntype{R}{>{\raggedleft\arraybackslash}X}
\newcolumntype{L}{>{\raggedright\arraybackslash}X}
\newcolumntype{C}{>{\centering\arraybackslash}X}

\caption{Synthetic DiD Analysis: Extensive Margin, Events in Over 60\% of Years}
\label{SDnD_CEDEExt_p60}
\begin{adjustbox}{max width=\linewidth, max height=\textwidth}\begin{tabular}{l|ccccccccc}

\toprule
{}&{(1)}&{(2)}&{(3)}&{(4)}&{(5)}&{(6)}&{(7)}&{(8)}&{(9)} \tabularnewline
\midrule \addlinespace[\belowrulesep]
\textbf{Panel A. Violence}&KLK Index&Anderson Index&&&&&&& \tabularnewline
\midrule Ceasefire \( \times \) FARC&--0.2329***&--0.3741***&&&&&&& \tabularnewline
&(0.0338)&(0.0840)&&&&&&& \tabularnewline
&&&&&&&&& \tabularnewline
Treated Municipalities&216&216&&&&&&& \tabularnewline
Potential Controls&727&727&&&&&&& \tabularnewline
Control Municipalities&590&627&&&&&&& \tabularnewline
&&&&&&&&& \tabularnewline
\midrule \textbf{Panel B. Economic}&Light Int.&Value Added (pc)&Built Up&Agr. Productivity&Share Urban&Firm Entry&Formal Empl.&KLK Index&Anderson Index \tabularnewline
\midrule Ceasefire \( \times \) FARC&--0.0071&--1.0999&--0.0020&0.1048&0.0002&--0.2330&--0.0111***&--0.0346***&--0.0258 \tabularnewline
&(0.0084)&(1.0281)&(0.0014)&(0.1853)&(0.0021)&(0.1818)&(0.0018)&(0.0117)&(0.0202) \tabularnewline
&&&&&&&&& \tabularnewline
Treated Municipalities&216&216&216&216&216&216&216&216&216 \tabularnewline
Potential Controls&711&711&711&711&711&711&711&727&727 \tabularnewline
Control Municipalities&610&633&576&632&638&629&625&631&624 \tabularnewline
&&&&&&&&& \tabularnewline
\midrule \textbf{Panel C. State Capacity}&Tax Revenue&Oper. Costs&Gov. Trans./Rev.&Fiscal Perf.&Admin. Perf.&KLK Index&Anderson Index&& \tabularnewline
\midrule Ceasefire \( \times \) FARC&--0.0324***&--0.0060**&0.5295&--0.0883**&--0.0690&--0.0987***&--0.0552*&& \tabularnewline
&(0.0073)&(0.0025)&(0.4474)&(0.0377)&(0.0524)&(0.0291)&(0.0291)&& \tabularnewline
&&&&&&&&& \tabularnewline
Treated Municipalities&207&207&207&207&207&216&216&& \tabularnewline
Potential Controls&704&704&704&704&704&727&727&& \tabularnewline
Control Municipalities&616&568&619&599&592&598&590&& \tabularnewline
\bottomrule \addlinespace[\belowrulesep]

\end{tabular} \end{adjustbox}
\\ \parbox{\linewidth}{\scriptsize \justifying \noindent \(Notes \): Estimated using Arkhangelsky et al. (2021) Synthetic Difference-in-Difference command, synthdid. Standard errors calculated using Jackknife. Considers only years from 2009 and before 2020. Defines the post period as 2015. KLK Index is a measure created following Kling, Liebman \& Katz (2007) that summarizes all the different outcomes variables. It is the (standardised) average of the z-scores of all dependent variables. Anderson index is based on Anderson (2008), and follows a similar procedure but attaching less weight to highly correlated variables (which bring relatively less new information to the index).}
\end{table}
\end{landscape}

\begin{landscape}
\begin{table}[h!] \centering
\newcolumntype{R}{>{\raggedleft\arraybackslash}X}
\newcolumntype{L}{>{\raggedright\arraybackslash}X}
\newcolumntype{C}{>{\centering\arraybackslash}X}

\caption{Synthetic DiD Analysis: Intensive Margin, Top 20\% Most Violent}
\label{SDnD_CEDEInt_p80}
\begin{adjustbox}{max width=\linewidth, max height=\textwidth}\begin{tabular}{l|ccccccccc}

\toprule
{}&{(1)}&{(2)}&{(3)}&{(4)}&{(5)}&{(6)}&{(7)}&{(8)}&{(9)} \tabularnewline
\midrule \addlinespace[\belowrulesep]
\textbf{Panel A. Violence}&KLK Index&Anderson Index&&&&&&& \tabularnewline
\midrule Ceasefire \( \times \) FARC&--0.2931***&--0.4228***&&&&&&& \tabularnewline
&(0.0400)&(0.0773)&&&&&&& \tabularnewline
&&&&&&&&& \tabularnewline
Treated Municipalities&153&153&&&&&&& \tabularnewline
Potential Controls&774&774&&&&&&& \tabularnewline
Control Municipalities&604&644&&&&&&& \tabularnewline
&&&&&&&&& \tabularnewline
\midrule \textbf{Panel B. Economic}&Light Int.&Value Added (pc)&Built Up&Agr. Productivity&Share Urban&Firm Entry&Formal Empl.&KLK Index&Anderson Index \tabularnewline
\midrule Ceasefire \( \times \) FARC&--0.0056&--1.7944&--0.0004&--0.0825&0.0067**&0.3894*&--0.0078***&--0.0182&--0.0033 \tabularnewline
&(0.0093)&(1.3941)&(0.0011)&(0.1617)&(0.0028)&(0.2013)&(0.0022)&(0.0146)&(0.0236) \tabularnewline
&&&&&&&&& \tabularnewline
Treated Municipalities&151&151&151&151&151&151&151&153&153 \tabularnewline
Potential Controls&759&759&759&759&759&759&759&774&774 \tabularnewline
Control Municipalities&639&671&625&675&673&669&665&672&655 \tabularnewline
&&&&&&&&& \tabularnewline
\midrule \textbf{Panel C. State Capacity}&Tax Revenue&Oper. Costs&Gov. Trans./Rev.&Fiscal Perf.&Admin. Perf.&KLK Index&Anderson Index&& \tabularnewline
\midrule Ceasefire \( \times \) FARC&--0.0305***&--0.0047&0.7371&--0.0274&--0.2572***&--0.1388***&--0.0825**&& \tabularnewline
&(0.0098)&(0.0033)&(0.5466)&(0.0433)&(0.0579)&(0.0334)&(0.0340)&& \tabularnewline
&&&&&&&&& \tabularnewline
Treated Municipalities&145&145&145&145&145&153&153&& \tabularnewline
Potential Controls&753&753&753&753&753&774&774&& \tabularnewline
Control Municipalities&665&633&660&644&628&623&624&& \tabularnewline
\bottomrule \addlinespace[\belowrulesep]

\end{tabular} \end{adjustbox}
\\ \parbox{\linewidth}{\scriptsize \justifying \noindent \(Notes \): Estimated using Arkhangelsky et al. (2021) Synthetic Difference-in-Difference command, synthdid. Standard errors calculated using Jackknife. Considers only years from 2009 and before 2020. Defines the post period as 2015. KLK Index is a measure created following Kling, Liebman \& Katz (2007) that summarizes all the different outcomes variables. It is the (standardised) average of the z-scores of all dependent variables. Anderson index is based on Anderson (2008), and follows a similar procedure but attaching less weight to highly correlated variables (which bring relatively less new information to the index).}
\end{table}
\end{landscape}

\clearpage

\subsubsection{Violations of Parallel Trend Assumption}

\begin{table}[h!] \centering
\newcolumntype{R}{>{\raggedleft\arraybackslash}X}
\newcolumntype{L}{>{\raggedright\arraybackslash}X}
\newcolumntype{C}{>{\centering\arraybackslash}X}

\caption{Slope and Bias of Pre-Trends -- Roth (2022)}
\label{Roth_baseline_slope}
\begin{adjustbox}{max width=\linewidth, max height=\textwidth}\begin{tabular}{l|cccc|cccc}

\toprule
& \multicolumn{4}{c|}{\textbf{Panel A. Extensive Margin, 60\%}} & \multicolumn{4}{c}{\textbf{Panel B. Intensive Margin, 20\%}} \tabularnewline & Slope & Uncond. Bias & Cond. Bias & Mean Post \(\hat{\beta}s\) & Slope & Uncond. Bias & Cond. Bias & Mean Post \(\hat{\beta}s\) \tabularnewline 
{}&{(1)}&{(2)}&{(3)}&{(4)}&{(5)}&{(6)}&{(7)}&{(8)} \tabularnewline
\midrule \addlinespace[\belowrulesep]
\midrule Forced Disappearances&0.019&0.056&0.064&--0.011&0.017&0.052&0.066&--0.012 \tabularnewline
Property Losses&0.07&0.211&0.211&--0.466&0.135&0.404&0.521&--0.272 \tabularnewline
Homicides&0.031&0.092&0.169&--0.045&0.022&0.067&0.098&--0.094 \tabularnewline
Mines&0.033&0.099&0.165&--0.221&0.042&0.127&0.184&--0.306 \tabularnewline
\textbf{And. Viol. Measures}&0.066&0.199&0.259&--0.298&0.073&0.218&0.402&--0.206 \tabularnewline
&&&&&&&& \tabularnewline
Nighttime Light Int.&0.019&0.057&0.066&--0.018&0.013&0.04&0.049&0.009 \tabularnewline
Agr. Productivity&0.225&0.676&0.551&0.001&0.166&0.499&0.522&--0.173 \tabularnewline
Share Urban Pop.&0.001&0.004&0.008&--0.001&0.001&0.003&0.006&0.007 \tabularnewline
Firm Entry&0.213&0.639&0.907&0.21&0.208&0.624&0.813&0.568 \tabularnewline
Formal Employment&0.003&0.009&0.009&--0.003&0.002&0.007&0.007&--0.004 \tabularnewline
\textbf{And. Econ. Measures}&0.022&0.066&0.073&--0.004&0.03&0.089&0.105&0.027 \tabularnewline
&&&&&&&& \tabularnewline
Tax Revenue&0.006&0.019&0.021&--0.019&0.009&0.026&0.03&--0.018 \tabularnewline
Functioning Costs&0.003&0.009&0.01&0.003&0.003&0.009&0.009&0.003 \tabularnewline
Ratio Gov. Trans. Revenue&0.389&1.167&1.065&--1.228&0.495&1.484&1.558&--0.289 \tabularnewline
Fiscal Performance&0.073&0.218&0.266&--0.04&0.053&0.16&0.196&--0.078 \tabularnewline
Admin Performance&0.109&0.326&0.392&--0.185&0.069&0.208&0.217&--0.243 \tabularnewline
\textbf{And. State Cap. Index}&0.04&0.119&0.165&0.061&0.031&0.093&0.121&--0.03 \tabularnewline
\bottomrule \addlinespace[\belowrulesep]

\end{tabular} \end{adjustbox}
\\ \parbox{\linewidth}{\scriptsize \justifying \noindent \(Notes \): Estimated following Roth (2022). Considers only years from  and before 2020. Defines the post period as . Mean Post $\hat\beta s$ is the mean of all the post-treatment coefficients. Anderson Index is a summary measure created following Anderson (2008) that summarizes all the different outcomes variables. It is the weighted average of the standardized outcomes, weighted by their inverted covariance matrix. Nighttime light intensity defined so that grid cells in the border of multiple municipalities are assigned in proportion to the share of the grid cell in each municipality (weighted). Agricultural productivity is defined as total tonnes produced of 271 agricultural crops divided by total area cultivated in hectares. Firm entry comes from RUES and is measured per 1000 inhabitants. Formal employment is measured as the average number of individuals paying contributions to healthcare, pension funds and workers' compensations across the year in the municipality per 18-60 years old, from PILA. Both tax revenue and operational costs are in millions of COP per capita. Ratio of transfers from the central government to total municipality revenue excludes transfers from the central government. These three measures have been winsorized for observations ten SDs away from the mean for the whole country. Fiscal performance is created by the National Department of Planning and measures how well the municipality spends its resources. Overall administrative performance is created by the National Department of Planning. It takes into account the local administration's efficiency and efficacy, compliance with rules, and the municipality's administrative capacity. These last two are standardised. And. State Cap Index summarizes the state capacity measures.}
\end{table}

\clearpage

\subsubsection{Extending Timeframe Until 2021}

\begin{landscape}
\begin{table}[h!] \centering
\newcolumntype{R}{>{\raggedleft\arraybackslash}X}
\newcolumntype{L}{>{\raggedright\arraybackslash}X}
\newcolumntype{C}{>{\centering\arraybackslash}X}

\caption{Comparison of Timeframes: Extensive Margin, Events in Over 60\% of Years}
\label{timeframes_CEDEExt_p60}
\begin{adjustbox}{max width=\linewidth, max height=\textwidth}\begin{tabular}{lccccccccccccccc}

\toprule
& Violence & Nighttime & Value & Share Urban & Agricultural & Firm & Formal & Economic & Economic & Tax & Operational & Gov. Trans. / & Fiscal & Administrative & State Cap. \tabularnewline  & Index & Light & Added (pc) & Population & Productivity & Entry & Empl. & Index & Index & Revenue (pc) & Costs (pc) & Revenue & Performance & Performance & Index \tabularnewline  & & & & & & &  & & (No Dissidents) & & & & & & \tabularnewline 
{}&{(1)}&{(2)}&{(3)}&{(4)}&{(5)}&{(6)}&{(7)}&{(8)}&{(9)}&{(10)}&{(11)}&{(12)}&{(13)}&{(14)}&{(15)} \tabularnewline
\midrule \addlinespace[\belowrulesep]
\midrule \textbf{Panel A. Diff-in-Diff}&&&&&&&&&&&&&&& \tabularnewline
Ceasefire \(\times\) FARC (2009--2019)&--0.289***&--0.007&--2.348*&--0.001&0.264&0.177&0.002&--0.004&--0.015&--0.009&0.005&--0.050&0.012&--0.058&0.067 \tabularnewline
&(0.104)&(0.021)&(1.208)&(0.004)&(0.491)&(0.329)&(0.004)&(0.038)&(0.043)&(0.011)&(0.005)&(0.539)&(0.108)&(0.090)&(0.054) \tabularnewline
Observations&2,827&2,827&2,313&2,827&2,827&2,827&2,827&2,827&2,211&2,822&2,822&2,815&2,823&2,827&2,827 \tabularnewline
&&&&&&&&&&&&&&& \tabularnewline
Ceasefire \(\times\) FARC (2009--2021)&--0.272***&--0.019&--2.221*&--0.002&--0.118&0.267&0.003&--0.013&--0.020&--0.011&0.007&0.603&--0.014&--0.029&0.057 \tabularnewline
&(0.079)&(0.021)&(1.262)&(0.006)&(0.548)&(0.345)&(0.004)&(0.040)&(0.046)&(0.012)&(0.006)&(0.768)&(0.100)&(0.086)&(0.049) \tabularnewline
Observations&3,341&3,341&2,827&3,341&3,341&3,341&3,341&3,341&2,613&3,336&3,336&3,329&3,337&3,341&3,341 \tabularnewline
&&&&&&&&&&&&&&& \tabularnewline
\midrule \textbf{Panel B. Diff-in-Disc}&&&&&&&&&&&&&&& \tabularnewline
IICA Treatment \(\times\) Post (2009--2019)&&0.031&0.692&--0.027**&--0.341&--1.064&--0.001&--0.022&--0.033&&&&&& \tabularnewline
&&(0.047)&(1.269)&(0.014)&(1.606)&(0.966)&(0.011)&(0.109)&(0.112)&&&&&& \tabularnewline
Observations&&1,859&936&1,067&1,100&1,584&1,166&1,353&1,397&&&&&& \tabularnewline
&&&&&&&&&&&&&&& \tabularnewline
IICA Treatment \(\times\) Post (2009--2021)&&--0.038&3.752&--0.036&0.453&0.204&--0.033&0.008&--0.030&&&&&& \tabularnewline
&&(0.054)&(2.622)&(0.024)&(1.449)&(1.106)&(0.020)&(0.105)&(0.113)&&&&&& \tabularnewline
Observations&&1,534&627&507&1,079&975&533&1,274&1,222&&&&&& \tabularnewline
\bottomrule \addlinespace[\belowrulesep]

\end{tabular} \end{adjustbox}
\\ \parbox{\linewidth}{\scriptsize \justifying \noindent \(Notes \): Results from estimating Equation \eqref{eq_did} are shown in Panel A, and from estimating Equation \eqref{eq_diff_in_disc} in Panel B. Standard errors clustered at the municipality level. Considers years between 2009 and 2019 in the upper half of each panel, and between 2009 and 2021 in the lower half of each panel. Defines the post period as 2015 in Panels A and C, and 2017 in Panel B. The first column uses a summary measure based on the different violence variables. The eight column uses a summary measure based on weighted nighttime light intensity, value added per capita (from DANE), share of urban population, agricultural productivity, formal employment and firm entry measures. The ninth column uses the same index, but excluding the municipalities in which FARC dissidents conducted any acts until 2019. The fifteenth column uses a summary measure based on tax revenue collected (per capita), operational costs (per capita), the ratio of transfers from the central government to total municipality revenue (excluding transfers from the central government), a measure of municipality's fiscal capacity, a meausre of municipality's administrative capacity, and a measure of municipality's compliance with the rules set out by the national government.}
\end{table}
\end{landscape}

\end{document}